# PARTIAL QUANTUM COHERENCE, ULTRASHORT ELECTRON PULSE STATISTICS, AND A PLASMON-ENHANCED NANOTIP EMITTER BASED ON METALLIZED OPTICAL FIBERS

by

Sam Keramati

A DISSERTATION

PARTIAL QUANTUM COHERENCE, ULTRASHORT ELECTRON PULSE STATISTICS, AND A PLASMON-ENHANCED NANOTIP EMITTER BASED ON METALLIZED OPTICAL FIBERS

Sam Keramati, Ph.D.

University of Nebraska, 2019

Advisor: Herman Batelaan

The present dissertation covers two related research projects. The first topic was initiated with the ultimate goal of observing quantum degeneracy in ultrashort free electron pulses. This constitutes a thorough theoretical analysis of the problem involving partial quantum coherence and spin polarization of the source in light of a path-integral treatment of the phenomenon of matter-wave diffraction-in-time. Subsequently, results of a trailblazing experiment, to be superseded by a Hanbury Brown-Twiss type conclusive test of free fermion antibunching with electrons, is reported. In this experiment, the statistical distribution of the emitted electrons is studied taking advantage of a double-detector coincidence detection technique. The utilized electron emitters are ultrafast photoemission tungsten nanotip needle sources which are known to procure large spatial coherence lengths. The emission statistics is found to be sub-Poissonian.

The second project involves introduction and full characterization of a novel laser-driven electron nanotip emitter based on metallized fiber optic tapers in which the emission mechanism is found to be assisted by surface plasmon resonance excitation as predicted prior to the design of such sources. It is shown that gold-coated fiber optic nanotips can emit electrons using both low-power continuous wave lasers as well as femtosecond pulsed lasers tuned to or near the surface plasmon resonance excitation wavelength of the system. The final chapter entails a proposal of a spin-polarized electron photoemitter based on the spin Hall effect for which such a fiber optic nanotip source is exploited.

The optically switchable spin-polarized tip source for the first project with the Pauli exclusion principle at its core is based on GaAs and is developed and characterized in-house by Mott polarimetry. Another novel spin-polarized emitter based on our proposed fiber optic nanotip sources therefore constitutes a confluence of the two research programs considered here and will wrap up the present dissertation. The reported findings in this work are of direct relevance to electron source and beam physics, ultrafast electron microscopy, and foundations of quantum mechanics in relation to the burgeoning field of free electron quantum optics.

*dedicated to my mother and father,*

*with gratitude*

## Acknowledgements

I joined Prof. Herman Batelaan's group in the summer of 2016. Looking back at the almost three and a half years which have passed since then, I feel enormously delighted and proud of being a member of this group and its achievements. In conjunction with that, I am deeply and graciously thankful of Prof. Batelaan for trusting me, investing in me, introducing me to multiple new things and laboratory techniques, giving me freedom to try and learn, and helping me shape my research career. I cannot overstate the professional support I received from Prof. Batelaan for which I will remain grateful the rest of my life.

Our experiments were performed in Prof. Kees Uiterwaal's lab where we share the ultrafast laser systems. I would like to express my sincere gratitude to Prof. Uiterwaal for letting me work in his lab just like his own students. In addition, it has been a great pleasure for me to receive the opportunity to collaborate with Prof. Uiterwaal, and his graduate student Joshua Beck, on the project presented in chapter four of this dissertation. Moreover, I am indebted to Prof. Uiterwaal for reading my dissertation, and prior to that my comprehensive exam report, as a committee member; his comments on both the scientific aspects and the writing style of my dissertation have noticeably improved the entire work.

One of the projects presented in this dissertation has relevance to the research conducted on spin-polarized electron sources in Prof. T. J. Gay's group. I received the opportunity to work in Prof. Gay's lab in relation with this research for a time span of several months where I gained invaluable experience on this topic. I am extremely thankful of Prof. Gay for this, and for the helpful discussions through which I have learned a lot. In addition, as a committee member, I have to express my gratefulness towards Prof. Gay for reading my dissertation as well as my comprehensive exam report. In spite of all the possible deficiencies in my authorship of the present work, I can firmly assert that all the comments I received from Prof. Gay on how to present research work have improved my efficiency and prudence in delivering the results to the scientific community. I would also like to thank Prof. Gay's students, Will Brunner for his collaboration, and William Newman and Karl Ahrendsen, for help during my time in Prof. Gay's lab.

I am genuinely thankful of Prof. David Pitts for attending my presentations as the outside-representative committee member, and the instructive questions he challenged me with.

I shall express my extreme gratitude towards Dr. Ali Passian of the Oak Ridge National Lab with whom we had a fruitful collaboration on the project of chapter four of the present dissertation. It has always been a pleasure for me to communicate the results of experiments and data analysis we performed at UNL with Dr. Passian, and virtually always and every time, I learned a few new things from him. Dr. Passian's positive attitude towards conducting professional research has become a great source of inspiration for me throughout the past few years.

I also thank Prof. Jeremy Armstrong from the University of Nebraska at Kearney for conversations on part of the project of chapter two.

I am thankful of Prof. Christian Binek for his support during my time at UNL before I joined Prof. Batelaan's group, and for the valuable experience I gained and the lessons I took in his and Prof. Shireen Adenwalla's lab.

I would like to take this chance and also thank my UNL friends for all their help and discussions on physics and beyond. Although I have to keep the list limited here, I genuinely need to express my indebtedness to Dr. Omid Zandi and Dr. Bahar Shahsavarani, both for all the discussions we had, and for their indispensable help during my transition time to the new environment where I had to learn how to pay the bills and get a driver's license. I also thank Zilin Chen for discussions and sharing instruments, and Dr. Peter Beierle, and Dr. Maaneli Derakhshani, for fruitful discussions on foundations of quantum mechanics.

I am grateful of all the wonderful staff of the physics department, including the electronic and instrument shops. It was not possible to complete the experiments without their expertise. In particular, I would like to thank Bob Rhynalds, the instrument shop manager, and Mike Thompson, who nicely machined various crucial parts for my experiments, invariably starting off my rough wiggly hand-sketches.

Last but not least, without the research environment provided by the people and the University of Nebraska, none of the scientific achievements presented here would have been possible to deliver. In addition to this, I would like to express my gratefulness towards the Maude Hammond Fling Fellowship program at the university of Nebraska which I was honored to be a recipient of in the academic year 2018-19. I am thankful of the National Science Foundation which has funded the majority of the work presented in this dissertation through Prof. Batelaan's research grant under the award number PHY-1912504. Part of the work was also funded by the Nebraska Research Initiative. I am grateful of the American Physical Society for granting me travel support to present our results in the APS DAMOP conferences twice.

I opened my acknowledgements by expressing my utmost gratitude towards Prof. Batelaan. I would like to point this out once more. I also add that as it goes without saying, in all the projects and their presented results, Prof. Batelaan's ideas are implicit without my explicit mentioning of them. In spite of this, I emphasize that any possible deficiencies and shortcomings are only those of mine, and I take complete responsibility for the content of this dissertation.

# Table of Contents

## Acronyms

AHE: Anomalous Hall Effect

ATE: Above-Threshold Emission

CC: Charge Current

CEM: Channel Electron Multiplier

CFD: Constant Fraction Discriminator

CM: Center of Mass

cps: counts per second

CW: Continuous-Wave

DIT: Diffraction-In-Time

ePPM: electron Point Projection Microscopy

FDTD: Finite Difference Time Domain

FE: Field Emission

FEQO: Free Electron Quantum Optics

FD: Fermi-Dirac (distribution)

FIB: Focused Ion Beam Milling

FN: Fowler-Nordheim (equation, emission, curve)

FONTES: Fiber Optic Nanotip Electron Source

GRIN: Graded Index

HBT: Hanbury Brown-Twiss (effect, interferometer)

HE: Hall Effect

HOM: Hong-Ou-Mandel (effect, interferometer)

ISHE: Inverse Spin Hall Effect

LED: Leading-Edge Discriminator

MCA: Multi-Channel Analyzer

MCP: Micro-Channel Plate (detector)

MCS: Multi-Channel Scalar

ML: Mode-Locked

MM: Multi-Mode

MPE: Multi-Photon Emission

MS: Mott Scattering

MWDIT: Matter-Wave Diffraction-In-Time

NIM: Nuclear Instrumentation Module

OAPM: Off-Axis Parabolic Mirror

O-FE: Optical Field Emission

OPA: Optical Parametric Amplifier

PEP: Pauli Exclusion Principle

PFE: Photo-Field Emission

PI: Path-Integral

PMT: Photo-Multiplier Tube

QE: Quantum Efficiency

QD: Quantum Degeneracy

QSHE: Quantum Spin Hall Effect

RLD: Richardson-Laue-Dushman (equation)

SC: Spin Current

SEM: Scanning Electron Microscopy(e)

S-FE: Schottky-assisted Field Emission

SHE: Spin Hall Effect

SNOM: Scanning Near-Field Optical Microscopy(e)

SO: Spin-Orbit (coupling, interaction)

SPES: Spin-Polarized Electron Source

SPM: Scanning Probe Microscopy(e)

SPR: Surface Plasmon Resonance

SQUID: Superconducting Quantum Interference Device

STM: Scanning Tunneling Microscopy(e)

TAC: Time-to-Amplitude Converter

TE: Thermionic Emission

T-FE: Thermally-assisted Field Emission

UED: Ultrafast Electron Diffraction

UEM: Ultrafast Electron Microscopy(e)

ULNES: Ultrafast Laser-driven Nanotip Electron Source

# List of Figures

# List of Tables

## Outline of the main results

The main results to be reported and discussed in detail throughout the present dissertation, following the introductory CH. 1, are summarized below.

**CH. 2**

2.a The problem of quantum degeneracy in ultrashort free electron pulses, which is manifest through the fermionic Hanbury Brown-Twiss effect, is studied implementing a path-integral approach to matter-wave diffraction-in-time.

2.b In particular, realistic *partially coherent* pulsed electron beams are modeled using quantum mixtures of coherent and incoherent contributions to the total density matrix and for an arbitrary degree of spin polarization. The problem is also tackled in a more formal manner – though as a first step – based on quantum decoherence theory.

**CH. 3**

3.a An *electron antibunching signal* of *1 part in 4 in magnitude* is reported on the detection plane using a femtosecond laser-driven tungsten nanotip photoemitter. This is significantly stronger than a previously reported electron antibunching signal of 1 part in 1000 using a high-brightness DC field-emission source from nearly two decades ago.

3.b Through an analytical calculation to characterize the *sub-Poissonian* nanotip source, accompanied by a computer simulation of the Poissonian counterpart of the same experiment, the electron-electron *Coulomb repulsion* in free space in ultrashort multi-electron pulses is identified as the antibunching mechanism. Other calculations show that we can approach the quantum degeneracy limit in upcoming experiments to study the Pauli force felt by free electron pairs.

**CH. 4**

4.a A new laser-driven nanotip electron source based on *metallized tapered optical fibers* is introduced, studied and characterized both optically and electronically. The emitter is back-illuminated and mechanically flexible.

4.b Fast continuous-wave electron emission is reported, and *photo-field emission* is identified as the underlying mechanism. The implemented laser powers are the *lowest* of their kind in the literature that have given rise to electron emission.

4.c Fast pulsed wavelength-dependent electron photoemission is reported using a femtosecond optical parametric amplifier as drive laser. *Above-threshold emission facilitated by surface plasmon resonance* excitation across a band in the visible spectrum is identified as the resonant emission mechanism. Multi-photon photoemission dominates the off-resonant domain.

**CH. 5**

5.a Spin-polarized sources of free electrons are directly relevant to the theoretical and experimental research work of CHs. 2&3. Given that ultrashort pulses of free electrons from coherent nanotip emitters are required to achieve enhanced antibunching signals investigated there, *ultrashort spin-polarized* sources of free electrons are to be developed. A proposal for one such nanotip source based on the *spin Hall effect*, which also exploits the fiber tip design of CH. 4, is put forward in the final chapter of this dissertation.

# CHAPTER 1: PROLEGOMENON TO PART I (CHs. 2&3), PART II (CH.4), AND THEIR CONFLUENCE (CH.5)

The present dissertation provides and discusses the results of the two major research projects which I worked on during my graduate studies in Prof. Herman Batelaan's group at the University of Nebraska-Lincoln (UNL). The present chapter gives a general introduction to these two related lines of research, with both of which falling under the umbrella of free-electron physics: ultrashort free-electron second-order matter-wave interferometry, and its underlying technology, coherent laser-driven nanotip electron sources. Additionally, I will give an overview of the relevant basic concepts throughout this chapter.

Part I, to which the next two chapters are dedicated, covers the first subject. The theoretical analysis of quantum degeneracy (QD) in an ultrashort partially coherent spin-polarized free-electron matter wave and its characteristic antibunching signal, studied through incorporation of the fermionic Hanbury Brown-Twiss (HBT) effect [1-3], is given in detail in CHAPTER 2. While the analysis of the fully coherent pulsed beams is straightforward, that of the realistic *partially coherent* beams had not been reported before this research. In addition, the recent advances in making optically switchable ultrashort spin-polarized tip sources based on GaAs in Prof. T. J. Gay's group here at UNL has been a further incentive to pursue this subject [4,5]. The reason is that as discussed in Ref. [3], observing the signature of the Pauli degeneracy force marks the prevalence of the quantum world where deviations from the apparent classical world are manifest. That is where electrons no longer seem to act like classical independent particles and therefore, their fermionic nature must be taken into account. A quantum mixture of independent '*classical'* electrons and quantum mechanically correlated *entangled* electrons, described by an antisymmetric total wavefunction, leads to partial quantum coherence.

The study of free electron QD has its significance rooted in both fundamental and practical aspects. From the fundamental point of view, QD tells us when and how the underlying quantum world of the free electrons ceases to remain hidden under the guise of a stack of quantum mechanically uncorrelated ‘classical’ particles. When that becomes the case, the Pauli exclusion principle (PEP) must be accounted for, just like in the electronic orbitals of atoms where no more than a pair of electrons – and that with opposite spins – are allowed to occupy an identical orbital. We normally ought not to be worried about this effect for the beams of *free* electrons, that is, electrons propagating in free space, under the condition that they are born independently of each other with sufficiently long time-intervals between successive single-electron emission events such that multi-electron interactions are negligible. This is particularly true for continuously emitting sources of electrons like the widely used DC field emission tunneling sources as well as the thermionic emission guns. With the advent of ultrashort laser-driven nanotip electron sources (ULNESs) over a decade ago [6-9] and the large amount of research undertaken on this topic, it is now possible to generate ultrashort multi-electron pulses with a non-negligible chance to contain quantum-mechanically correlated electrons. QD then quantifies how likely it is for the electrons to fall within a single coherence volume, which is the jargon used to indicate whether the various relevant quantum mechanical effects (PEP, matter wave diffraction and interferences of various orders, etc.) can be observed or not. Beams of ultrashort free electron pulses thus seem to be providing a means for the experimental study of quantum foundations and photon-electron reciprocal control and interaction in the burgeoning field of free electron quantum optics (FEQO) [9], which can be seen as the electronic version of the deeply-researched field of photon quantum optics in free space, that has proved to be a trailblazer for quantum information science over the last two decades of the twentieth century. FEQO may someday culminate in reliably reproducible production of on-demand entangled free electron pairs. For the time being, we want to know how ‘*quantum*’ our ultrashort pulses of free electrons are.

The initial impetus of the research on ULNESs was primarily technological which provides a second reason why research on free electron QD has been of interest. The main advantage of nanotip sources lies in their significantly larger transversal (spatial) coherence lengths. In fact, the electrons in such point-like sources are generated within a very small volume with a radius on the order of 1-10 nm that encompasses only a few hundreds to a few millions of atoms. With their natural emission timing due to the well-defined repetition rate of the driving pulsed laser, these sources have been used in ultrafast electron diffraction (UED), as well as to make so-called molecular movies in time-resolved scanning electron microscopy (SEM) and transmission electron microscopy (TEM) schemes. Ultrafast electron microscopy (UEM) [10,11] is mainly performed in three modes of operation as discussed in Ref. [12]. In the stroboscopic mode, single- or few-electron pulses are used to form the time-resolved image over a relatively long experiment time. In the repeated mode, a larger number of electrons, on the order of $10^4$ electrons per pulse, is used to construct the image at a faster pace compared with the stroboscopic mode, but at a cost of a reduced spatiotemporal resolution. This is due to the Coulombic interactions among the electrons of the bunch which is detrimental to the final resolution. In fact, such space charge repulsive effects not only broaden the pulse duration, but also contribute to the energy spread and beam divergence [13-15]. The loss of coherence in this case is more significant in the low-energy nonrelativistic regime where the typical interaction time is longer. For the more commonly used flat photo-guns, with a plethora of applications from electron microscopes to particle accelerators [16], the quantum coherence length and time is too short to make the Pauli degeneracy force a crucial factor to consider. For nanotip sources designed to give in a considerable spatial coherence, a conclusive characterization of the contributions of the various Coulomb and Pauli force effects has yet to occur. For the single-shot imaging and diffraction mode, which is the last of the main three modes of UEM, $10^5$-$10^8$ electrons per bunch is typical to achieve spatial and temporal resolutions on the order of 10 nm and 10 ns, respectively, although subpicosecond resolution is demonstrated to be within reach [12,15]. Here, in order to reduce the total exposure time as well as to gain a better contrast in the extracted images, a larger number of electrons are required to exist in each pulse, and based on the above discussion, understanding the electron repulsion effects both due to the PEP and the Coulomb interaction is demanded in order to fully characterize such multi-electron pulses.

These two incentives were the basis of the research that led to Part I of the present work. The detailed theoretical analysis of the former will follow in CHAPTER 2.

The first experimental results on this topic are provided and analyzed in CHAPTER 3. Here, the experimental setup involves everything except for an electrostatic lens which is pivotal in exposing quantum effects by expanding the coherence volume through postselection of the coherently illuminated area. The significance of this experiment, however, is twofold. First, it is the natural precursor of the final free electron HBT experiments by ensuring that our coincidence detection electronics is indeed functional. It also turned out to be markedly helpful in regards with identification and subsequent suppression of the various noise sources and spurious effects before installing the electrostatic lens. All these proved highly advantageous in the design and construction process of a new high-vacuum chamber built in-house to hunt for the signatures of the Pauli degeneracy pressure in ultrashort free electron pulses. A discussion on the design of this new chamber and the experiment to follow will also constitute a portion of CHAPTER 3.

The second equally important significance of the experiment whose results and analysis are elucidated in CHAPTER 3 is that it provides a method to experimentally characterize the statistical randomness of the stream of electrons emitted from our source – a tungsten (W) nanotip needle. It is taken for granted in the community of electron physicists that the photoemission process by itself is statistically random, i.e. that which follows a Poissonian distribution for the consecutive emission time intervals [3]. In our experiment, we empirically inspected whether this is the case on the detection plane as well. Our method, based on utilization of a double-detector in a coincidence measurement scheme, is practical to deploy to characterize the particle statistics of virtually any free electron source, from photocathode emitters in various types of electron accelerators to ULNESs in UED and UEM applications and beyond.

Part II which is covered in CHAPTER 4, is on the development, physics, and characterization of a new electron nanotip source which works based on surface plasmon resonance (SPR) excitation of the system in the noble metal coating of tapered optical fibers using ultrafast drive lasers, and photo-field emission (PFE) using low-power lasers.

The significance of the nanotip sources was pointed out throughout the above discussion on the topic of Part I. It is therefore advantageous to characterize such laser-driven emitters. This is particularly true of the fiber-optic-based nanotip electron source (FONTES) presented in CHAPTER 4. The reason behind this emphasis is twofold. On the one hand, it is shown that such sources can be operated in the continuous-wave (CW) mode using low power CW lasers which can, interestingly, emit photoelectrons in the long wavelength regime, that is, with the photon energy of the driving laser being smaller than the work function of the metal. This is normally not expected for the nanotip sources which require illumination with femtosecond laser pulses that enjoy strong electric fields at each pulse peak. Such fields with appropriate polarizations are further intensified due to the geometrical shape of the metallic nanotip and can thus give rise to the nonlinear effect of multiphoton absorption for the electrons to overcome the potential barrier of the metal. A host of emission mechanisms including thermionic and thermally-assisted effects are also possible, at least in principle, which will be overviewed later in this chapter. To our knowledge, such a photoemission effect has not been reported for low-power (sub-mW) CW lasers.

The second reason why this new source is worthy of attention is that it can be operated in the pulsed mode as well. Although ULNESs in their modern form have now been around for nearly two decades, the FONTES proposed, studied, and characterized in this dissertation possesses a key advantage in contrast with the regular nanotip sources which is the alignment-free delivery of the emitted electrons anywhere inside the vacuum chamber. The only thing the user would need to do is to couple the laser light into the fiber conveniently from outside the vacuum chamber. The electron emitting nanotip at the metallized tapered end of the fiber can be positioned and oriented as needed and even raster scanned on top of a sample surface for potential time-resolved scanning tunneling microscopy (STM) applications. This is in stark contrast with how the common ULNESs are operated using alignment-sensitive optics to tightly focus the incoming laser beam on the nanometric emitting tip.

Despite being seemingly two separate research topics presented in a pair of distinct parts in three chapters, ULNESs are at the heart of this dissertation. In addition, as the PEP is a quantum effect that is intimately sensitive to the electron spin as will be discussed in detail in CHAPTER 2, spin-polarized electron sources (SPESs) are a key in segregating the effects of the Coulomb repulsion and that of the PEP for it is only the latter which responds to a '*spin knob*'. We use GaAs-based electron sources which can give rise to ultrashort electron pulses with a spin polarization of over 10% [4,5]. More details on this type of source will be given later on, but for now it should be clear why SPESs are important for FEQO aside from their significance in other fields of science and technology. CHAPTER 5 is a proposal for a novel SPES based on the spin Hall effect (SHE). The ultimate design is based on a FONTES coated with a material like W that can reportedly give rise to SHE. CHAPTER 5 is thus a confluence of Parts I&II.

In the remainder of this chapter, Section I.1 provides a more detailed introduction to Part I, followed by an overview of the fermionic HBT effect in Section I.2. Section II.1 serves as an introduction to the topic of Part II. Lastly, an illustrative overview of the electron photoemission physics and the role of the SPR in it when applicable provided in Section II.2 will close the present chapter.

## I.1 Quantum coherence and the Hanbury Brown-Twiss effect

The observation of electron degeneracy pressure for a CW free electron beam with a high estimated degeneracy was reported in 2002 [3]. The free electron HBT effect was exploited to detect signatures of QD, and the observed characteristic feature, namely electron antibunching, has been claimed to be expressible by Coulomb repulsion rather than by electron QD in a subsequent theoretical analysis of the problem [17] which is also pointed out by the authors of a different article who performed a similar experiment using a TEM column [18]. Currently, the production of degenerate free electron beams is an open problem. The question on what limits the density of electron pulses, degeneracy pressure or Coulomb interaction, is both relevant for quantum mechanics and for applications such as making snap-shot movies of molecular interactions as pointed out in the opening to this chapter.

In the free-electron HBT experiment of Ref. [3], the nanotip electron source that produced a continuous emission of electrons, was claimed to have provided a world-record degeneracy of $1.6\times10^{-4}$, while femtosecond pulsed nanotip electron sources are projected to be capable of offering an electron degeneracy in excess of $10^{-1}$ *per pulse*. It will be discussed in detail throughout this and the next chapter that *degeneracy* in the context of unbound systems such as beams of free electrons is defined similar to that in bound systems – atoms and solid-state crystals, as representative examples. For instance, for fully occupied electronic states up to the Fermi level in a metal at a certain temperature, the system is fully degenerate below the Fermi level and non-degenerate elsewhere. Accordingly, for free electron beams, *degeneracy* is quantified as the number of electrons per coherence volume; the latter is defined by the longitudinal and transverse coherence lengths. We should also note that the concept of coherence in matter-wave interferometry is defined similarly to that in optical interferometry. It is therefore possible to characterize the coherence volume by, say, taking advantage of double-slit interference. Subsequently, from the average number of particles in a known beam segment, the degeneracy can be evaluated. For free fermions such as electrons, the maximum value of degeneracy is *one* in compliance with the PEP. The reason is that for electrons within a coherence volume, the only remaining quantum number that must differ for them to be able to occupy the same state, namely the same coherence volume, is the azimuthal spin quantum number and there are only two of them for the spin-1/2 electrons. For ultrashort electron pulses, the HBT antibunching signal, which, as will be shown, is a manifestation of QD, is therefore predicted in this research to be stronger by orders of magnitude compared with the CW experiment of Ref. [3]. Moreover, in a 1D approach, it has been shown that the Coulomb interaction and electron degeneracy both contribute to the free-electron HBT antibunching effect with equal proportions for pulsed nanotip electron sources [19].

A difficulty in the experiments is that the QD will not attain the maximum possible value of unity in practice. At such a value a fully coherent description of the quantum mechanical wavefunction would be adequate. However, at smaller values the electron sources and pulses are only *partially* coherent. To address the question how partial coherence influences the dynamics of two-electron wavepackets and their experimental antibunching signature, a theoretical framework is developed which will be discussed in detail in CHAPTER 2. We will conclude that the coincidence techniques can be used to attempt the observation of antibunching for particle-pair mutual detection, even when the electron pulse is partially coherent to the extent as expected for current pulsed nanotip electron sources. Our theoretical approach is intended to make estimates of the electron count rate and the coincidence signals as functions of time for design parameters of table-top experiments.

Historically, R. Hanbury Brown and R. Q. Twiss (HBT) demonstrated that intensity fluctuations of photons emitted from independent thermal light sources show positive correlations when two detectors are located close to each other; for detectors outside of the same coherence volume no correlations were observed [1,2]. The HBT stellar interferometer is shown in Fig. 1-1 schematically. The resultant photon bunching, which gives rise to an enhancement in the photon coincidence counts on two detectors within a single coherence volume, can both be justified by classical intensity fluctuation theory as well as quantum two-photon interference theory [20]. The former attributes the enhancement in photon counts to an increase in the normalized second order correlation function, which is defined as

$$g^{(2)} = \frac{\langle I_1 I_2 \rangle}{\langle I_1 \rangle \langle I_2 \rangle}. \qquad \text{(1-1)}$$

Here, $I_1$ and $I_2$ are time-dependent intensities detected by the two detectors, labeled *1* and *2*, respectively, and the brackets indicate averaging over time. We will get back to the significance of this definition in section I.2.2. The latter approach provides a full quantum optical treatment of the problem based on the bosonic nature of the photons [21].

Identical bosons tend to bunch due to the enhancement in their second order correlation function rooted in the two-particle interference with a positive sign, while identical fermions tend to antibunch because of the PEP which inhibits simultaneous phase-

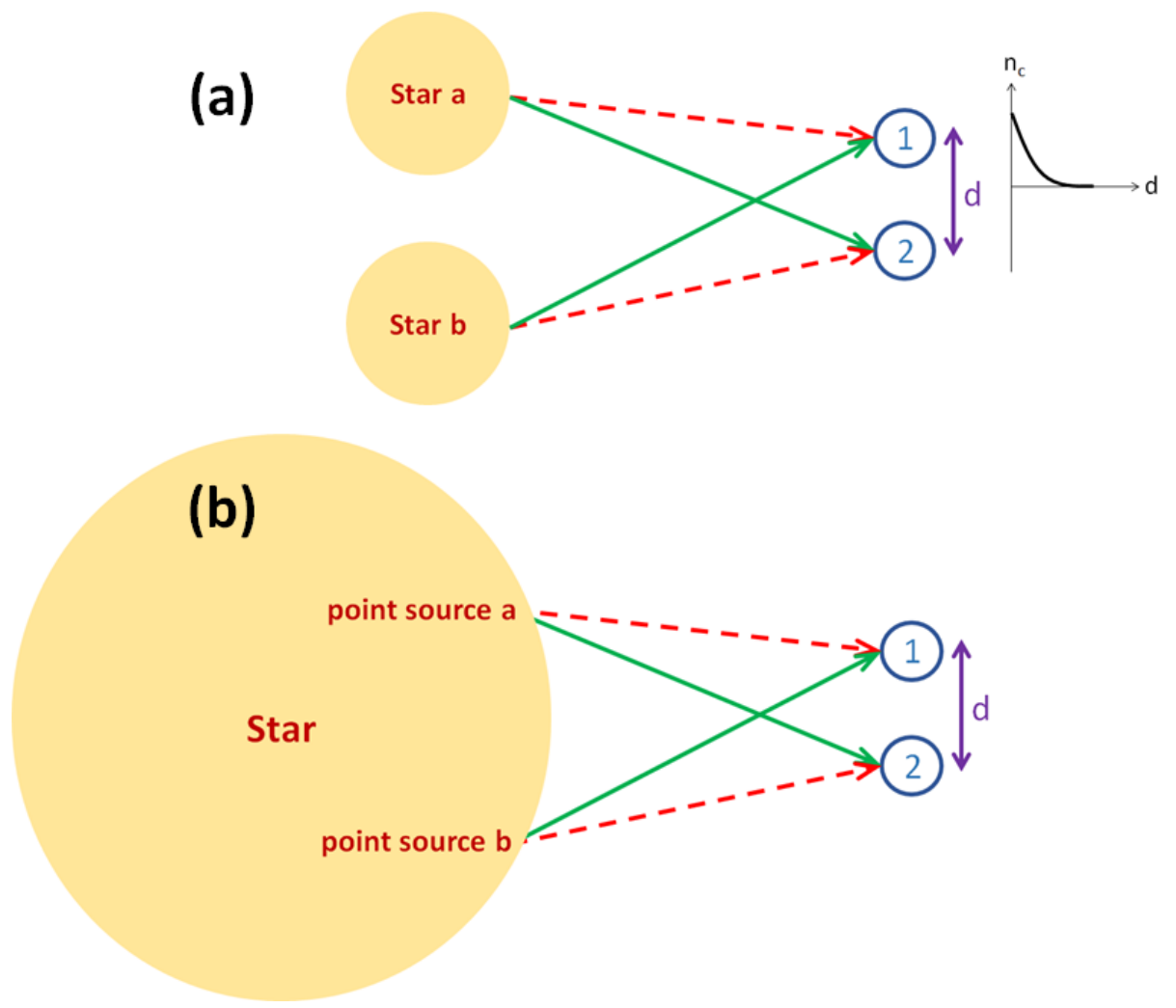

**Fig. 1-1 | Hanbury Brown-Twiss (HBT) stellar interferometer.** The HBT interferometer was invented in 1956 by Robert Hanbury Brown and Richard Twiss. The main advantage of this interferometer over the then-common Michelson interferometer which used to be exploited to determine the angular size of distant stars is that, firstly, it is more compact. Secondly, by correlating *intensities* from independent detectors (rather than fields as in Michelson interferometer) it could bypass stringent stability requirements noting to the fact that the average intensity is not sensitive to atmospheric fluctuations. **(a)** Two possible two-photon trajectories, from (binary) stars *a* and *b* to two detectors labeled *1* and *2* located near each other on the Earth, are depicted by green solid arrows and red dashed arrows (more on this will follow in the discussion of Fig. 1-2). Intensity interference fringes are observed by varying the detectors separation *d*. Here, the fringes constitute the number of mutual detection events at the detector pair in coincidence with each other labeled $n_c$ (within an appropriately set window). This was justified by Roy Glauber in 1963 to be due to the bosonic nature of photons because of which the photons emitted from incoherent thermal sources tend to clump up (bunch) and arrive within a coherence volume *in coincidence* more often than a *random* stream of photons, that is, when the detector pair are separated by one coherence length. **(b)**

The effect is not restricted to double-stars. In fact, any pair of point-like sources on a single star can give rise to the HBT bunching signal. When integrated over the star surface, a similar effect will be observed. The angular size of the star can be determined this way when its distance from the Earth is known. This is explained later on in this chapter in the discussion related to Fig. 1-8.

space occupation of any coherence volume. The latter is indeed the case for a spatial wavefunction representing a two-particle interference with a negative sign. Strictly speaking, it is the symmetry of the spatial wavefunction of the identical particles under consideration that determines whether the corresponding statistical distribution should be bunched or antibunched. Clearly, for an electron pair prepared in a pure state with an antisymmetric total wavefunction, a spin-singlet state couples with a symmetric spatial wavefunction representing a two-particle interference with a positive sign. This indeed gives rise to an electron bunching effect. It is therefore important to note that particle bunching or antibunching *per se* are not inherent to the bosonic or fermionic nature of the streams of identical particles.

HBT-type correlations have been used to probe quantum mechanical particle statistics for various kinds of massive identical particles. Antibunching of thermal neutrons [22], antibunching in a Fermi gas of $^{40}K$ atoms [23], and the HBT effect for $^{3}He$ and $^{4}He$, with the former being a fermion (showing antibunching) and the latter a boson (demonstrating bunching) [24], were reported about a decade ago. The HBT effect with electrons in semiconductor devices typically consisting of electron sources, drains, and mesoscopic beam splitters was also observed [25-27]. In such experiments, high phase-space degeneracies, i.e. the number of electrons per phase-space-cell volume [19], are available and the Coulomb repulsion between electrons is screened by space charge effects. For free electron beams, as pointed out earlier, a conclusive demonstration of the HBT effect remains a challenge. A tour-de-force experiment was reported in 2002 [3]. A reduction of $1.26\times10^{-3}$ in the coincidence count rate of free electrons in a continuous beam emitted from a high-brightness cold field-emission tip was demonstrated. However, at least one later analysis claimed that such small signals could have been dominated by Coulomb interactions between the free electrons in the beam casting doubt on the original claim that a direct signature of the PEP had been observed [17-18].

## I.2 Overview of the fermionic Hanbury Brown-Twiss effect

This section is meant to provide an introductory overview of the topic of Part I. Since the HBT effect employed in this research as a means of probing the quantum correlations in the ultrashort pulses of free electrons is an example of *second-order* matter-wave interferometry, where *two* independent detectors are used to probe the coincidences, an introduction to first- and second-order interferometry is given before discussing the free electron HBT effect as a specific example. Numerous sources, from books on quantum optics, quantum field theory, and quantum information theory, to research and review articles on both optical and matter-wave interferometry exist in the literature for an introduction to the topic under consideration here. Among them, Refs. [21,28-32] provide an inclusive list directly used in the following two brief introductory sections on the treatment of interferometric phenomena based on the first and second order correlation functions. Let us first juxtapose the first- and second-order interferometry schematically.

The two interferometry schemes are pictorially compared in Fig. 1-2. The well-known Young's double-slit interference [33,34] which is a prime example of the more commonly taught and discussed first-order effect is depicted in panel (a). In optics, the first-order interferometry hinges on the interference of the electric field of light beams within a coherence length, while the second-order interferometry – perhaps counterintuitively at first sight – deals with the position-

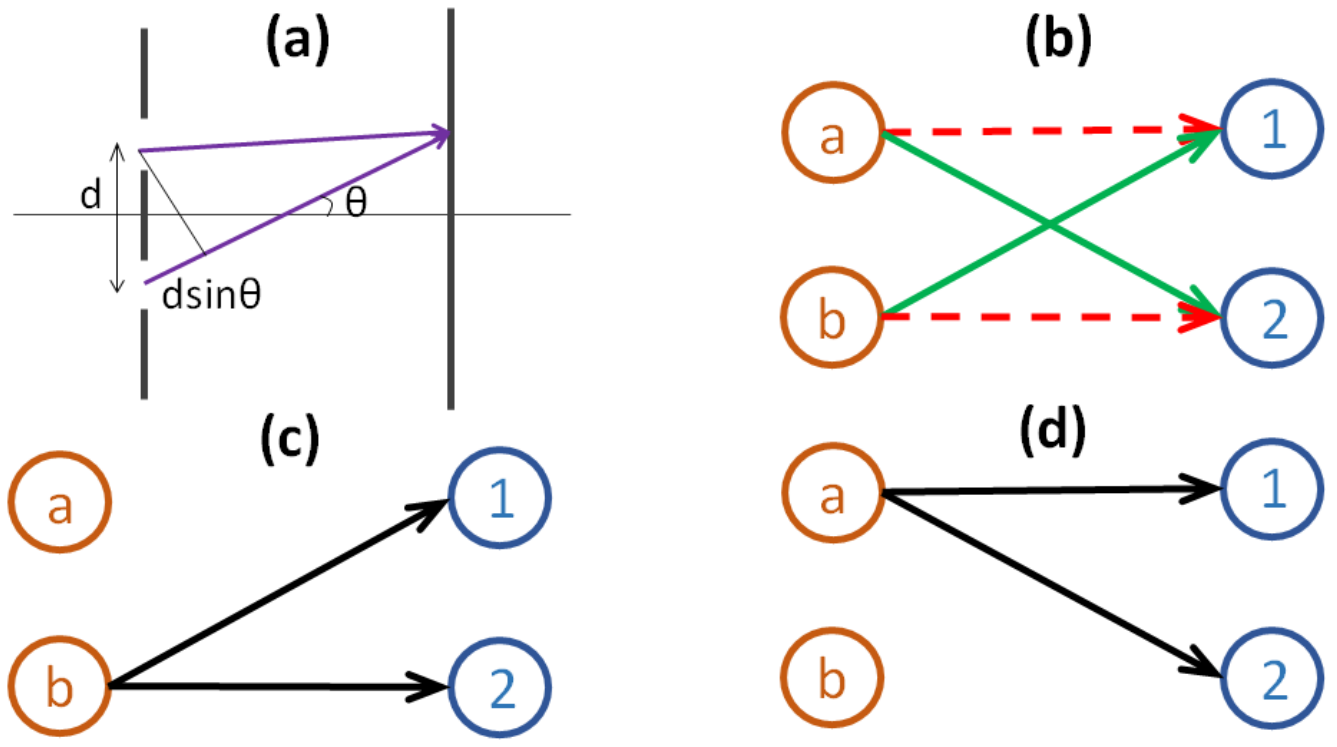

**Fig. 1-2 | Schematic comparison between the double-slit and HBT interferometry. (a)** The phase difference between indistinguishable paths from the two slits which constitute the *interfering alternatives* gives rise to the first order quantum interference. **(b)** The general HBT setup involves two independent sources *a* and *b*, and two detectors *1* and *2* in the second-order interference. The green solid arrows and the red dashed arrows show the pair of *indistinguishable* two-particle trajectories as the pair of interfering alternatives. The general scheme of the figure is taken from Ref. [21]. **(c,d)** The two possible *distinguishable* combinations of two-particle trajectories which add up incoherently without contributing to the periodic interference pattern are depicted for comparison. These obviously do not form mutual alternatives. To observe this in a different way, as will be elucidated further later on in this chapter, for coherent electron pairs for which the two-particle state is required to be antisymmetric, the normalized total wavefunction in panel (b) becomes $(\varphi_{a1}\varphi_{b2}-\varphi_{a2}\varphi_{b1})/\sqrt{2}$ in terms of the single-electron wavefunctions each starting from one of the sources and striking one of the two detectors. This wavefunction is clearly antisymmetric under the exchange of any two source or detector labels. Considering the cases represented in panels (c,d), the hypothetical normalized wavefunction $(\varphi_{a1}\varphi_{a2}-\varphi_{b1}\varphi_{b2})/\sqrt{2}$ for the total state, would not be antisymmetric under the exchange of the detector labels. This shows that we cannot write down a total antisymmetric wavefunction for this pair of alternatives, just as

anticipated from the above-mentioned observation that they are indeed a pair of distinguishable two-electron paths which contribute incoherently to the final probability distribution – as opposed to contributing with an antisymmetric two-electron pure state like the indistinguishable alternatives of panel (b).

and time-dependent periodic pattern in the correlation of the intensity fluctuations of independent beams striking a pair of detectors. By correlation, we refer to the temporal average of the product of the intensities delayed by a fixed amount of time in space which can be measured experimentally in order to extract the corresponding oscillatory pattern. While in the double-slit experiment, the '*which-way*' knowledge will wash-out the interference pattern formed by indistinguishable particle paths according to the standard interpretation of quantum mechanics [35,36], in a second-order interferometry experiment as in that of the HBT effect schematically shown in Fig. 1-2(b), it is the presence of a pair of indistinguishable two-particle trajectories that gives rise to the periodic pattern in the correlated intensities. The remaining two possibilities shown in panels (c) and (d) are of course distinguishable. To envision this, one may note that, say, in Fig. 1-2(c), the photon pair detected at the two detectors labeled *1* and *2* are both originated at a single point source *b*; loosely speaking, the observer would only *see* one of the sources through her detectors which constitutes an unambiguously *distinguishable 'history'* as compared with panel (d) where the independent source *a* is being observed. In Fig. 1-2(b), in contrast, there is no way to tell whether a photon registered at the detector *1* or *2* was originated at source point *a* or *b* so long as the detectors are coherently illuminated, i.e. positioned within one coherence volume of the incident particle stream.

### I.2.1 First-order interferometry

Consider the positive and negative frequency components of the electric field operator denoted with $\hat{E}^{(+)}$ and $\hat{E}^{(-)} = \left(\hat{E}^{(+)}\right)^*$, respectively, with the former (latter) being proportional to the annihilation (creation) operator, $\hat{a}$ ($\hat{a}^\dagger$). The time averaged intensity of light at the space-time coordinate $(\vec{r},t)$, say, on the detection plane is therefore dictated by $\langle \hat{E}^{(-)}(\vec{r},t)\hat{E}^{(+)}(\vec{r},t)\rangle$. The first-order (amplitude) correlation function takes such terms at the same or at two different space-time coordinates, e.g. at the pair of pinholes in the Young's interference of Fig. 1-2(a), to express their degree of mutual coherence:

$$G^{(1)}(\vec{r}_1,t_1;\vec{r}_2,t_2)=\langle \hat{E}^{(-)}(\vec{r}_1,t_1)\hat{E}^{(+)}(\vec{r}_2,t_2)\rangle. \tag{1-2}$$

For such different space-time points, this gives the cross-correlation terms which form the periodic interference pattern. In the Young's experiment, for instance, it can be shown that the average intensity at the coordinate $(\vec{r},t)$ on the screen is given by

$$\begin{aligned} G^{(1)}(\vec{r},t;\vec{r},t) &= G^{(1)}(\vec{r}_1,t_1;\vec{r}_1,t_1)+G^{(1)}(\vec{r}_2,t_2;\vec{r}_2,t_2) \\ &\quad +G^{(1)}(\vec{r}_1,t_1;\vec{r}_2,t_2)+G^{(1)}(\vec{r}_2,t_2;\vec{r}_1,t_1). \end{aligned} \tag{1-3}$$

We also make the observation that at fixed points in space this relation can be utilized to discuss the interference of radiation or matter waves in time.

The main mathematical properties of the correlation function are as follows:

$$\begin{aligned} &G^{(1)}(\vec{r}_1,t_1;\vec{r}_2,t_2)=\left(G^{(1)}(\vec{r}_1,t_1;\vec{r}_2,t_2)\right)^*, \\ &G^{(1)}(\vec{r},t;\vec{r},t)\geq 0, \\ &\left|G^{(1)}(\vec{r}_1,t_1;\vec{r}_2,t_2)\right|^2 \leq G^{(1)}(\vec{r}_1,t_1;\vec{r}_1,t_1)G^{(1)}(\vec{r}_2,t_2;\vec{r}_2,t_2). \end{aligned} \tag{1-4}$$

The latter is the analogue of the Schwarz inequality in the linear algebra of two state vectors which sets an upper bound on the cross-correlation function and the fringe visibility thereof. In other words, the equals case corresponds to mutually coherent space-time points. In addition, it can be shown that for such points, the first order cross-correlation function factorizes to a product of two fields such that corresponding to the upper bound of the above Schwarz inequality, in the double-slit experiment, we get

$$\left|G^{(1)}\left(\vec{r}_1,t_1;\vec{r}_2,t_2\right)\right|^2=\left|\hat{E}\left(\vec{r}_1,t_1\right)\right|^2\left|\hat{E}\left(\vec{r}_2,t_2\right)\right|^2, \tag{1-5}$$

as expected for the superposition of two fully coherent beams. An important point to consider here is that for stationary light sources in the steady state, the cross-correlation function depends on the time interval $\tau=t_1-t_2$, rather than the individual times which implies

$$G^{(1)}\left(\tau\right)=\hat{E}^*\left(t_1\right)\hat{E}\left(t_2\right). \tag{1-6}$$

The only possible solutions to this equation, however, is monochromatic waves. In other words, the monochromaticity requirement to realize fully coherent interference with maximum visibility is a consequence of the steady state condition of the source [32]. Lastly, the normalized first order correlation function is defined as

$$g^{(1)}\left(\vec{r}_1,t_1;\vec{r}_2,t_2\right)=\frac{G^{(1)}\left(\vec{r}_1,t_1;\vec{r}_2,t_2\right)}{\left[G^{(1)}\left(\vec{r}_1,t_1;\vec{r}_1,t_1\right)G^{(1)}\left(\vec{r}_2,t_2;\vec{r}_2,t_2\right)\right]^{1/2}}. \tag{1-7}$$

As an example, to demonstrate the significance of all these definitions, it can be shown that in the Michelson interferometer with a stationary source, the fringe visibility is

$$V=\frac{I_{\max}-I_{\min}}{I_{\max}+I_{\min}}=\left|g^{(1)}\left(\tau\right)\right|, \tag{1-8}$$

where

$$g^{(1)}(\tau)=\frac{\left\langle \hat{E}^{(-)}(\vec{r},t)\hat{E}^{(+)}(\vec{r},t+\tau)\right\rangle}{\left\langle \hat{E}^{(-)}(\vec{r},t)\hat{E}^{(+)}(\vec{r},t)\right\rangle}. \tag{1-9}$$

In general, for fully coherent light $g^{(1)}=1$, while for fully incoherent light $g^{(1)}(\tau)=\delta(\tau)$, the Dirac delta-function. These are the two extremes of the realistic *partially* coherent situations. The finite width of $g^{(1)}$ corresponds to a finite frequency distribution in the Fourier components of the radiation field which is caused by various collision (homogeneous, Lorentzian) and Doppler (inhomogeneous, Gaussian) broadening mechanisms in the source. The width of the normalized correlation function (Gaussian or Lorentzian) therefore gives the coherence time $\tau_c$ of the source and the emanated light. The function $g^{(1)}(\tau)$ is maximum and equal to one at $\tau=0$ and approaches zero asymptotically as $\tau\to\infty$.

### I.2.2 Second-order interferometry

The second-order interferometry is concerned with intensity correlations using a pair of independent detectors, rather than those pertinent to the field amplitude constructed using one detector or a screen to detect one particle at a time. The HBT effect using the radiofrequency waves emitted from a distant star (Sirius) and detected by a pair of nearby antennas on the Earth provides a historic example [2] (see Fig 1-1). Compared with the then-prevalent Michelson stellar interferometer exploited to measure the angular size of the stars, the Hanbury Brown-Twiss interferometer was much more compact. More noticeably, the HBT intensity interferometer was clearly not prone to atmospheric turbulence as it no longer relied on the optical field interference and was therefore far less affected by mechanical stability of its constituent optical components [31]. The elucidation of the quantum mechanical origins of what later on became known as two-photon correlation interferometry (and its variations, e.g. intensity interferometry) nonetheless required extensive intellectual debates and scientific deliberations in the years following Hanbury Brown's and Twiss' discovery of higher order *quantum* interference. It is an interesting historical incident that this was partly rooted in Paul Dirac's misinterpretation of the quantum mechanical concept of interference. In his pioneering and widely-read textbook on quantum mechanics [37, p.9], after discussing the common single-photon interference in the Michelson interferometer, he states, "*[e]ach photon then interferes only with itself. Interference between two different photons never occurs*". Indeed, it is the abstract wavefunctions, namely, single- and multi-particle probability amplitudes in standard quantum mechanics, that may or may not interfere with themselves, not the physical particles and objects in 'real' 4D space-time [32].

Two-particle interferometry, as it turned out in 1960's, is expressible for both identical bosons and identical fermions based on their respective symmetrization postulate. In the case of light, the classical theory can still explain the HBT effect and, up to a certain level of contrast (< 50%), the Hong-Ou-Mandel (HOM) effect [38]. For massive bosons, and specifically for identical fermions with their distinct quantum statistics, the classical theory fails irreparably making use of the quantum theory imperative to construct the correct distributions of interfering particles. For a quantum mechanically correlated (i.e. *mutually* coherent, namely, entangled) identical pair, the two indistinguishable two-particle paths of Fig. 1-2(b) thus contribute to the mutual detection probability density with a negative (positive) sign for the fermions (bosons) which share the same coherence volume upon striking the double-detector. The more realistic case of the *partially* coherent HBT effect with identical electron pairs is the topic of CH.2.

Although we will stick with calculation and measurement of probability density and probability mass functions – as they are directly determinable quantities and easier to deal with by intuition – we wrap up this section with a summary of the main properties of the more abstract generalization of the same concepts, namely, the second-order correlation function. The relation between the two sets of quantities is straightforward; quantum probabilities are proportional to the light intensities, and for non-relativistic massive particles, probability amplitudes replace the electric field operators in the general picture. So long as our quantum system is in the non-relativistic realm, the Schrödinger equation and path integrals [35] are often natural choices to compute the probabilities of various dynamical observables. A brief discussion of the differences between the bosonic and fermionic field operators in the canonical quantization scheme, in the context of the spin statistics theorem and the PEP can be found in Ref. [39].

The normalized second-order correlation function is defined after an obvious extension to Eq. (1-9). This gives

$$\begin{aligned} g^{(2)}(\tau) &= \frac{\left\langle \hat{E}^{(-)}(\vec{r},t)\hat{E}^{(-)}(\vec{r},t+\tau)\hat{E}^{(+)}(\vec{r},t)\hat{E}^{(+)}(\vec{r},t+\tau)\right\rangle}{\left\langle \hat{E}^{(-)}(\vec{r},t)\hat{E}^{(+)}(\vec{r},t)\right\rangle^2} \\ &= \frac{\left\langle \bar{I}(t)\bar{I}(t+\tau)\right\rangle}{\left\langle \bar{I}(t)\right\rangle^2}, \end{aligned} \tag{1-10}$$

where in the second line, $\bar{I}$ is the average intensity of the radiation field. In the photon counting mode, this expression becomes equivalent to

$$g^{(2)}(\tau) = \frac{\left\langle \hat{N}(t)\hat{N}(t+\tau)\right\rangle}{\left\langle \hat{N}(t)\right\rangle^2}, \tag{1-11}$$

with $\hat{N} = \hat{a}^\dagger \hat{a}$ being the number operator. For the instantaneous particle number

$$n(t) = \bar{n} + \Delta n(t), \qquad \langle \Delta n \rangle = 0, \tag{1-12}$$

with constant mean value $\bar{n}$ one readily obtains

$$g^{(2)}(\tau) = 1 + \frac{\langle \Delta n(t) \Delta n(t+\tau) \rangle}{\bar{n}^2}. \tag{1-13}$$

Here, $g^{(2)}(\tau)$ is seen to correlate the fluctuations in the particle flux density separated by a time interval $\tau$. This, in turn, implies that $g^{(2)}(\tau = 0)$ can be equal to 1, larger than 1, or smaller than 1. It is equal to one for random photon distributions which obey the *Poissonian* statistics. This is in fact the case for the coherent output beam of a CW laser light. For a *super-Poissonian* distribution, the temporal and spatial separation between successive photons downstream the beam is reduced compared with the random Poissonian distribution (photon *bunching*) such that the second term in Eq. (1-13) is now positive. A larger spacing than that of the random distribution (photon *antibunching*), in contrast, leads to a reduction in $g^{(2)}(0)$ to a value below 1. While the former is indeed the case for chaotic and thermal light, the latter is purely non-classical. An example is single-photon sources like diamond-based nitrogen-vacancy centers which are capable of giving rise to antibunched photon streams. It can be shown that for a given Fock state $|n\rangle$ at $\tau = 0$,

$$g^{(2)}(\tau = 0) = \frac{\left\langle \left(\hat{a}^{\dagger}(t)\right)^2 \left(\hat{a}(t)\right)^2 \right\rangle}{\left\langle \hat{a}^{\dagger}(t)\hat{a}(t)\right\rangle^2} = \frac{n(n-1)}{n^2}. \tag{1-14}$$

For a true single-photon emitter and an HBT interferometer which is widely implemented in nanophotonics for this purpose, at any arbitrary time $t$, $n = 0 \ or \ 1$, both of which give $g^{(2)}(0) = 0$ [40]. Although this is an example of boson antibunching for an artificial source, as pointed out earlier, boson bunching is what normally occurs in nature. For electrons, *antibunching* is prevalent; for a spin-polarized source it is dominant as the spatial wavefunction of the electron pair is antisymmetric for spin-parallel states. Even for an unpolarized source, the spin-triplet state that couples with an antisymmetric orbital state occurs 3 times as frequently as the spin-singlet state and thus electron antibunching remains dominant.

Finally, an important connection exists between the first- and second-order correlation functions for chaotic light known as the Siegert's relation [31],

$$g^{(2)}(\tau)=1+\left|g^{(1)}(\tau)\right|^2, \tag{1-15}$$

which concatenates this section with the preceding one.

### I.2.3 Quantum coherence, degeneracy pressure, and free electron statistics

The term *coherence* begs for a bit of clarification. I, for one, used to be misled by the varied and interrelated usages of this word under different contexts in physics for a while. We just saw above (though without complete quantum optical proof) that the coherent beam of a CW laser gives the ideal random photon statistics for which, for instance, deviation of $g^{(2)}$ from 1 at $\tau=0$ is zero. But does this make of the coherent beam a more 'classical' source than a thermal source? Is it not a legitimate expectation to be able to get to *see more* of the quantum world when something is *coherent* than not? Is quantum coherence an identical concept to the more familiar classical wave coherence? It is time to elucidate these, making sure the QD referred to in the present dissertation is precisely defined from now on, and introduce the statistical tools we will need to put to use in the subsequent two chapters.

We restrict ourselves to the massive nonrelativistic matter waves that are pertinent to this research on low energy electron beams. As a bonus, the contrast between this case and that of the classical optics, or even quantum optics with its meticulously defined cross-correlation functions, will be more nuanced without losing generality which should serve the purpose of this brief introductory section.

Qualitatively speaking, just as in optics with radiation fields, quantum coherence quantifies the capability of matter waves to take part in various interference phenomena with non-vanishing fringe visibility. The first-order correlation function is intended to do exactly that. In single-particle interference, say, a double-slit experiment with one electron incident at a time [41], so long as the paths taken are not localized – even in principle, regardless of our own knowledge of it – large transverse coherence lengths are attainable upon adequate suppression of the decoherence channels. The HBT interference, on the other hand, involves two electrons landing at a detector pair in a given time interval (sub-nanosecond or higher). Here, *quantum coherence* needs to be refined to ensure that identical mutually coherent electrons (thus expressed by a total antisymmetric state) do not transition to the deterministic *incoherent* 'classical' world in our analysis. A potentially confusing element in this discussion is the role of entanglement. In the quantum decoherence theory [36,42-45] which is at the core of a formal treatment of the partially coherent electron pulses in Section 2.2, entanglement of single electrons with their environment undermines the first order coherence of the electron system in, say, a double-slit experiment. This is because the entanglement to the environmental states effectively acts as a *which-way* detector. Take, for instance, the superposition state

$$\left|\psi\right\rangle = c_1\left|e_1\right\rangle\left|\psi_1\right\rangle + c_2\left|e_2\right\rangle\left|\psi_2\right\rangle, \tag{1-16}$$

in which $\left|e_{1,2}\right\rangle$ are a pair of *distinguishable* environmental states, for which $\left\langle e_1 \middle| e_2\right\rangle = 0$, entangled with the electron states at slits *1* and *2* of a Young's experiment. Clearly, the interference pattern is now washed-out entirely leaving behind only incoherent sums with constant probabilities. This is an archetypal example surrounding the quantum decoherence theory.

Now let us get back to the quantum mechanically correlated electron pair. First of all, we may use the terms *quantum mechanically correlated* and *mutually coherent* interchangeably. These are actually relative concepts in the sense that we can quantify a degree of quantum correlation or coherence for quantum systems. *Entanglement*, on the other hand, is attributable to a *pure* multi-particle state only; it is pure therefore the particles involved are mutually quantum coherent such that the corresponding multi-particle density operator satisfies the purity test, $\hat{\rho}^2 = \hat{\rho}$. We now also precisely know what we mean by *quantum mechanically uncorrelated* and *mutually incoherent*. Such uncorrelated or incoherent pairs of identical particles are effectively *independent* from each other and act as if they are classical particles. In the wavepacket picture – when implemented with caution as a useful model while knowing its ontological limitations due to the abstract nature of quantum states, and epistemic approximations such as those imposed by the Heisenberg uncertainty principle – this corresponds to non-overlapping wavefunctions pictorially shown in Fig. 1-3. Send an ensemble of two such electrons at a time towards the double-detector of an appropriate HBT interferometer and they will keep $g^{(2)}$ equal to unity unless we manage to make non-random distributions from them employing other means at hand, such as the 'classical' Coulomb interaction. In that case, their quantum memory is given off irreversibly to the environment just like the case of everyday macroscopic objects, i.e. stacks of microscopic particles in innumerable configurations, which are never found in superpositions of different locations – at least until now.

The quantum degeneracy (QD) $\delta$ for spin-1/2 fermions is defined as

$$\delta = \frac{nV_c}{2V}, \tag{1-17}$$

where $n$ is the average number of particles per volume $V$ of the beam. From the PEP, we know that each coherence volume $V_c$ can be occupied by a maximum of two identical fermions which sets the upper bound of unity for $\delta$. We will refer to this extreme case as *fully degenerate*. For a general fermion with spin $s$, the factor 2 in the denominator shall be replaced with $2s+1$. For nanotip sources such as chemically etched tungsten needles, the point-like source procures a large spatial coherence in the direction transverse to the beam. As can be grasped from Fig. 1-4, in the cylindrically symmetric approximation, the longitudinal coherence alone establishes the beam degeneracy.

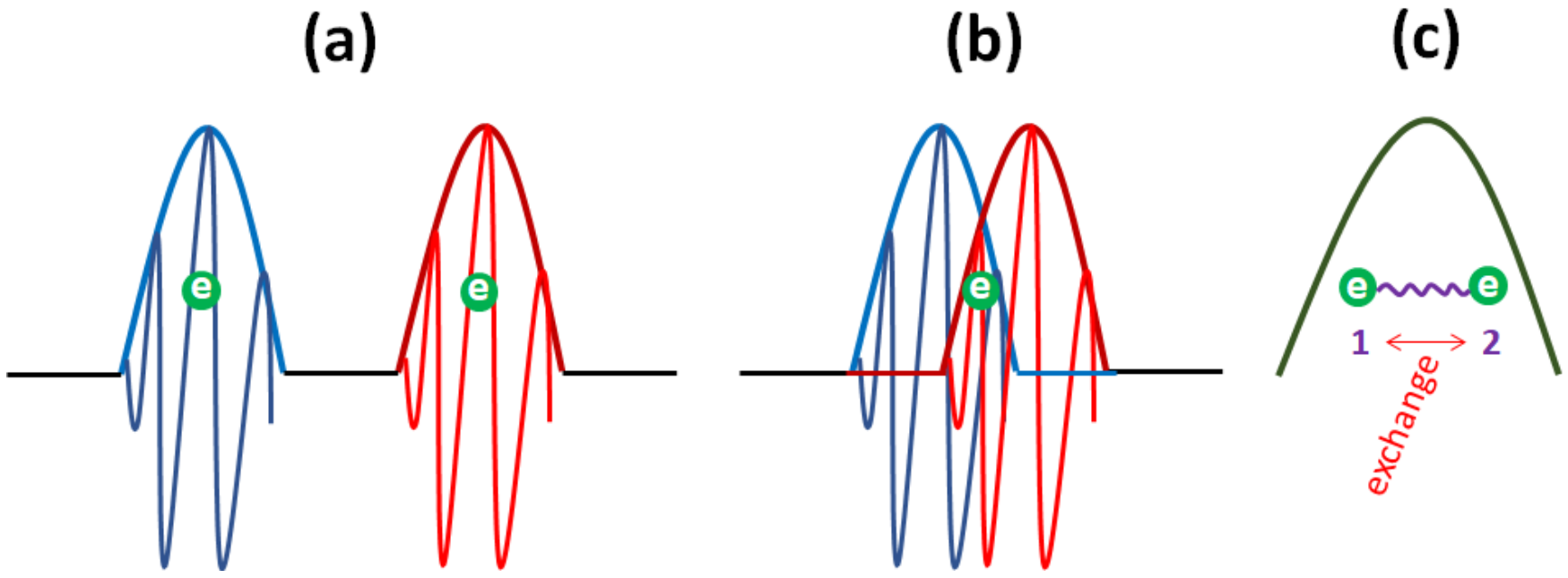

**Fig. 1-3 | Overlapping and nonoverlapping wavepackets. (a)** A pair of *nonoverlapping* single-electron wavepackets corresponding to two quantum mechanically *uncorrelated* electrons are illustrated schematically. The standard deviation of the position operator in the state represented by such individual wavepackets, which are practically nonzero over a finite span of the position space, gives the spatial range of localization of the occupying electron upon measurement of this observable. Such a pair of electrons are *mutually incoherent* while the single-electron position states are *pure* and thus expressible by a wavefunction. **(b)** Partially *overlapping* wavefunctions are depicted in this panel. It is crucial to keep in mind that such a superposition state pictured in this manner is occupied by a single electron only as in the double-slit interference which is a first-order phenomenon. The overlap region corresponds to the finite coherence volume of a *partially coherent* single-electron matter wave where quantum interference can happen. **(c)** A two-electron wavepacket carrying a pair of *entangled* electrons can be thought of as occupying a finite *mutual coherence volume* where the total wavefunction is antisymmetric under the exchange symmetry operation. The wiggly line is solely to denote existence of *quantum correlation* between the pair. It is not possible to represent this wavepacket on the position coordinates of one of the electrons alone unlike in panel (a) where a common coordinate is used for the 'classical' particles involved. In CH.2, we will show how to construct the partially coherent *two-electron* analogue to (b) using probabilistic sums of (a) and (c) combinatorically.

The record-high value of degeneracy for a CW beam so far, to our knowledge, is reported in Ref. [3], which is the first attempt to detect HBT correlations for free electrons, as mentioned earlier. The authors report a $\delta = 1.6\times10^{-4}$. Our remedy to improve the signal-to-noise ratio in order to enhance the HBT antibunching effect caused by QD pressure is to employ ultrashort free electron pulses for four reasons:

1. **Enhanced degeneracy per pulse:** A fraction of the 50-fs pulses carry two electrons (we neglect the smaller fraction of three-electron pulses) while the coherence time is roughly estimated at 1-10 fs. The chance for two such electrons to be born within one coherence time is thus several orders of magnitude higher than that in the CW beam. This heralds an enhanced QD per pulse, approaching unity under realistic conditions.
2. **Time-resolved character:** On top of this, due to the inherent timing of the emission at the well-defined repetition time of the incident laser pulses (13.2 ns), all the mutual detection events that register

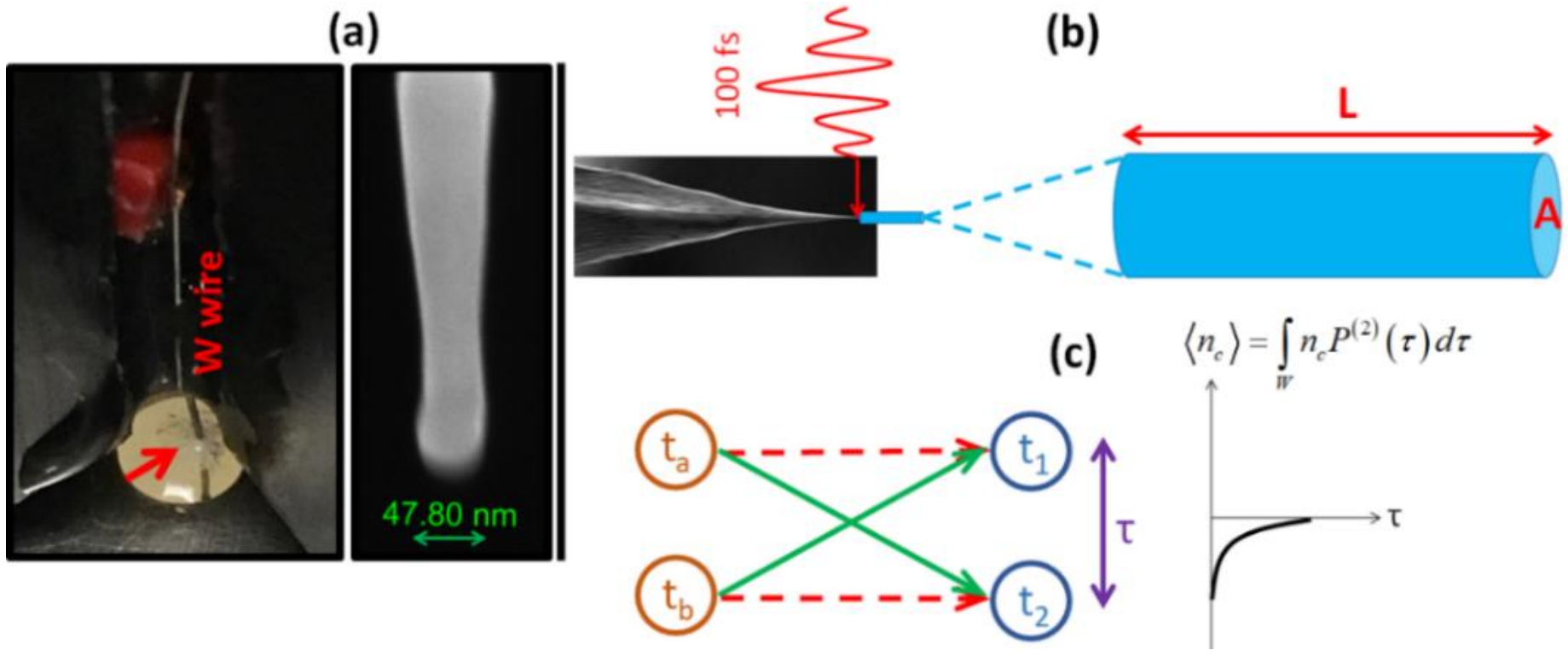

**Fig. 1-4 | Large transverse coherence lengths achieved by point-like nanotip emitters. (a)** High-resolution SEM image of a typical electro-chemically etched tungsten nanotip needle electron photoemitter fabricated by lamella drop-off method [46] is shown (right). The bubbles in the KOH solution (left, red arrow) show the region where the W wire is being etched. Upon detachment, the tip falls down with the apex heading upwards. The sharp tip of this point-like electron source is seen to have a radius of curvature of approximately r = 25 nm. Noting to the nearest-neighbor distance of d = 2.74 Å for crystalline W with bcc lattice parameter of a = 3.16 Å [47, p.21], the emission hemisphere contains about $(2\pi/3)\cdot(r/d)^3 = 1.6\times10^6$ atoms which is tinier than the Avogadro number by over 17 orders of magnitude. **(b)** The tungsten tip, located inside a high-vacuum chamber, is brought into the tight focus of a femtosecond laser beam with parallel (on-axis) polarization, pulse duration of approximately 100 fs, repetition time (rate) of 13.2 ns (76 MHz), and an estimated focal spot size of 9 μm, using a 3D-micro-stage-mounted flexible bellows, giving rise to ultrafast pulsed electron photoemission. Due to the point-like nature of the source, the degree of transverse coherence can safely be taken as complete across the beam cross-section, $A_c \approx A$, in the first approximation. The electron pulse degeneracy, defined in Eq. (1-17), is therefore determined by the longitudinal coherence properties, noting that $V_c/V = T_c/T_{pulse}$, where $T_c$ and $T_{pulse}$ denote the coherence time and pulse duration, respectively. **(c)** The HBT diagram for the emitted ultrashort

electron pulses is depicted. The emphasis here is that we now deal with point-like temporal sources, denoted with $t_a$ and $t_b$, corresponding to emission *times* of an electron pair originated from the spatially point-like nanotip. The detection times are labeled *1* and *2*. The fermionic HBT effect will then be manifest by a decrease in the expectation value of the coincidence counts $n_c$ as a function of the detection time delay $\tau = t_1 - t_2$ which can be calculated from the integral of the two-electron probability density function $P^{(2)}(\tau)$ of the problem over a preset detection window (compare with Fig. 1-1).

within 13.2 ns of each other are distinctively due to the two electrons of a single pulse. This is in contrast to the CW experiment with no such natural timing where it is not possible to directly identify the genuine correlated-pair events from *chance coincidences* of uncorrelated electrons. The problem here is that the best coincidence timing resolutions attainable with the existing detection electronics is on the order of 10 ps which is 3 orders of magnitude longer than typical longitudinal coherence times of existing electron guns. This implies that an HBT signal reduced by 3 orders of magnitude across a coincidence detection window (which itself can be 3 orders of magnitude wider than the actual signal width) is expected. Put simply, in Ref. [3], despite the record-high degeneracy, it is not possible to directly access the femtosecond coincidence time interval of the quantum-correlated electrons which makes it impossible to exclude the mentioned chance coincidences deterministically. This brings the third advantage of utilizing ultrashort electron pulses to the fore.

3. **Diffraction-in-time:** It will be shown in the next chapter that electron pulses broaden in time during propagation. Similar to the generation of a diffraction pattern from a pinhole which forms over the spatial extent of a screen, the ultrashort pulses of electrons born within a short time interval diffract out in time leading to (arrival time) temporal diffraction patterns that become progressively broader for longer distances and lower kinetic energies. More significantly, it will also be demonstrated that the HBT antibunching signal also becomes temporally broader, accordingly. While in the CW experiment of Ref. [3] the 32.5-fs phenomenon is masked by the 26-ps resolution, in our simulation based on realistic parameters of our setup, a 10-fs correlation gives rise to a few-ns HBT signal at the detector – about 2 orders of magnitude in excess of the coincidence detection resolution time.

4. **Fast spin-knob:** As pointed out earlier, we have access to an optically switchable ultrashort spin-polarized electron source whose polarization is determined by a home-built Mott polarimeter [4,5]. While the Coulomb interaction is insensitive to spin polarization, the Pauli degeneracy pressure that gives rise to the fermionic HBT antibunching signal is indeed spin-dependent. Implementing our *all-optical spin-knob* we anticipate conclusive demonstration of QD for free electrons. Our method, scrutinized in CH.3 to empirically determine the particle statistics of the beam of a typical unpolarized tungsten nanotip needle source in the regime insensitive to QD pressure, provides a general machinery to experimentally characterize the (non)randomness of any electron source.

An illustrative example adopted from Ref. [47] is discussed in Fig. 1-5 which demonstrates how the QD pressure of the Pauli force can be quantified in a simple bound atomic system.

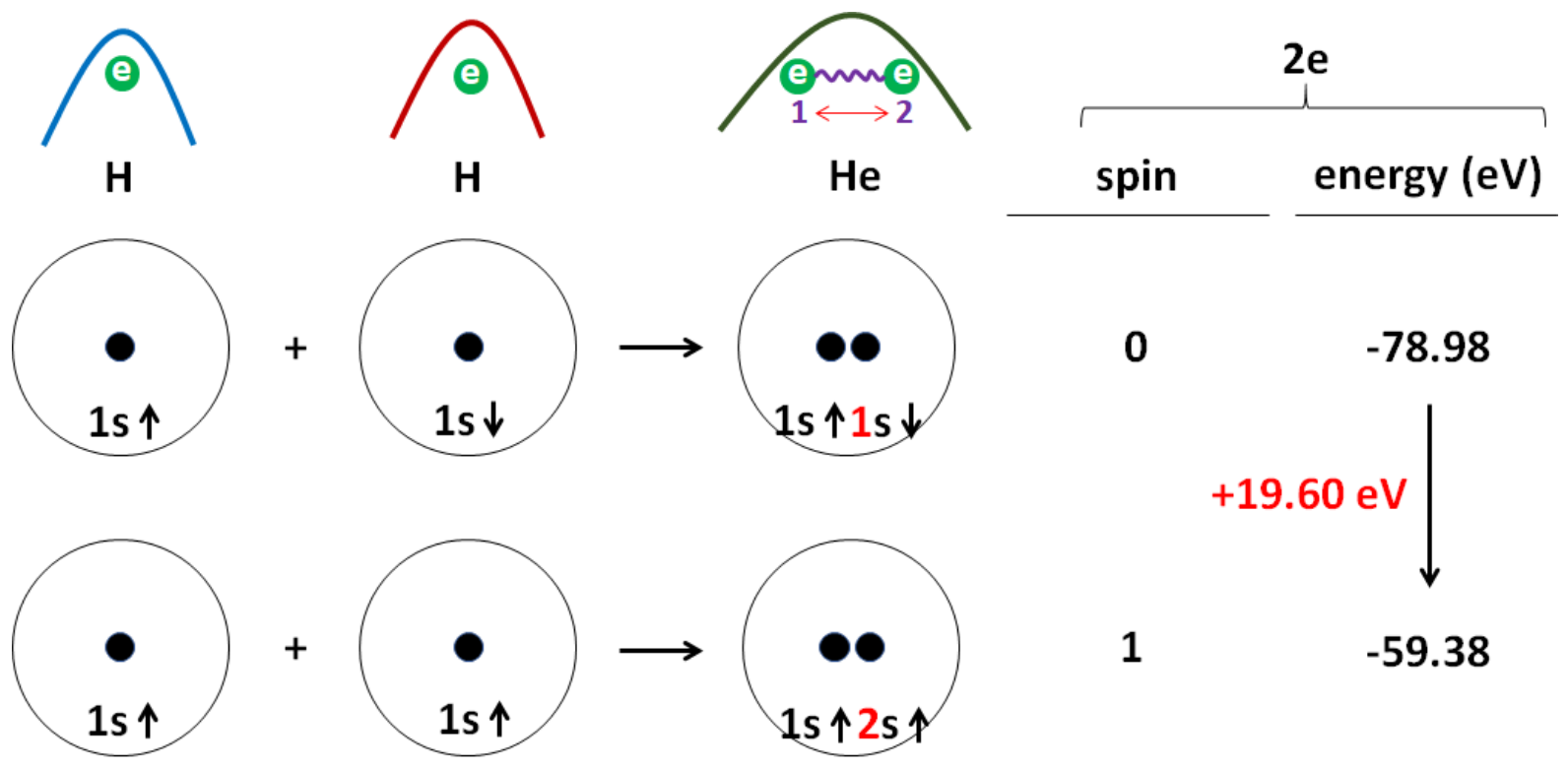

**Fig. 1-5 | Illustration of the repulsive quantum degeneracy pressure.** Two hydrogen atoms are brought to extreme proximity with each other such that their now overlapping electron clouds form a helium-like bound state with an identical Coulomb potential as in the real He atom. In the absence of spin-flip mechanisms, occupation of the same s-orbital is hindered by the *Pauli exclusion principle* for spin-parallel electrons which promotes one of the identical electrons of the pair from the 1s orbital of H to the 2s orbital of He, instead of the 1s orbital of the latter that is the case for antiparallel electron spins. The subtracted total energy of the entangled spin pair is thus attributed to the repulsive *Pauli force*. In principle, the Pauli force and *quantum degeneracy pressure* for identical fermions can be calculated by similar considerations for the unbound state of a (mutually) coherent free electron pair carried by an ultrashort pulse envelope. The repulsive energy of the protons (black circles) is neglected. (Adopted from Ref. [47, p.57].)

### I.2.4 Hanbury Brown-Twiss coincidence interferometry with free electrons

"*When alternatives cannot possibly be resolved by an experiment, they always interfere.*" – quoted from the quantum mechanics textbook by Feynman & Hibbs [35, p.14]. I will discuss how this is at the core of the HBT effect under discussion

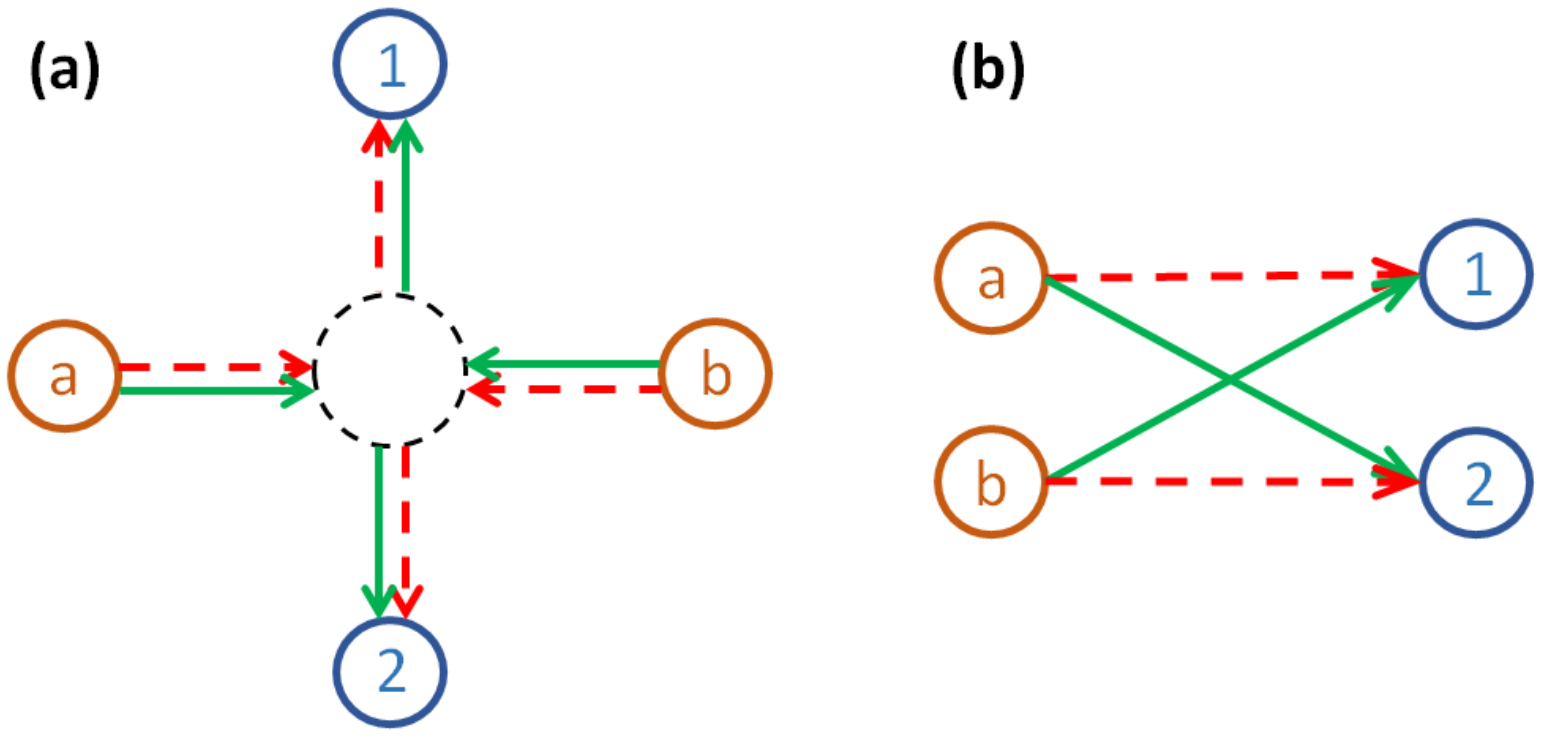

**Fig. 1-6 | Interfering alternatives in the scattering of identical particles compared with the Hanbury Brown-Twiss effect. (a)** In this example taken from Feynman & Hibbs [35, pp.14-16], two particles, distinguishable or indistinguishable, start off the source points *a* and *b*, respectively. After scattering in the central region, the mutual detection probability for the detection points *1* and *2* are calculated. I have argued in the text that the authors' analysis is partly incorrect. The correct manner to put forward a sound description for identical electrons in this setting is that in the overlap region where scattering occurs, the electrons become entangled. It is then possible to calculate the final probabilities from the total antisymmetric state describing the indistinguishable pair. **(b)** In the HBT diagram discussed earlier in Fig. 1-2(b), the situation is similar. Here, it is not a scattering process which renders the total wavefunction antisymmetric. Rather, the illumination of the double-detector is to be made *coherent* by postselection. (Panel (a) is partly adopted from [35, p.15].)

here after commenting on an example worked out in the above text following this excerpt.

The authors make an illustration of their point by an example considering a 90-degree scattering of two nuclei in the center of mass frame. Their schematic setup is juxtaposed with our HBT diagram in Fig. 1-6. Due to the 90-degree geometry, the probability amplitude $\phi(1,2;a,b)$ for the nucleus emitted from source *a* (*b*) to be detected at *1* (*2*) is equal to the exchanged amplitude $\phi(2,1;a,b)$. The total detection probability for such a distinguishable pair of nuclei is

$$\left|\phi(1,2;a,b)\right|^2 + \left|\phi(2,1;a,b)\right|^2 = 2p. \tag{1-18}$$

For indistinguishable nuclei like two alpha particles, however, the total mutual detection probability becomes

$$\left|\phi(1,2;a,b) + \phi(2,1;a,b)\right|^2 = 4p, \tag{1-19}$$

for the pair of interfering alternatives. There is no way, for instance, to determine whether the particle detected at *1* was originated at the source point *a* or *b* in this case, hence no two-particle *which-way* information attainable in principle. "*If electrons scatter electrons,*" the authors continue, "*the result is different in two ways. First, ... spin is not changed to first approximation for scattering at low energy. The spin carries a magnetic moment. At low velocities the main forces are electrical, owing to charge, and the magnetic influences make only a small correction, which we neglect. So if the electron from a* (A in the original text) *has spin up and the electron from b* (B in the original text) *has spin down, we could later tell which arrived at 1 by measuring its spin. If up, it is from a; if down, from b. The scattering probability is then ... 2p in this case. If, however, electrons at both a and b start with spin up, we cannot later tell which is which and we would expect ...* $\left|\phi(1,2;a,b) - \phi(2,1;a,b)\right|^2$. *In our case of* $90^\circ$ *scattering* $\phi(1,2;a,b) = \phi(2,1;a,b)$, *so this is zero.*" [35, pp.15-16]

Part of this argument, in my opinion, is flawed; here is why. Electrons are identical fermions. The electrons involved in the described scattering process are either indistinguishable or not, *regardless of their spins being parallel or antiparallel*, just as in the above discussion of Fig. 1-5. If they are not distinguishable for whatever reason that confers distinguishable *'histories'* on them, even in principle, that is, irrespective of the experimenter's consciousness of it, then the mutual probability is *2p* as if they were independent 'classical' particles. For indistinguishable electrons, however, one must antisymmetrize the total two-electron probability amplitudes obtained by the tensor product of the spatial amplitudes $\phi$ and the total spin state $\chi$ which gives

$$\psi_{tot} = \begin{cases} \phi_S \otimes \chi_{AS} \\ \phi_{AS} \otimes \chi_S \end{cases}, \tag{1-20}$$

for the total two-electron wavefunction. The symmetric (S) and antisymmetric (AS) spatial components are given by

$$\begin{aligned} \phi_S &= \left[\phi(1,2;a,b) + \phi(2,1;a,b)\right]/\sqrt{2}, \\ \phi_{AS} &= \left[\phi(1,2;a,b) - \phi(2,1;a,b)\right]/\sqrt{2}, \end{aligned} \tag{1-21}$$

and the symmetric and antisymmetric spin states are the familiar spin-triplet and -singlet sets. It is therefore by no means correct to rule out the possibility of a symmetric spatial (orbital) state for such indistinguishable electrons with antiparallel spins (with probability *4p*) precisely because they are *indistinguishable* in the quantum mechanical sense of the word. In the absence of spin-flip mechanisms, starting off identical spin-parallel electrons at *a* and *b*, the mutual detection probability at *1* and *2* in this geometry is zero. However, for spin-antiparallel electrons, the pair do not suddenly metamorphose into distinguishable particles giving a mutual detection probability of *2p*. The formalism of standard quantum mechanics requires us to write down the total wavefunction in the form given by Eq. (1-21) in the scattering region, for indistinguishable particles of any spin orientation, where the particles start to become entangled through isolated interactions. This is exactly how we are going to go about the free electron HBT problem in CH.2. For a brief review on how the notions of identicality and distinguishability were inconsistently defined by some authors of quantum mechanics textbooks the reader may consult with Ref. [48].

There is an interesting resemblance between the phenomenology of the scattering problem described above and that of the free electron HBT effect. In the scattering of indistinguishable electrons, it is indeed the interaction in the proximity region that supports entanglement. In the HBT setup, in contrast, there is no scattering between the identical particle pair. If so, then what makes seemingly independent photons originated from points separated by gigantic spans on distant stars indistinguishable here on Earth? Strikingly enough, the photons may have even been emitted at totally different times (say, with respect to the lab frame). The answer is *through expansion of the mutual coherence volume by postselection* which is explained in Fig. 1-7. One of the main aspects of this notion is conceived straightforwardly when contrasted with *preselection*. In the context of our experiments, preselection corresponds to the preparation of the initial state by picking a narrow beam, through a pinhole in the double-slit experiment, and by triggering electron emission from the point-like apex of a nanotip in the electron photoemission experiments. The initial state is thus forced to have a narrow $\Delta x$, corresponding to a comparatively large $\Delta p_x \sim \hbar/\Delta x$, in accordance with the Heisenberg's position-momentum uncertainty relation. This indeed corresponds to a longer transverse coherence length that can now encompass an appreciable segment of the far-field screen which thereby supports formation of multiple fringes in the first-order interference exemplified by the Young's experiment. Postselection then involves further 3D-spatial expansion of the coherence volume *after* the emission of the electrons and *before* they land at either of the two detectors of the HBT setup. In our experiment with ultrashort free electron pulses, the electrons in each pulse are born at the nanotip apex within about 50 fs of each other. For temporal separations exceeding a coherence time of 1-10 fs, the particles are distinguishable and for realistic parameters, the postselection mechanism through an electron quadrupole lens cannot bring about a coherent HBT effect. This is actually ideal for practical purposes related to source characterization – its degree of coherence in this case. For electrons born within a coherence time, on the other hand, an electrostatic quadrupole lens can be exploited to further expand the spatial coherence volumes over the beam segments without undermining the mutual correlation properties (as it acts unitarily by turning on an external field). This enables coherent illumination of the macroscopically-sized detection region.

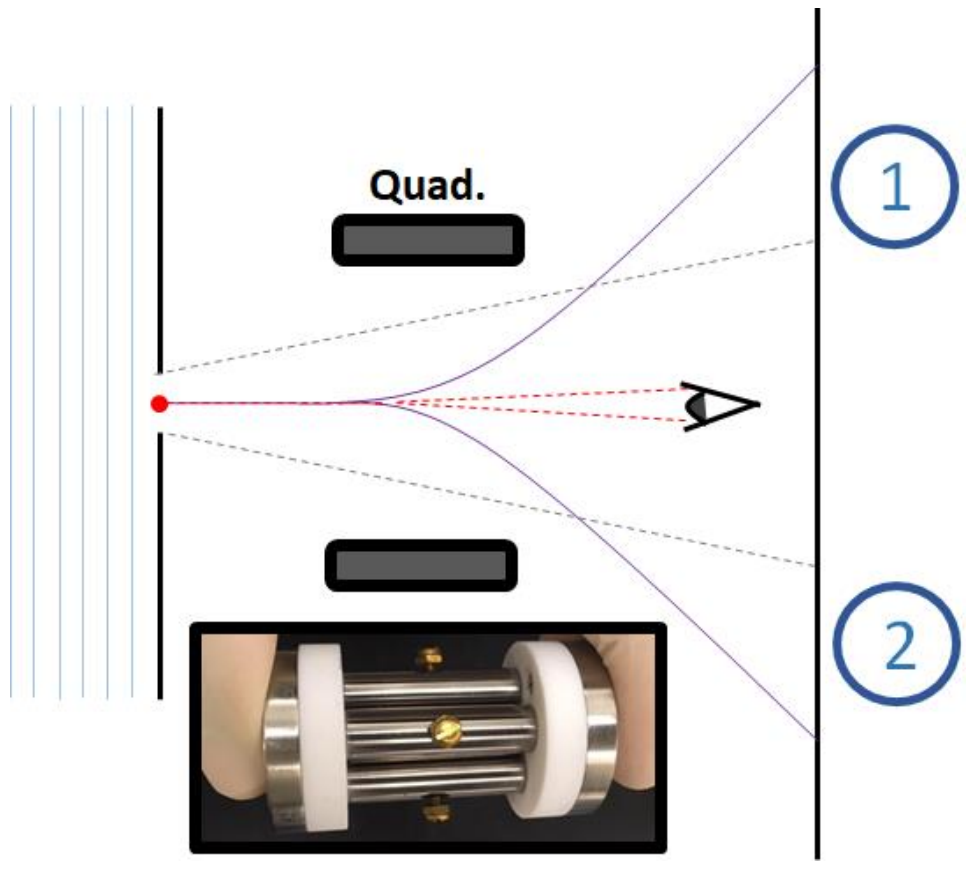

**Fig. 1-7 | Postselection of the coherence volume by an electrostatic lens.** The incoming plane wave is *preselected* to a narrow cross-section for an improved spatial coherence (grey dashed lines). The narrow slit represents the nanometric point-like electron emitter. For macroscopically separated detectors *1* and *2*, the illumination can still not be coherent in practice. The spatial coherence volume may be further expanded transversely (purple solid lines) using an electrostatic quadrupole lens (Quad.) shown in the inset. The detector pair is now within a range of large $\Delta p_x$ corresponding to a small $\Delta x$ . This is equivalent to a virtual emitter (red dot and dashed lines to the virtual eye) that is yet much smaller in size compared with the preselected slit. The count rate per detector will consequently decrease, however, for sufficiently long data acquisition times, one may still record an adequate number of *correlated* electron pairs. Depending on the setup geometry, appropriate DC voltages on the order of 10 V are typically sufficient for the four cylindrical electrodes. See Fig. 3-24 for more discussion.

In general, the van Cittert-Zernike theorem in optics, which states that partial coherence may be produced through propagation of an incoherent beam in free space [29, p.103], is intimately related to this problem [49]. Application of this theorem, in principle, enables determination of the virtual source size based on the lateral coherence length which can be measured by first-order interferometry and corresponds to the transverse length on the far-field screen where the interference fringes are visible. For nanotip electron emitters, the virtual source size of a tungsten nanotip measured in an interferometry experiment using carbon nanotubes, was recently estimated at 0.65 nm which is an order of magnitude smaller than the *geometrical* size of the implemented nanometric source [50]. As explained in this work, the van Cittert-Zernike theorem gives the effective virtual source size,

$$r_{eff} = \frac{\lambda_{dB} l}{\pi \xi_{\perp}}, \qquad (1\text{-}22)$$

where $\lambda_{dB}$ is the de Broglie wavelength of the interfering electron, $l$ is the distance between the carbon nanotube biprism (effectively a double-slit) to the screen, and $\xi_{\perp}$ is the lateral span of the far-field interference pattern on the screen [50]. In general, while we know from the HBT interferometry in astronomy that placing the detectors close enough with each other is in principle sufficient, for electron sources we cannot rely on the free space propagation alone. The key difference is that the degeneracy is much higher for optical beams by many orders of magnitude. For a 1-mW HeNe laser, for instance, the degeneracy is reported at about $10^4$ while for a ruby laser it is calculated to be $10^{16}$ [51, p.93]. For electrons, we observed earlier that the maximum value for $\delta$ is 1 and that the highest reported number is only $10^{-4}$. For ultrashort electron pulses, we anticipate larger values, even those approaching unity, as discussed earlier. For thermal neutrons, as another example, the situation is even worse and the reported degeneracies are on the order of $10^{-10}$ [51, p.93] although long coherence times on the order of 10 ns – 6 orders of magnitude longer than that of electron sources – are reported to have been achieved [22].

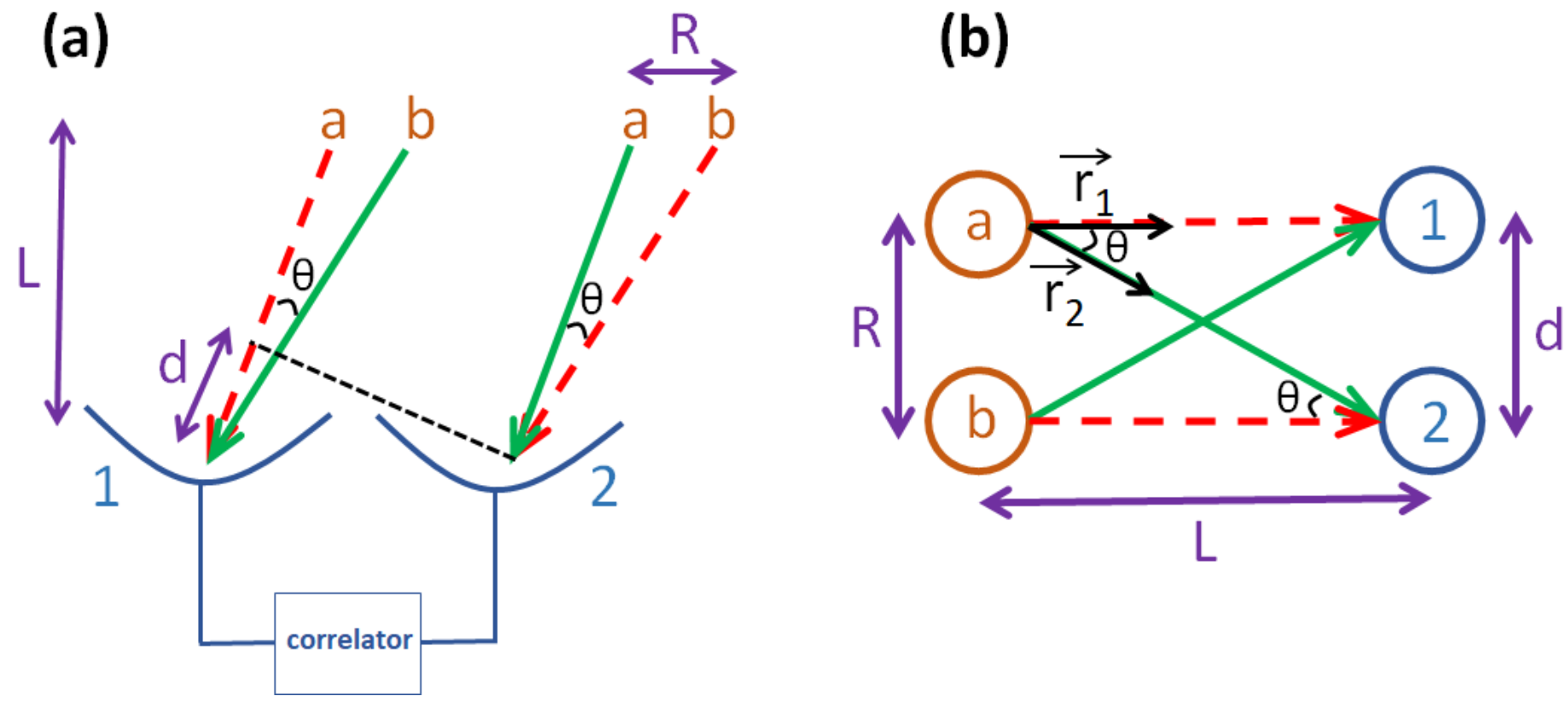

**Fig. 1-8 | The Hanbury Brown-Twiss intensity interferometer.** **(a)** The pair of radio-frequency antennas *1* and *2* generate photocurrents to be correlated upon mutual detection of the incoming light from two distant stars labeled *a* and *b*. **(b)** The HBT schematic diagram is sketched and the relevant parameters are defined. (Panels (a) and (b) are adopted from Ref. [51] and Ref. [21], respectively.) For light, the joint probability distribution which is proportional to $g^{(2)}$ can be recorded as a function of $d$ to exhibit a photon bunching peak when $d = 0$ as discussed in Fig. 1-1. For electrons, when $R = 0$ and $d = 0$, the source can be pulsed, and the electrons can arrive at different times (see Fig. 1-4). In that case, we will find that interference can be observed as a function of the time interval between the detection of the two electrons, $\tau = t_2^f - t_1^f$. This will be elucidated further in CH.2.

Before we discuss in more detail the CW electron antibunching experiment of Ref. [3], it is instructive to mathematically show how the periodic pattern in the cross-correlation function is formed in the double-star HBT stellar interferometer (see Fig. 1-1), as this will equip us to more thoroughly understand the problem.

The stellar interferometer devised by Hanbury Brown and Twiss is sketched in Fig. 1-8(a) [51]. The corresponding HBT schematic diagram is also shown in panel (b). All the relevant parameters are defined in this figure. We are now ready to calculate the second-order correlation function for this problem. The sources *a* and *b* are two distant starts. As can be seen from panel (a), the phase shift is given by the distance *d*. To observe the HBT effect, the radio-frequency detectors need to be positioned sufficiently close to each other for this distance not to fall out of one transverse coherence length for at least one interference fringe to emerge.

The second-order correlation function for the HBT interferometer of Fig. 1-8(a) is given as a function of *d* in Ref. [21] without full proof. Here, we derive the result.

Consider the spherical waves

$$\begin{aligned} S_a &= \alpha \exp\left[ ik\left|\vec{r}-\vec{r}_a\right| + i\varphi_a \right] \big/ \left|\vec{r}-\vec{r}_a\right|, \\ S_b &= \beta \exp\left[ ik\left|\vec{r}-\vec{r}_b\right| + i\varphi_b \right] \big/ \left|\vec{r}-\vec{r}_b\right|, \end{aligned} \tag{1-23}$$

emitted from the stars *a* and *b*, respectively. The polarization is ignored, and $\varphi_a$ and $\varphi_b$ are random phases. The total wave amplitude at the detector *1* is therefore

$$A_1 = \frac{1}{L}\left\{ \alpha \exp\left[ ikr_{1a} + i\varphi_a \right] + \beta \exp\left[ ikr_{1b} + i\varphi_b \right] \right\}. \tag{1-24}$$

From here, we can calculate the intensity at this detector,

$$I_1 = \frac{1}{L^2}\Big(|\alpha|^2 + |\beta|^2 + \alpha^*\beta \exp\left\{i\left[k\left(r_{1b} - r_{1a}\right) + \varphi_b - \varphi_a\right]\right\} \\ + \alpha\beta^* \exp\left\{-i\left[k\left(r_{1b} - r_{1a}\right) + \varphi_b - \varphi_a\right]\right\}\Big). \tag{1-25}$$

Similarly, the intensity at the second detector is

$$I_2 = \frac{1}{L^2}\Big(|\alpha|^2 + |\beta|^2 + \alpha^*\beta \exp\left\{i\left[k\left(r_{2b} - r_{2a}\right) + \varphi_b - \varphi_a\right]\right\} \\ + \alpha\beta^* \exp\left\{-i\left[k\left(r_{2b} - r_{2a}\right) + \varphi_b - \varphi_a\right]\right\}\Big), \tag{1-26}$$

and thus, we can calculate the correlation function:

$$\begin{aligned}
I_1 I_2 = \frac{1}{L^4}\Big\{ & |\alpha|^4 + |\beta|^4 + 2|\alpha|^2|\beta|^2 \\
& + \left(|\alpha|^2 + |\beta|^2\right)\alpha^*\beta \exp\left\{i\left[k\left(r_{1b} - r_{1a}\right) + \left(\varphi_b - \varphi_a\right)\right]\right\} \\
& + \left(|\alpha|^2 + |\beta|^2\right)\alpha\beta^* \exp\left\{-i\left[k\left(r_{1b} - r_{1a}\right) + \left(\varphi_b - \varphi_a\right)\right]\right\} \\
& + \left(|\alpha|^2 + |\beta|^2\right)\alpha^*\beta \exp\left\{i\left[k\left(r_{2b} - r_{2a}\right) + \left(\varphi_b - \varphi_a\right)\right]\right\} \\
& + \left(|\alpha|^2 + |\beta|^2\right)\alpha\beta^* \exp\left\{-i\left[k\left(r_{2b} - r_{2a}\right) + \left(\varphi_b - \varphi_a\right)\right]\right\} \\
& + \alpha^*\beta\alpha^*\beta \exp\left\{i\left[k\left(r_{1b} - r_{1a}\right) + \left(\varphi_b - \varphi_a\right)\right]\right\} \\
& \quad \times \exp\left\{i\left[k\left(r_{2b} - r_{2a}\right) + \left(\varphi_b - \varphi_a\right)\right]\right\} \\
& + \alpha\beta^*\alpha\beta^* \exp\left\{-i\left[k\left(r_{1b} - r_{1a}\right) + \left(\varphi_b - \varphi_a\right)\right]\right\} \\
& \quad \times \exp\left\{-i\left[k\left(r_{2b} - r_{2a}\right) + \left(\varphi_b - \varphi_a\right)\right]\right\} \\
& + |\alpha|^2|\beta|^2 \exp\left\{i\left[k\left(r_{1b} - r_{1a}\right) + \left(\varphi_b - \varphi_a\right)\right]\right\} \\
& \quad \times \exp\left\{-i\left[k\left(r_{2b} - r_{2a}\right) + \left(\varphi_b - \varphi_a\right)\right]\right\} \\
& + |\alpha|^2|\beta|^2 \exp\left\{-i\left[k\left(r_{1b} - r_{1a}\right) + \left(\varphi_b - \varphi_a\right)\right]\right\} \\
& \quad \times \exp\left\{i\left[k\left(r_{2b} - r_{2a}\right) + \left(\varphi_b - \varphi_a\right)\right]\right\}\Big\}.
\end{aligned} \tag{1-27}$$

After carrying out the multiplications, all the exponential terms vanish upon a subsequent averaging in time *except* the two terms originated from the last four lines. This leaves us with

$$
\begin{aligned}
\langle I_1 I_2 \rangle &= \frac{1}{L^4}\Big\langle \Big\{ |\alpha|^4 + |\beta|^4 + 2|\alpha|^2|\beta|^2 \\
&+ |\alpha|^2|\beta|^2 \exp\left[ik\left(r_{1b} - r_{1a} - r_{2b} + r_{2a}\right)\right] \\
&+ |\alpha|^2|\beta|^2 \exp\left[-ik\left(r_{1b} - r_{1a} - r_{2b} + r_{2a}\right)\right]\Big\}\Big\rangle \\
&= \frac{1}{L^4}\Big\langle \Big\{ |\alpha|^4 + |\beta|^4 + 2|\alpha|^2|\beta|^2\left(1 + \cos\left[k\left(r_{1a} - r_{2a} - r_{1b} + r_{2b}\right)\right]\right)\Big\}\Big\rangle .
\end{aligned}
\tag{1-28}
$$

Noting that from the Eqs. (1-25) & (1-26),

$$
\langle I_1 \rangle \langle I_2 \rangle = \frac{1}{L^4}\left\langle |\alpha|^4 + |\beta|^4 + 2|\alpha|^2|\beta|^2 \right\rangle, \tag{1-29}
$$

we obtain the desired expression,

$$
g^{(2)}(d) = \frac{\langle I_1 I_2 \rangle}{\langle I_1 \rangle \langle I_2 \rangle} = 1 + 2\frac{\left\langle |\alpha|^2 \right\rangle \left\langle |\beta|^2 \right\rangle}{\left(\left\langle |\alpha|^2 \right\rangle + \left\langle |\beta|^2 \right\rangle\right)^2} \times \cos\left[k\left(r_{1a} - r_{2a} - r_{1b} + r_{2b}\right)\right], \tag{1-30}
$$

for the second-order correlation function which is periodic with respect to the detectors' separation *d*. To see this more clearly, let us take advantage of the small-angle approximation, $L \gg R$. In this limit,

$$
\left[\left(r_{1a} + r_{2b}\right) - \left(r_{2a} + r_{1b}\right)\right] \sim \left|\vec{r}_1 - \vec{r}_2\right| \sim d\sin\left(\theta\right) \sim d\theta \sim d\frac{R}{L}, \tag{1-31}
$$

where the position vectors $\vec{r}_1$ and $\vec{r}_2$ were defined in Fig. 1-8(b). Plugging this into Eq. (1-30), we readily arrive at the conclusion that

$$g^{(2)} - 1 \propto \cos\left(\frac{2\pi Rd}{\lambda L}\right). \tag{1-32}$$

By varying the detectors' separation $d$ within one coherence length, it is possible to determine the periodicity of the second-order coincidence fringes. Assuming the distance $L$ is known from other measurements (e.g. stellar spectroscopy), one can then take advantage of the above equation to find the separation distance between the two (binary) stars. In addition, the same experiment can be performed using the light originated from a single star as a broad source (see Fig. 1-1). The angular size of the star can therefore be measured in the same way, and again, if the distance $L$ is known, one will be able to determine the diameter of the star by varying the separation distance of the two independent detectors here on Earth.

The probability mass function of the Poissonian distribution which governs the particle statistics of a true random source is given by

$$P(n) = \frac{\lambda^n}{n!} e^{-\lambda}, \tag{1-33}$$

in which $P(n)$ is the probability to detect $n = 0,1,2,...$ particles within a fixed time interval and $\lambda$ is a characteristic constant real number. The normalization condition is

$$\sum_{n=0}^{\infty} P_n = 1, \tag{1-34}$$

the mean value is

$$\sum_{n=0}^{\infty} nP_n = \lambda \sum_{n=1}^{\infty} \frac{\lambda^{n-1}}{(n-1)!} e^{-\lambda} = \lambda \sum_{m=0}^{\infty} \frac{\lambda^m}{m!} e^{-\lambda} = \lambda, \tag{1-35}$$

and it is straightforward to show by means of similar calculations that the standard deviation of this discrete distribution is equal to $\sqrt{\lambda}$ . The characteristic parameter $\lambda = r\tau$ for a stream of particles at the rate of $r$ , where $\tau$ denotes the time interval between successive particles. To find the continuous distribution function of the time intervals between consecutive particles in the beam denoted with $I_1(\tau)$ , one must note that firstly, for a truly random source (one that obeys the Poisson distribution) the absolute times are not significant and only time intervals are pertinent (Markov process). Secondly, the probability to detect a particle after a time interval $\tau$ has passed from the detection of the previous particle is equal to the probability of detecting no particles within the same time interval times the probability of detecting one particle within $(\tau, \tau + d\tau)$ [52]. This gives

$$I_1(\tau)d\tau = e^{-r\tau} \times rd\tau \Rightarrow I_1(\tau) = re^{-r\tau}, \tag{1-36}$$

which is a normalized probability density function,

$$\int_0^\infty I_1(\tau)d\tau = 1. \tag{1-37}$$

Noting to the fact that in the coincidence correlation measurement of the HBT experiment, the measured quantity is the distribution of the electron counts as a function of the time interval between the mutual detection events, the exponential distribution of Eq. (1-36) is precisely what we need to compare with the experimentally constructed histogram of the temporal separation between mutual detection events. Such a histogram is indeed produced in the CW free electron HBT experiment of Ref. [3].

In their experiment [3], the coherence volume is expanded by a quadrupole lens as explained earlier in this section. By correlating the detection counts of the detector pair, the histogram of the arrival times is constructed for mutual detection events within a 26-ps window which sets the finite resolution. The authors report a coherence time of 32.5 fs, a total current of 1.5 μA, and a coherent count rate (with the quadrupole turned on) of $4.7\times10^{9}$ cps. By changing the illumination from incoherent to coherent, a relative reduction of $1.26\times10^{-3}$ with a signal-to-noise ratio of 3 is reported for the total number of coincidences over an acquisition time $t_{acq}$ = 30 h; that is a reduced ratio of coincidence counts of about $10^{3}$ in $5\times10^{5}$ parts or 1 in 500.

Despite being an ingenious pioneering experiment, the authors forgo any consideration of Coulomb repulsion between mutual electrons that could also give rise to apparent antibunching signals. This has raised some controversies over the validity of their interpretation of the results as fermion antibunching.

Let us analyze the results of Ref. [3] based on the Poissonian arrival time distribution. For the observed reduction in the coincidence counts to genuinely be due to QD pressure, the number of electrons born within one coherence time of the source must be consistent with the reported signal. For $r = 4.7\times10^{9}\,cps$ and the coherence time $\tau_c = 32.5\times10^{-15}\,s$, the mean value $\lambda_c$ corresponding to time intervals of duration $\tau_c$ is

$$\lambda_c = r\tau_c = 1.5275\times10^{-4}. \tag{1-38}$$

From this we get

$$\begin{cases} P_0 = e^{-\lambda_c} = 99.984\% \\ P_1 = \lambda_c e^{-\lambda_c} = 1.5272\times10^{-2}\% \\ P_2 = \dfrac{\lambda_c^{\,2}}{2} e^{-\lambda_c} = 1.1664\times10^{-6}\% \end{cases} . \tag{1-39}$$

Therefore,

$$N_2^{tot} = P_2 r t_{acq} = 5.92\times10^6. \tag{1-40}$$

For the sake of comparison, $N_1^{tot} = 7.75\times10^{10}$ is the total number of electrons in this experiment which are separated from their nearest neighbors by a time interval longer than one coherence time. The effect of interest corresponds to a time interval of approximately 30 fs, while the resolution is about 30 ps. An order of magnitude of 3 in the coincidence rate is therefore given up due to the lack of perfect resolution. Were all the $N_2$ electrons able to give rise to coincidence clicks, one would anticipate an HBT antibunching signal strength of approximately $6\times10^3$ which is already satisfactorily within the same order of magnitude as the observed signal. In CH.3, we will argue that from combinatoric considerations, only about 50% of the $N_2$ detection events may be anticipated to ultimately give rise to mutual coincidence counts. This would bring our estimate still closer to the observed signal of Ref. [3].

The quadrupole lens acts as a *coherence-knob* in this experiment. However, by altering the 'classical' trajectories of the electrons inhomogeneously, it may alter the effect of the Coulomb interaction as well. The novel experiment of Ref. [3], therefore, cannot be interpreted as providing *conclusive* evidence in support of the observation of QD pressure in free electron beams.

## II.1 A laser-driven nanotip electron source based on tapered optical fibers

Ultrafast laser-driven photoemission electron sources based on nanotip needles have been studied for over a decade and have attracted a lot of attention due to their potential to act as point-like sources of ultrashort pulsed electrons with a few-femtosecond pulse duration [7-9]. These, together, produce a larger spatial coherence (due to their being point-like in size) along with the ability to be implemented in time-resolved electron microscopy (due to their being an ultrafast pulsed source).

Achieving a sufficiently strong and spatially coherent electron source – a nanoscale-size source with temporal resolution – is a holy grail for free electron physics and technology. The reason is rooted in the widely envisioned applications as a source for UED, UEM [10,11,19], time-resolved STM, electron nano-patterning, and in combining electron probes with scanning near-field optical microscopy (SNOM).

Chemically-etched metallic nanotip needles are the most common samples used to achieve ultrashort pulsed electron emission with sizable spatial coherence lengths upon irradiation with ultrafast laser pulses [46]. In such experiments, the femtosecond laser beam is sent into the high vacuum chamber where the nanotip source is mounted. The laser beam, often the near-IR output of a Ti:Sapphire oscillator, is subsequently focused onto the apex of the nanotip needle. As a result, due to the enhanced electric field at the focal spot on the sharp nanometric apex, the probability of occurrence of nonlinear effects is no longer negligible. In fact, ultrashort pulsed beams of photoelectrons have been achieved under these conditions governed by multiphoton absorption mechanism. While a single IR photon does not have sufficient energy to liberate an electron from the potential barrier of the metallic source, several photons can cooperate nonlinearly to achieve this which can happen when the electric field is strong enough, thus making the use of ultrafast laser pulses imperative.

Our proposed and successfully demonstrated technology in Part II benefits from two main advantages that the conventional scheme overviewed above lacks: *i*) Our laser-driven source is alignment-free due to the fact that we deliver the laser power to the metallic tip emitter in a completely different manner. In fact, our technique bypasses all the complications in focusing the laser beam on the apex of the tip located inside the vacuum chamber which is typically done using parabolic mirrors. *ii*) We have demonstrated that we can achieve photoelectron emission not only implementing ultrashort pulsed lasers, which in turn give rise to short pulsed electron emission, but also using CW lasers which makes it a bimodal source. Our sources are shown to give rise to electron photoemission in the long-wavelength domain using sub-mW CW lasers. We have evidence that the electron emission is due to surface plasmon resonance (SPR) excitation in the nanotip metallized (gold-coated) fiber taper by the driving laser field.

Our technology will *i*) Enable convenient alignment-free operation of the bimodal (CW and pulsed) electron tip source. *ii*) It will enable production of a low-cost plug-and-play photoelectron source noting to the low price of commercial CW diode lasers available at various different wavelengths compared with a sophisticated and costly ultrafast laser system. *iii*) It will enable convenient integration of photoemission sources in various devices like STM and SEM, and wherever raster scanning a photoemission source is needed or preferred. An example of this is in nanopatterning using the emitted electron beam, in which case both CW and pulsed modes of operation (hence the term bimodal) will be available as discussed above.

A succinct overview of the electron emission mechanisms and a brief discussion of the role of surface plasmons in enhancing the emission yield in the following section will wrap up this chapter.

## II.2 Overview of the electron photoemission physics

In CH.4, we will endeavor to identify the electron emission mechanism in our proposed fiber optic nanotip source (FONTES) and will thus consider the various assisted and non-assisted photoemission mechanisms to compare with our experimental results. The goal of this section is to provide a descriptive overview of the main electron emission mechanisms, specifically those pertinent to nanotip emitters, in a mostly qualitative manner for readers less familiar with electron field emission physics.

Numerous excellent resources exist in the literature after nearly a century and a half of experimentation in electron field emission physics, assisted with vigorous modeling and theoretical analysis, and encompassing legendary figures like Nobel laureates Albert Einstein and Max von Laue, as well as inspired inventors like Thomas Edison – to name a very few. In what follows, I have taken advantage of Ref. [53] which is a recent textbook on this topic, Ref.'s [7,8,54] as contemporary review articles which also discuss the characteristics of various electron emission mechanisms, and Ref. [55] in which a brief summary of the major governing equations of several of the mechanisms is given along with the essential considerations to design plasmon-enhanced photocathodes. The characteristic brightness of various electron photoemission sources which are or of potential to be used in UED and UEM applications is reviewed and critically analyzed in Ref. [12]. Lastly, an example on how to calculate the maximum current density in photoemission sources using the Child-Langmuir law in the context of space-charge limited emission theory can be found in Ref. [56].

Electrons can be excited to the vacuum level through several different mechanisms and their combinations alike. These are sketched schematically in Fig. 1-9. Heating, for instance, can raise the Fermi level to near the vacuum state thereby providing enough energy for the nearby electrons to escape the metal through thermal emission, also known as *thermionic emission*. Electrons can also tunnel through sufficiently narrow potential barriers with appreciably high probabilities to give rise to *cold field emission*. An applied DC electric field can narrow down the potential barrier by giving it a triangular shape, noting to the linear dependence of the electric field to the electrostatic potential. Thermally-assisted field emission is therefore another possibility that can be realized, provided that the thermal energy is adequate to promote the electrons to previously unoccupied levels in the proximity of the vacuum level, so that field emission can lead to emission current from the narrower segment of the barrier where tunneling is more probable. Electron *photoemission* is another well-known mechanism to liberate electrons. In the short-wavelength domain, a single photon is sufficient to knock out an electron from the metal (or semiconductor) as the photon energy is larger than the work function. In the long-wavelength domain, it is still possible to achieve electron emission through the nonlinear effect of multiphoton absorption whose probability of occurrence can become high enough for intense femtosecond pulsed lasers. This is particularly true for tight focusing of such beams onto the apex of a metallic nanotip, often in the near-IR domain, where the photon energy is several times smaller than the work function of most metals. The conical shape of the nanotip and the *lightening rod effect* (larger electron density around the sharp apex) together with the parallel-polarization (relative to the tip axis) of the focused femtosecond pulses are known to boost the local electron field for various multiphoton emission processes to become dominant. The potential barrier (solid black line) in all of the nine panels of Fig. 1-9 is assumed to be triangular which is the case in the presence of a negative external bias on the emitter noting to the linear dependence of the electric potential with the electrostatic field as mentioned above. Since in most of the experiments with laser-driven sources we exert such a bias, often as an accelerating field and not necessarily to assist with the emission process significantly, this is taken to be the case here although, in principle, it is required only in mechanisms involving DC (no laser) field emission (FE) tunneling.

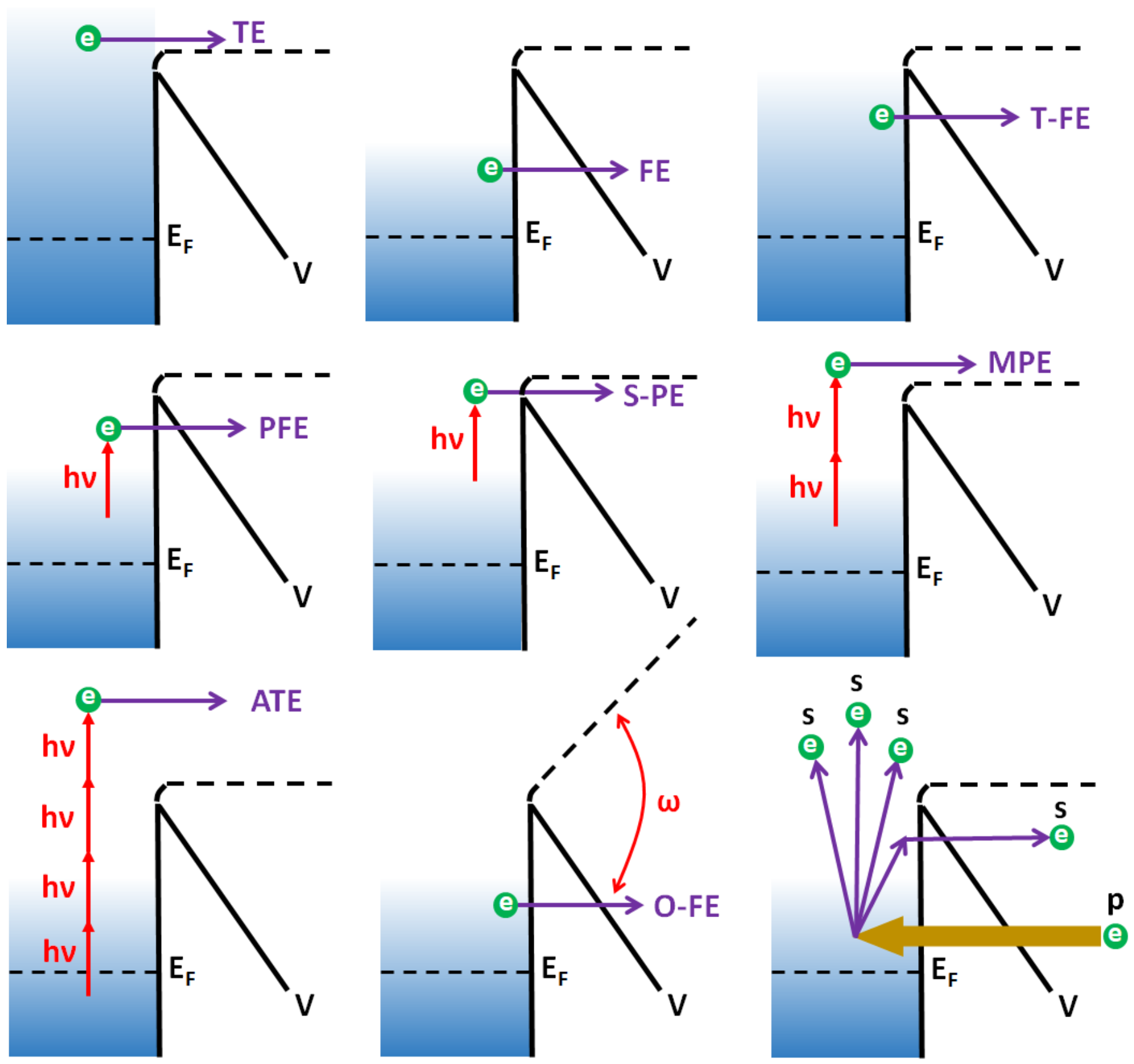

**Fig. 1-9 | Illustration of various electron emission mechanisms.** The blue region indicates the valence band of the metal to the left of the vacuum. The Fermi energy (i.e. the Fermi level at 0 K) is denoted with $E_F$. At no applied electrostatic field, the potential barrier follows the black dashed line. Upon application of an external DC bias field, which can be further enhanced by surface plasmon resonance excitation where possible, the potential barrier starts to assume a triangular shape. In addition, the peak potential will also be slightly lowered, a phenomenon called Schottky lowering, which is explained in the literature using the image charge model. All of the emission mechanisms illustrated here, except for field emission (FE) tunneling, may in principle happen in the absence of any external DC bias fields which leaves the potential barrier rectangular in shape. **(top, left)** Thermionic emission (TE) is shown after raising the Fermi level (highest

occupied level) from that at room temperature to above the vacuum level by thermal heating. The heating process can be through an applied current flowing in a metallic wire or by a beam of laser irradiating the surface of the emitter, just to name a few. **(top, middle)** Cold field emission (FE) by tunneling into the vacuum through the barrier. **(top, right)** Thermally-assisted FE (T-FE) is tunneling from a higher level after an initial promotion by heating. **(middle, left)** Photo-field emission (PFE) is tunneling after promotion to a higher level upon absorption of a photon. **middle, middle)** Schottky-assisted photoemission (S-PE). For a photon energy larger than the work function at room temperature single-photon photoemission (i.e. Einstein's photoelectric effect) can occur. This is one of its variants in which the Schottky lowering is also playing a role in decreasing the barrier height. **(middle, right)** Multiphoton photoemission (MPE) can occur when the laser intensity is large enough to support the nonlinear effect of multiphoton absorption. **(bottom, left)** Above-threshold emission (ATE) is similar to MPE except that the intensity is even larger here to support higher order emission in which the liberated electron accrues an appreciable kinetic energy as well. **(bottom, middle)** Optical FE (O-FE) occurs at very high intensities and for very short laser pulses on the order of 1 fs in duration which encompass only a few light cycles. The potential barrier oscillates with the laser carrier (as opposed to the pulse envelope) phase frequency without photon absorption. During the fraction of a cycle where the potential barrier narrows down FE may thus occur. **(bottom, right)** Secondary electron (s) emission upon impact of an energetic primary electron (p) that has penetrated the potential barrier from the vacuum side.

# CHAPTER 2: ULTRASHORT PARTIALLY COHERENT SPIN-POLARIZED ELECTRON MATTER WAVE: TOWARDS QUANTUM DEGENERACY

This is the first of the two chapters dedicated to Part I. An introduction to this topic was given in CH.1, sections I.1 and I.2, with which the readers' familiarity is assumed.

## 2.1 Partial quantum coherence: A path integral approach based on maximal quantum mixtures

The electronic HBT effect can be described by the second-order correlation function introduced in section I.2.2. However, this is not the only possible function to use to characterize the two-particle interference [57]. The probability distribution $p(n,T)$ for $n$ counts to be registered by a single detector in a time interval $T$ , as well as the joint detection probability density distribution $P(\tau)$ for the time delay $\tau$ to occur between two successive detection events are the other suitable functions. In our analytical model, it is most convenient to obtain the two-electron joint detection probability on a detector located in the far-field of the emitter as a function of the detection time delay which is $P(\tau)$. This is specifically the case for the nanotip emitters with relatively large transverse coherence. Non-relativistic quantum mechanics suffices for our demonstration of the enhancement in the HBT effect with pulsed electron beams compared with continuous beams. In addition, we neglect multi-particle interference effects that involve more than two electrons since they are rare for the experimental scenario considered here which involves low particle numbers in each pulse and short coherence times. We consider the wavepacket propagation along the beam axis. This is justified noting to the fact that in the experimental arrangement in Ref. [3], a quadrupole lens is used to expand the coherence volume appropriately in the transverse plane. This in effect trivializes the physics in the perpendicular directions, conveniently leaving the implications of the PEP pertinent to the longitudinal axis which makes the 1D treatment valid.

The two-electron wavefunction is the tensor product of a spatial part $\varphi(\vec{r}_1,\vec{r}_2)$ and a spinor $\chi(s,m)$, where $\vec{r}_1$ and $\vec{r}_2$ are the position vectors of the two electrons labeled with $1$ and $2$, $s$ gives the total spin eigenvalue, and $m$ gives the eigenvalue of the spin operator in the *z*-direction. The symmetric triplet states are denoted by $\chi_S(s=1, m=1,0,-1)$, and the antisymmetric singlet state by $\chi_{AS}(s=0\,, m=0)$. According to the spin statistics theorem, the total wavefunction of the two electrons

$$\psi(x_1,x_2;s,m)=\varphi(\vec{r}_1,\vec{r}_2)\chi(s,m), \tag{2-1}$$

must be antisymmetric. Consequently, an antisymmetric spatial wavefunction $\varphi_{AS}(\vec{r}_1,\vec{r}_2)$ should be used with $\chi_S$ and a symmetric one $\varphi_S(\vec{r}_1,\vec{r}_2)$ with $\chi_{AS}$. The joint detection probability of a pair of identical electrons is then given by

$$P_{S,AS}(\vec{r}_1,\vec{r}_2)=\left|\varphi_{S,AS}(\vec{r}_1,\vec{r}_2)\right|^2, \tag{2-2}$$

where the normalized symmetric and antisymmetric spatial wavefunctions are

$$\varphi_{S,AS}(\vec{r}_1,\vec{r}_2)=\frac{1}{\sqrt{2}}\left[\varphi_1(\vec{r}_1)\varphi_2(\vec{r}_2)\pm\varphi_1(\vec{r}_2)\varphi_2(\vec{r}_1)\right], \tag{2-3}$$

with the (minus) plus sign for the (anti-)symmetric function. The subscripts in the single-electron wavefunctions $\varphi_i$ with $i=1,2$ denote the particle exchange number. Substituting Eq. (2-3) in Eq. (2-2), one obtains

$$P_{S,AS}\left(\vec{r}_1,\vec{r}_2\right)=\frac{1}{2}\Big\{\left|\varphi_1\left(\vec{r}_1\right)\right|^2\left|\varphi_2\left(\vec{r}_2\right)\right|^2+\left|\varphi_1\left(\vec{r}_2\right)\right|^2\left|\varphi_2\left(\vec{r}_1\right)\right|^2$$

$$\pm 2\,\mathrm{Re}\left[\varphi^*{}_1\left(\vec{r}_1\right)\varphi^*{}_2\left(\vec{r}_2\right)\varphi_1\left(\vec{r}_2\right)\varphi_2\left(\vec{r}_1\right)\right]\Big\}, \qquad (2\text{-}4)$$

for the joint detection probability density of the electron pair. As pointed out earlier and will become clearer shortly, it is the two-electron interference term given by the overlap of the single-particle wavefunctions in $\varphi_{AS}$ that gives rise to the HBT antibunching effect. For an electron source which coherently emits singlet pairs $\chi_{AS}$ (associated with $\varphi_S$), constructive interference will resultantly occur that resembles the familiar boson bunching effect. This is not a violation of PEP since such electron pairs with antiparallel spins do not occupy the same phase-space cell. A fully spin-polarized and fully coherent electron source emitting two spin-up electrons at a time with $\chi_S\left(s=1,m=1\right)$ associated with $\varphi_{AS}$, would thus be expected to give rise to an HBT antibunching effect with perfect contrast, i.e. with zero joint detection probability density for complete coincidence of the electron pairs. The question is, how will the partial coherence of a more realistic electron source affect the measurable HBT contrast of the joint detection probability density. To this end, we need to introduce partial quantum coherence to our analysis.

Consider a pair of electrons with a coherence time $T_c$ carried by a pulse with duration $T_{pulse} \geq T_c$. The limit $T_{pulse} \rightarrow \infty$ will asymptotically reach the continuous beam condition that was experimentally studied in the past [3]. Fig. 2-1 shows the discretization of the time axis into equally sized intervals. Each coherence time interval $T_c$ is split into two equal intervals each with duration $a=T_c/2$. Every such interval then corresponds to a phase-space cell that can bear up to one electron according to the PEP. This way, a pair of electrons in any two neighboring intervals are quantum mechanically *correlated*, namely *coherent* in the sense explained in CH.1, and quantum mechanically *uncorrelated*, namely *incoherent*, otherwise. As pointed out in [58, p.182], *"it is possible to regard a "realistic" beam of particles either as a mixture of plane-wave states (monoenergetic free particle states) or as a mixture of wave-packet states"*. The second approach is what is implemented here. Each wavepacket is thus considered a plane wave restricted within a time interval with duration $a$ as defined above, and is box-normalized accordingly.

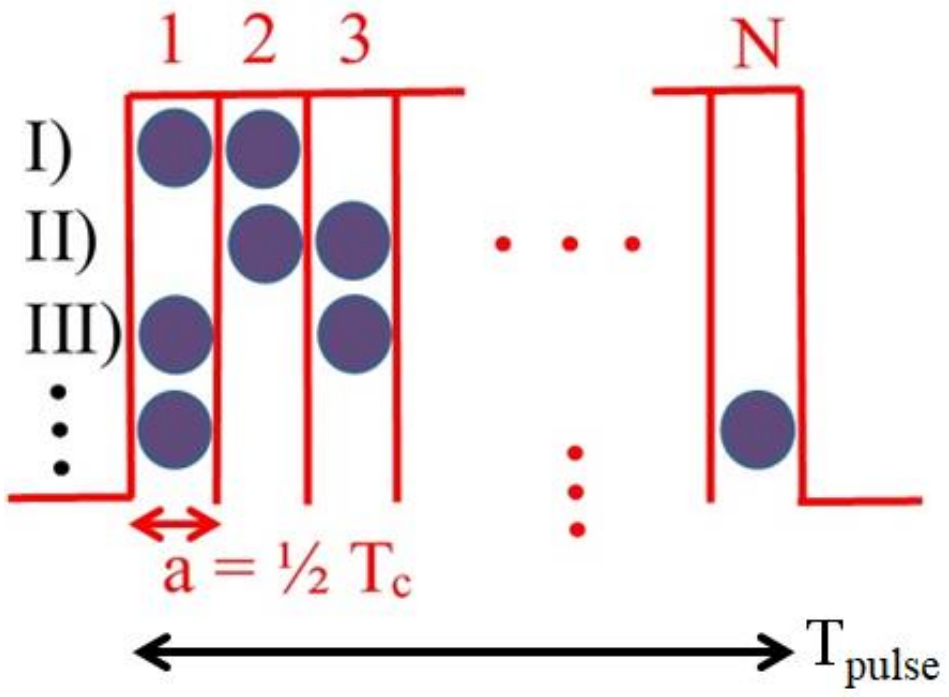

**Fig. 2-1 | Temporal distribution of a pulsed two-electron source.** The pulse duration $T_{pulse}$ is divided into $N$ equal time segments of duration $a$ and the coherence time $T_c$ extends over two such segments. Two electrons separated by a time interval shorter than $T_c$ are in a two-particle pure state corresponding to an antisymmetric total wavefunction. During the fraction of time when the separation time of the electron pair is longer than the coherence time, they contribute incoherently to the final joint detection probability density. A more realistic beam can be treated as a quantum mixture of all these possible configurations.

In our simulations, propagation to a far-field detector at a fixed distance from the source was performed using the path-integral method to be discussed in the following section. It should be noted that the propagation part of the present problem falls under the context of the diffraction in time (DIT) of matter waves. While in the case of diffraction in space, the sources, and the detector screen are extended objects in 3D space, for DIT the point source and the point detector are held fixed at certain positions. This way, the detectors labeled with *1* and *2* in our HBT diagram of CH.1 (see Fig. 1-7) correspond to a pair of mutual detection times, temporally separated by a time interval $\tau$ , for a point-like detector located at a fixed position in space. An electron pulse with a finite duration is also assumed to have been produced at the point-like source which resembles a point-like matter wave shutter that opens for a short time to spit out an electron pulse. Analogously, the electrons are pictured to move through a temporal slit and subsequently diffract in time monitored by a point detector. Our present HBT problem is an extension of this scenario. We therefore look for the joint detection probability density $P(\tau)$ as a function of the two detection times $t_1$ and $t_2$ with $\tau = t_2 - t_1$, rather than as a function of detection positions as in Eq. (2-4). For the case I) in Fig. 2-1, where the two electrons are emitted within one coherence time interval, the two-particle coherent density matrix elements are given by

$$\rho_{S,AS}\left(t_1,t_2;t_1',t_2'\right) = \left\langle t_1,t_2 \middle| \varphi_{S,AS} \right\rangle \left\langle \varphi_{S,AS} \middle| t_1',t_2' \right\rangle = \varphi_{S,AS}\left(t_1,t_2\right)\varphi^*{}_{S,AS}\left(t_1',t_2'\right), \qquad (2\text{-}5)$$

with

$$\varphi_{S,AS}\left(t_1,t_2\right) = \frac{1}{\sqrt{2}}\left[\varphi_1\left(t_1\right)\varphi_2\left(t_2\right) \pm \varphi_1\left(t_2\right)\varphi_2\left(t_1\right)\right], \qquad (2\text{-}6)$$

as the symmetric and antisymmetric coherent spatial wavefunctions. On the other hand, for the case III) in Fig. 2-1, where the two electrons are emitted outside one coherence time window, the elements of the corresponding incoherent density operator are, as a result, given by

$$\begin{aligned}\rho_{incoh}\left(t_1,t_3;t_1',t_3'\right) &= \frac{1}{2}\langle t_1,t_3|\varphi_p\rangle\langle\varphi_p|t_1',t_3'\rangle+\frac{1}{2}\langle t_1,t_3|\varphi_p{}^P\rangle\langle\varphi_p{}^P|t_1',t_3'\rangle \\ &= \frac{1}{2}\varphi_p\left(t_1,t_3\right)\varphi_p^*\left(t_1',t_3'\right)+\frac{1}{2}\varphi_p{}^P\left(t_1,t_3\right)\varphi_p^{P*}\left(t_1',t_3'\right),\end{aligned} \tag{2-7}$$

where the product wavefunctions are

$$\begin{cases}\varphi_p\left(t_1,t_3\right)=\varphi_1\left(t_1\right)\varphi_2\left(t_3\right),\\ \varphi_p{}^P\left(t_1,t_3\right)=\varphi_1\left(t_3\right)\varphi_2\left(t_1\right).\end{cases} \tag{2-8}$$

No coherent terms are present for this part of the total density matrix. From the total partially coherent density operator $\rho$, the final time-dependent joint detection probability density is thus formally written as

$$\begin{aligned}P\left(t_1,t_2\right)&=Tr\left(\rho\left|t_1,t_2\right\rangle\left\langle t_1,t_2\right|\right)=\int\langle t_1',t_2'|\rho|t_1,t_2\rangle\langle t_1,t_2|t_1',t_2'\rangle dt_1'dt_2'\\ &=\int\rho\left(t_1',t_2';t_1,t_2\right)\delta\left(t_1-t_1'\right)\delta\left(t_2-t_2'\right)dt_1'dt_2'=\rho\left(t_1,t_2;t_1,t_2\right).\end{aligned} \tag{2-9}$$

The two-electron problem consists partly of the propagation of electron pulses from the point source to the point detector. Before tackling the two-electron problem, the path integral treatment of the matter wave diffraction in time (MWDIT) is demonstrated and explained for single-electron pulses in the following section.

### 2.1.1 Matter-wave diffraction-in-time

The problem under focus here consists of two main parts: *i*) Writing down an appropriate initial two-electron state, and *ii*) Propagating the state to a detector located at the far-field limit of the emitter. The first step was discussed above. The simulation results after the second step will be given in the subsequent section. Here, an overview of the path-integral (PI) method used for the propagation problem in the context of MWDIT is given and explained in detail.

The MWDIT is sketched schematically in Fig. 2-2. When the shutter opens for a time interval $a$, the electrons start to emit toward a detector held fixed at a sufficiently long distance $D$ from the source (to satisfy the far field approximation) which can register the arrival time of each emitted electron. This is exactly the case for the emission of an electron pulse with duration $a$ that embeds one electron. The observed diffraction pattern which indicates wavepacket broadening can conveniently be computed using the PI method where the probability amplitude is obtained by integrating the accumulated complex phase $e^{iS/\hbar}$, with $S$ being the action defined by the temporal integral of the Lagrangian – here that of the free space – over all possible phase-space trajectories.

The free-space PI kernel for path number $k$ is given by

$$K\left(t_k^i;t_k^f\right)=\sqrt{\frac{m_e}{i2\pi\hbar\left(t_k^f-t_k^i\right)}}\exp\left[\frac{im_eD^2}{2\hbar\left(t_k^f-t_k^i\right)}\right], \tag{2-10}$$

between an initial and a final time denoted by superscripts $i$ and $f$, respectively, $m_e$ is the electron mass and $D$ is the constant distance between the source and the detector. Same as in the classical optics, the wavelets propagated through the paths *1* and *2* originated at the edges of the wide slit in Fig. 2-2 must arrive at the detector in-phase at time $T$ as the condition for constructive interference. The accumulated phase through each of these two paths is determined by the exponential term in Eq. (2-10). Also, for a plane wavefunction at the source, the initial phase is given by $\omega t$ up to a constant phase, where $\omega$ is the angular frequency. The condition of constructive interference at time $T$ will thus enable us to determine the expression for $\omega$ consistent with this requirement that we can subsequently use in our simulations. The constructive interference condition therefore becomes

$$\xi_2^0+\xi_2-\xi_1^0-\xi_1=0, \tag{2-11}$$

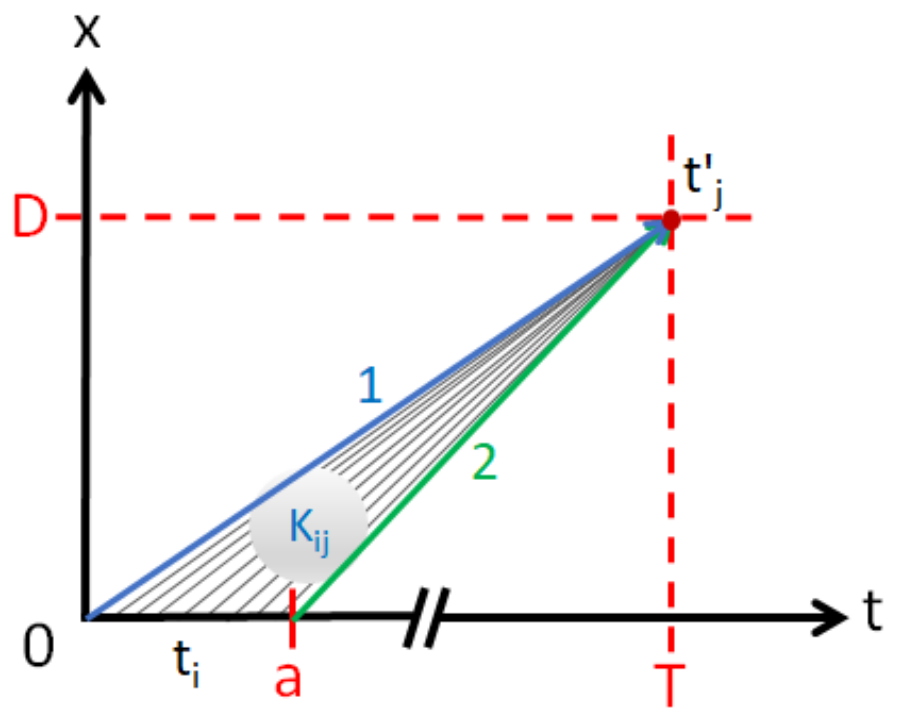

**Fig. 2-2 | Schematics of matter wave diffraction-in-time.** The $x$-$t$ diagram for propagation in 1D space is given. When the shutter is open for a time $a$, electrons are emitted toward a detector held fixed at a sufficiently long distance $D$ from the source (to satisfy the far-field approximation) which can register the arrival time of the emitted electron(s). For the emission of a one-electron pulse with duration $a$, the diffraction pattern can be computed using the path integral method. The probability amplitude is obtained by integrating the accumulated complex phase $e^{iS/\hbar}$, with $S$ being the action defined by the temporal integral of the Lagrangian over all possible phase-space trajectories. A few representative paths starting at an initial time $t_i$ and ending at a detection time $t'_j = T$ are sketched here. The modulus square of the Feynman kernel, $K_{ij}$, gives the probability for an electron to go from $t_i$ to $t_f$ along the path.

where $\xi_i$ with $i = 1,2$ is the accumulated phases over each of the two paths. The superscript $0$ indicates the initial phase on the source. Plugging in the correct expression for each phase function as described above we obtain

$$\frac{m_e D^2}{2\hbar T}\left[\frac{1}{1+\dfrac{a}{T}} - 1\right] - \omega a = 0, \qquad \text{(2-12)}$$

as the exact phase equation. In the far-field limit $a \ll T$ which gives

$$\omega = \frac{-m_e D^2}{2\hbar T^2}. \tag{2-13}$$

For the sake of comparison, for the more familiar case of diffraction in space, the corresponding wave-number obtained by the same method is derived as

$$\kappa = \frac{m_e \Delta}{\hbar \tau}, \tag{2-14}$$

where $\tau$ is the constant detection time and $\Delta$ is the point on the screen where the diffraction peak will show up after a sufficiently large number of single-electron detection events. The far-field approximation that leads to Eq. (2-14) is expressed as $\alpha \ll \Delta$, where $\alpha$ is the spatial width of the source.

It is also instructive to find an expression for the first zeros of the diffraction pattern as a consistency check to make sure that our simulation results comply with it. The condition for destructive interference at time $t_0 \gg a$ is given by

$$\xi_2^0 + \xi_2 - \xi_1^0 - \xi_1 = \pm 2\pi, \tag{2-15}$$

as in optics for a wide source. The reader less familiar with interferometry must note that in this case, for every trajectory originated from one half of the source, there is another trajectory started at a point on the second half which arrives at the first zero with an absolute phase difference of $\pi$ relative to the first path, hence a destructive interference. It follows from Eq. (2-15) that the times of the first zeros are given by

$$t_0 = \frac{1}{\sqrt{\frac{\pm 4\pi\hbar}{m_e a D^2} + \frac{1}{T^2}}}, \tag{2-16}$$

where the minus (plus) sign corresponds to the first zero on the leading (trailing) edge of the diffraction pattern. Again, for comparison with the more familiar case of the matter wave diffraction in space, the first zeros of the diffraction in that case are obtained, following a similar procedure, as

$$x_0 = \Delta \pm \frac{2\pi\hbar\tau}{m_e\alpha}. \quad \text{(2-17)}$$

The double-slit experiment in time can also be treated similarly. Two shutters unblocking an incoming single-electron matter wave packet in the absence of decoherence is equivalent to the case where one electron is prepared in an equal superposition of the time interval *1*, corresponding to the slit number 1, and the time interval *2*, corresponding to another slit, labeled 2, with a negligible detection probability elsewhere. When the slits are separated by a time interval, typically on the order of the slit duration in practice to maintain the transvers coherence of the superposition state, a temporal interference pattern can be observed by measuring the arrival times of electrons at the detector for an adequately large number of detection events. One must keep in mind that the double-slit experiment is a single-particle interference event. For this to be the case for MWDIT in the presence of a physical synchronous double-shutter, the overlap between the emitted electrons from the source must be negligible which is the case for low degeneracies. In fact, in the reported experiments to date with CW sources, each electron arrives at the double-slit long before its subsequent electron is born at the source [59].

Representative simulation results of the single- and double-slit electron diffraction and interference in time are shown in Fig. 2-3. An interesting feature of the MWDIT pattern is that unlike in the more familiar case of the light and matter-

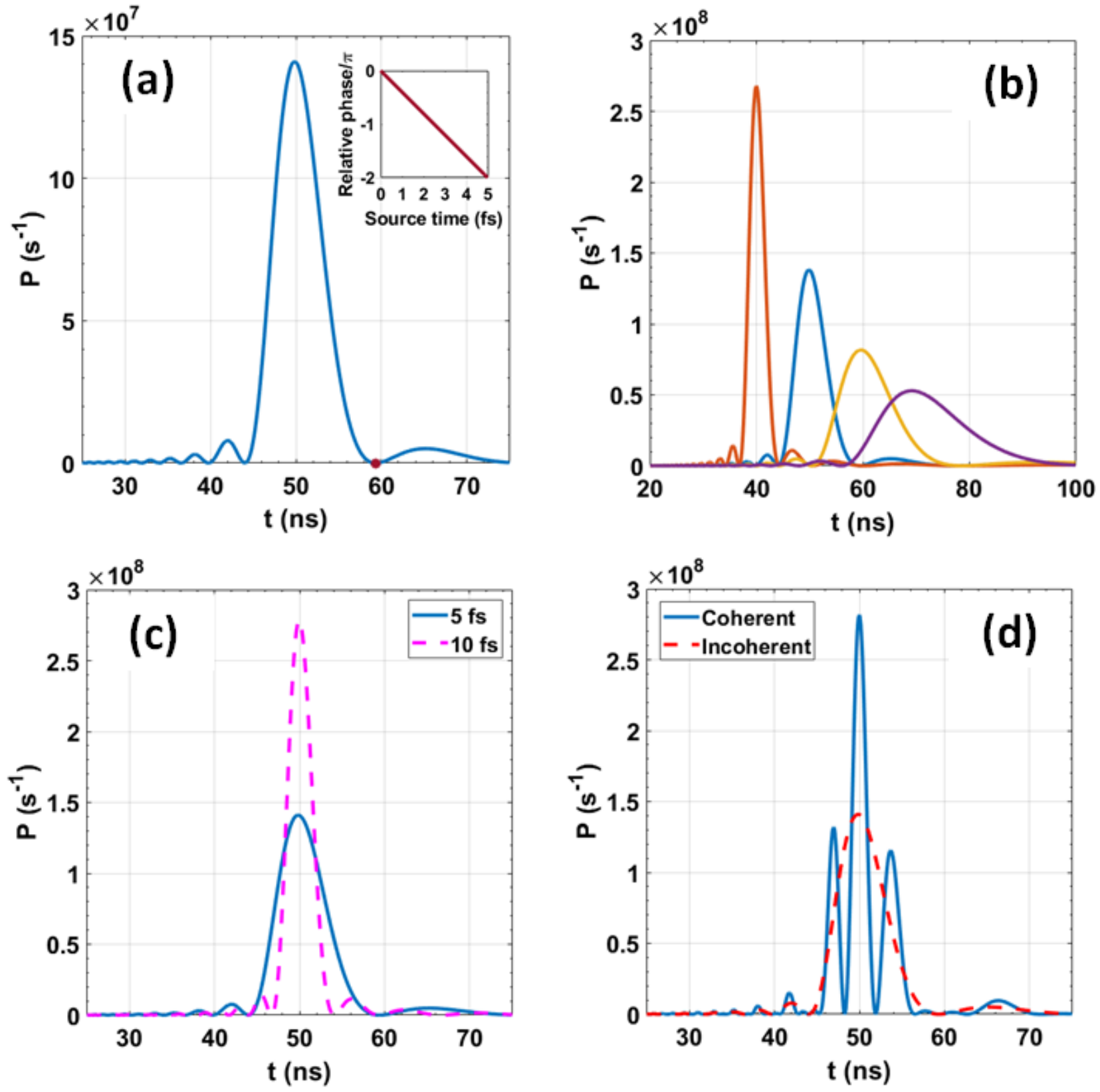

**Fig. 2-3 | Single-particle diffraction-in-time**. This is an overview of the matter wave diffraction-in-time (MWDIT) showing the single-electron detection probability density as a function of the detection time. **(a)** Single-slit electron diffraction shows pulse broadening for $a = 5$ fs and $D = 5$ cm. The inset shows the relative phase of the trajectories originating from different points on the source and arriving at the first zero at the detector position. The first zero is indicated by a (red) dot at $t = 59$ ns on the trailing edge of the diffraction pattern. The total phase difference, which is computed relative to the phase from the trajectory originated at $t = 0$, is $-2\pi$ as indicated in the text. In addition, unlike the more familiar case of diffraction in space, the MWDIT pattern is non-symmetric in time. The reason is that the electrons on the trailing edge take a longer time to travel the same distance $D$ as compared with those on the leading edge. In other words, they have smaller momenta and consequently longer de

Broglie wavelengths and a beat pattern with a longer period is the result. **(b)** From left to right; faster electron packets broaden less than slower ones when traveling the same distance from the source. **(c)** The computed patterns for two different source durations are compared. Shorter wavepackets broaden faster, consistent with Heisenberg's uncertainty relation. **(d)** Coherent and incoherent contributions for double-slit diffraction are shown. All parameters are the same as in panel (a) with the center-to-center slit temporal separation being equal to $2a = 10$ fs.

wave diffraction in space, the MWDIT pattern is non-symmetric in time. The reason is that the electrons on the leading edge are slower than those which land on the trailing edge and thus take a longer time to travel the same distance *D*. In other words, they have smaller momenta and consequently longer de Broglie wavelengths consistent with this observation. We must recall here that in light optics, the first zeros of diffraction are proportional to the wavelength; the longer the wavelength, the wider the diffraction peak. This is also apparent from the Eqs. (2-16) & (2-17); while the first zeros are located at an equal distance from the peak of the diffraction in the latter, those of the former are not symmetrically distributed around the time of the main peak. Another feature to take note of from these same pair of equations is that clearly, for a particle more massive than electron, like neutron, the diffraction peak is narrower under identical conditions, consistent with the claim that it is generally harder to observe quantum interference for larger particles as they tend to keep it localized, or in other words, delocalize at a slower pace compared with lighter particles. The single-electron double-slit pattern is also shown in Fig. 2-3(d), using the same parameters as in Fig. 2-3(a), with the center-to-center time separation of the temporal slits being equal to twice each slit duration. The normalized coherent probability density is obtained by coherent addition of the electron wavefunction corresponding to each slit, while the normalized incoherent pattern, which is also identical to that of a single-electron single-slit DIT in the far-field limit under the same experimental conditions, is obtained by adding the wavefunctions probabilistically, that is, by adding the separate probabilities as we would justifiably do for 'classical' particles. This would in fact be the case for an actual experiment with electrons but with quantum mechanically uncorrelated shutters; that way, the electrons are no longer born in a superposition state of the pair of shutter time intervals. Rather, each electron is born in an equal quantum mixture of the two intervals giving rise to the incoherent detection probability which is identical to that of the single-electron MWDIT in panel (a) of the same figure.

### 2.1.2 Simulation and Analysis: Two-electron partial coherence

The symmetric $P_S(\tau)$ and antisymmetric $P_{AS}(\tau)$ components of the normalized joint detection probability density $P(\tau)$ are made up of an equal mixture of coherent contributions $P_{coh}(\tau)$ associated with Eq. (2-5), and incoherent terms $P_{incoh}(\tau)$ corresponding to Eq. (2-7). The difference is that while the coherent part of $P_S(\tau)$ is symmetric, that of $P_{AS}(\tau)$ is antisymmetric. We thus denote the coherent contribution to $P_S(\tau)$ with $P_{coh,S}(\tau)$ and the coherent contribution to $P_{AS}(\tau)$ with $P_{coh,AS}(\tau)$. Noting to the fact that for the $N$ intervals shown in Fig. 2-1, where

$$N = \frac{2T_{pulse}}{T_c}, \tag{2-18}$$

there are a total of $(N-1)$ coherent contributions and $(N-2)(N-1)/2$ incoherent possibilities, the functions $P_{S/AS}(\tau)$ are expressed as

$$P_{S/AS}(\tau) = \frac{2}{N} P_{coh,S/AS}(\tau) + \frac{N-2}{N} P_{incoh}(\tau). \tag{2-19}$$

The normalized antisymmetric probability density function $P_{AS}(\tau)$ for the fully coherent case of $N = 2$ is shown in Fig. 2-4(d) along with the real part of its single-particle components in panel (a), the corresponding probabilities (not normalized) in panel (b), and the real part of the constituent two-electron wavefunctions in panel (c). The coherence time is 10 fs. The point detector is 5 cm away from the point source, and the anticoincidence time is 50 ns.

Fig. 2-5(a) shows the symmetric (dashed lines) and antisymmetric (solid lines) normalized joint detection probability densities of Eq. (2-19) for several different values of pulse duration $T_{pulse}$ for the same parameters as above, including a coherence time of 10 fs. All of the parameters are consistent with the experimentally achievable values. While $P_{coh,S}(\tau)$ couples with the spin-singlet probability density of its corresponding electron pair, $P_{coh,AS}(\tau)$ goes with the spin-triplet probability density. Clearly, and as pointed out earlier in CH.1, it is the spatially antisymmetric contributions corresponding to the spin-parallel states which give rise to the HBT dip. Now, noting to the fact that for a completely spin-polarized source for which the electron pair are born in a symmetric spin state, the spatial part of the coherent contribution to the total density matrix is antisymmetric, we conclude that the antisymmetric curves $P_{coh,AS}(\tau)$ in Fig. 2-5(a) are indeed the final electronic HBT joint detection probability densities for completely spin-polarized sources with different degrees of coherence. The fully degenerate case with a maximum degeneracy of $\delta = 1$ for which the pulse duration is equal to the coherence time is therefore seen to giving rise to the perfect HBT dip contrast for which the joint probability density is identically zero at the time of the would-be perfect coincidence – the *anticoincidence time* in the present context. This is identical to what was shown above in Fig. 1-4(d). The fully spin-polarized but non-degenerate pulses which are partially coherent consequently give rise to partial HBT dips which do not go all the way down to zero at the anticoincidence time as can be seen from the rest of the antisymmetric joint probability densities for the

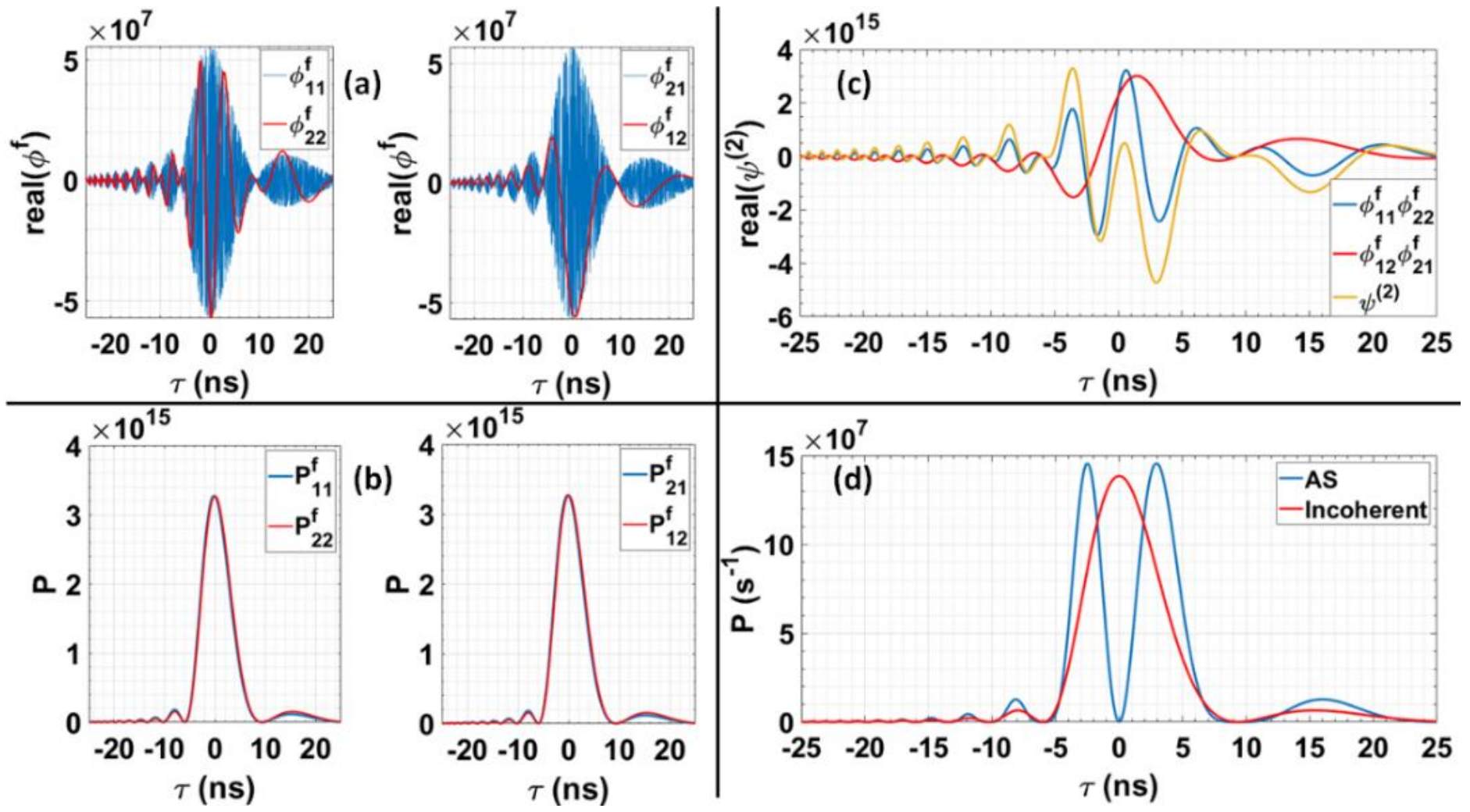

**Fig. 2-4 | The antisymmetric coherent two-electron probability density function and its constituents.** The final (f) single-electron wavepackets are denoted with $\varphi_{ij}$ ; the first index (1 or 2) labels the two source intervals of the fully coherent pulse with $N = 2$ corresponding to the source points $a$ and $b$ in the HBT diagrams of CH. 1. The second index (1 or 2) denotes the two detection times. One detection time is fixed at the so-called *anticoincidence time* of 50 ns, while the other is varied around this time such that the final two-electron (spatial) wavefunction is expressed as $\psi_{\tau}^{(2)} = \left[ \varphi_{a,T}^{(1)} \varphi_{b,T+\tau}^{(1)} - \varphi_{a,T+\tau}^{(1)} \varphi_{b,T}^{(1)} \right] \Big/ \sqrt{2}$ . The superscripts denote the number of particles in the packet. **(a)** The real part of the four single-electron wavepackets involved in this problem are compared. **(b)** The single-electron detection probability corresponding to these functions resemble each other in the far-field approximation as expected. The imaginary part of the wavefunctions which also enter the probability functions are not shown for brevity. **(c)** The real part of the final two-electron wavefunction along with its product components are compared. **(d)** The normalized HBT probability density for a pair of spin-parallel electrons as a function of the mutual detection time delay is shown and contrasted with the corresponding incoherent pattern (had the electron pair ceased to become entangled at the source).

When coherent, the electrons are hindered to click at the same time due to the quantum degeneracy pressure (see Fig. 1-5). The coherence time is 10 fs. The detector is separated from the source by 5 cm, and the anticoincidence time that corresponds to the minimum of the HBT antibunching dip in this panel, as well as the maxima of the peaks in panel (b), is 50 ns. This corresponds to a low single-electron kinetic energy of 2.84 eV.

representative pulse durations in panel (a). On the other hand, an unpolarized electron source is represented by a quantum mixture of the one spin-singlet and the three spin-triplet states [19]. Therefore, the total joint detection probability density for a completely unpolarized source is derived as

$$\begin{aligned} P(\tau) &= \frac{2}{N}\left(\frac{1}{4}P_{coh,S} + \frac{3}{4}P_{coh,AS}\right) + \frac{N-2}{N}P_{incoh} \\ &= \frac{1}{2N}P_{coh,S}(\tau) + \frac{3}{2N}P_{coh,AS}(\tau) + \frac{N-2}{N}P_{incoh}(\tau). \end{aligned} \tag{2-20}$$

Fig. 2-5(b) shows the normalized joint detection probability density $P(\tau)$ for various unpolarized sources obtained using the symmetric and antisymmetric components of panel (a) of the same figure. Fig. 2-6 subsequently demonstrates how the antisymmetric and quantum degenerate joint detection probability density of Fig. 2-5(a) (the solid blue curve) varies with increasing the electron kinetic energy. The kinetic energy is $KE = m_e (D/T)^2 / 2$ where $D$ is the fixed source-to-detector distance and $T$, the anticoincidence time, is identically the average flight time of the electrons.

The quantum degeneracy $\delta$ was defined in CH.1. It was pointed out that for nanoemitters with large spatial coherence the longitudinal coherence and pulse duration take control of setting the value of this quantity (see Fig. 1-4). Let us add a few additional points directly relevant to the present problem.

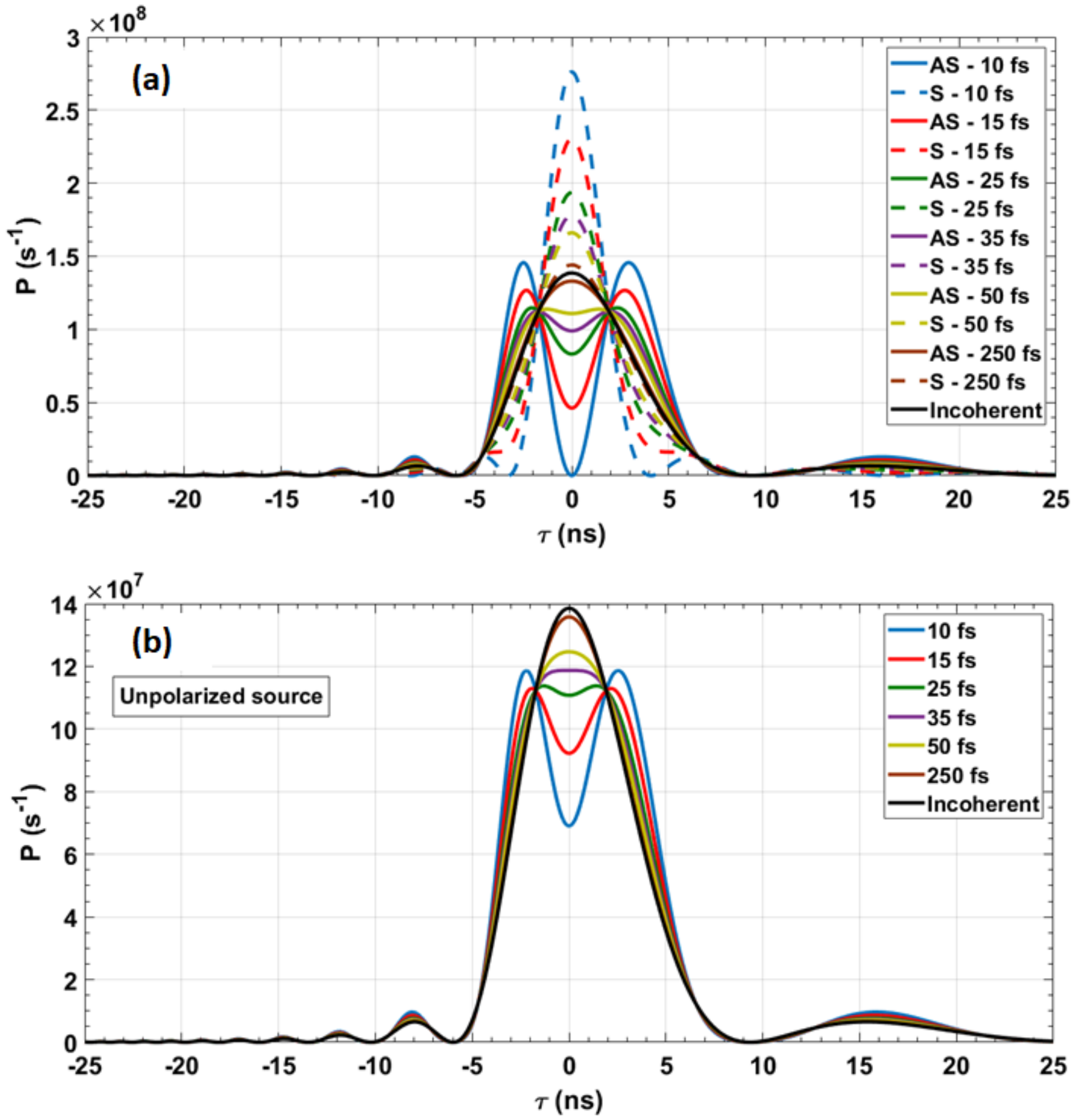

**Fig. 2-5 | Simulation results of the HBT effect for ultra-short pulsed electron beams with arbitrary degrees of coherence.** The two-electron joint detection probability density is plotted as a function of the mutual detection time delay. **(a)** The symmetric *S* and antisymmetric *AS* components of the normalized joint electron detection probability density for several different pulse durations are shown separately. The antisymmetric curves are also the final joint probability distributions when the pulsed source is completely spin-polarized. The incoherent contribution in the far field is also shown. **(b)** For an unpolarized source, the joint probability density is given by a quantum mixture of the symmetric and antisymmetric components in accordance with Eq. (2-20). In general, some other interesting features like the flat-top

probability distribution for the pulse duration of 32 fs in the present case, due to which mutual detections within an absolute delay time of roughly 1.5 ns are all equally probable deserve future scrutiny.

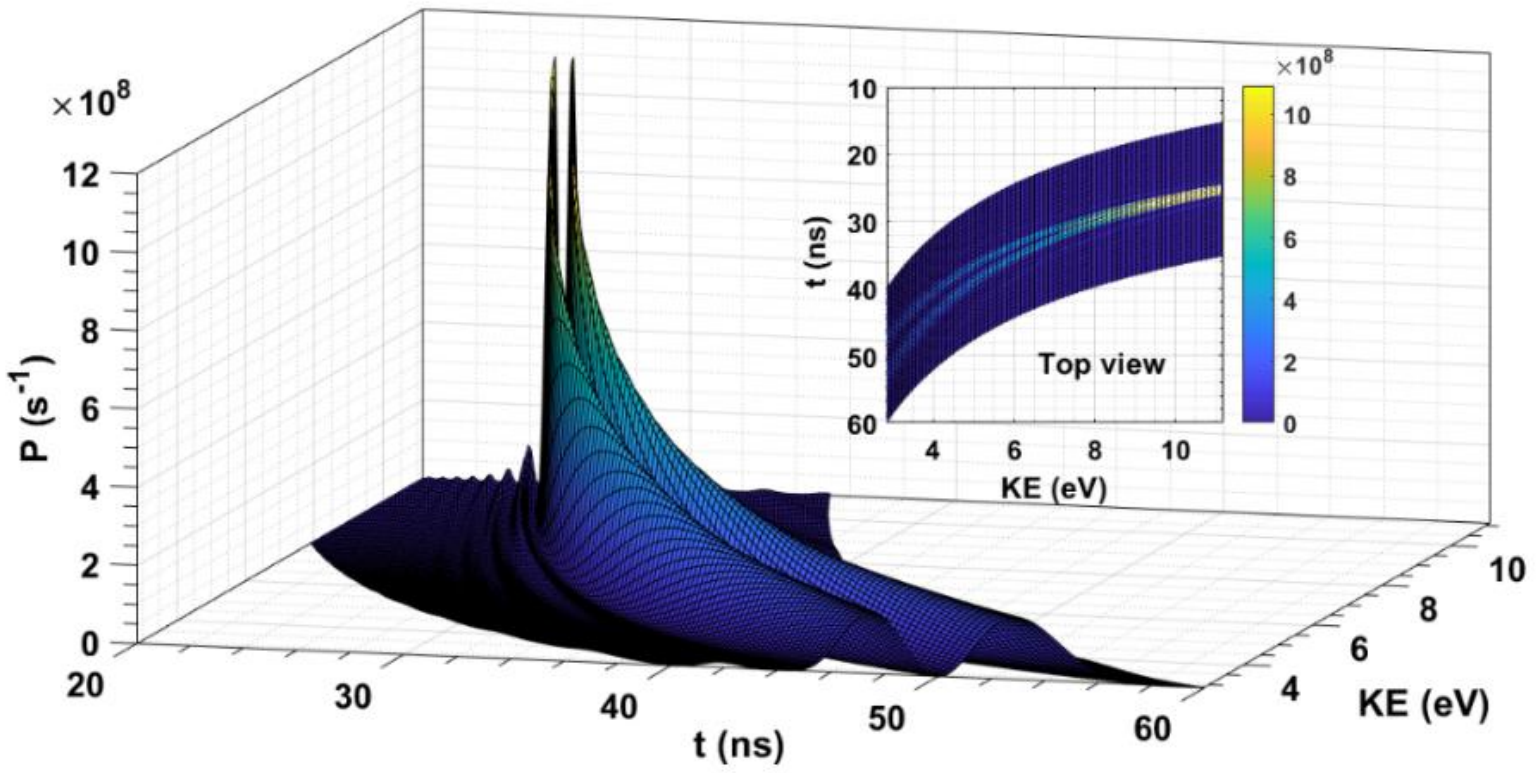

**Fig. 2-6 | Variation of the HBT joint probability density with the average electron kinetic energy.** The degenerate polarized two-electron joint probability density is given as a function of the average electron kinetic energy, $m_e\left(D/T\right)^2/2$, where $D$ is the source-to-detector distance, $T$ is the average flight time of the electrons, and $t = T + \tau$. The curve with the lowest kinetic energy of $2.8$ eV (with $T = 50$ ns) overlaps the solid blue line in Fig. 2-5(a). The HBT dip becomes narrower for progressively increasing values of the kinetic energy (for a fixed distance $D$). The diffraction-in-time broadening of electron pulses helps to push the width of the HBT dip into the nanosecond regime as outlined in CH.1, where detection is straightforward. This is a feature predicted for pulsed electron sources, but not for continuous electron beams. The top view is depicted in the inset for better visibility. As can be noted swiftly, the HBT dip becomes narrower for gradually increasing values of the kinetic energy.

The coherence volume is the space across which the electrons are quantum mechanically correlated and can thus interfere with each other. The size of this volume can be estimated using the energy-time uncertainly relation. The physical volume is determined by the experiment, for example, a cylinder with a radius related to the diameter of the aperture emitting the electrons and a length related to the electron velocity and pulse duration. Assuming cylindrical symmetry which is appropriate for nanotip electron sources with large transverse coherence as compared with, say, flat photocathodes, the number of electrons in the coherence volume can be written as $\delta = \left( nA_c l_c / 2Al \right)$, where $A$ is the cross-sectional area of the beam and $l$ is the length of the volume in question. For most of the present chapter, we limit ourselves to the regime of two electrons per pulse, that is $n = 2$. The coherence volume in this case is therefore set by $l_c / l = T_c / T_{pulse}$, leading to

$$\delta = n \frac{T_c}{2T_{pulse}}. \tag{2-21}$$

This equation thus links the degree of temporal coherence of the pulse $T_c / T_{pulse}$ to the degeneracy. To produce the results shown in Fig. 2-5, as mentioned above, the coherence time $T_c$ is taken to be 10 fs, and the pulse duration $T_{pulse}$ is varied around 50 fs, both consistent with experimentally achieved values. The lack of complete coherence for $T_{pulse} > T_c$, along with the symmetric wavefunction components for an unpolarized electron source, reduce the contrast as demonstrated in Fig. 2-5(b). In general, arbitrary degrees of coherence and polarization can be treated on the same grounds, and according to our results, both a higher degree of quantum coherence and a higher degree of spin polarization in a beam of two-electron pulses enhance the fermionic HBT antibunching signal.

The HBT contrast can be defined as

$$C_{HBT} = \frac{P_{incoh}^0 - P^0}{P_{incoh}^0 + P^0}, \tag{2-22}$$

in which the superscript zero indicates that the functions $P_{incoh}\left(\tau\right)$ and $P\left(\tau\right)$ are evaluated at $\tau = 0$ which coincides with the anticoincidence time $T$ defined earlier. The contrast expressed in this way equals unity when the beam is fully coherent $\left(T_{pulse} = T_c\right)$ and fully spin-polarized. The contrast is zero in the absence of coherence which is the 'classical' realm. Substituting Eq. (2-20) in Eq. (2-22), the HBT contrast for an unpolarized source is derived as

$$C_{HBT}^{unpol} = \frac{P_{incoh}^{0} - \frac{1}{4}P_{coh,S}^{0}}{\left(N-1\right)P_{incoh}^{0} + \frac{1}{4}P_{coh,S}^{0}} = \frac{1}{2N-1}, \tag{2-23}$$

where $N$ was previously given in Eq. (2-18) in terms of the characteristic pulse parameters. In the last step, use has been made of Eq. (2-4) in the time-domain which gives $P_{coh,S}^{0} = 2P_{incoh}^{0}$. For completely polarized two-electron emission there is no symmetric component which subsequently increases the HBT contrast to

$$C_{HBT}^{pol} = \frac{1}{N-1}. \tag{2-24}$$

As an example, consider the special case of a degenerate beam with $T_{pulse} = T_c$. In that case $C_{HBT}^{pol} = 1$ as expected using Eqs. (2-18) & (2-24), whereas for an unpolarized source we get $C_{HBT}^{unpol} = 1/3$ using Eq. (2-23) instead of the latter. This shows a reduction in the contrast by a factor of 3 compared with a similar but polarized beam and identifies a measurable effect in the free electron HBT experiments using controllable ultrafast spin-polarized pulsed electron sources [4,5]. For $T_{pulse} \gg T_c$ (or equivalently $N \to \infty$), which corresponds to long pulse durations approaching the CW limit, the HBT contrast is small and often negligible even though the spin polarization can still make a difference, noting that

$$\lim_{N\to\infty}\frac{C_{HBT}^{pol}}{C_{HBT}^{unpol}}=\lim_{N\to\infty}\frac{2N-1}{N-1}=2. \tag{2-25}$$

The magnitude of the HBT dip is given by the difference between the coincidence rate at $\tau=0$ and the coincidence rate for incoherent illumination of the detectors. This quantity is in fact what was sought after in experiment [3]. The count rate *R* is related to the probability density *P* by $R=P\times t_w\times f$, where $f$ is the repetition rate and $t_w$ the coincidence time window. Starting from Eq. (2-20), the difference in the joint detection probability density at the anticoincidence time is obtained as

$$\Delta P^{unpol}=\frac{2}{N}P_{incoh}^{0}-\frac{1}{2N}P_{coh,S}^{0}=\frac{T_c}{T_{pulse}}\left[P_{incoh}^{0}-\frac{1}{4}P_{coh,S}^{0}\right]=\frac{T_c}{2T_{pulse}}P_{incoh}^{0}, \tag{2-26}$$

for an unpolarized beam, and

$$\Delta P^{pol}=\frac{T_c}{T_{pulse}}P_{incoh}^{0}=2\Delta P^{unpol}. \tag{2-27}$$

For example, for a repetition rate of $f=80$ MHz, a coincidence time window of $t_w=1$ ns (appropriate for the parameters used in Fig. 2-5), an unpolarized electron beam, and a pulse duration of $T_{pulse}=50$ fs, the difference in the count rate would be $\Delta R^{unpol}=\Delta P^{unpol}\times t_w\times f\sim 1.5\times 10^{6}$ cps. This is, however, unrealistic as we have to take into account that not every pulse will be a two-electron pulse.

We assume a Poissonian probability distribution for the number of electrons in each pulse as outlined in CH.1, Eq. (1-33). We define $\lambda=\langle R_e\rangle/f$, where $R_e$ is the electron emission rate, for the integer number of emitted electrons found in each pulse, *n*. For example, for typical mid-range emission rates of $10^6$ electrons per second [60,61], one finds $\lambda \sim 0.01$. For $n = 2$ this gives $\Delta R^{unpol} \sim 100$ cps. For a quadrupole setting that gives an expansion of the solid angle of a factor of 100, a more realistic count rate of $\Delta R^{unpol} \sim 1$ cps is found [9]. For the CW experiment of Ref. [3], the difference in the coincidence counts was $\sim 1000$ for a measuring time of $T_{acq} = 30h \sim 10^5 s$, and a coincidence time window of $t_w = 1$ ps (for each channel detected).

| $T_{\text{pulse}}$ (fs) | $C_{\text{HBT}}^{\text{pol}}$ | $C_{\text{HBT}}^{\text{unpol}}$ | $C_{\text{HBT}}^{\text{pol}} / C_{\text{HBT}}^{\text{unpol}}$ |
|---|---|---|---|
| **10** | 1.0000 | 0.3350 | 2.985 |
| **15** | 0.5000 | 0.2009 | 2.488 |
| **25** | 0.2500 | 0.1116 | 2.240 |
| **35** | 0.1667 | 0.0772 | 2.159 |
| **50** | 0.1111 | 0.0528 | 2.104 |
| **250** | 0.0204 | 0.0101 | 2.019 |

**Table 2-1 | The calculated values of the HBT contrast for different pulse durations.** The contrast is given for the simulation results of Fig. 2-5 as a function of the electron pulse duration. The coherence time is 10 fs. The marginal deviation of $C_{HBT}^{unpol}$ from $1/3$ for a degenerate pulse is due to using the far-field approximation. For (an) a (un)polarized electron pulse the contrast is given in the second (third) column. The fourth column shows that spin polarization gradually becomes less significant in improving the contrast as the pulse duration grows towards the continuous beam regime. The ratio approaches the value derived in Eq. (2-25) for long pulses.

The HBT contrasts corresponding to the joint detection probability distributions, including the polarization dependence illustrated in Fig. 2-5, are collected and shown in Table 2-1.

### 2.1.3 Partially coherent multi-electron pulses

We can generalize our model to include pulses with more than two electrons in them. Here we reserve the notation $T_{pulse}$ for the shortest time interval which includes two electrons on average and denote the actual electron pulse duration with $\Delta T_e$. The maximum number of electrons in such a pulse is given by

$$\eta_{\max} = 2\frac{\Delta T_e}{T_c}, \tag{2-28}$$

which satisfies $T_{pulse} = T_c$. For $2 \le \eta \le \eta_{\max}$,

$$T_{pulse} = \left(\frac{\eta_{\max}}{\eta}\right) T_c = \frac{2\Delta T_e}{\eta}, \tag{2-29}$$

where Eq. (2-28) is used in the last step. This modifies the expression for the number of intervals $N$, formerly given in Eq. (2-18) for two-electron pulses, as

$$N = \frac{2T_{pulse}}{T_c} = \frac{4\Delta T_e}{\eta T_c}. \tag{2-30}$$

Assuming a Poissonian probability distribution for the number of electrons in each pulse, the average reduction in the joint detection probability density at the anticoincidence time, using Eqs. (2-26) & (2-30), is obtained as

$$\left\langle \Delta P^{unpol} \right\rangle = \sum_{\eta=2}^{\eta_{\max}} \left[ \frac{\eta}{\eta_{\max}} \times \frac{P_{incoh}^{0}}{2} \times P_{Poisson}(\eta) \right]. \tag{2-31}$$

### 2.1.4 Double-slit diffraction with two-electron pulses: What to expect?

In the free electron HBT effect, the antibunching signal is created by quantum mechanically correlated electrons whose composite pure state is spatially antisymmetric in compliance with the spin statistics theorem. On the other hand, the double-slit interference effect is a familiar single-particle – first-order – quantum interference. Can we predict the probability density on the screen in a double-slit experiment with quantum degenerate, or more realistically partially quantum coherent two-electron pulses currently available? Will it be an interplay between the first- and higher-order interference effects, or will one interference order dominate when the experiment is performed in practice? And why?

Where the main constructive interference fringe is observed for the double-slit setup as shown earlier in Fig. 2-3(d), an HBT dip occurs under similar experimental conditions as in Fig. 2-5. The former is the single-particle detection probability density at a certain time, while the latter, is that of the joint detection of an electron pair at the same time. Fig. 2-7 compares the (two-electron) fermionic HBT effect with that of the (one-electron) double-slit experiment under identical experimental conditions. For electrons outside the mutual coherence volume the situation is resembled by two independent shutters whose individual diffraction patterns contribute incoherently to the final effect. Two electrons in a single coherence volume then contribute to the HBT dip. In our treatment of free electron pulses scrutinized in this chapter so far, no double-slits cross the beam path. The only effects present are therefore incoherent contributions mimicking the single-slit single-electron diffraction pattern, and the antibunching signal due to coherent electron pairs. One may thus envision spatial and temporal double-slit experiments with degenerate pulsed electron beams manifesting an interplay between coherent single-particle interferences and the two-electron HBT effect. The incoherent electrons in our model will then undergo double-slit interference with maximum peak heights of up to twice the incoherent diffraction pattern at the HBT anticoincidence time which will reduce the antibunching signal even further. From the perspective of the single-electron interferences, however, it is the very presence of two-fermion correlation that diminishes the first-order fringe visibility. The question begging an answer is thenceforth whether the double-slit will decohere the mutually coherent electron pairs or not. If it does it with no exceptions, no HBT signal shall be observed. Any hint of antibunching will thus indicate that some coherent pairs of electrons are not affected by the double-slit. Another possibility is the coherent electron pair in each ultrashort pulse act as a single boson with *twice* the mass of one electron undergoing a double-slit arrangement and giving rise to single-particle interference patterns while the uncorrelated electrons in each such pulse seek to generate single-electron interference patterns which are broader by a factor of two. Fig. 2-8 demonstrates aspects of this latter conjecture.

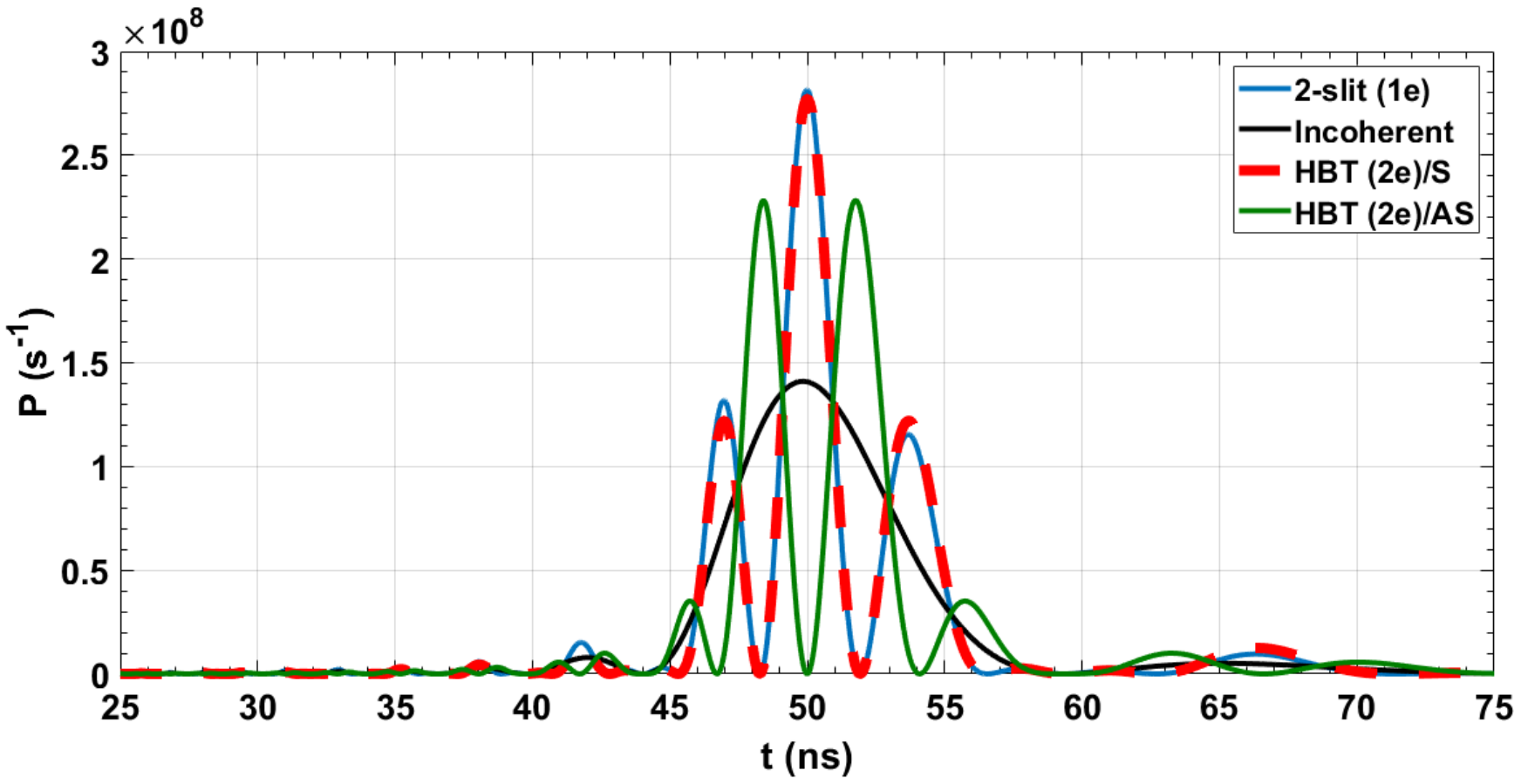

**Fig. 2-7 | One-electron probability contrasted with two-electron probability.** A comparison is made between the single-electron probability density in a double-slit experiment and the symmetric and antisymmetric components of the two-electron joint detection probability density in the fermionic HBT effect for the same experimental conditions. The numerical values of the relevant parameters are identical to those which led to Fig. 2-3(d) and Fig. 2-5(a).

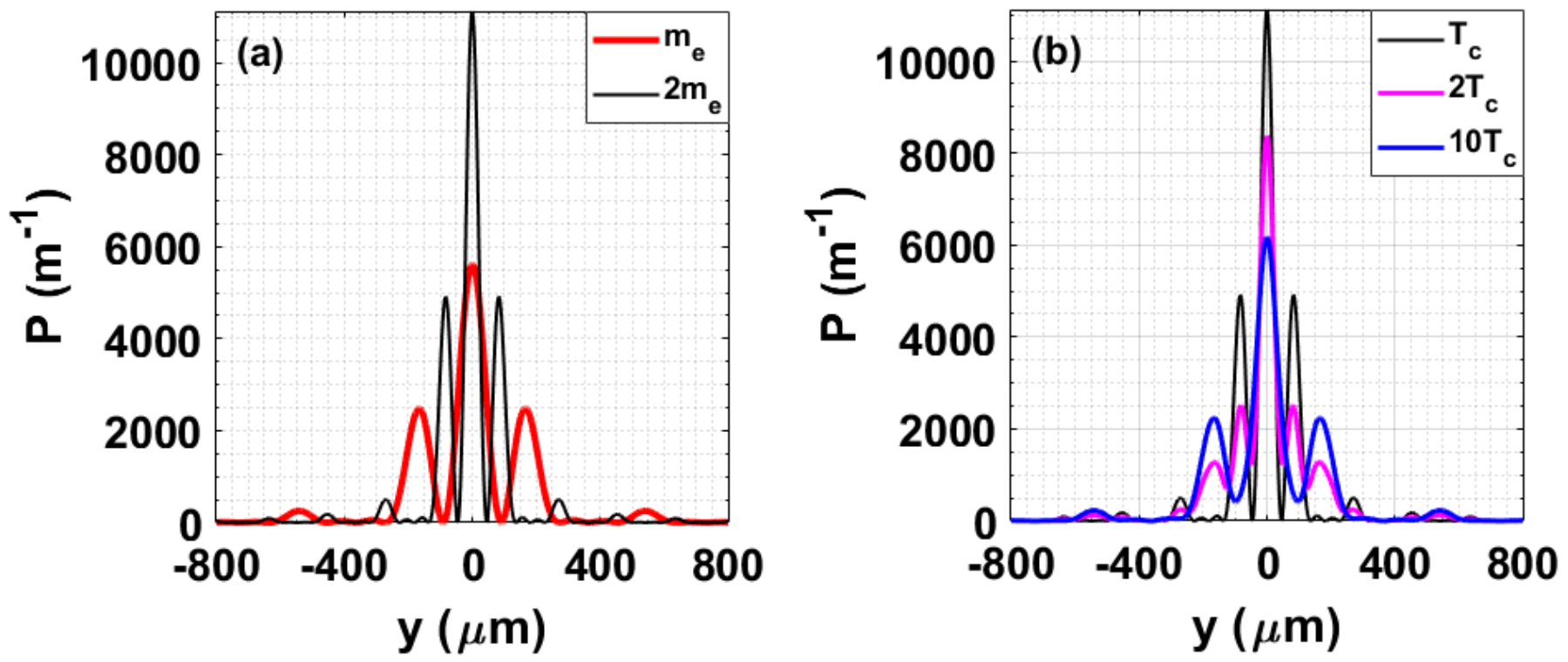

**Fig. 2-8 | A conjecture: What if the entangled electron pair acted as a single twice-as-massive boson in the double-slit experiment?** One of the conjectures discussed in the present section is demonstrated in this figure. The physical parameters are identical to those of Fig. 2-7. As elaborated on in the main text, two-electron pulses propagating in free space will lead to HBT dips with various degrees of contrast depending on their degrees of coherence. Now what happens if a spatial double-slit is placed in the path of the two-electron pulsed beam? One possibility is that when the two electrons are within one coherence time they act as a single boson with twice the mass of one electron while two uncorrelated electrons will each contribute to the regular single-electron interference pattern. The two contributions are compared in panel **(a)**. Keeping up with this conjecture and as shown in panel **(b)** for three different pulse durations, for a completely coherent pair, the bosonic pattern is retrieved, whereas for partially coherent beams the single-electron interference pattern will tend to progressively dominate as the degree of coherence diminishes with increasing the ratio between the two-electron pulse duration and the coherence time. Not overviewed in this figure is the other conjecture discussed in the text according to which the fermionic HBT effect would persist to show up despite the presence of a double-slit in the path of the beam.

### 2.1.5 Temporally point-like sources and matter-wave beating in time

It is instructive to compare the main realistic results for finite temporal slits which we have discussed so far with those of the ideal temporally point-like source points at times *0* and $t_s$. The rest of the parameters are left unchanged. We will first make the comparison computationally, and then analytically. It will be shown that a beat-note-like pattern in time that was implicit in the presented results becomes more manifest.

A comparison between the finite-slit and point-source simulation results is given in Fig. 2-9. An interesting feature is that the effect of the symmetrization type seems to be a phase shift of π in the two-particle joint probability which was not possible to grasp from the wide-source results due to their decaying envelope. For the incoherent case, the energy of the two electrons is completely uncertain in accordance with the energy-time uncertainty relation which gives an identical detection probability density across the entire detection window.

Let us derive the observed phase shift of π in the computational results analytically. The final single-electron states can be written down using Eqs. (2-10) & (2-13) for the pair of point-sources in time under consideration. Since we are interested in calculating the phase, we will write the wavefunctions up to a normalization factor for brevity without any loss of generality. We use the same notation in labeling the wavefunctions as before. The four propagated wavepackets are therefore,

$$
\begin{aligned}
\varphi_{11}^{f} &= \exp\left[\frac{im_e D^2}{2\hbar T}\right],\\
\varphi_{12}^{f} &= \exp\left[\frac{im_e D^2}{2\hbar\left(T+\tau\right)}\right],\\
\varphi_{21}^{f} &= \exp\left[\frac{-im_e D^2 t_s}{2\hbar T^2}\right]\exp\left[\frac{im_e D^2}{2\hbar\left(T-t_s\right)}\right],\\
\varphi_{22}^{f} &= \exp\left[\frac{-im_e D^2 t_s}{2\hbar\left(T+\tau\right)^2}\right]\exp\left[\frac{im_e D^2}{2\hbar\left(T+\tau-t_s\right)}\right],
\end{aligned}
\tag{2-32}
$$

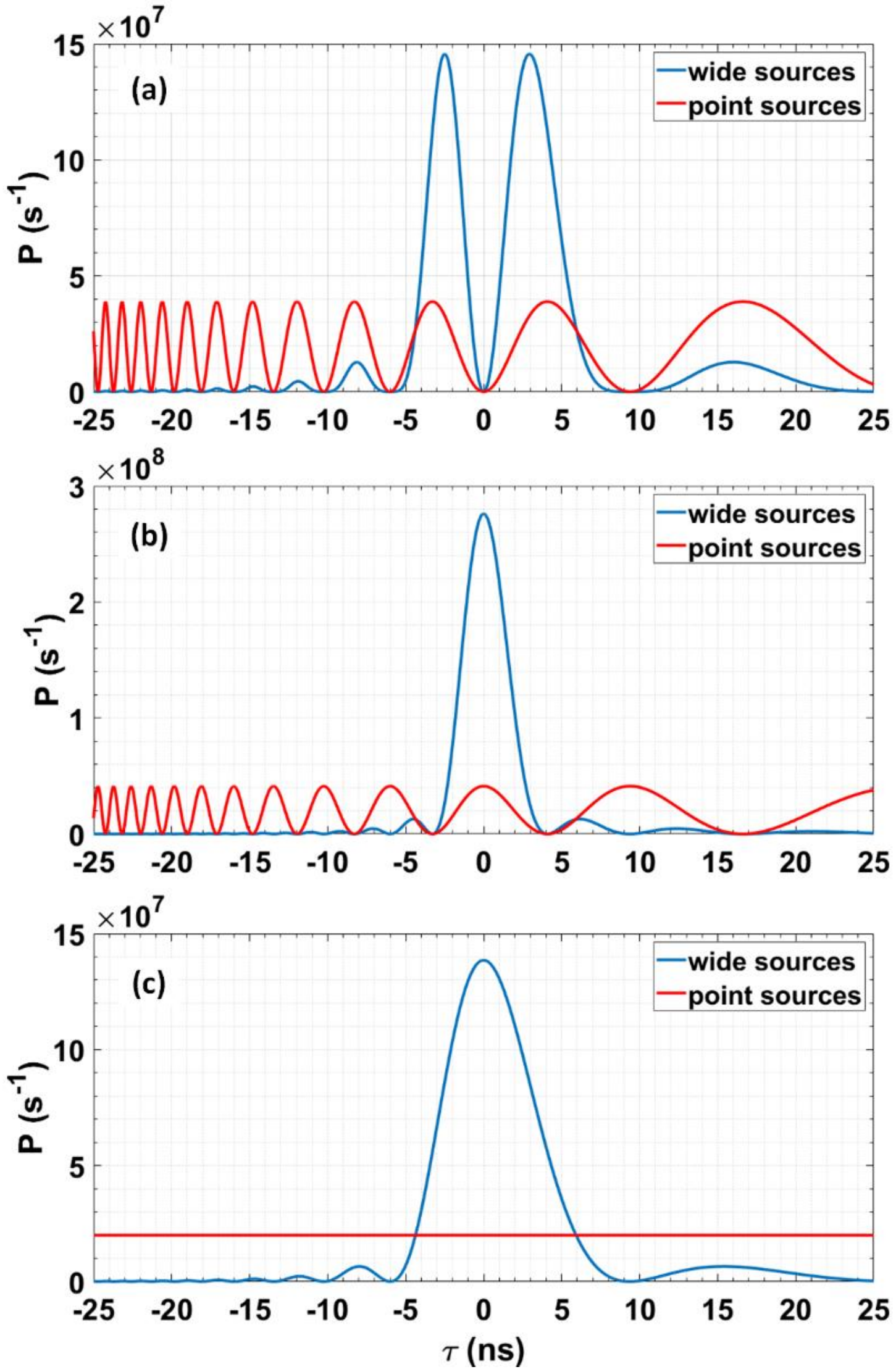

**Fig. 2-9 | Comparison between the finite-slit and temporal point-source results. (a)** The normalized antisymmetric mutual probability density for the pair of finite coherent slits (pulses) is contrasted with its corresponding normalized point-source pattern with the source pair at time 0 and 5 fs while the wide slits are extended from [0,5fs] and [5fs,10fs] as before. **(b)** The symmetric distributions are juxtaposed. **(c)** The incoherent pattern, resembling that of the single-electron diffraction in time for the

finite pulses is compared with the (box-) normalized probability density for the pair of point-sources. As expected, the detection probability density is constant everywhere noting to the fact that the electron energy is completely uncertain in this case as commanded by the energy-time uncertainty relation.

---

and the final two-electron wavefunction is

$$\psi_f^{AS} = \varphi_{11}^f \varphi_{22}^f - \varphi_{12}^f \varphi_{21}^f. \tag{2-33}$$

Combining Eqs. (2-32) & (2-33), we arrive at the expression

$$\begin{aligned}\psi_f^{AS} &= \exp\left[\frac{im_e D^2}{2\hbar T}\right]\exp\left[\frac{-im_e D^2 t_s}{2\hbar\left(T+\tau\right)}\right]\exp\left[\frac{im_e D^2}{2\hbar\left(T+\tau-t_s\right)}\right] \\ &\quad -\exp\left[\frac{im_e D^2}{2\hbar\left(T+\tau\right)}\right]\exp\left[\frac{-im_e D^2 t_s}{2\hbar T}\right]\exp\left[\frac{im_e D^2}{2\hbar\left(T-t_s\right)}\right] \\ &\equiv \exp\left(i\theta_1\right) - \exp\left(i\theta_2\right),\end{aligned} \tag{2-34}$$

in which the phases denoted with $\theta_1$ and $\theta_2$ are defined as

$$\begin{aligned}\theta_1 &\equiv \frac{m_e D^2}{2\hbar}\left[\frac{1}{T} - \frac{t}{\left(T+\tau\right)^2} + \frac{1}{T+\tau-t_s}\right], \\ \theta_2 &\equiv \frac{m_e D^2}{2\hbar}\left[\frac{1}{T+\tau} - \frac{t}{T^2} + \frac{1}{T-t_s}\right].\end{aligned} \tag{2-35}$$

With these definitions, the antisymmetric probability density function, up to a normalization factor, becomes

$$P_f^{AS} = \left|\psi_f^{AS}\right|^2 = 4\sin^2\left(\frac{\theta_1-\theta_2}{2}\right). \tag{2-36}$$

Similarly,

$$P_f^S = \left|\psi_f^S\right|^2 = 4\cos^2\left(\frac{\theta_1 - \theta_2}{2}\right), \tag{2-37}$$

is derived as the symmetric joint probability density function. This proves that the phase difference is indeed equal to π.

Noting to the fact that in the far-field approximation, $t_s \ll T$, we estimate

$$\begin{aligned}\frac{1}{T - t_s} &\approx \frac{1}{T},\\ \frac{1}{T + \tau - t_s} &\approx \frac{1}{T + \tau}.\end{aligned} \tag{2-38}$$

This therefore enables us to write down the phase difference as

$$\Delta\phi \equiv \frac{\theta_1 - \theta_2}{2} \approx \frac{m_e D^2}{4\hbar}\left[\frac{t_s}{T^2} - \frac{t_s}{\left(T + \tau\right)^2}\right], \tag{2-39}$$

which can be plugged into Eqs. (2-36) & (2-37) to obtain the desired expressions for the antisymmetric and symmetric joint probability densities, respectively, up to a normalization factor. At the anticoincidence time corresponding to $\tau = 0$, the phase $\Delta\phi = 0$, which is an exact expression as can be seen upon direct evaluation of Eq. (2-35) at $\tau = 0$. The periodicity $\tau_0$ of the two-electron wavepackets which is the characteristic time of the problem is also readily calculated from the condition

$$\frac{m_e D^2}{4\hbar}\left[\frac{t_s}{T^2} - \frac{t_s}{\left(T + \tau_0\right)^2}\right] = \pi. \tag{2-40}$$

On the right-hand-side, only the positive sign is physical noting to the fact that the left-hand-side is positive-definite for nonvanishing $\tau_0$. This leads to

$$\tau_0 = -T + \sqrt{\frac{1}{\frac{1}{T^2} - \frac{4\pi\hbar}{m_e D^2 t_s}}}. \tag{2-41}$$

The positive sign on the right-hand-side ensures $\tau_0$ is positive. The expressions obtained in Eqs. (2-36), (2-37), and (2-39) are plotted and shown in Fig. 2-10 (a-c). The matter wave beat-note pattern in Fig. 2-10(d) is constructed by introducing a short time delay $\delta T$ between the average flight times of the spin-triplet and -singlet electron pairs. While in practice this is a challenging task to realize, it is instructive to demonstrate the possibility of creating such temporally beating joint probability density distributions. As can be seen in this panel, for a 1-ns delay, a pair of distinct envelopes in the blue beat-note pattern appear for shorter detection time delays. It should be noted that the effect is visible as we are considering temporally point-like electron sources which approximate nanotip photo-emitters. For wide sources the effect would be obscured (see Fig. 2-9). For an spin-unpolarized beam no beat-note pattern would be observed (black line in Fig. 2-10 (c,d)).

It is also instructive to compare the one- and two-electron probabilities just as we did earlier in Fig. 2-7 for wide sources. The single-electron double-slit pattern for a pair of point-like sources is plotted along with its corresponding antisymmetric two-electron probability density in Fig. 2-11. Inspecting the final two-electron wavefunctions of Eq. (2-32) and the time-dependent normalization factor from the PI kernel previously given in Eq. (2-10), it is straightforward to show that neglecting the short time duration of the point-like sources at *0* and $t_s$ compared with the detection time window, for the two-electron HBT case, the probability density is modulated by a factor $1/Tt$, where $t$ is the (final) detection time, while in the single-electron double-slit setting, in which the pattern is formed

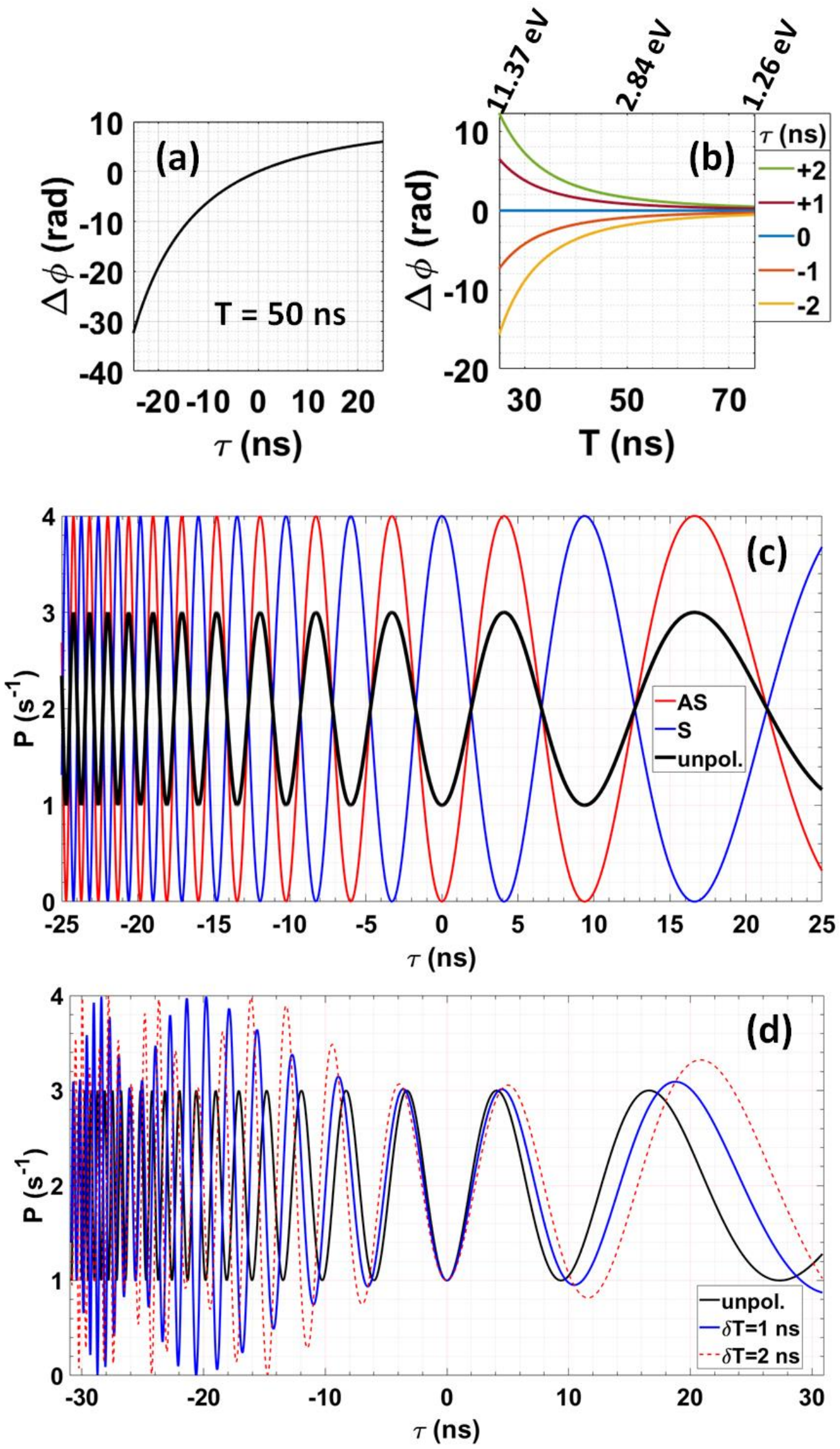

**Fig. 2-10 | Matter-wave beating-in-time. (a)** The phase difference $\Delta\phi$ in radians defined in Eq. (2-39) is plotted as a function of the mutual detection time delay. As before, the anticoincidence time is T = 50 ns, and D = 5 cm. The pair of point sources are separated by $t_s$ = 5 fs. **(b)** The same phase difference is plotted as a function of the anticoincidence time (corresponding

to different kinetic energies shown in eV) at five different values of the delay time $\tau$. As discussed in the main text, the phase difference is zero at the anticoincidence time which is an exact result. Larger phase differences at higher kinetic energies correspond to more rapid oscillations, while at lower kinetic energies the oscillations in time slow down, making it more favorable for observation purposes (see Fig. 2-6). **(c)** The symmetric and antisymmetric normalized joint detection probability densities corresponding to the phase difference depicted in panel (a) are shown and compared. Clearly, the two oscillations are out of phase by a half cycle. In general, the progressively decreasing slope of $\Delta\phi$ is a consequence of the gradually increasing de Broglie wavelengths for lower momenta coupled with delayed arrival times. The effect therefore is a consequence of *interfering alternatives* with a continuum of de Broglie wavelengths. The black line is the corresponding joint detection probability density mixture for an unpolarized source. **(d)** To produce a matter-wave beat-note-in-time distribution, a short time delay $\delta T$ between the average flight times of the triplet and singlet pairs is introduced. The beating pattern is manifest.

---

by the superposition of the single-electron amplitudes $\varphi_{11}^{f}$ and $\varphi_{21}^{f}$, the probability density decays as $1/t$. In the former, $1/Tt \approx 1/T^2$, which is ultimately absorbed in the final box-normalized probability density function.

### 2.1.6 Error analysis

In the computer simulations, the probability density distribution for the incoherent cases $(i < j, j \neq i+1)$, for $i = 1, 2, ..., N-2$ are all approximated by the same diffraction pattern. This is justified by the far-field approximation, because the source time difference for different values of $j$ is very small compared to the time-of-flight. To estimate the error that is introduced by this approximation, a sample of joint detection probability distributions is calculated with and without the approximation. Let us consider the case $N = 3$. The joint probability is given by

$$P(\tau)=\frac{1}{3}P_{coh}^{1,2}(\tau)+\frac{1}{3}P_{coh}^{2,3}(\tau)+\frac{1}{3}P_{incoh}^{1,3}(\tau), \tag{2-42}$$

where the superscripts denote the interval numbers $i$ and $j$, respectively. Eq. (2-42) is approximated with

$$P(\tau)=\frac{2}{3}P_{coh}^{1,2}(\tau)+\frac{1}{3}P_{incoh}^{1,3}(\tau), \tag{2-43}$$

as in the far-field approximation, $P_{coh}^{2,3}(\tau)\approx P_{coh}^{1,2}(\tau)$. Similarly, for $N>3$ we use the same approximation for the coherent terms, $(i<j,\, j=i+1)$, for $i=1,2,...,N-1$. The incoherent terms are approximated with $P_{incoh}^{1,3}(\tau)$ as above. In all such cases,

$$P_{coh,incoh}^{i,j}(\tau)=P_{coh,incoh}^{1,2or3}(\tau)+e_{coh,incoh}^{i,j}, \tag{2-44}$$

where $e_{coh,incoh}^{i,j}=\max\left(e_{coh,incoh}^{i,j}(\tau)\right)$ is the upper limit of the error. Table 2-2 shows the relative errors in the maximum value of $P_{incoh}(\tau)$ for simulation results in a number of examples using parameters of Fig. (2-5).

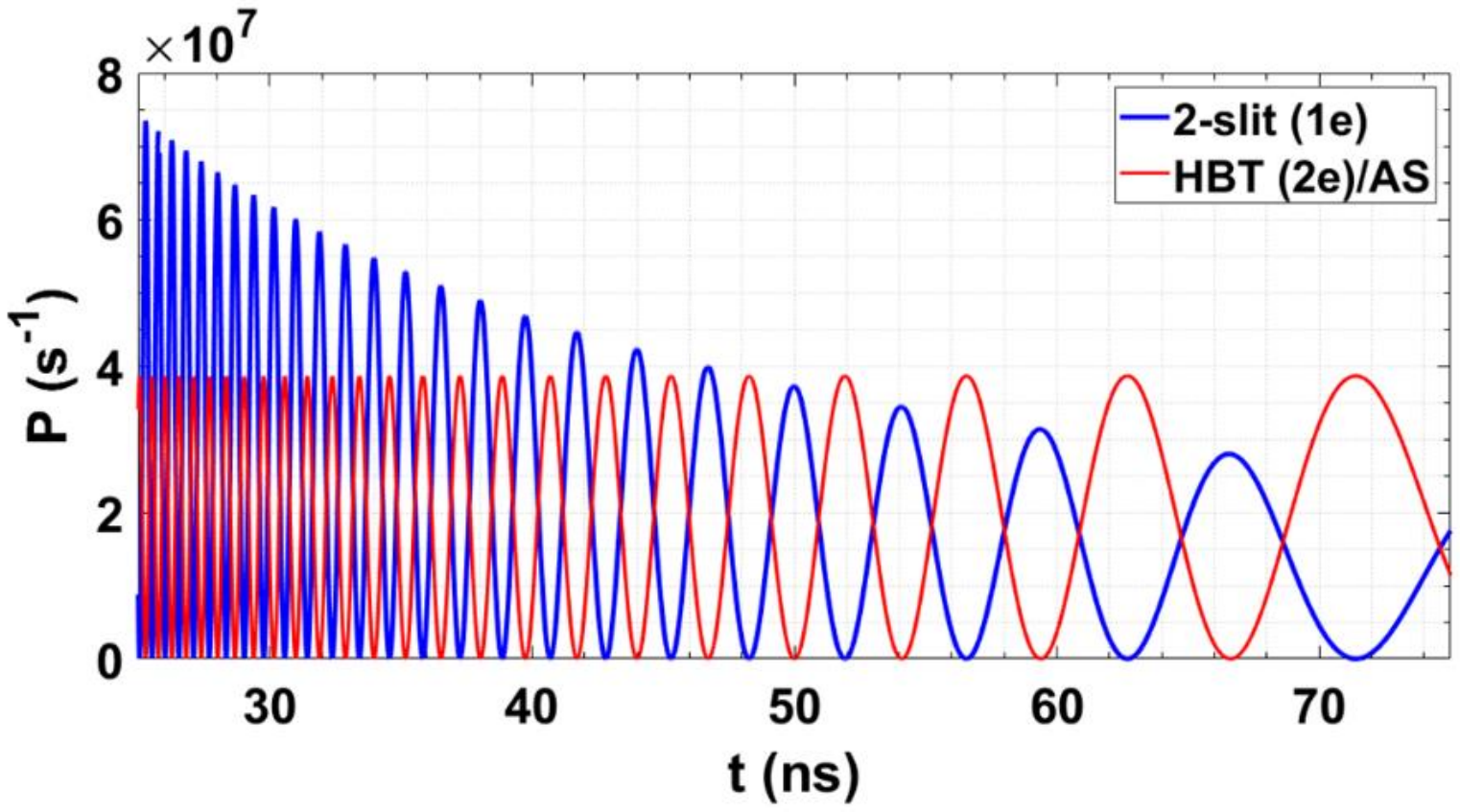

**Fig. 2-11 | Comparison between the point-source double-slit pattern and the antisymmetric HBT probability density.** For both cases, the pair of point sources are at $t_s$ = 0 and $t_s$ = 10 fs. As before, D = 5 cm, and T = 50 ns for the HBT effect.

| i, j | Relative error |
|---|---|
| **1,3** | 0 |
| **1,10** | $1.4\times10^{-6}$ |
| **8,10** | $2.1\times10^{-6}$ |
| **1,50** | $7.2\times10^{-6}$ |
| **1,100** | $1.5\times10^{-5}$ |
| **98,100** | $3.0\times10^{-5}$ |

**Table 2-2 | Computational errors are negligible.** Examples of the error in the maximum value of $P_{incoh}(\tau)$ relative to the case used as a valid approximate.

## 2.2 Partial quantum coherence: A formal approach based on the quantum decoherence theory

The initial partially coherent state described in Fig. 2-1 is given in the density matrix formalism. It is possible to relate this state, in principle, to quantum decoherence theory [36,42-45], as we will illustrate below with one example state constructed for that purpose. The basic idea is to write an example state that is fully coherent, where, upon tracing over all the unobserved terms, we recover the partially coherent state that we used as our initial state, thereby justifying the claim that our approach is consistent with the decoherence theory.

For the example state, it is sufficient to consider a pure entangled state with well-defined symmetries for the two electrons on the source, which is itself entangled with a two-level environmental state, as required by the PEP. The entangled three-particle pure state is described by the density operator

$$\rho_{12E} = |\psi_{12E}\rangle\langle\psi_{12E}|, \tag{2-45}$$

where the subscripts denote the particle exchange number of the two electrons and the *environmental* particle $E$ . The partially coherent two-electron state is found by tracing out the environmental states,

$$\rho_{12} = Tr_E \rho_{12E}. \tag{2-46}$$

To obtain the spatial state involved in the HBT effect, we trace over the unobserved the two-particle spin states $\{|\uparrow\downarrow\rangle, |\downarrow\uparrow\rangle\}$ ,

$$\bar{\rho}_{12} = Tr_s \rho_{12}. \tag{2-47}$$

In our simulations of the HBT effect this is an example of the type of state that is used as our initial state. The electron spin-up (-down) is denoted with $\uparrow$ ( $\downarrow$ ). The orbital states, $|t_a\rangle$ , $|t_b\rangle$ , and $|t_c\rangle$ are three single-electron orbitals at the source corresponding to $N = 3$ intervals (see Fig. 2-1). Unless the tensor-product sign $\otimes$ is explicitly used, factorizable states such as $|t_a\rangle \otimes |t_b\rangle$ are simply written as $|t_a t_b\rangle$ . In all such cases, the particle exchange number is 1 (2) for the left (right) eigenvalue.

Let us now consider the following normalized entangled state:

$$\begin{aligned} |\psi_{12E}\rangle = \frac{1}{4}\Big\{ & \Big[\big(|t_a t_b\rangle + |t_b t_a\rangle\big) \otimes \big(|\uparrow\downarrow\rangle - |\downarrow\uparrow\rangle\big) \otimes |g_E\rangle\Big] \\ & + \Big[\big(|t_b t_c\rangle + |t_c t_b\rangle\big) \otimes \big(|\uparrow\downarrow\rangle - |\downarrow\uparrow\rangle\big) \otimes |g_E\rangle\Big] \\ & + \Big[\big(|t_a t_c\rangle + |t_c t_a\rangle\big) \otimes \big(|\uparrow\downarrow\rangle - |\downarrow\uparrow\rangle\big) \otimes |g_E\rangle\Big] \\ & + \Big[\big(|t_a t_c\rangle - |t_c t_a\rangle\big) \otimes \big(|\uparrow\downarrow\rangle + |\downarrow\uparrow\rangle\big) \otimes |h_E\rangle\Big]\Big\}, \end{aligned} \tag{2-48}$$

which is a pure composite state, antisymmetric under the exchange of the two electrons. The kets $\left|h_E\right\rangle$ and $\left|g_E\right\rangle$ indicate the two basis (pointer) states of the environmental particle. We represent $\rho_{12}$, given by Eq. (2-46), on the basis states

$$\left\{\begin{matrix} \left|t_a t_b;\uparrow\downarrow\right\rangle, \left|t_a t_b;\downarrow\uparrow\right\rangle, \left|t_b t_a;\uparrow\downarrow\right\rangle, \left|t_b t_a;\downarrow\uparrow\right\rangle, \left|t_a t_c;\uparrow\downarrow\right\rangle, \left|t_a t_c;\downarrow\uparrow\right\rangle, \\ \left|t_c t_a;\uparrow\downarrow\right\rangle, \left|t_c t_a;\downarrow\uparrow\right\rangle, \left|t_b t_c;\uparrow\downarrow\right\rangle, \left|t_b t_c;\downarrow\uparrow\right\rangle, \left|t_c t_b;\uparrow\downarrow\right\rangle, \left|t_c t_b;\downarrow\uparrow\right\rangle \end{matrix}\right\}, \tag{2-49}$$

in order from left to right (top to bottom) for each row (column) of the matrix. Consequently, we obtain

$$\rho_{12} \doteq \frac{1}{16}\begin{pmatrix} 1 & -1 & 1 & -1 & 1 & -1 & 1 & -1 & 1 & -1 & 1 & -1 \\ -1 & 1 & -1 & 1 & -1 & 1 & -1 & 1 & -1 & 1 & -1 & 1 \\ 1 & -1 & 1 & -1 & 1 & -1 & 1 & -1 & 1 & -1 & 1 & -1 \\ -1 & 1 & -1 & 1 & -1 & 1 & -1 & 1 & -1 & 1 & -1 & 1 \\ 1 & -1 & 1 & -1 & 2 & 0 & 0 & -2 & 1 & -1 & 1 & -1 \\ -1 & 1 & -1 & 1 & 0 & 2 & -2 & 0 & -1 & 1 & -1 & 1 \\ 1 & -1 & 1 & -1 & 0 & -2 & 2 & 0 & 1 & -1 & 1 & -1 \\ -1 & 1 & -1 & 1 & -2 & 0 & 0 & 2 & -1 & 1 & -1 & 1 \\ 1 & -1 & 1 & -1 & 1 & -1 & 1 & -1 & 1 & -1 & 1 & -1 \\ -1 & 1 & -1 & 1 & -1 & 1 & -1 & 1 & -1 & 1 & -1 & 1 \\ 1 & -1 & 1 & -1 & 1 & -1 & 1 & -1 & 1 & -1 & 1 & -1 \\ -1 & 1 & -1 & 1 & -1 & 1 & -1 & 1 & -1 & 1 & -1 & 1 \end{pmatrix}. \tag{2-50}$$

The dot in $\doteq$ indicates that the right-hand side is representation-dependent. Using Eq. (2-47), the spin-averaged two-electron density matrix represented on the set of basis states given by

$$\left\{\left|t_a t_b\right\rangle, \left|t_b t_a\right\rangle, \left|t_a t_c\right\rangle, \left|t_c t_a\right\rangle, \left|t_b t_c\right\rangle, \left|t_c t_b\right\rangle\right\}, \tag{2-51}$$

is finally computed as

$$\bar{\rho}_{12} \doteq \frac{1}{8}\begin{pmatrix} 1 & 1 & 1 & 1 & 1 & 1 \\ 1 & 1 & 1 & 1 & 1 & 1 \\ 1 & 1 & 2 & 0 & 1 & 1 \\ 1 & 1 & 0 & 2 & 1 & 1 \\ 1 & 1 & 1 & 1 & 1 & 1 \\ 1 & 1 & 1 & 1 & 1 & 1 \end{pmatrix}. \qquad (2\text{-}52)$$

Clearly, the sub-space $\{|t_a t_c\rangle, |t_c t_a\rangle\}$ has lost its coherence in this process while both $\{|t_a t_b\rangle, |t_b t_a\rangle\}$ and $\{|t_b t_c\rangle, |t_c t_b\rangle\}$ sub-spaces are remained coherent which is what we desired for. The generalization to longer pulses with $N > 3$ is straightforward. Lastly, the reader must note that the example state put forth in Eq. (2-48) is not the only possible entangled state to serve our purpose. It is, in fact, possible to write down other distinguishable states – consisting of an equal or more terms – which could still support the density matrix in Eq. (2-52). Here, the goal was to explore how partial quantum coherence could *formally* be modeled following considerations of the quantum decoherence theory.

# CHAPTER 3: VERIFICATION OF PARTICLE STATISTICS IN ULTRASHORT PULSED ELECTRON BEAMS USING THE COINCIDENCE TECHNIQUE

In the literature on ultrashort laser-driven free electron sources, it is often – and at least to my knowledge always – assumed that the electron emission is random, that is, the number of electrons emitted per incident laser pulse obeys the Poissonian distribution. Moreover, randomness of the photoelectrons distribution on the detection plane, specifically at low emission rates, is presumed. Taking advantage of our double-detector coincidence detection setup, we aimed to verify or refute the validity of this claim for an electro-chemically etched tungsten nanotip needle source, to our knowledge for the first time. Additionally, this has helped us to identify the background signal and noise effects in our upcoming free electron HBT experiment after addition of a quadrupole lens to the setup in a newly designed vacuum chamber. We have also developed a statistical machinery for the purpose of analyzing the coincidence detection spectra using both analytical and computational considerations. The Poissonian probability mass function given earlier in Eq. (1-33) and its ensuing discussion in CH.1 will be at the core of the data analysis of the present chapter.

In the upcoming sections, we will first describe our experimental setup. Next, our experimental results will be presented and discussed in detail. An analytical statistical formulation of the realistic problem, that involves the possibility of the production of multi-electron pulses and finite detection probabilities, will demonstrate that the photoelectron statistics of the tungsten nanotip needle source on the detection plane is in fact sub-Poissonian. The outcome of the analytical analysis will subsequently be put to meticulous scrutiny through combinatoric statistical considerations in trying to better illustrate the problem. Prior to that, the supporting results of a computer simulation carried out using SIMION™ and performed by undergraduate student Will Brunner from Prof. T. J. Gay's group will also be given and discussed. The output raw data of the SIMION™ simulations which constitute the detection times at two detectors were subsequently visualized using histograms of electron arrival times by the author of the present dissertation. A short overview of our newly-designed vacuum chamber, which will host the follow-up free electron HBT experiments after incorporating a diverging electrostatic lens in the existing machinery, will close this chapter.

## 3.1 Experimental setup

To perform coincidence experiments, we first need to characterize the experimental setup with two core parameters: *i*) channel width, and *ii*) resolution. Before explaining what these quantities are and how we can determine their numerical values for our experiment, let us first consider the experimental setup and its components, sketched schematically in Fig. 3-1. We use Dr.Sjuts™ channel electron multipliers (CEMs) model KBL-510 as electron detectors. Channeltrons function by generation of a cloud of secondary electrons (see Fig. 1-8) from their interior lead-glass coating forming a continuous dynode. The operational gain, defined as the integrated output pulse charge (in number of secondary electrons) generated per incident primary electron, is on the order of $10^7$. For the CEM output signal width of about 10 ns, this corresponds to a current of roughly $10^{-4}$ A flowing across the 50-Ω termination of, say, an oscilloscope input port. This gives rise to a voltage signal that is approximately -5 mV in amplitude. The finite signal width of 10 ns with the voltage rise time of about 1 ns, as visualized on an oscilloscope, is a restriction due to the finite dead time of the CEM; upon generation of a cloud of secondaries in the channeltron funnel after incidence of an incoming primary, the detector becomes insensitive to the detection of a subsequent primary electron for a time span of about 1 ns. This is indeed consistent with the observation of a finite signal rise time on the order of 1 ns. The dead time is due to two effects. Firstly, the second incoming electron is repelled by the existing electron cloud in the detector funnel. Secondly, the secondary electrons hinder generation of more secondaries across this time span as they occupy most of the otherwise available vacuum state. This is therefore the time required for the cloud of secondaries to evacuate the CEM funnel. We will soon see how the coincidence technique can, remarkably, procure few-ps time resolutions (as in the CW HBT experiment of Ref. [3]), but first let us observe how the *detection resolution* is determined. The reader must distinguish between this quantity and the *detector dead time* of ~ 1 ns. The former gives us the typical pulse-height histogram width generated by a multichannel analyzer (MCA) under the condition that the incident count rate is low enough not to saturate the detector. The dead time is therefore not pertinent there.

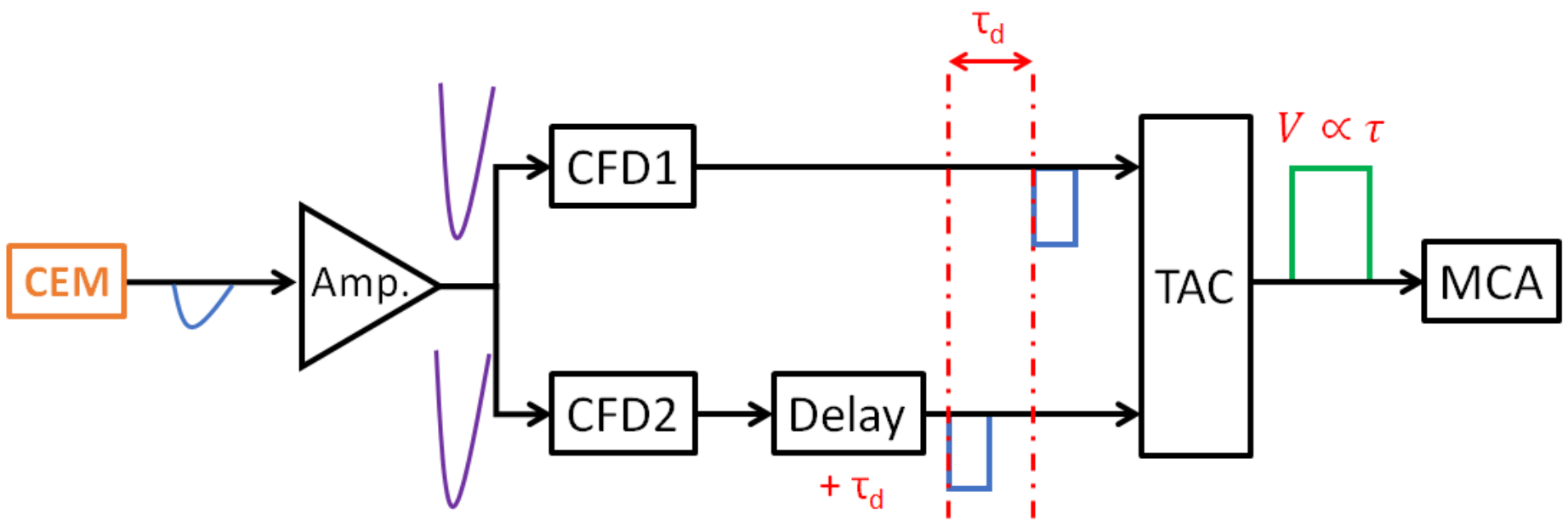

**Fig. 3-1 | Electron detection resolution measurement setup.** Schematic setup for measuring the detection resolution is illustrated in this figure. A negative voltage signal is sent out from the channel electron multiplier (CEM) detector on the left. The amplified signal is split and fed into two separate signals using an equal length of identical coaxial cables (black arrows) to ensure identical propagation times. Two channels of a constant fraction discriminator (CFD) subsequently filter out the background noise upon receipt of the amplified signals at the same time within the margin of error. One of the output signals is then delayed by a known amount of time (a few ns) with respect to the other signal. The time-to-amplitude converter (TAC) is a device that outputs a positive square signal whose height is proportional to the time delay between the inputted *start* and *stop* signals where the former precedes the latter in arrival time. Using a multichannel analyzer (MCA) a histogram of the TAC output voltage heights is generated which can subsequently be translated into the corresponding mutual time delay histogram for the split signals. This will be a peak on the mean time delay with a certain width due to the energy and velocity distributions of the incoming electrons. The width defines the resolution of the setup. When the CEM is replaced with a standard low-noise low-jitter pulse generator, the peak will narrow down significantly which in turn gives the channel width of the MCA, namely, the minimum detectable time interval under conditions compatible with an artificial low-noise source. The time delay $\tau$ equals $\tau_d$ plus any time-dependent noise and jitter.

The detection resolution in an experiment with a laser-driven tungsten nanotip source in our setup, measured as explained in Fig. 1-3, is approximately 50 ps which is very close to the (presumably FWHM) value of 26 ps reported in the experiment of Ref. [3]. The result is shown in Fig. 3-2(a). From the computational results of CH.2, we anticipate effects on the time-scale of 1 ns or more. Therefore, this amount of resolution is totally adequate for our purpose. We use ORTEC$^{TM}$ Model VT-120 low-noise preamplifiers with a constant gain of 200. An ORTEC$^{TM}$ Model 935 constant fraction discriminator (CFD) with four identical quads enables background noise removal with a very low systematic error on the order of 1 ps. The constant delay module is a stand-alone circuit and does not contribute to the time jitter and noise noticeably. Lastly, the time-to-amplitude converter, ORTEC$^{TM}$ Model 567, has a FWHM time resolution of less than 6 ps (obtained as 5 ps in excess of the 0.01% of the full scale) for a time window of 100 ns which is frequently used in the experiments of this chapter. To witness the role of the discriminator noise as part of the whole setup which ultimately sets the resolution, for the time spectrum of Fig. 3-2(b), the CFD in the *start* signal line was replaced with a leading-edge discriminator (LED), ORTEC$^{TM}$ Model 436.

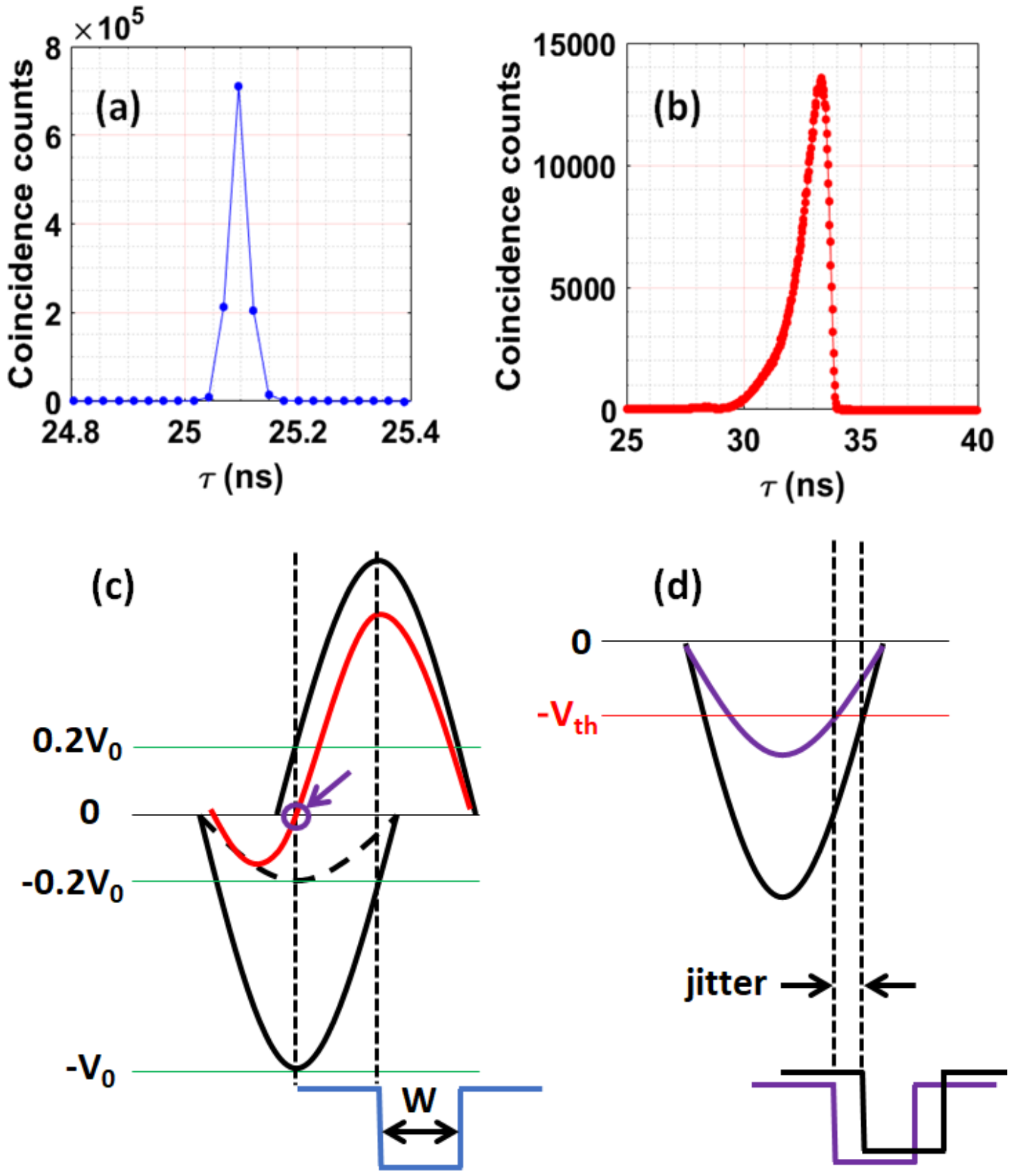

**Fig. 3-2 | The detection resolution is a property of the entire setup as a whole. (a)** In an experiment with pulsed photoelectrons of a laser-driven tungsten nanotip needle source the FWHM resolution is obtained as ~ 50 ps. **(b)** For the CFD of the *start* line replaced with a leading-edge discriminator (LED) while leaving everything else, including the electron source count rate, unaltered (note that the approximate peak areas are comparable), the spectrum broadens asymmetrically by a factor of ~ 40. **(c)** The function of the CFD with a *constant fraction* factor of 20% is illustrated schematically. A negative signal of amplitude $-V_0$ is inputted to the CFD. The dashed black signal shows its attenuated version reduced by the constant fraction factor. The positive black signal is, on the other hand, its inverted and shifted version such that when added with the attenuated signal, the *zero-crossing* point (purple arrow) always coincides with the peak of the signal

to a high accuracy and precision, *irrespective of the signal height*, with a systematic error on the order of 1 ps. The negative logic signal with adjustable width *W* is then outputted at the time corresponding to 20% of the signal height measured from this precisely defined zero-time point. The time jitter, known as *walk*, of the CFD is therefore very low, on the order of 1 ps. The signal width *W* of the ORTEC™ Model 935 CFD is adjustable from a few nanoseconds to over 100 ns so that in the so-called *updating mode* of operation, a second signal inputted to the CFD within this window only stretches the output without generating a new signal count. This is highly favorable as it makes it possible to exclude unwanted side-peaks due to RF signal impedance mismatch in the coaxial cables and connectors with bypassing the requirement to further raise $V_{th}$, which would progressively suppress a fraction of the real detector signals. In the *blocking mode* of operation, such a second input signal does not give rise to a new output pulse without stretching the output logic signal. **(d)** Contrary to CFDs, the LEDs suffer from a systematic time jitter (walk) on the order of 1 ns or higher. Here, the output signal, a logic pulse or in general a Nuclear Instrumentation Module (NIM) standard signal, is outputted at the time corresponding to the intersection of the signal with the threshold voltage level $V_{th}$. The corresponding time, however, varies for typically around 1 ns for signals with varying amplitudes. In CFDs, as explained in (c), the $V_{th}$ setting does not affect the timing of the output and this is precisely why their time jitter is noticeably lower.

---

This type of discriminator generally suffers from a higher level of systematic noise compared with the state-of-the-art CFDs. The rest of the conditions were left unchanged. Clearly, the FWHM resolution is undermined in this case and is now about 2 ns (deteriorated by a factor of ~ 40).

The channel width is the minimum bin size that the TAC and the MCA together can provide. MCAs offer a few fixed number of bins in powers of 2, in our case 512, 1028, and 2048. We use the maximum bin number of 2048 which gives the shortest bin size. The bin duration is determined by the TAC window noting to the fact that its positive output pulse height is proportional to the time delay between the start and stop channels. The maximum pulse height of 10 V represents the maximum time delay set by the TAC window. For example, with the minimum mutual detection window of 50 ns, the MCA channel (bin) number 2048 corresponds to a delay of 50 ns which gives a channel width of 24.4 ps in theory. In practice, it is highly preferred to *measure* this quantity by replacing the CEM signal in the resolution-measurement setup of Fig. 3-1 with a low-noise signal generator, that itself gives a very narrow time spectrum peak spanning over merely one or two MCA bins, and subsequently, implement a known amount of external time delay to compare with the resultant shift in the peak position. The measured MCA channel widths for a few different values of the TAC window are given in Table 3-1. The external delay was held at 4 ns.

| TAC window | Measured channel width | Calculated channel width |
|---|---|---|
| **50 ns** | 26.5 ps | 24.4 ps |
| **100 ns** | 53.0 ps | 48.8 ps |
| **200 ns** | 108.1 ps | 97.7 ps |
| **500 ns** | 250 ps | 244.3 ps |
| **1 µs** | 500 ps | 488.5 ps |
| **10 µs** | 4 ns | 4.8 ns |
| **100 µs** | N/A | 48.8 ns |

**Table 3-1 | MCA channel widths for various different TAC windows.** The measured and calculated values are compared. For TAC windows below 1 µs, the measured values are more reliable. For wider windows, the available low-noise external delay of 126 ns (here 4 ns is used for all rows) ceases to be sufficient to measure the channel width up to a reliable number of significant figures.

A few SEM images of the tungsten tip used in the remainder of the experiments in the present chapter are shown in Fig. 3-3. The apex radius of curvature is seen to be crudely ~ 50 nm. This is followed by the laser-electron coincidence spectra in Fig. 3-4 vindicating time-of-flight consistency in our setup for the detected electrons; $V_{tip}$ is the tip electrostatic bias potential that sets the electron kinetic energy, and $V_f$ is the negative bias voltage applied to the narrow rectangular margin of the front face of the detector to further suppress the unwanted secondary electrons created in the chamber upon impact of the energetic primaries with the chamber walls and the other components. The three peaks are successively labeled *A*, *B*, and *C* on Fig. 3-4(b). Qualitatively speaking, the results show that the electrons at the lower kinetic energy of 50 eV are slower which is expected. To inspect this more quantitatively, let us compare the peaks labeled *A* and *C* which are separated by $t_C - t_A = 3.90\, ns$. The tip-to-detector distance is approximately *5 cm* in our setup inside the high-vacuum chamber (~ $2\times10^{-7}$ Torr). Forgoing the effects $V_f$ might have on the kinetic energy of the incoming electrons would lead to an underestimation of this time interval by 0.4 ns. On the contrary, if we take the difference potential $V_{tip} - V_f$ as a possibly better approximation of the kinetic energy, we arrive at 4.9 ns for the calculated $t_A - t_C$ which is worse in terms of agreement with the observed value. For the middle peak labeled *B* one cannot consider the latter approximation as that would be zero. Coupled with a qualitative consideration, it therefore appears that the front negative bias $V_f$ which is applied to the outer margin of the CEM face is effective in excluding the low-energy stray secondaries without slowing down the more forward-going electrons significantly. The majority of these electrons, defined by an aperture, therefore reach the detection region noting that the peak areas are comparable.

To investigate the effect of $V_f$ more quantitatively, the electrostatic potential distribution inside a CEM with its front rectangular margin held at a non-zero (negative) potential is calculated. A schematic sketch of the CEM along with how the relevant parameters are defined are shown in Fig. 3-5. From there, we can readily express the circumference of the funnel as a function of *x* as

$$\begin{aligned}
&W(x) = 2L_y(x) + 2L_z(x) \\
&L_y(x) = a_0 + \kappa(x - x_0) \Rightarrow \kappa = (a - a_0)/(L - x_0) \\
&L_z(x) = b_0 + \lambda(x - x_0) \Rightarrow \lambda = (b - b_0)/(L - x_0) \\
&\Rightarrow W(x) = 2L_y(x) + 2L_z(x) \\
&\qquad = 2(a_0 + b_0) + 2\left[\frac{a - a_0 + b - b_0}{(L - x_0)}\right](x - x_0) = 2\left[\alpha + \beta(x - x_0)\right] \\
&\alpha \equiv a_0 + b_0\ ; \quad \beta \equiv \frac{a - a_0 + b - b_0}{(L - x_0)}.
\end{aligned} \tag{3-1}$$

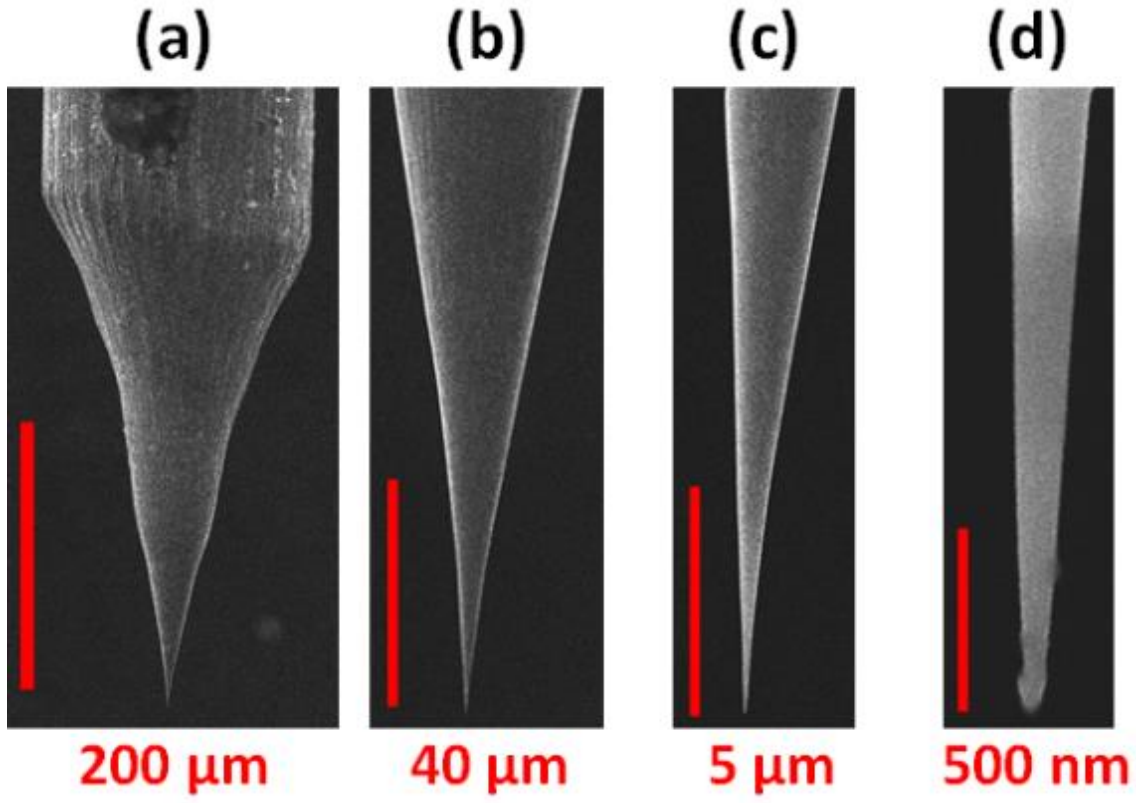

**Fig. 3-3 | SEM images of the tungsten nanotip.** This chemically etched nanotip needle source was used in the experiments to be discussed in the rest of this chapter. The actual length corresponding to each scale bar is denoted under the corresponding image. From panel (d), the geometric apex radius of curvature is crudely estimated at 50 nm.

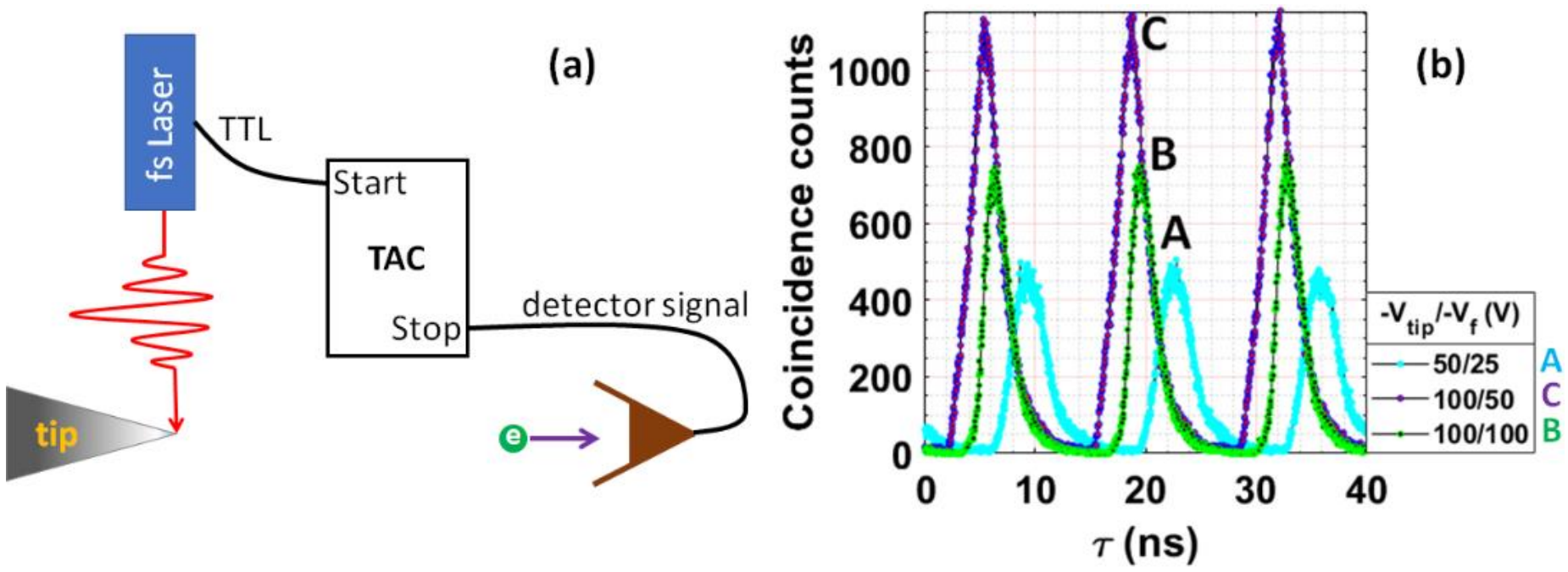

**Fig. 3-4 | Laser (start)-electron (stop) coincidence time spectra. (a)** The setup is shown schematically. A femtosecond laser TTL signal (at 76 MHz) opens the TAC window (set to 50 ns in the present case). A subsequent detector signal fed into the stop input leads to a coincidence count on the MCA (see Fig. 3-1). **(b)** The results are compared for three different combinations of the tip voltage $V_{tip}$ and the negative bias voltage $V_f$ applied to the margin of the front face of the detector in order to expel low-energy secondary electrons generated inside the chamber. Peak *A* corresponds to $(-V_{tip},-V_f) = (50,25)$, peak *B* to (100,100), and peak *C* to (100,50), all in unit of volts.

On the rectangular cone,

$$dR = \rho\frac{dx}{A} = \rho\frac{dx}{W(x)t} = R_s\frac{dx}{W(x)}, \tag{3-2}$$

where $R$ is the resistance of the wall, $\rho$ is its resistivity, $t$ is its thickness, and $R_s = \rho/t$ is its sheet resistance. We therefore arrive at the expression

$$\begin{aligned}
R_{cone}(x) &= \int_{x_0}^{x} dR = \frac{1}{2}R_s\int_{x_0}^{x}\frac{dx}{\alpha + \beta(x - x_0)} = \frac{1}{2}R_s\frac{1}{\beta}\ln\left(\alpha + \beta(x - x_0)\right)\Big|_{x_0}^{x} \\
&= \frac{R_s}{2\beta}\ln\frac{\alpha + \beta(x - x_0)}{\alpha} = \frac{R_s}{2\beta}\ln\left[1 + \delta(x - x_0)\right],
\end{aligned} \tag{3-3}$$

as the resistance of the cone. In the final step, $\delta \equiv \beta/\alpha$ defined in Eq. (3-1). For the cylindrical tube with diameter $D$, $\pi D = 2(a_0 + b_0) \Rightarrow D = 2\alpha/\pi$. Consequently,

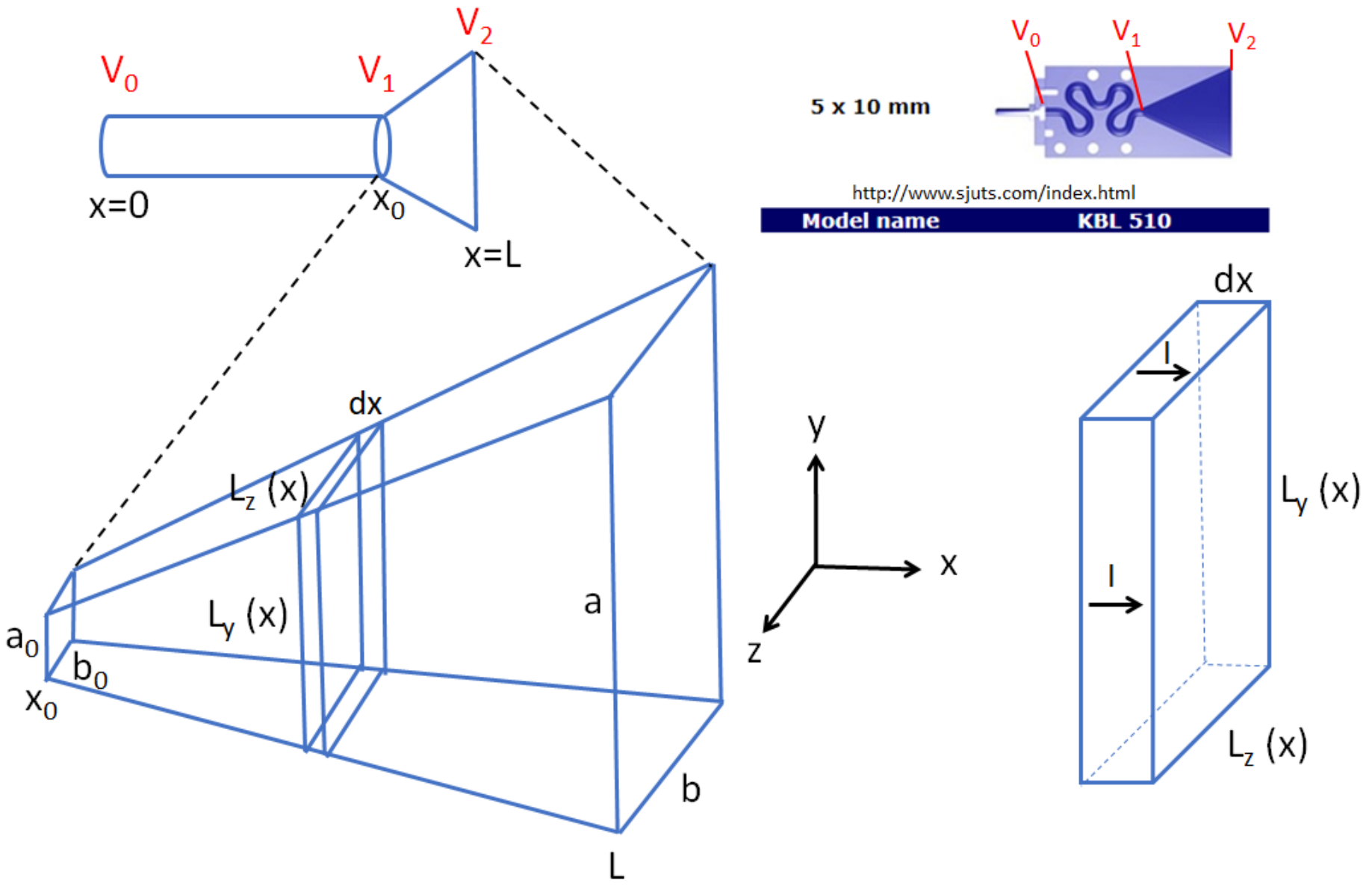

**Fig. 3-5 | Electrostatic potential distribution inside the CEM.** A cutaway view of the Dr.Sjuts™ CEM Model KBL-510 used in this research is shown on the top right. The input funnel is a rectangular cone, 5×10 mm$^2$ at the opening mouth. The front electrode is negatively biased at $V_f = V_2$, and a positive high voltage $V_0$ is applied to the back of the CEM. We approximate the curly tube, that is made in this shape to suppress ion feedback, with a straight cylinder. The coordinate system and the parametrized dimensions are trivially defined in the bottom-left section. The funnel is discretized in the x-direction. One such 1D-infinitesimal element is depicted at the bottom-right corner. The structure is hollow, and a current *I* may flow across its walls which is made of a secondary-emissive lead-glass with a high wall resistance of $2\times10^8$ Ω and a typical operational gain on the order of $10^7$.

$$R_{tube}(x) = \int_0^x \frac{\rho dx}{\pi D t} = \frac{R_s x}{2\alpha} \tag{3-4}$$

is the tube resistance. To find the position-dependent potential across the CEM, we note that from the Dirichlet boundary condition and Eqs. (3-3) & (3-4),

$$V_2 = V_0 - \frac{R_s x_0}{2\alpha} I - \frac{R_s I}{2\beta} \ln\left[1 + \delta\left(L - x_0\right)\right]. \tag{3-5}$$

We define

$$K \equiv \frac{R_s I}{2} = \frac{V_0 - V_2}{\frac{x_0}{\alpha} + \frac{1}{\beta}\ln\left[1+\delta\left(L - x_0\right)\right]} = \frac{V_0 - V_2}{\gamma + \frac{1}{\beta}\ln\left[1+\delta\left(L - x_0\right)\right]}, \tag{3-6}$$

in which $\gamma \equiv x_0/\alpha$. For $x_0 \le x \le L$, the electrostatic potential of the cone becomes

$$V_{cone}\left(x\right) = V_0 - K_\alpha x_0 - K_\beta \ln\left[1+\delta\left(x - x_0\right)\right], \tag{3-7}$$

where $K_\alpha \equiv K/\alpha$ and $K_\beta \equiv K/\beta$. For $0 \le x \le x_0$,

$$V_{tube}\left(x\right) = V_0 - K_\alpha x. \tag{3-8}$$

The variation of the electrostatic potential on the CEM interior walls as a function of the position coordinate $x$ is plotted and shown in Fig. 3-6. In order to better visualize the effect of $V_f$ on the incoming electrons, the CEM mouth is pixelated as depicted to the left of Table 3-2. The kinetic energy of the incoming photoelectrons is typically set between a few eV which is desirable for the HBT experiment (see Fig. 2-6) to 100 eV. It is therefore grasped from Table 3-2 that the electrons entering the CEM from the region enfolded by the 8$^{th}$ pixel immediately feel the positive potential and get accelerated into the tube. Depending on their energy and angle of incidence, some electrons will be decelerated before detection which contributes to the finite detection resolution previously discussed in Fig. 3-2. The low-energy stray electrons in the chamber, which for the most part

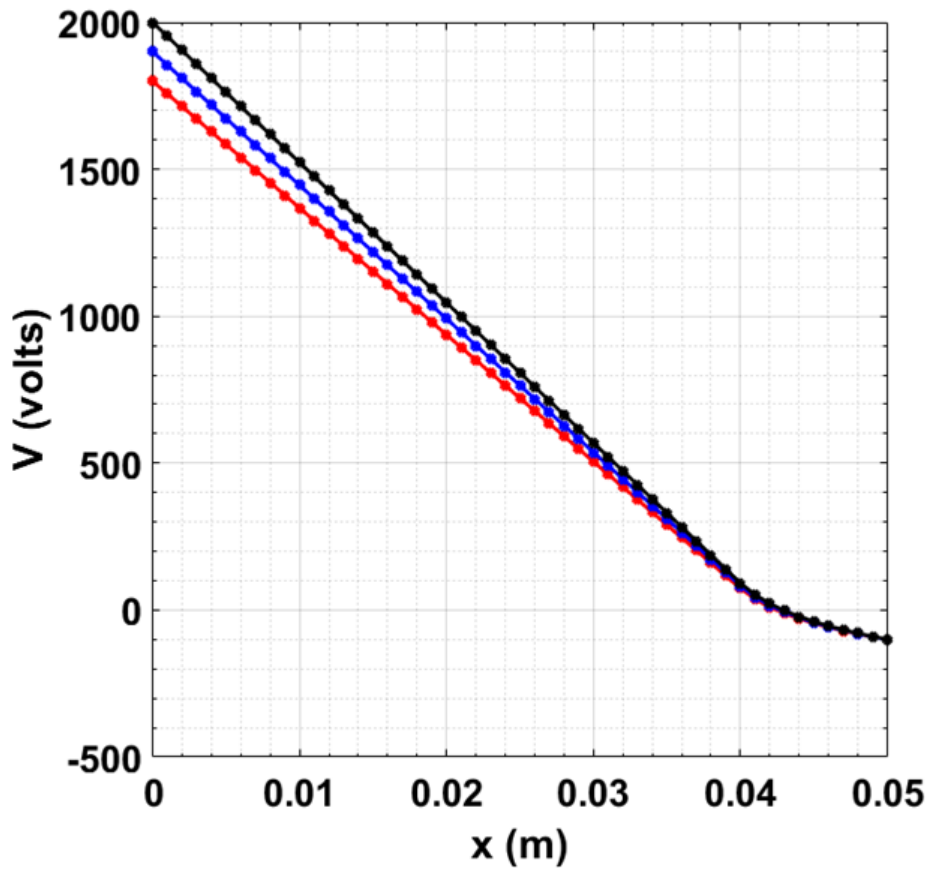

**Fig. 3-6 | Variation of the electrostatic potential on the CEM interior walls with its axial position.** Eqs. (3-7)&(3-8) are used to plot the potential on the tube and the cone walls, respectively. Three typical operational values of the applied positive high-voltage $V_0$ are taken for comparison. The numerical values are:

$a_0 = 2\,mm, b_0 = 1\,mm, a = 10\,mm, b = 5\,mm, x_0 = 4\,cm, L = 5\,cm$

and $V_2 = -100\,V$ .

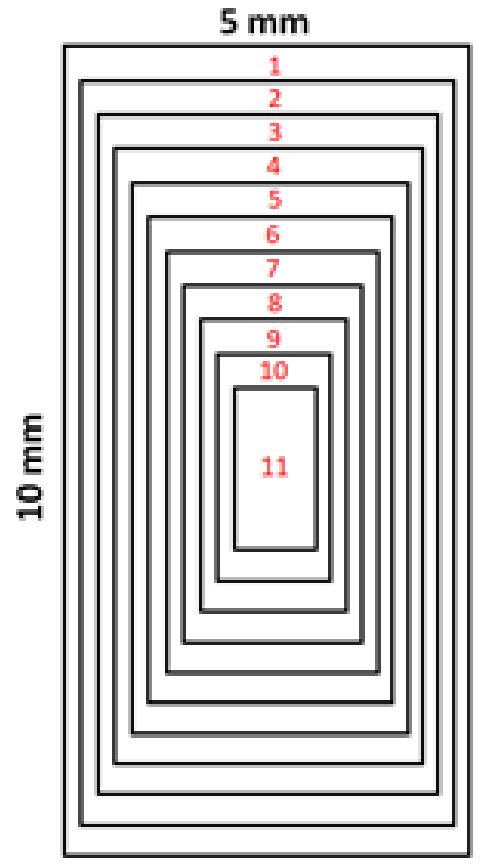

| n | $x_n$ (mm) | $a_n$ (mm) | $b_n$ (mm) | V (volts) |
|---|---|---|---|---|
| 1 | 10 | 10.0 | 5.0 | -100.0000 |
| 2 | 9 | 9.2 | 4.6 | -91.0034 |
| 3 | 8 | 8.4 | 4.2 | -81.1878 |
| 4 | 7 | 7.6 | 3.8 | -70.3892 |
| 5 | 6 | 6.8 | 3.4 | -58.3883 |
| 6 | 5 | 6.0 | 3.0 | -44.8836 |
| 7 | 4 | 5.2 | 2.6 | -29.4435 |
| 8 | 3 | 4.4 | 2.2 | -11.4189 |
| 9 | 2 | 3.6 | 1.8 | +10.2328 |
| 10 | 1 | 2.8 | 1.4 | +37.3488 |
| 11 | 0 | 2.0 | 1.0 | +73.6530 |

$x_n$ is the x-position of the n'th pixel relative to the innermost one (#11).

**Table 3-2 | Pixelated rectangular cone of the CEM mouth.** The electrostatic potential for each pixel, which are rectangular ribbons save for the deepmost pixel number 11 that is a complete rectangle, constitute the 5$^{th}$ column of the table. $V_0$ = 1800 V, typically suitable for a fresh channeltron.

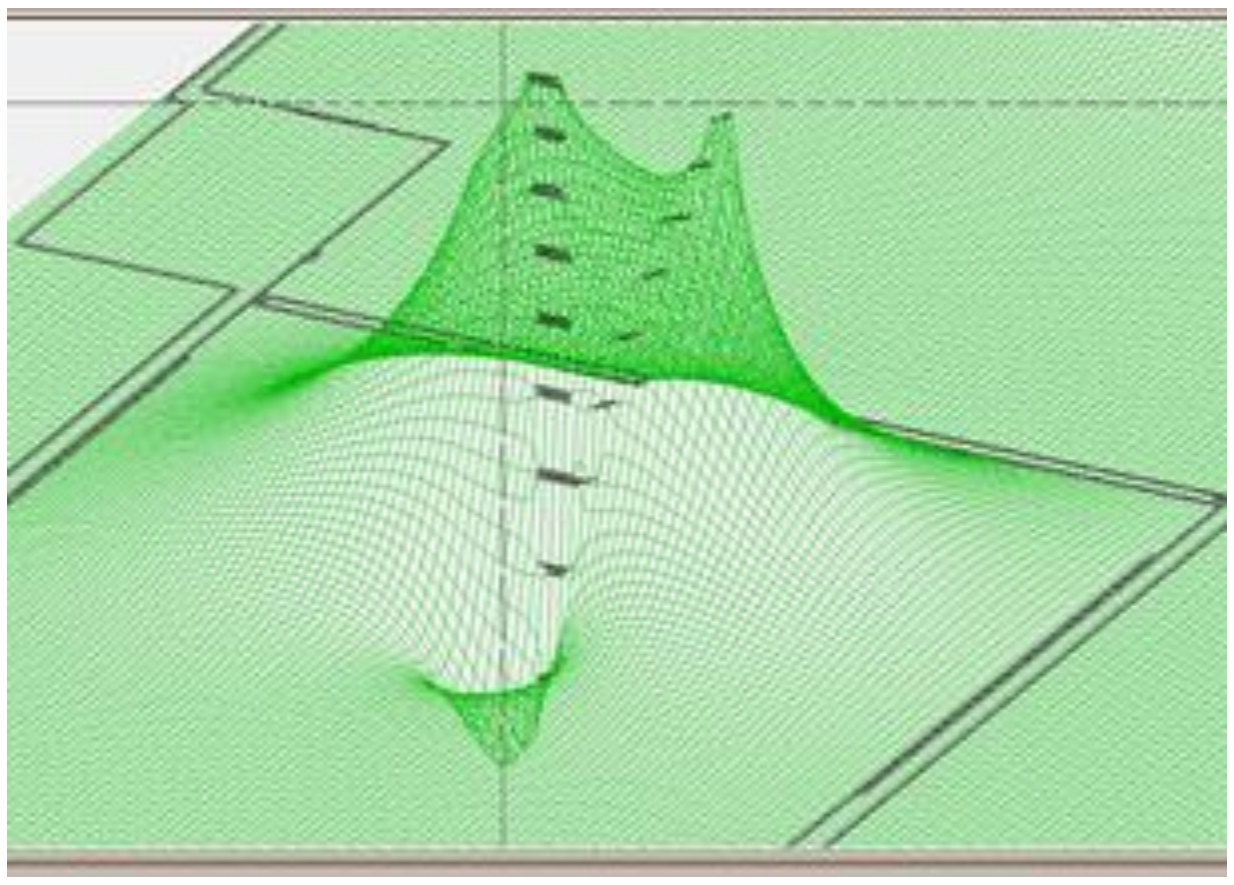

**Fig. 3-7 | SIMION8.1™ simulation of the electrostatic potential map inside the CEM.** The analytical values of the potential on the CEM walls from Table 3-2 were used as the boundary condition. A higher point represents a more negative potential. The simulation was performed by Will Brunner from Prof. T. J. Gay's group.

constitute randomly generated secondary electrons after collision of the primary photoelectrons with the various components in the chamber, will thus have a higher chance to be expelled. This, indeed, proved to be highly efficient to suppress the uncorrelated background counts in practice. A SIMION8.1™ simulation of the potential distribution inside the CEM, with the calculated values of Table 3-2 inputted as the boundary potential, is illustrated in Fig. 3-7.

We have so far considered the effect of $V_f$ in detail. On the back end of the channeltron, the positive high voltage $V_0$ is applied through a voltage divider. This is illustrated and explained in Fig. 3-8.

Our two-CEM setup for coincidence detection experiments is schematically shown in Fig. 3-9. The contraption enables us to measure the time delay between mutual detection events with a resolution set by the MCA channel width which can be as low as approximately 25 ps, rather than the single-detector dead time of 1 ns as explained earlier. This is in fact one main advantage of employing the double-the signal height will be reduced. Additionally, for a capacitance that is too large, the stored energy is large which may damage the pre-amplifier upon discharge. Lastly, the RC time-constant should not be too large, or the electron flow across the 1-MΩ resistor could be hindered due to which the anode potential will drop, and it may no longer be able to collect all the secondaries, leading to a decreased signal height once more.

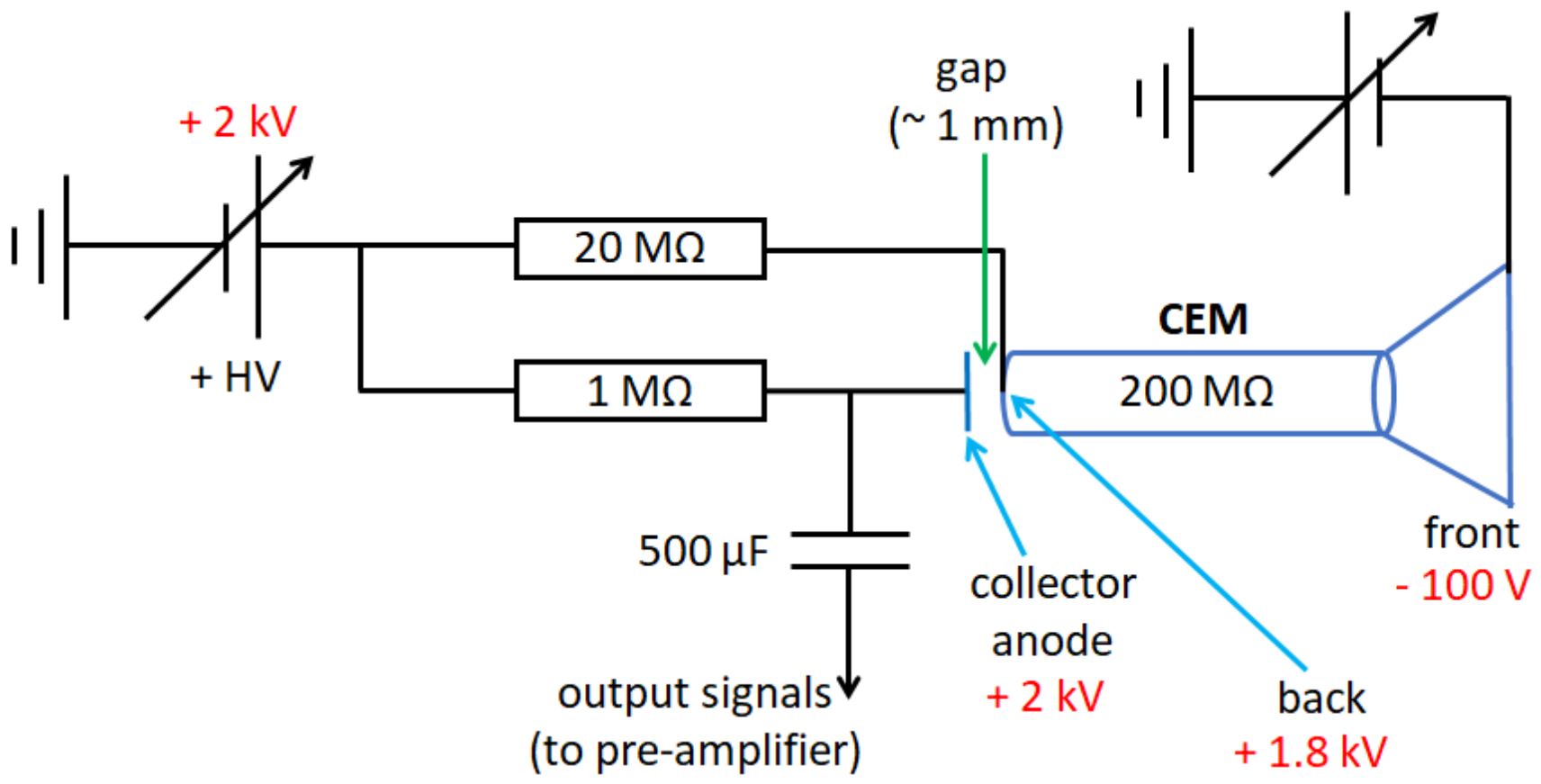

**Fig. 3-8 | CEM electronic circuit in the pulse-counting mode.** CEMs can detect single electrons. This is crucial for free electron technologies in almost the same way that single-photon detection has been important for quantum optics. For currents > 10 pA a picoammeter should to be used. CEMs can be wired for current measurement as well, but the signal itself would appear at a high voltage. Leakage currents and background DC levels due the electronics would be hurdles to overcome in that case. When the pulse-counting mode is suitable, the issue can be greatly ameliorated taking advantage of the circuit shown here. The 20-MΩ resistor is in series with the CEM itself. This forms a voltage divider across which an electrostatic potential of roughly 2.1 kV is applied in the present example. The back end of the CEM is therefore at 1.8 kV approximately. The collector anode, however, is at 2 kV. The cloud of secondary electrons that emerge at the end of the tube are thus accelerated toward the anode due to the resultant potential difference of about 200 V. Given that in this mode of operation, instead of a direct current individual pulses of secondaries are counted, a capacitor is used in the signal line which removes the DC level due to the electronic equipment, and current leakage to the grounded vacuum chamber is no longer an issue. The chosen capacitance and the resistance of the 1-MΩ resistor are typical. In general, the impedance of the capacitor must be several times lower than that of the coaxial cable used to deliver the signals to outside of the vacuum chamber, otherwise

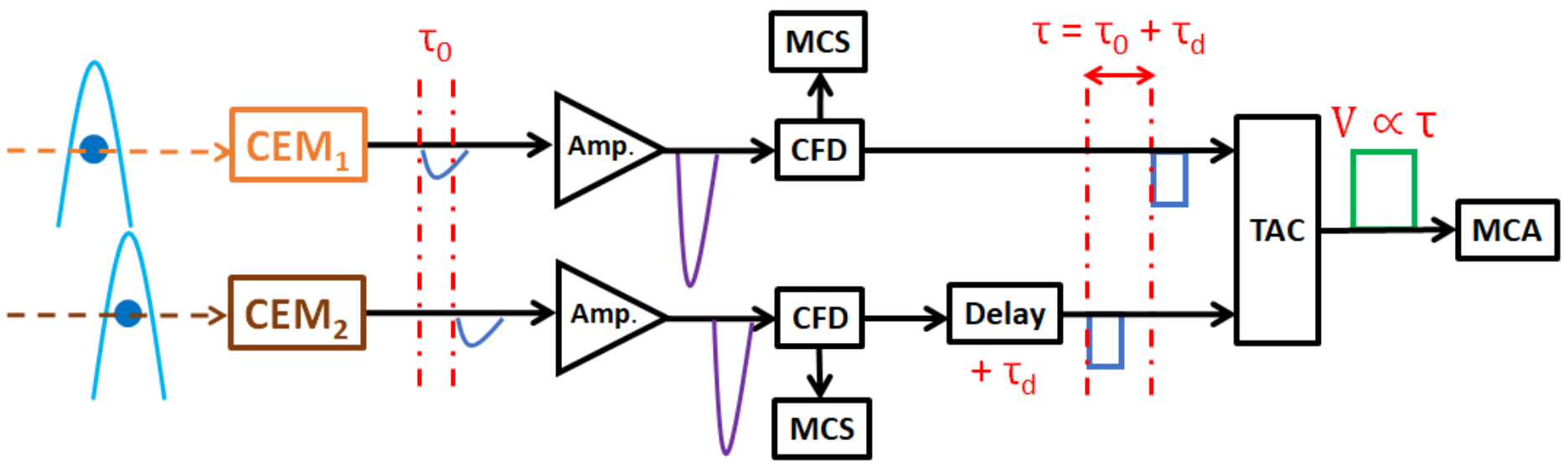

**Fig. 3-9 | Schematics of the coincidence detection setup.** Two electrons (blue circles) land at two independent chnneltrons with a time interval $\tau_0$. With the implemented external delay, it is possible to conveniently define the zero-delay time by shifting the signals relative to each other by a constant known amount. The count rates are registered in real-time using a pair of multichannel scalers (MCSs) which output bar-plots of the variation of the detection counts over the experiment time.

detector scheme. The second main advantage is indeed the fact that the double-detector set-up is sensitive to the spatial distribution of the photoelectrons in the direction transverse to the beam which is illustrated in Fig. 3-10. Later in this chapter, it will be shown that the particle statistics of the incoming beam can be inferred from the two-detector coincidence spectra. Pictures of the home-built cubic chamber [46], and the double-detector assembled and mounted by the author of the present dissertation, are shown in Fig. 3-11.

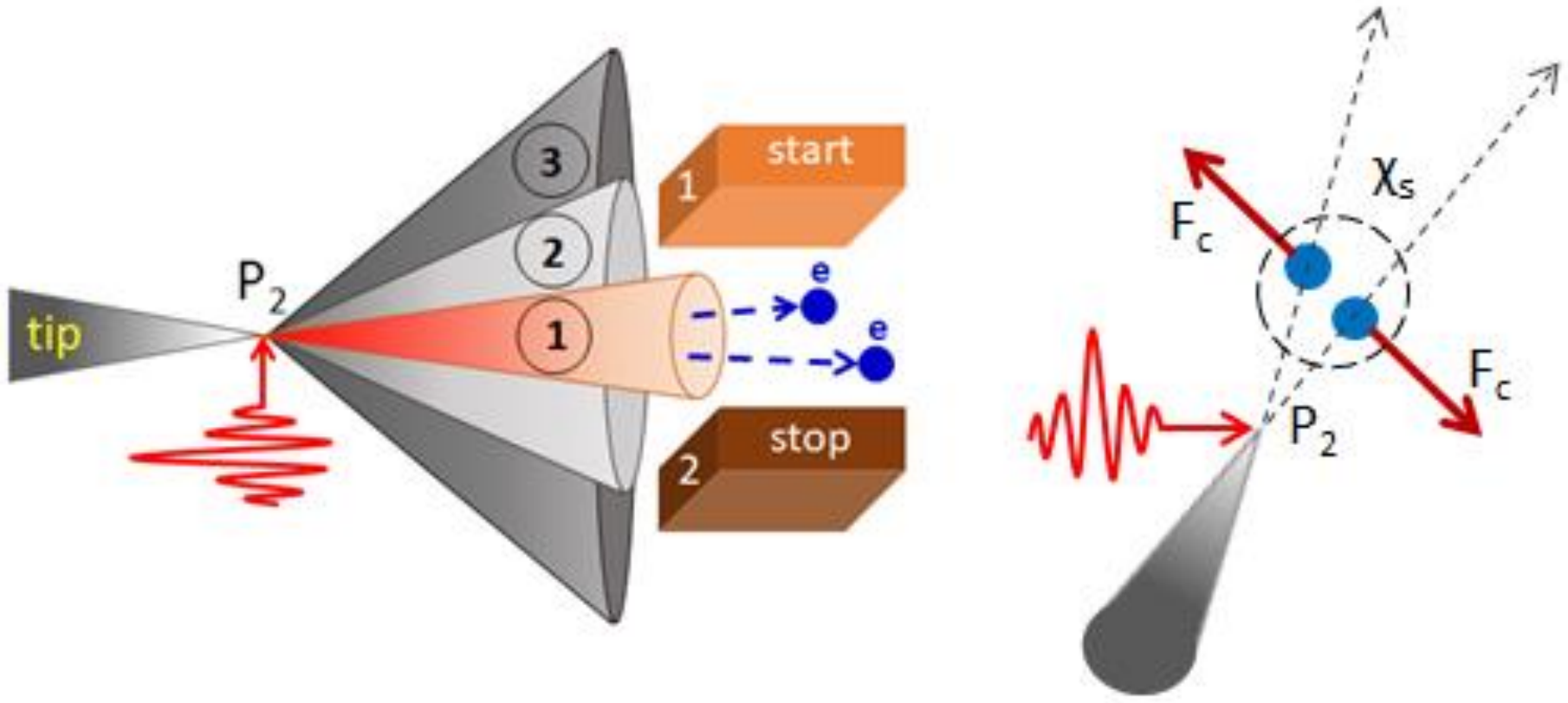

**Fig. 3-10 | The double-detector set-up provides information on the spatial distribution of the photoelectrons on the detection plane.** A pair of photoelectrons are generated through a multiphoton absorption process with probability $P_2$ upon incidence of a fs laser pulse on the nanotip apex. The spatial separation of the electrons on the detection plane depends on *i*) the photoemission momentum and angular distribution, *ii*) the initial separation of the electrons that sets the strength of the Coulomb interaction felt by them, and *iii*) the size of the coherence volume on the detection plane that sets the strength of the Pauli force that also tends to increase the separation of the electron pair by allowing only spin-singlet pairs in the same spatial orbital (right). In the example on the left side, the distance between the start and stop detectors are such that the electron pairs in regions 1 and 3 do not strike any of the detectors while those in region 2 contribute to the ensuing coincidence spectrum.

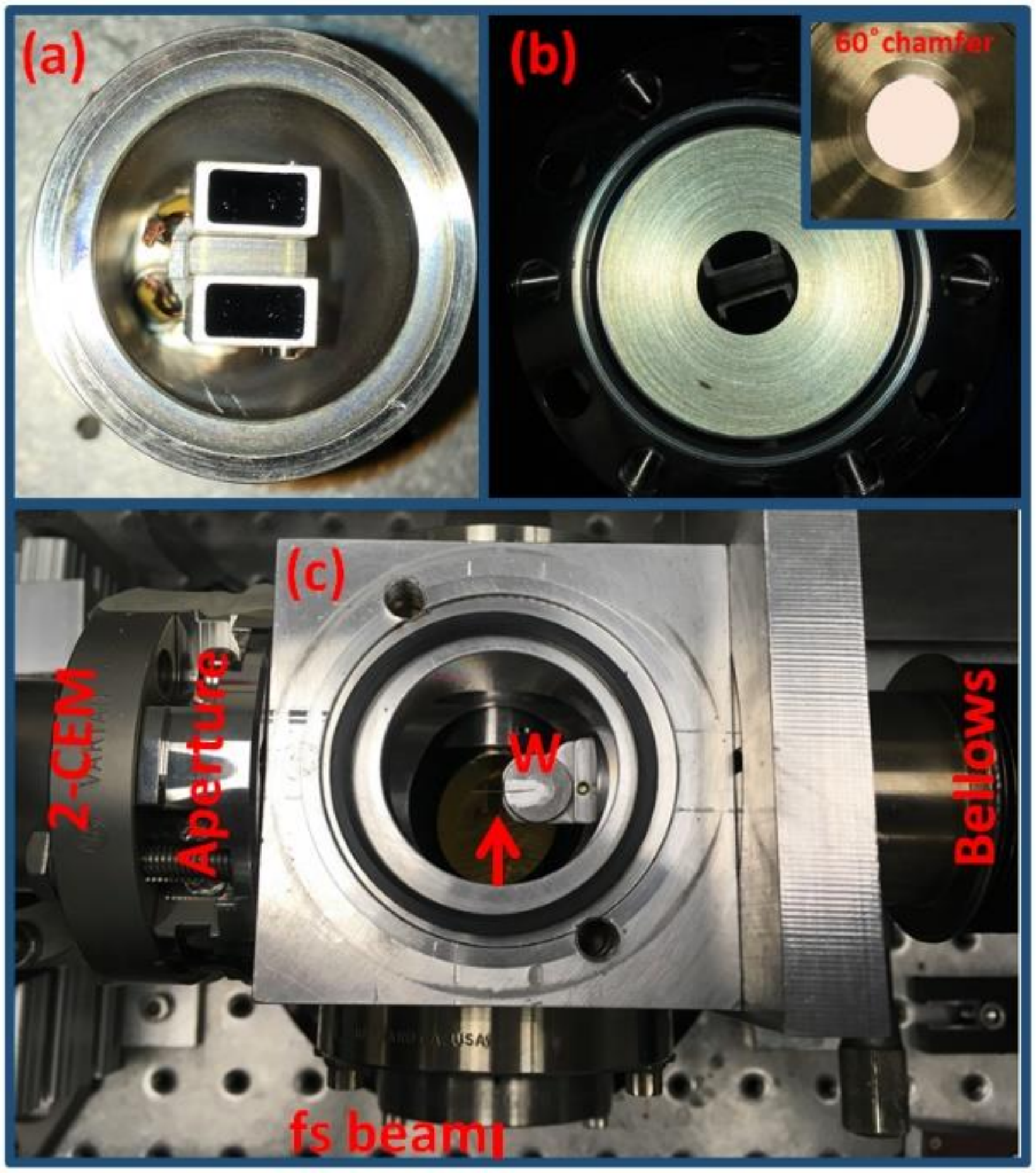

**Fig. 3-11 | The double-detector combination. (a)** The front face of the CEM pair is shown. The rectangular opening of each detector is 5×10 $mm^2$. **(b)** A brass aperture 1.4 cm in diameter is positioned at a distance of 1.4 cm from the detector faces to block the stray secondaries. The inset shows the 60-degree chamfer on its rear face. **(c)** Top view of the cubic vacuum chamber is shown. A tungsten tip is mounted on top of an off-axis parabolic mirror (OAPM) that focuses the incoming femtosecond pulsed laser beam, with linear polarization parallel to the tip axis, onto its apex aberration-free and without pulse stretching as would be the case due to chromatic dispersion in curved lenses. The focal length is

1 inch. The focal spot diameter is estimated at 8.6 μm using the Gaussian Rayleigh length formula and the knife-edged full-width-at-1/e of 2.95 mm for the incident beam. The tip is scanned across the focal area using the flexible bellows on the right. The aperture and the pair of detectors are mounted on the left. During the experiments, the tip is at a distance of approximately 5 cm from the detectors.

## 3.2 Experimental results

We introduced the Poissonian probability distribution in Eq. (1-33) and its ensuing discussion where we statistically analyzed the results reported in Ref. [3]. For a true random source, as explained in CH.1, the histogram of successive time intervals obeys the exponential distribution,

$$I_1(\tau) \propto e^{-r\tau}, \tag{3-9}$$

up to a normalization factor given by the particle rate *r*. Later on, we will provide reason why the double-detector scheme can still generate the desired distribution despite the fact that in the ideal world, the time interval between two such successive electrons is supposedly the detection time delay of a single detector sitting in front of the source. In practice, however, the single-electron scheme does not serve our purpose for two reasons. First of all, the 1-ns dead time of the CEM hinders detection of electrons separated by less than a nanosecond. Secondly, the coincidence technique would not be suitable to resolve the time interval using only one detector. Take, for instance, the schematic setup of Fig. 3-1. We opened this chapter by a discussion of how the resolution time of the setup can be measured using this contraption. We observed that each CEM signal is split into two identical signals, after low-noise amplification, to be fed into the TAC start and stop channels and thus the coincidence time spectral resolution peaks of Fig. 3-2 were obtained accordingly. Therefore, even if the finite dead time were not a restrictive issue, one would need to manage to send every other amplified output signal of a detector to a separate coaxial line. For these reasons, the double-detector coincidence setup depicted in Fig. 3-9 is the best bet to exploit to construct the time interval histogram of a stream of electrons thereof coupled with prudent analytical considerations and supported by statistical and combinatoric deliberations. The reader is hereby reminded of the second goal behind this experiment. As elaborated on in CH.1, the setup is identical to that of the ultrashort free electron HBT experiment save for a quadrupole lens. The results are therefore anticipated to be those of the negligibly small coherence volumes in the context of matter wave HBT experiment, corresponding to incoherent illumination of the double-detector assembly outlined in Ref. [3], where the antibunching signal is expected to be observed as a reduced peak.

The anticipated particle statistics for a femtosecond laser-driven nanotip emitter is illustrated in Fig. 3-12, Generation of 3 electrons or higher per incident laser pulse is negligible for low-brightness sources. For a true random source, the probabilities obey that of the Poissonian distribution of Eq. (1-33). It is important to note that in our setup as described in detail in the previous section, the flight time of the photoemitted electrons is shorter than the laser repetition time. At 100 eV, for instance, the time of flight is 4.9 ns shorter than the repetition time. At 40 eV, the flight time is approximately equal to the repetition time. This implies that the Coulomb interaction is present in our problem only when a two-electron pulse is flying towards the detection plane. This makes inclusion of the repulsive Coulomb force more straightforward than in an analogous CW experiment. As illustrated in Fig. 3-12, a fraction of the pulses escapes detection even after surviving the aperture’s blockade (see Fig. 3-10). Also, the order of detection is crucial in giving rise to coincidence clicks. All these beg for statistical and combinatoric considerations.

A coincidence spectrum for a TAC window of 50 ns is shown in Fig. 3-13. The coincidence counts about the zero time-delay, which correspond to electrons

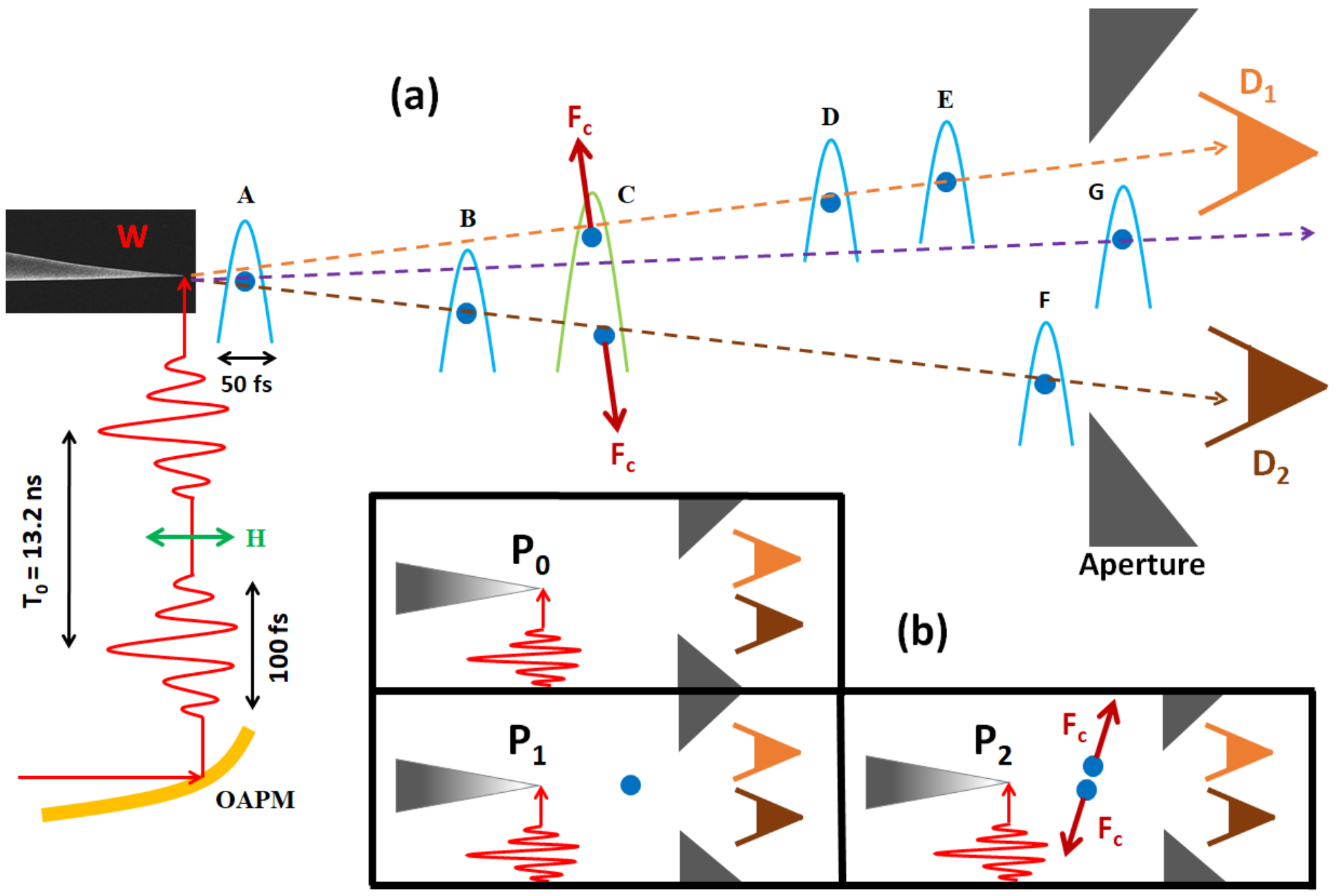

**Fig. 3-12 | Schematics of the electron photoemission statistics from a nanotip source. (a,b)** The photoemitted electrons from a tungsten nanotip propagate towards the double-detector after crossing the aperture. Some of the laser pulses generate no electrons (with probability $P_0$), while the rest may liberate one ($P_1$) or two ($P_2$) electrons as shown in (b); at low brightness, larger numbers of electrons per pulse are progressively smaller and negligible. This is why some of the successive electron pulses like *A* and *B* are separated by an interval longer than the repetition time of the horizontally polarized (H) femtosecond laser beam, $T_0$. To the right, the electron pulse *G* is going to miss the detectors, while *F* and *E* will give rise to a coincidence click separated by a value around $-T_0$ assuming the $D_1$ output signal is fed into the *start* input of the TAC and that an appropriate amount of external time delay is used in the *stop* line to adequately shift the time of zero-delay. A two-electron pulse in which the radial repulsive Coulomb force $F_c$ may be non-negligible is also shown at *C*. In our setup, we only have *one* electron pulse propagating towards the detection plane at a time, as in panel (b), due to the fact that the flight time is shorter than the laser repetition time. Coulomb repulsion in this

problem is therefore present when such pulses carrying more than a single electron are emitted and subsequently undertake free-space propagation. Another possibility not considered here is the Coulomb attraction between the emitted electrons and the positive ion left behind at the nanotip emitter which may affect the observed dynamics despite electron-hole recombination times on the order of 10 fs in metals.

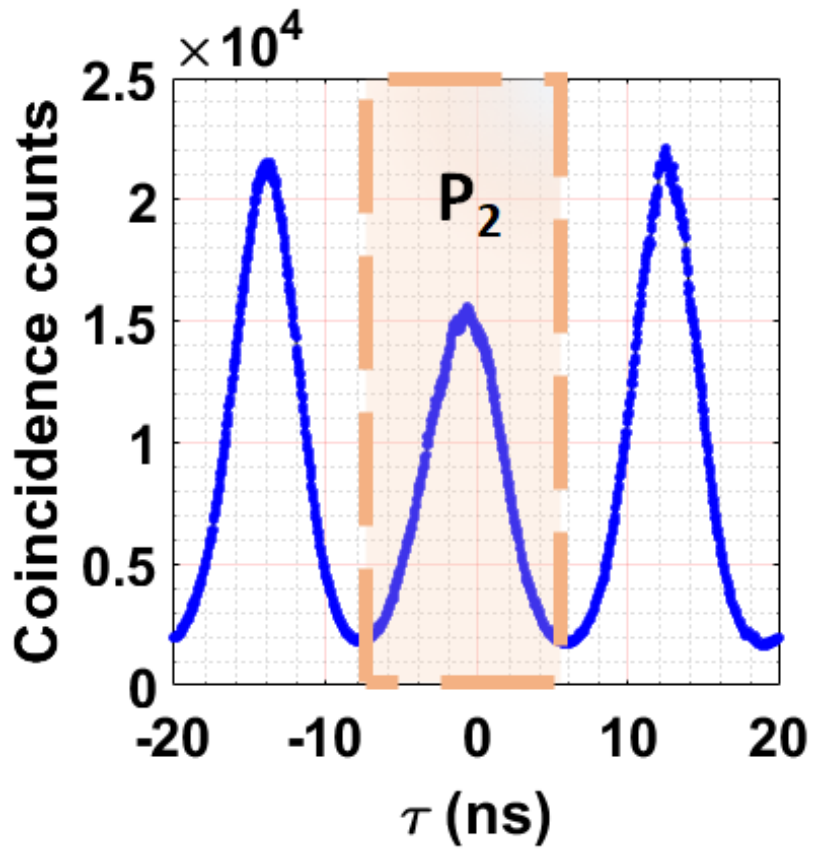

**Fig. 3-13 | Experimental result of a mutual detection coincidence spectrum.** The spectrum is obtained at a TAC window of 50 ns, $V_{tip}$ = -100 V, and $V_f$ = -50 V. Strikingly enough, the center peak about the zero time-delay is shorter than the other two. The coincidence points on this peak (the shaded area) which are separated by time intervals shorter than one repetition time (13.2 ns) can only be given rise to by the electrons carried in multi-electron pulses. We can therefore make the definitive claim that the *zero-delay peak* represents $P_2$-events, ignoring $P_3$ and higher orders. The data acquisition time was 15 min.

separated by less than one repetition time, can only be due to pulses carrying two (or more) electrons (after appropriate background reduction due to secondaries using an aperture (see Fig. 3-6) and a negative applied voltage bias $V_f$). Armed with the observation that we now have a means to access direct information about the two-electron events in the photoemission process, we are now a major step closer to our goal. We thus attempt to inspect this observation methodically.

A set of coincidence spectra at three different ranges of TAC detection window are contrasted in Fig. 3-14. Here, $V_{tip}$ = -100 V, as in Fig. 3-13, but the front bias is further increased to $V_f$ = -100 V (see the discussion of Fig. 3-4) to achieve a further reduction in the observed coincidence background level. The corresponding MCS spectra for both the $CEM_1$ (start) and $CEM_2$ (stop) detectors, taken in real-time over the span of the MCA data acquisition process, is also shown with bin widths of 1 s. As a reminder, the MCS alone cannot resolve the two-electron events for which the separation time is shorter than or on the order of the detector dead time at first place.

Before discussing the particle statistics, let us first show that the observed peak width which is one repetition time on top of the coincidence background level cannot be justified by a back-of-the-envelope calculation of the Coulomb broadening in the center-of-mass (CM) frame of an emitted electron pair.

For a kinetic energy of $E = 100\,eV$, the electron speed is approximately $v_0 = 6\times10^6\ m/s$. This gives $x_0 = v_0 t_0 = 150\,nm$ for the mean electron spatial separation at the source corresponding to one-half of the electron pulse duration which is approximately 50 fs. Conservation of energy in the CM frame can thus be written as

$$\frac{kq^2}{x_0} = \frac{(\Delta p_c)^2}{2\mu}, \qquad (3\text{-}10)$$

in which $\mu = m_e/2$ is the reduced mass of the electron, $k = 9\times10^9\ N\cdot m^2/C^2$ is the Coulomb constant, $q$ is the electron charge, and $\Delta p_c$ is the relative momentum of the electron pair due to Coulomb interaction. We therefore obtain

$$\Delta p_c = \sqrt{\frac{2\mu kq^2}{x_0}} = 3.74\times10^{-26}\ kg\cdot m/s. \qquad (3\text{-}11)$$

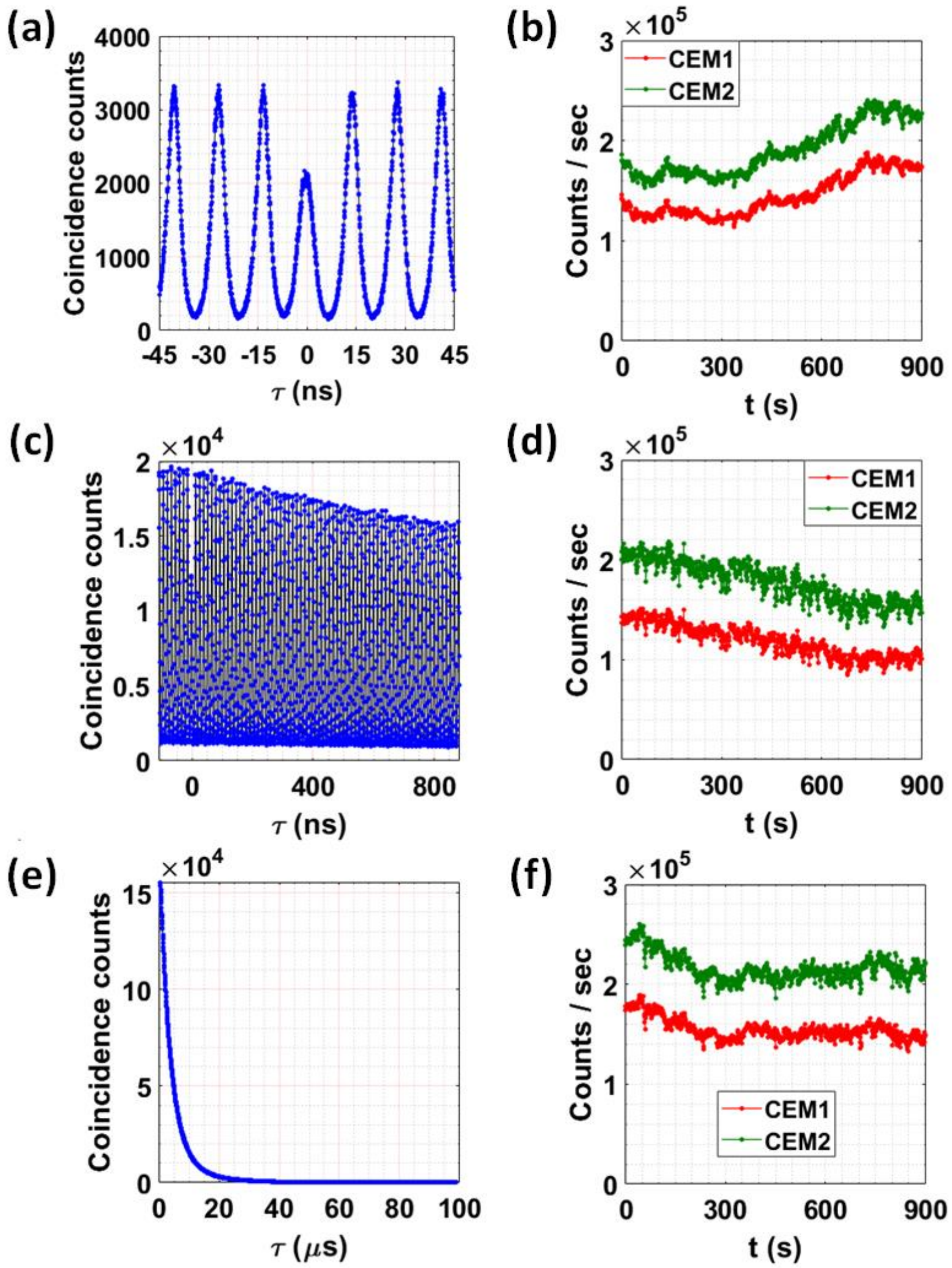

**Fig. 3-14 | Coincidence spectra for three different detection windows are compared.** Panels **(a,c,e)** show the MCA coincidence spectra while the corresponding MCS count rate spectra are depicted in panels **(b,d,f)**. The TAC window is, from top to bottom, 100 ns, 1 μs, and 100 μs. Refer to Table 3-1 for the corresponding channel widths. CEM1(2) is the start (stop) detector. No external time delay is used to generate the time

spectrum of panel (e). The shorter $P_2$-peak is standing out in (a) and (c). Data in panels (e,f) were taken starting 1 min after the end of (a,b) acquisition, and followed by (c,d), again separated with 1 min. In all panels, $V_{tip} = V_f$ = -100 V, and the 100-fs laser pulse energy is 1.58 nJ at 76 MHz and peak wavelength of 800 nm.

From here, the relative speed is calculated as

$$\Delta v_c = \frac{\Delta p_c}{\mu} = 8.2 \times 10^4\ m/s. \tag{3-12}$$

For a flight time of 8.3 ns with the above kinetic energy over the distance of 5 cm, we arrive at

$$\Delta t_c = t_f \cdot \frac{\Delta v_c}{v_0} \approx 113\ ps, \tag{3-13}$$

for an estimate of the Coulomb broadening. In addition to this, the broadening due to the energy spread of the W nanotip source which is $\Delta E \approx 1\,eV$ should also be taken into account. Noting to the fact that

$$\frac{\Delta v_E}{v_0} = \frac{\Delta t_E}{t_f} = \frac{\Delta E}{2E}, \tag{3-14}$$

a broadening of

$$\Delta t_E \approx 41\ ps, \tag{3-15}$$

is estimated in this case. The upper limit of the total broadening originated from Coulomb interaction and energy spread is therefore anticipated to be

$$\Delta t_{\max} \approx \sqrt{\Delta t_c^2 + \Delta t_E^2} \approx 120\ ps, \tag{3-16}$$

in our experiment. As an example, for a (constant) transverse velocity of $v_0/10$, this maximum time separation (for electrons emitted at the apex on the transverse plane) corresponds to a transverse spatial spread of 72 μm which is roughly 2 orders of magnitude shorter than the separation distance of 5 mm between the nearest edges of the two detectors. At any rate, the peak widths in the MCA spectra do not seem to be justifiable by these broadening mechanisms alone. They are, however, consistent with the laser-electron coincidence spectra of Fig. 3-4.

## 3.3 Theory and data analysis

Consider the double-detector set $D_1$ and $D_2$ which were schematically shown in Fig. 3-12. We observed in the same figure that in realistic experiments, we need to take the possibility of consecutive detections in one detector of the coincidence setup into account. In addition, we opened the present chapter by a consideration of the finite dead time of each detector which leads to suppression of a fraction of genuine detector signals. Moreover, the finite detector resolution (not to be confused with that of the detection electronics discussed in Fig. 3-2), with an upper bound of approximately 10 ns as judged by the observed detector pulse duration using an oscilloscope, is longer than the dead time. We therefore attempt to make sure that our analytical machinery will take care of practical constraints such as these before tackling the main problem. The final goal in this section is to find the relative height of the $P_2$-peak (or better speaking, zero-delay peak) that is given rise by multi-electron pulses for a true Poissonian distribution for realistic (i.e. imperfect) detectors and propagation constraints. We will then be able to compare our experimental results of Fig. 3-14 with the Poissonian prediction and conclude the type of particle statistics of the nanotip source from there. We undertake this task through five *observations* as outlined below (*A* stands for *Analytical*).

**A1.** The detection system, involving a detector and the detection electronics as a whole, has a finite *dead (recovery) time* $t_d$. Consequently, *n* primary electrons striking a detector during this time interval are counted as one event. To begin with, we will find the first-order correction in the rate due to this limitation.

**A2.** We will then derive an expression for the number of successive detection counts $N(n\tau_0)$ in the coincidence setup, namely, mutual detection events separated in time by an average interval $n\tau_0$, that is, $n\tau_0 \pm \tau_0/2$, where $n$ is any positive or negative integer, and $\tau_0 = 13.2\,ns$ is the pulse repetition time.

**A3.** Next, we will find a relation between single-detector counts and double-detector coincidence counts for uniform illumination of the detection plane. This enables us to understand how to correctly formulate the two-detector coincidence probability at each detector pair based on the single-event probabilities.

**A4.** Exploiting the identical notation for double-detector events as in A2, we will show that for the Poissonian distribution, ignoring $P_3$- and higher-order photoemission events, $N(1\tau_0)/N(0\tau_0) = e^{-\lambda}$. This indeed predicts a $P_2$-peak taller than the rest of the decaying $P_1$-dominated peaks unlike with our experimental results of Fig. 3-13 and Fig. 3-14(a).

**A5.** Lastly, we will consider the general realistic situation with no multi-electron events ignored. Here, we will show that for a large number of incident laser pulses $N_p \gg 1$ which in fact applies to our experiments performed at a laser repetition rate of 76 MHz, $\lim_{N_p \to \infty} \left( N(1\tau_0)/N(0\tau_0) \right) = 1$ for low emission rates. The ratio is larger than 1 as the emission rate increases. This sets the Poissonian level higher than our observed zero-delay peak height on the coincidence spectra of Fig. 3-13 and Fig. 3-14(a). This will prove that our implemented femtosecond laser-driven tungsten nanotip needle source is *sub-Poissonian* on the detection plane.

**Observation A1.** For $\lambda = rt$, the Poissonian probability mass function of Eq. (1-33) becomes

$$P_n(r) = \frac{(rt)^n}{n!} e^{-rt}, \tag{3-17}$$

where *r* is the particle rate. In general, we must note that although we are hunting for deviations from the Poissonian distribution and we do not want to assume that the source is truly random, such departures are negligible in the total counts which are dominated by the single-electron $P_1$-events in our experiments where the count rate $R \ll 1/\tau_0$ (= 76 MHz, the laser repetition rate). We thus expect to encounter digressions from random distribution on the $N_2$-level, and we safely ignore $N_3$ and higher-order terms. With this in mind, for average electron detection count $\bar{N}_d \ll 1,$ striking one detector over its dead time $t = t_d$ (or resolution time if it is longer), we use the Poissonian probability density to evaluate the first-order correction to the source rate $r$ to find the approximate value of the observed detection rate $r_d$ over the dead time $t_d$. The low-counts condition implies $dp = rdt$ from which we conclude,

$$\begin{aligned} \bar{N}_d &= \sum_{n\geq 1}^{\infty} P_n(r) = 1 - P_0 \\ &= 1 - e^{-rt_d} = 1 - \left[1 - rt_d + \frac{1}{2}(rt_d)^2 - ...\right] \\ &\approx rt_d - \frac{1}{2}(rt_d)^2, \end{aligned} \tag{3-18}$$

which is identically the approximate probability for at least one electron to land on the detector face over its dead time. The approximation made in the last line is valid noting to the fact that $rt_d \ll 1$. The corrected detection rate, up to first order, therefore becomes

$$r_d \approx r - \frac{1}{2} r^2 t_d . \tag{3-19}$$

**Observation A2.** Consider a number $N_p$ of the laser pulses irradiating the nanotip over the experiment time. We discretize the time axis onto $N_p$ intervals accordingly, each with duration $\tau_0$, such that $N(n\tau_0)$ is the number of detected coincidence events which are separated by an average time interval $n\tau_0$. For example, $N(1\tau_0)$ is the number of coincidence events separated by approximately one repetition time, namely, those constituting the tall peak centered about $+1\tau_0$ in Fig. 3-14(a). It is important to note that in our formulation of the problem, we correctly take into account successive detections in one detector which do not lead to *coincidence* counts as well. We first take note of the fact that,

*N(nτ0) = [probability to detect at least 1 electron in the start detector]*

×*[probability to detect zero electrons in the stop detector over interval (n-1)τ0]*

×*[probability to detect at least 1 electron in the stop detector]*

×*[total number of time intervals separated by nτ0],* (3-20)

which we formally denote as

$$N(n\tau_0) = P_A \times \left(\tilde{P}_B\right)^{n-1} \times P_B \times \left(N_p - n\right). \tag{3-21}$$

For the detected $n\tau_0$-events we get,

$$N(n\tau_0) = \left[1 - P_0(r_A)\right] \times \left[P_0(r_B)\right]^{k-1} \times \left[1 - P_0(r_B)\right] \times \left(N_p - n\right), \tag{3-22}$$

where $r_A$ ($r_B$) denotes the rate on the *start* (*stop*) detector. At sufficiently low emission rates,

$$1-P_0\left(r_{A,B}\right)=1-e^{-r_{A,B}\tau_0}\approx r_{A,B}, \tag{3-23}$$

and we therefore conclude that

$$N\left(n\tau_0\right)\approx r_A\tau_0\times e^{-r_B(n-1)\tau_0}\times r_B\tau_0\times N_p, \tag{3-24}$$

for $0<n\ll N_p$. It is important to note that the rates $r_A$ and $r_B$ are *detection* rates. From this discussion, we may take the lesson that the exponential decay in the coincidence counts for cumulatively longer time delays is actually governed by the *stop channel rate* as can be grasped from the factor $r_B\left(n-1\right)$ in the exponent.

**Observation A3.** Consider uniform illumination of a double-detector set with individual detectors labeled *A* and *B*, respectively. For identical detectors, the face (sensitive) area *S* is the same for both. Now let us also consider a third detector *C* with properties identical to each of the detectors of the two-detector assembly except for its face area which is twice as large as that of the detector *A* (or *B*). Take an identical point source positioned at the same distance from the center of the double-detector combination and the detector *C* in each case. This way, everything is identical in the two experiments except that $S_C=2S_A=2S_B$. For uniform illumination, this implies that $r_C=2r_A=2r_B$. The question is, whether the correct coincidence term in the double-detector scenario for single-electron pulses is $P_1^AP_1^B$, or is it $P_1^AP_1^B+P_1^BP_1^A=2P_1^AP_1^B$? In the trivial notation employed here, $P_1^A$ ($P_1^B$) is the probability to detect one electron at *A* (*B*). As before, we neglect $N_3$ and higher-order emission events.

To answer this question, we compare the two settings for emission of an electron pair. Without any loss of generality, we assume that the detection face for the identical detectors under consideration is large enough so that the beam divergence encompasses the entire sensitive area; in other words, we assume that all the emitted electrons get detected with 100% efficiency. Imperfect detection efficiencies due to propagation losses and finite dead times will be accounted for in our final exact analysis. For random emission, the probability to detect *two* electrons at *C* at a source rate *r* is given by

$$P_2^C\left(r\tau_0\right)=\frac{\left(r\tau_0\right)^2}{2!}e^{-r\tau_0}, \tag{3-25}$$

using Eq. (3-17). In contrast, assuming the coincidence term is given by the product $P_1^A P_1^B$ (without an additional factor of 2), we observe that in the double-detector setting,

$$\begin{aligned} P_2^{A\&B} &= P_1^A\left(\frac{r}{2}\tau_0\right)P_1^B\left(\frac{r}{2}\tau_0\right)+P_2^A\left(\frac{r}{2}\tau_0\right)P_0^B\left(\frac{r}{2}\tau_0\right)+P_0^A\left(\frac{r}{2}\tau_0\right)P_2^B\left(\frac{r}{2}\tau_0\right) \\ &=\left[\left(\frac{r}{2}\tau_0\right)e^{-(r/2)\tau_0}\right]^2+2\times\frac{1}{2!}\left(\frac{r}{2}\tau_0\right)^2 e^{-r\tau_0}=\frac{1}{2}r^2{\tau_0}^2e^{-r\tau_0}=P_2^C\left(r\tau_0\right). \end{aligned} \tag{3-26}$$

The correct coincidence term hence does not contain an extra factor of 2; it represents one electron going to *A* and the other striking *B*, regardless of the permutation order.

We now inspect the case of two-electron pulses emitted from the source. The probability for both electrons to be detected at *C* is $P_2^C\left(r\tau_0\right)$; perfect detection efficiency is assumed as before. Here, we cannot use the Poissonian probability $P_1$ which pertains to single-electron pulses as two-electron pulses are being under consideration, exclusively. We take resort to combinatorics: Once the electron pair is emitted, there is $(1/2)^2 = 25\%$ chance for both of them to strike either the detector *A* or *B*. These two possibilities, with a total probability of 50%, do not give rise to any coincidence counts. The remaining possibilities, with the probability of 50%, which constitute the two permutations of the case where one electron gets detected at each of the two detectors *A* or *B*, therefore lead to coincidence clicks. Note that we are considering a two-electron pulse emitted at a time, exclusively. We will consider the combinatoric generalization to this involving one- and two-electron pulses after the exact analytical determination of the predicted Poissonian coincidence distribution.

**Observation A4.** We are now ready to tackle the main problem. We do that first ignoring $P_3$- and higher-order photoemission events. Consider the double-detector of the previous observation. Let us assume that the detector *A* (*B*) provides the *start* (*stop*) input to the TAC. We denote the probability for each emitted electron to strike the detector *A* (*B*) after propagation from the source to the detection plane with $\varepsilon_A$ ($\varepsilon_B$).

We proceed, having the results of the previous three observations in mind. The number of counts on the $P_2$-peak is

$$N(0\tau) = P_2(\varepsilon_A\varepsilon_B + \varepsilon_B\varepsilon_A)N_p = 2P_2\varepsilon_A\varepsilon_B N_p. \qquad (3\text{-}27)$$

For single-electron-dominated coincidence counts about $+1\tau_0$ we get

$$N(1\tau) = P_1\varepsilon_A P_1\varepsilon_B(N_p - 1), \qquad (3\text{-}28)$$

and similarly,

$$N(2\tau) = P_1\varepsilon_A P_0 P_1 \varepsilon_B (N_p - 2),$$
$$\vdots \tag{3-29}$$
$$N(n\tau) = P_1\varepsilon_A P_0^{n-1} P_1 \varepsilon_B (N_p - n) \propto (N_p - n) e^{-(n-1)\lambda}.$$

The last line justifies the exponential decay in Fig. 3-14(e). We conclude that

$$\frac{N(0\tau)}{N(1\tau)} = \frac{2\frac{\lambda^2}{2} e^{-\lambda} \varepsilon_A \varepsilon_B N_p}{\left(\lambda e^{-\lambda}\right)^2 \varepsilon_A \varepsilon_B (N_p - 1)} \xrightarrow{N_p \gg 1} e^{\lambda}. \tag{3-30}$$

This shows that the $P_2$-peak for a Poissonian distribution is, in dramatic contrast to our experimental result of Fig. 3-14(a), taller than the first peak to its right by a factor of $e^{\lambda} = e^{r\tau_0}$ in this approximation.

**Observation A5. The general case.** We now take into account the realistic possibility of multi-electron generation per incident fs laser pulse. We now derive an expression for $N(n\tau_0)$ with *n* being an integer. We first derive the desired result for non-zero *n* and then consider the $n = 0$ case. For the former, we start off from Eq. (3-21).

The probability to detect an emitted photoelectron at the detector *A*, is

$$P_A = \left[P_1\varepsilon_A + P_2\varepsilon_A + \ldots + P_k\varepsilon_A + \ldots\right] = \sum_{n=1}^{\infty} P_n \varepsilon_A = \varepsilon_A (1 - P_0) = \varepsilon_A \left(1 - e^{-\lambda}\right), \tag{3-31}$$

where $P_n$ is the Poissonian probability given in Eq. (1-33). For detector *B*, only the subscript *A* should be replaced with *B*. These give the first and the third terms in Eq. (3-21). For the second term we have

$$\left(\tilde{P}_B\right)^{n-1} = \left[P_0 + P_1\left(1-\varepsilon_B\right) + P_2\left(1-\varepsilon_B\right)^2 + \ldots + P_k\left(1-\varepsilon_B\right)^k + \ldots\right]^{n-1}$$
$$= \left[\sum_{k=0}^{\infty} P_k\left(1-\varepsilon_B\right)^k\right]^{n-1} = \left\{\sum_{k=0}^{\infty} \frac{\left[\lambda\left(1-\varepsilon_B\right)\right]^k}{k!} e^{-\lambda}\right\}^{n-1} \tag{3-32}$$
$$= \mathrm{e}^{-(n-1)\lambda} \times \mathrm{e}^{(n-1)\lambda(1-\varepsilon_B)} = \mathrm{e}^{-(n-1)\lambda\varepsilon_B} = P_0^{(n-1)\varepsilon_B},$$

in which, quite obviously, the term $\left(1-\varepsilon_B\right)$ is the probability for an emitted electron to not strike the detector *B*. Plugging Eqs. (3-31,32) into Eq. (3-21) we find that

$$N\left(n\tau_0\right) = \varepsilon_A\varepsilon_B\left(1-e^{-\lambda}\right)^2 e^{-(n-1)\varepsilon_B\lambda}\left(N_p - n\right). \qquad n \neq 0 \tag{3-33}$$

To predict the number of coincidence events $N\left(n\tau_0\right)$ as a function of time delay $\tau = n\tau_0$, we inspect the dependence on *n*. For a data acquisition time of 900 s as in Fig. 3-14, the number of laser pulses $N_p \simeq 7\times10^{10} \gg n$. In Fig. 3-14(e) that corresponds to an expanded coincidence window of 100 μs, $n < 10^4$ and the factor $N_p - n$ in Eq. (3-33) is nearly constant. The remaining *n*-dependent factor is $e^{-n\varepsilon_B\lambda} = e^{-n\varepsilon_B r\tau_0}$ (see Eqs. (3-17) & (3-25)) which decays exponentially. In Fig. 3-14(e) the individual delay peaks are not resolved as the timing bin nearly 50 ns (see Table 3-1)). For the zero-time delay peak with $n = 0$, we find that

$$N\left(0\tau_0\right) = \left[2P_2\varepsilon_A\varepsilon_B + 6P_3\varepsilon_A\varepsilon_B + \cdots + n\left(n-1\right)P_k\varepsilon_A\varepsilon_B + \cdots\right] \times N_p$$
$$= \varepsilon_A\varepsilon_B N_p \sum_{n=2}^{\infty} n\left(n-1\right)P_k, \tag{3-34}$$

in which all the possible permutations for $n \geq 2$ emitted electrons to strike the detector pair are taken into account. The reader should be reminded that this peak is formed by multi-electron pulses exclusively hence $n = 1$ is excluded. To evaluate the summation on the RHS of Eq. (3-34), we note that for Poissonian distributions,

$$\sigma_n = \langle n^2 \rangle - \langle n \rangle^2 = \langle n \rangle = \sum_{n=0}^{\infty} nP_n = \lambda \Rightarrow \langle n^2 \rangle = \lambda + \lambda^2. \qquad (3\text{-}35)$$

Subsequently,

$$\sum_{n=2}^{\infty} n(n-1)P_n = \sum_{n=0}^{\infty} n(n-1)P_n = \langle n^2 \rangle - \langle n \rangle = \lambda + \lambda^2 - \lambda = \lambda^2. \qquad (3\text{-}36)$$

From Eqs. (3-34) & (3-36) we therefore deduce that,

$$N(0\tau_0) = \varepsilon_A \varepsilon_B N_p \lambda^2. \qquad (3\text{-}37)$$

Comparing Eqs. (3-33) & (3-37), we conclude that for a Poissonian distribution, the ratio of the content of the peak centered at $\tau/\tau_0 = 0$ to that of the peak centered at $\tau/\tau_0 = 1$ is

$$R_{01} \equiv \frac{N(0\tau_0)}{N(1\tau_0)} = \frac{\lambda^2}{\left(1-e^{-\lambda}\right)^2} \xrightarrow{0<\lambda\ll 1} 1. \qquad (3\text{-}38)$$

In other words, at sufficiently low emission rates, the zero-delay peak is expected to have about the same height as its neighboring tall peak at $+1\tau_0$. As the emission rate increases, more and more multi-electron pulses are created which leads to an enhanced contribution to the zero-delay peak. Consequently, the height of this peak (to be precise its content) increases beyond that of its neighboring peaks making it even taller than them as can be seen in Fig. 3-15. The fact that in Fig. 3-14(a), the zero-delay peak is shorter therefore demonstrates that the photoelectrons distribution on the detection plane is indeed *sub-Poissonian* unambiguously by a

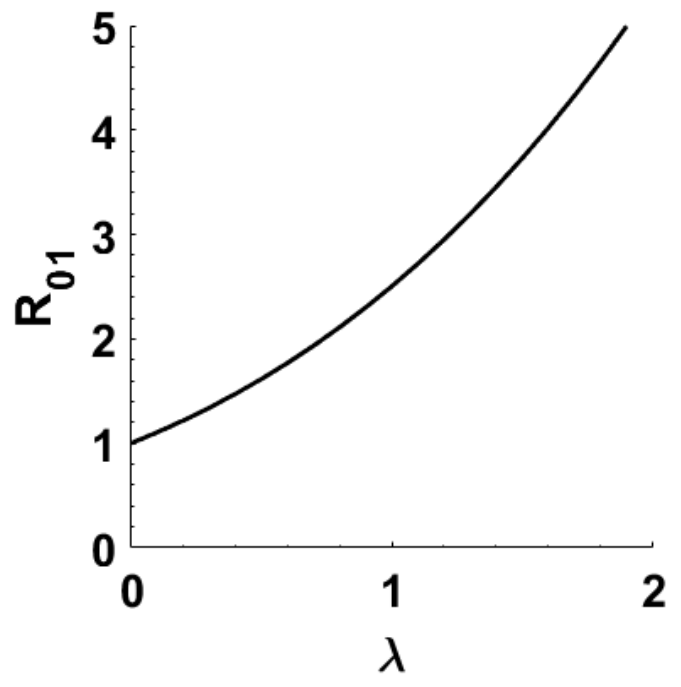

**Fig. 3-15 | The ratio of the zero-delay peak content to that of the +1$\tau_0$ peak.** $R_{01}$ from Eq. (3-38) is plotted against the mean value of the distribution, λ. The ratio is always greater than 1 and increases by increasing λ monotonically.

(negative) relative ratio

$$V \equiv \frac{N^{\exp}(0\tau_0) - N^{\exp}(1\tau_0)}{N^{\exp}(1\tau_0)} = -24.4\%, \tag{3-39}$$

for an estimated $N^{\exp}(0\tau_0) = 1.9963\times10^5$ and $N^{\exp}(1\tau_0) = 2.6403\times10^5$.

This is also significant from a different perspective: Non-random electron beams may lead to new technologies such as correlative electron microscopy and on-demand single-electron sources, aside from the production of quantum degenerate free electron beams (see CH. 2 and the corresponding introductory discussions in CH. 1). While the first report on such beams [3] (using a high-brightness CW source) involved production of an antibunching signal on the order of *1 part in 1000*, here we have demonstrated such a signal, this time on the order of *1 part in 4*. The approximately 250-fold enhancement achieved in the present research may help pave the way for the novel free electron technologies mentioned above. The effect was robust and always observed for several different nanotips.

Let us now consider the antibunching mechanism. It is fair to assume that the (multi-photon) photoemission process itself is Poissonian. If so, then we are left with two reasonably possible mechanisms to explain the non-randomness of the photoelectrons distribution on the detection plane: *i*) the so-called Pauli force, and *ii*) the e-e Coulomb repulsion. We first investigate if quantum degeneracy that leads to Pauli force can justify at least a fraction of the observed antibunching signal in the present experiment.

The HBT dip contrast $C_{HBT}$ was studied in detail for 1D propagation in CH. 2 (see Eqs. (2-23) & (2-24)). The magnitude of the antibunching signal, namely the dip $D_{HBT}$ can also be readily calculated as

$$D_{HBT} = \frac{P^0_{incoh} - P^0}{P^0_{incoh}} = 1 - \frac{P^0}{P^0_{incoh}} = \frac{1}{N}, \tag{3-40}$$

from Eq. (2-20) and the same considerations that led to Eq. (2-23). The quantity $N$ was previously defined in Eq. (2-18). Straightforward generalization of this result to 3D propagation yields,

$$D_{HBT} = \frac{1}{4\frac{\Delta t_e}{\tau_c}\frac{X_{tip}}{X_c}\frac{Y_{tip}}{Y_c}}, \tag{3-41}$$

in which $\tau_c$, $X_c$, and $Y_c$ are the coherence time and the transverse coherence lengths, $\Delta t_e$ is the electron pulse duration, and $X_{tip}$ and $Y_{tip}$ are the estimated sizes of the nanotip in the transverse (x and y) directions. The minimum energy-time uncertainty relation limits the coherence time to, $\tau_c = \hbar/2\Delta E = 0.66\ fs$ for $\Delta E = 0.5\ eV$. The coherence lengths, on the other hand, can be estimated from the position-momentum uncertainty relation. The angular width of the distribution of the electrons that reach the detection plane is

$$\gamma = \tan^{-1}\left(\frac{\det_{wx} + X_{\det}/2}{L}\right), \tag{3-42}$$

where $X_{\det}$ is the distance between the detectors centers and $\det_{wx}$ is the detectors opening size. The transverse ($x$) component of the uncertainty in the linear momentum therefore becomes

$$\Delta p_x = 2p\gamma. \tag{3-43}$$

The momentum is $p = m_e v$ for 100 eV electrons travelling with speed $v = 6\times10^6\ m/s$. Consequently, $X_c = \hbar/2\Delta p_x$, and $Y_c = X_c$ is assumed. The emission site lateral widths are estimated as $X_{tip} = 2R_{tip}\sin\gamma$, where $R_{tip}$ is the nanotip emission site radius of curvature, and $Y_{tip} = X_{tip}$ is assumed. For these parameters the effect of quantum degeneracy turns out to be negligible in the present experiment as can be seen in Fig. 3-16. This was in fact expected given that the coherence volume on the detection plane was not expanded (see Fig. 1-7).

Detailed analysis of the problem based on a simulation involving the classical e-e Coulomb repulsion is underway. We can still indirectly verify or refute our remaining conjecture that it is the Coulomb interaction that has caused electron antibunching. To this goal, we have simulated the coincidence experiment starting from a Poissonian distribution of photoelectrons emerging from a nanotip with a similar energy spread as in our experiment, except that the Coulomb interaction was turned off. In the subsequent section, we will show that the outcome is in fact in complete compliance with our analytical calculation that led to Eq. (3-38), demonstrating that in the absence of e-e Coulomb interaction, we would have observed an identical height for the zero-delay peak compared with that of its nearest neighbor counterpart.

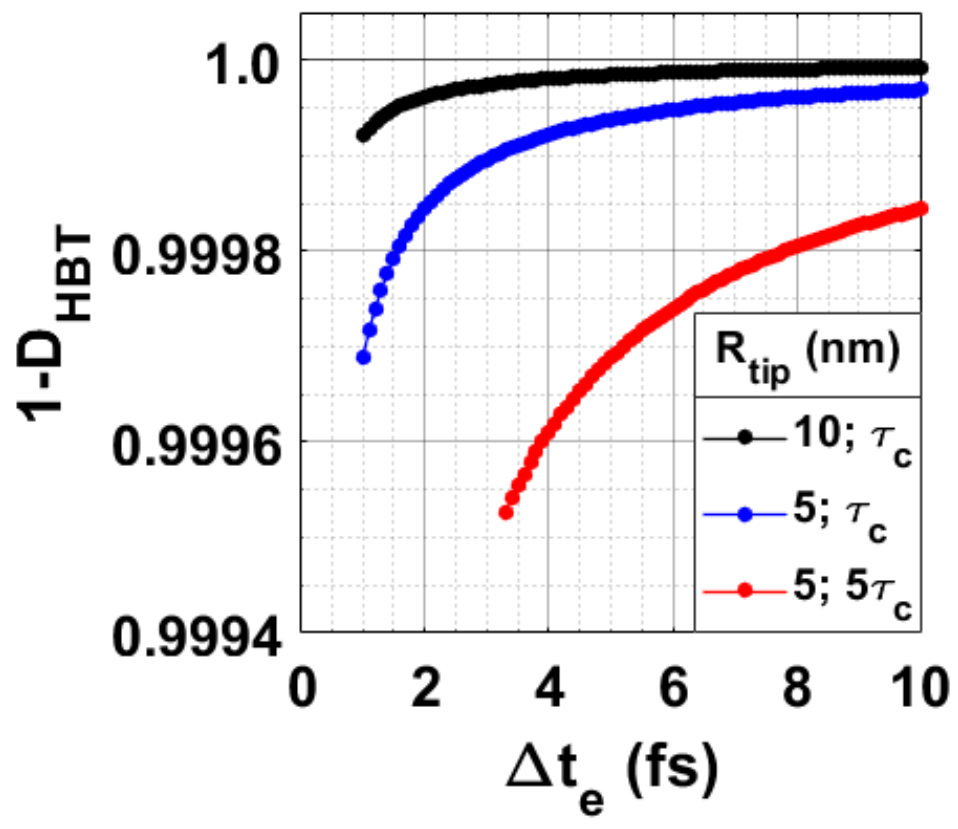

**Fig. 3-16 | The quantum degeneracy dip as a function of electron pulse duration.** Three different combinations of nanotip emission site radius of curvature and coherence time are compared. Even a sharp tip and a coherence time 5 times the estimated value (red line) cannot justify the unprecedentedly large antibunching signal of 1 part in 4 observed in the present experiment.

### 3.3.1 Computer simulation of Poissonian distributions

We used the software package SIMION™ to inspect the shape of the coincidence histogram that can be generated from its output raw data. The setup was defined in SIMION™ and the simulations were performed by undergraduate student Will Brunner from Prof. T. J. Gay's group here at UNL. We ignored the Coulomb interaction in the two-electron pulses for the reason explained above. The physical dimensions were taken to match those of our setup which consists of the tip emitter, an aperture, and the double-detector. A kinetic energy of 100 eV, and an energy spread of 1 eV were considered. The electrons that cross the aperture are produced within a cone of emission angles encompassing the aperture hole.

The computation setup in SIMION™ is shown in Fig. 3-17. We first made sure a CW random emission expectedly gives rise to Poissonian distribution on the detection plane. This turned out to be true which also indicates that as anticipated, the mutual detection setup can in fact be used to study the particle statistics. We then moved on to a pulsed point source which is more relevant to our problem and thus we consider it here. The physical parameters of the ultrashort pulses are the same as the realistic situation, namely, 50-fs pulses with the repetition time of 13.2 ns. Let us first explain how a histogram of the mutual coincidence detection times is generated through the MATLAB™ code written by the author of the present dissertation to analyze the raw output data of SIMION™. I forgo the initial trivial steps which involve, for instance, removal of the (redundant) information on the electrons which have escaped detection. For symmetrically mirrored detectors with respect to the *x*-*z* plane, which detector has clicked is determined from the sign of the *y*-component of the detection point leaving the rest of the code with a column of detection times and a corresponding column of +1's and -1's, denoting detection at the top or the bottom detector, respectively. The algorithm must subsequently ensure that none of the detection times are used twice to generate the histogram of the coincidence events as in the realistic situation. Additionally, assuming $D_1$ is *start* and $D_2$ is *stop*, the correct sign for each coincidence event should also be determined.

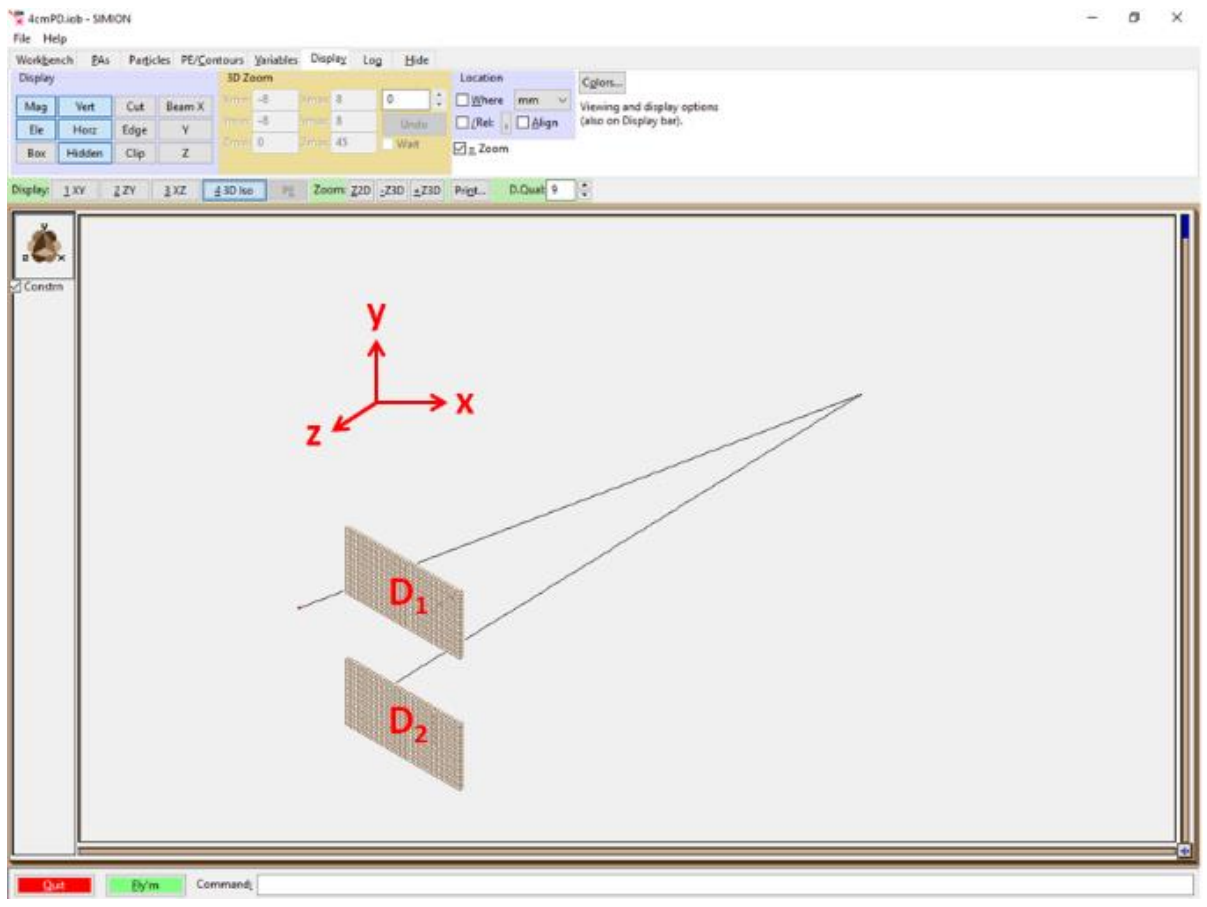

**Fig. 3-17 | The experimental setup defined in SIMION™.** A pair of representative electrons are shown to have been emitted from the point source on the right. The two rectangles are the detectors. As can be seen in this representative example, one of the electrons of the two-electron pulse is detected by the bottom detector while the other one is left undetected. The simulation is performed by Will Brunner from Prof. T. J. Gay's group.

| | t (μs) | Sign | Raw Coincidences | Real Coincidences |
|---|---|---|---|---|
| **1** | 0.00677514600000000 | -1 | -2 | -2 |
| **2** | 0.191603011000000 | 1 | 0 | 0 |
| **3** | 0.336792193000000 | 1 | 2 | 2 |
| **4** | 0.666773003000000 | -1 | -2 | 0 |
| **5** | 1.26080163100000 | 1 | 0 | 0 |
| **6** | 1.40605183800000 | 1 | 2 | 2 |
| **7** | 1.44560354800000 | -1 | -2 | 0 |
| **8** | 1.82837141500000 | 1 | 2 | 2 |
| **9** | 1.93401291600000 | -1 | 2 | 0 |
| **10** | 2.47516686700000 | -1 | | |

**Table 3-3 | Algorithm to generate coincidence histograms from the raw SIMION™ data.** Refer to the main text for an explanation of the computing procedure.

The first 10 detection times of a representative data set are listed in the second column of Table 3-3. Next, the MATLAB™ code finds which detector has clicked for each detection time; +1 for the top detector, and -1 for the bottom detector. In this example, we should achieve a negative coincidence point for the detection times $t_1$ and $t_2$. Noting that up to this point $t_2$ is already used once, the next coincidence event should be between $t_3$ and $t_4$ which is positive in this case, then $t_5$ and $t_6$, etc. The key observation again is that $t_4$ cannot click with $t_5$ as it is already used with $t_3$. The first element of the column labeled Raw Coincidences is obtained by subtracting the second element of column Sign from its first element, the second element is similarly Sign (3) – Sign (2), and so on. A closer look at the table solves the problem; we cannot allow consecutive 2's and -2's as those exclusively correspond to illegal repetitions. The MATLAB™ code then finds such repetitions, and in each occasion, it replaces a repeated 2 or -2 with a zero. Now we can work with the last column; the 2's correspond to positive coincidence events, and -2's to negative coincidences. Take for instance the -2 on the first row. The associated coincidence time interval is $-(t_2 - t_1)$ as expected. The 2 in the third row correctly gives $+(t_4 - t_3)$. Lastly, all the zeros in the final column correspond to *no* coincidences either due to actual successive detections on a single detector as for the zero in the fifth row, or due to avoiding repeated usage of the same detection time as for the zero in the fourth row, or both, as for the zero on the second row.

In a simulation of a pulsed experiment, two particle rates of $r = 10^7$ cps and r/2 crossing the aperture were considered. At a laser repetition rate of 76 MHz, the first corresponds to

$$\begin{cases} P_0 = 86.84\% \\ P_1 = 12.25\% \\ P_2 = 1 - P_0 - P_1 = 0.91\% \end{cases} . \qquad (3\text{-}44)$$

Similarly, at the reduced rate of r/2 we get

$$\begin{cases} P_0 = 93.42\% \\ P_1 = 6.36\% \\ P_2 = 1 - P_0 - P_1 = 0.22\% \end{cases} . \qquad (3\text{-}45)$$

The number of zero-, one-, and two-electron pulses inputted to SIMION™ in a random order can therefore be inputted from Eqs. (3-44) & (3-45) to generate the desired Poissonian emission. The resulting histograms are shown in Fig. 3-18 for

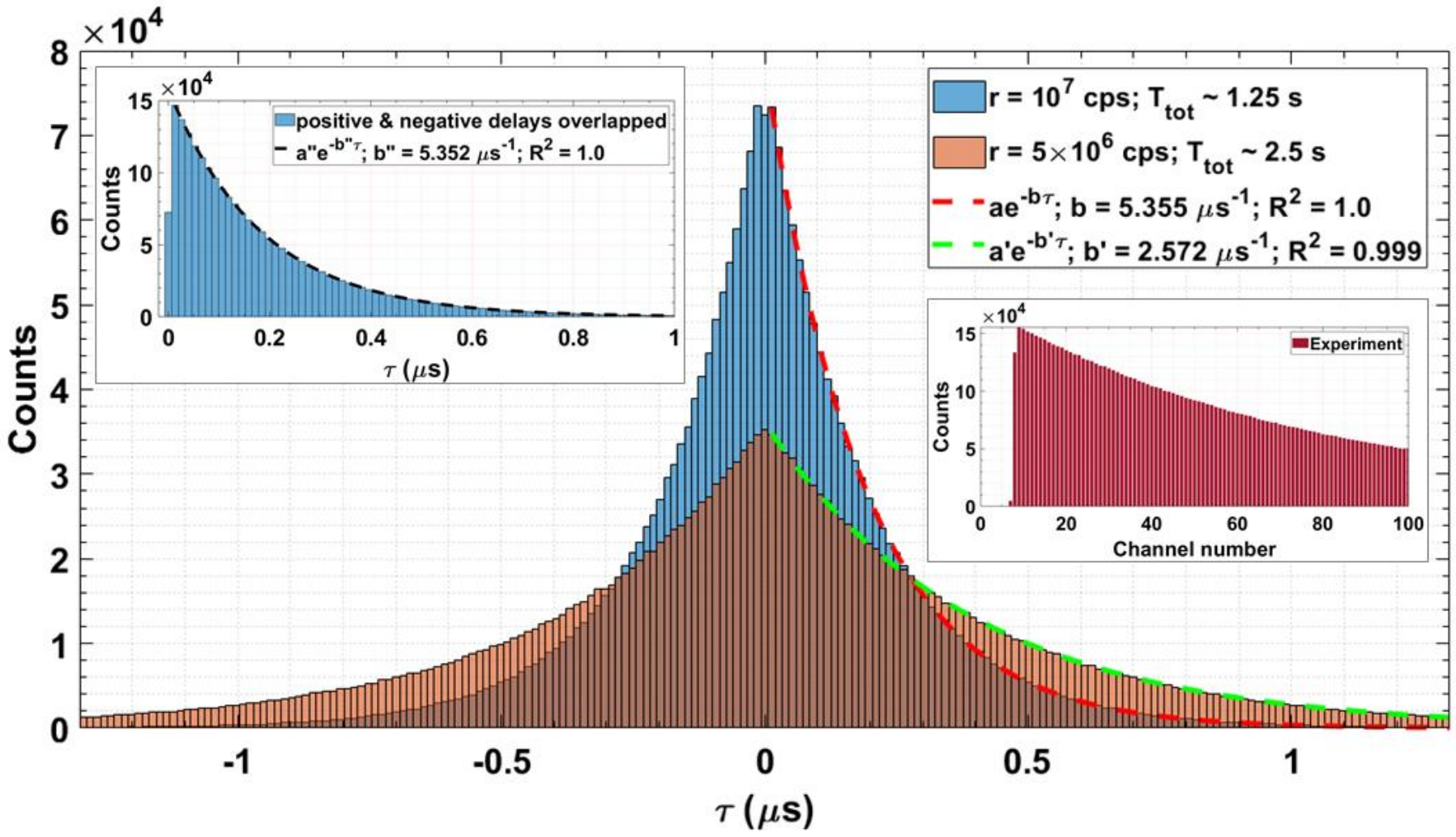

**Fig 3-18 | The computed coincidence histogram resulted from a simulation of the mutual detection events.** A Poissonian pulsed source with the approximation $P_2 = 1 - P_0 - P_1$ (due to computational resource constraints) is taken for this simulation at two different particle rates indicated in the legend box on the top-right corner. $T_{tot}$ is the exposure time in each case. The zero-delay peak $P_2$ at the higher particle rate is slightly shorter than its neighboring tall peaks. This is precisely because of the above approximation implemented here which is not appropriate for such a high rate. For an analogous sub-Poissonian distribution, this peak would be even shorter than this. For the second histogram at a lower rate, the deviation from the Poissonian distribution caused by the implemented approximation is much less dramatic. Consequently, the $P_2$-peak is correctly observed to be taller than its side-peaks. Both histograms are seen to decay perfectly consistent with the exponential trend of a random particle stream. The respective coincidence rates $b$ and $b'$ are obtained by curve-fitting, and as expected from the corresponding particle rates denoted with $r$, their ratio follows that of the latter. The inset on the top-left corner is identically the blue histogram corresponding to the electron stream at a higher rate with the negative-delay bins superimposed on their positive-delay twins. As a result, each $N_1$ bin-height is approximately twice higher in this plot. The lesson to take from here is, the fact that we fail to

observe 50% of the coincidence events when no external delay is used (as in Fig. 3-14(e)) does not affect the *observed coincidence rate* pertaining to the real situation which is of course expected as doubling all the bin heights doubles the constant *a* only and not the coincidence rate *b*. The inset on the right-middle is the experimental histogram showing the first one hundred MCA channels of Fig. 3-14(e). The channel width is 48.8 ns (see Table 3-1) which encompasses 3.7 repetition times. Nonetheless, we can still observe the pair of reduced bins at the beginning (left end) of the dynamic range of the MCA reflecting, in part, the $N_2$ contributions.

---

a bin width of 13.2 ns. The key observation is the exponential decay of the coincidence rate. It is noteworthy to point out that in both of these simulation results, approximately 50% of the electrons crossing the aperture get detected by one of the two detectors. Among them, about 35% ultimately give rise to both positive and negative coincidence counts leading to a total coincidence counts of approximately 16% for the particle rate crossing the aperture, or about 8% for positive-delay counts only.

We then attempted to simulate the Poissonian equivalent of our main experimental results of Fig. 3-14(a,e). As explained after Fig. 3-16, this was to confirm if the observed antibunching effect is caused by e-e Coulomb repulsion in multi-electron pulses. A virtual aperture at the distance of 3.6 cm from the tip apex was considered and 100 eV electrons with a FWHM of 1 eV were emitted at isotropic random angles from a point source subtended by this aperture. The SIMION™ 8.1 program simulated 70 s of experiment time. An electron pulse lasting 50 fs was generated every $\tau_0 = 13.1617\, ns$. The number of electrons per pulse was determined based on the Poissonian probability distribution with the count rate through the aperture of $r = 7.7095 \times 10^5\ cps$. The time at which each electron was emitted during the 50 fs pulse duration was determined randomly from a uniform temporal distribution function: 0-, 1-, 2-, and 3-electron pulses were taken into account. As before, the SIMION$^{TM}$ simulation was performed by Will Brunner from Prof. T. J. Gay's group and the raw data were subsequently analyzed through a MATLAB$^{TM}$ code by the author of the present dissertation as explained in Table 3-3.

The tall peaks of Fig. 3-14 (a) have a width of approximately 5 ns. The typical CEM dead time is about 0.5 ns which cannot justify this amount of broadening. A finite detector resolution time due to an angular distribution of the incident electrons, however, provides a reasonable explanation: It takes a longer time for an electron that has entered the CEM mouth at a sharp angle relative to the x-axis (see Fig. 3-5) to leave its mark on the collector plate at the CEM end by a cloud of secondaries than a more forward-propagating electron. The laser-electron coincidence peaks of Fig. 3-4 show a width of less than 3 ns which is consistent noting that in that case only one CEM is involved. A set of 4 Gaussian random number generators all with mean value 0 were therefore used in the MATLAB$^{TM}$ code to fit the histogram widths and the local minima to the experimental values. Two of these Gaussian profiles had a standard deviation of 1.1 ns while that of the other two was 2.0 ns. The peak heights were collectively adjusted by varying the histogram bin widths. The result is shown in Fig. 3-19.

### 3.3.2 A combinatoric afterword to analytical calculations

Now that we have demonstrated the experimental results and have drawn analytical conclusions to determine the particle statistics of the electron source – or any particle source, the goal is reached. Nevertheless, let us inspect the problem on pure combinatoric considerations as it seems instructive to look at the data from that angle as well.

A set of possibilities for one- and two-electron pulses in the double-detector setting are depicted schematically and discussed in Fig. 3-20. Our goal is to find the fraction of emitted two-electron pulses which ultimately contribute to the

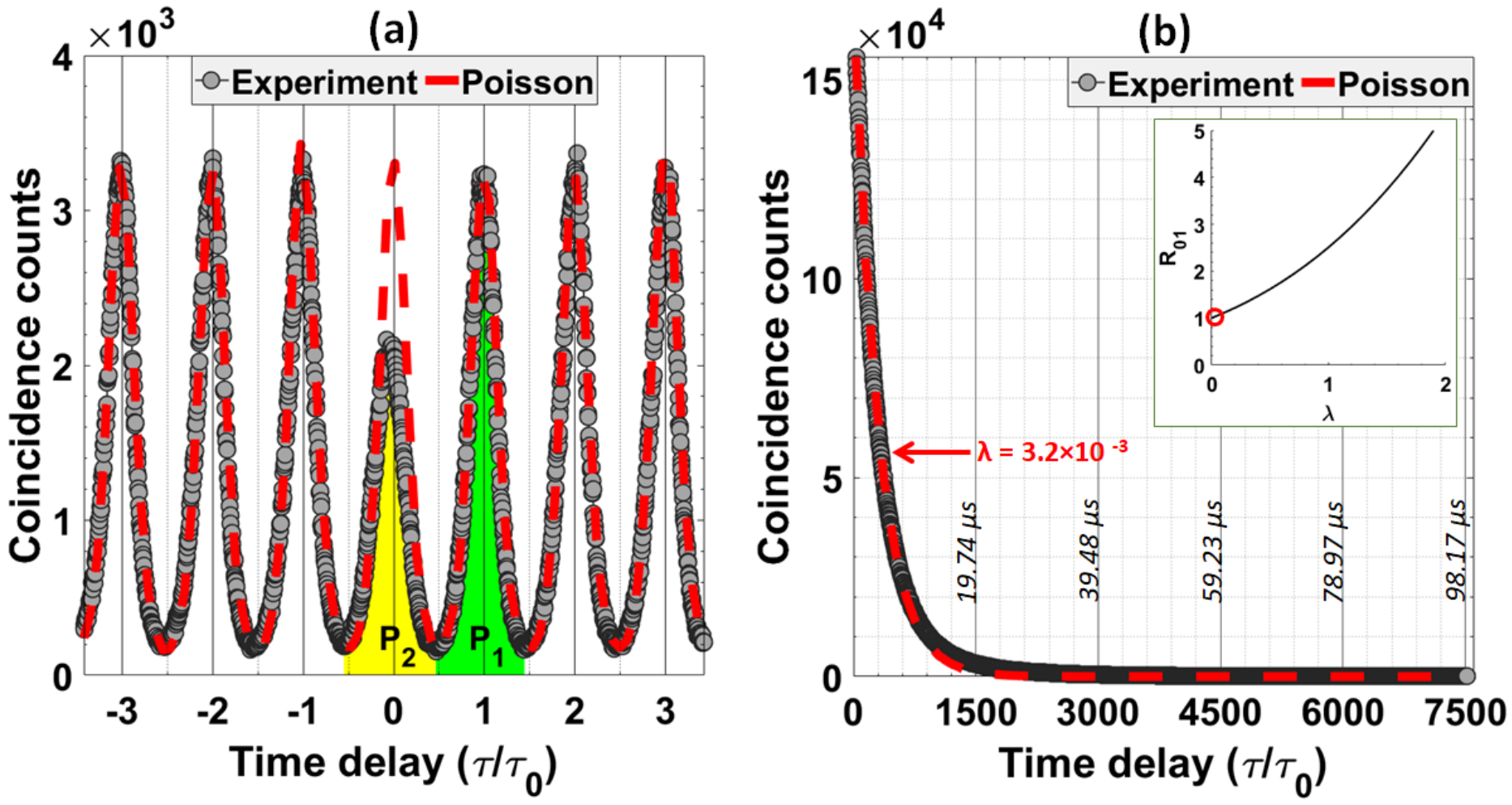

**Fig. 3-19 | A Poissonian distribution would not lead to an antibunching signal.** The grey circles are the same experimental data points of Fig. 3-14 (a,e) with the repetition time $\tau_0$ = 13.1617 ns (76 MHz). The Poissonian equivalent of the identical physical situation, as can be seen in **(a)**, predicts no antibunching signal. The count rate of the photoelectrons crossing the aperture is such that the exponential decay of panel **(b)** is reproduced satisfactorily. The fact that the height of the zero-delay peak (yellow area) is predicted to be the same as that of the first tall peak (green area) is indeed consistent with our earlier analytical result of Eq. (3-38). To emphasize this, the location of the mean value of the distribution $\lambda = 3.2\times10^{-3}$ is marked by a red arrow in (b). The same Fig. 3-15 is shown here in the inset. For such small values of λ within the red circle, equal heights are in fact expected in (a). It was shown earlier in Fig. 3-16 that quantum degeneracy cannot justify the large antibunching signal of 1 part in 4 observed in (a) for a coincidence window of 100 ns. The only other viable hypothesis to explain this observation is thus e-e Coulomb repulsion. The SIMION™ simulation was performed by Will Brunner from Prof. T. J. Gay's group. The coincidence spectra (red dashed lines) were generated through a MATLAB™ code explained in Table 3-3 by the author of the present dissertation.

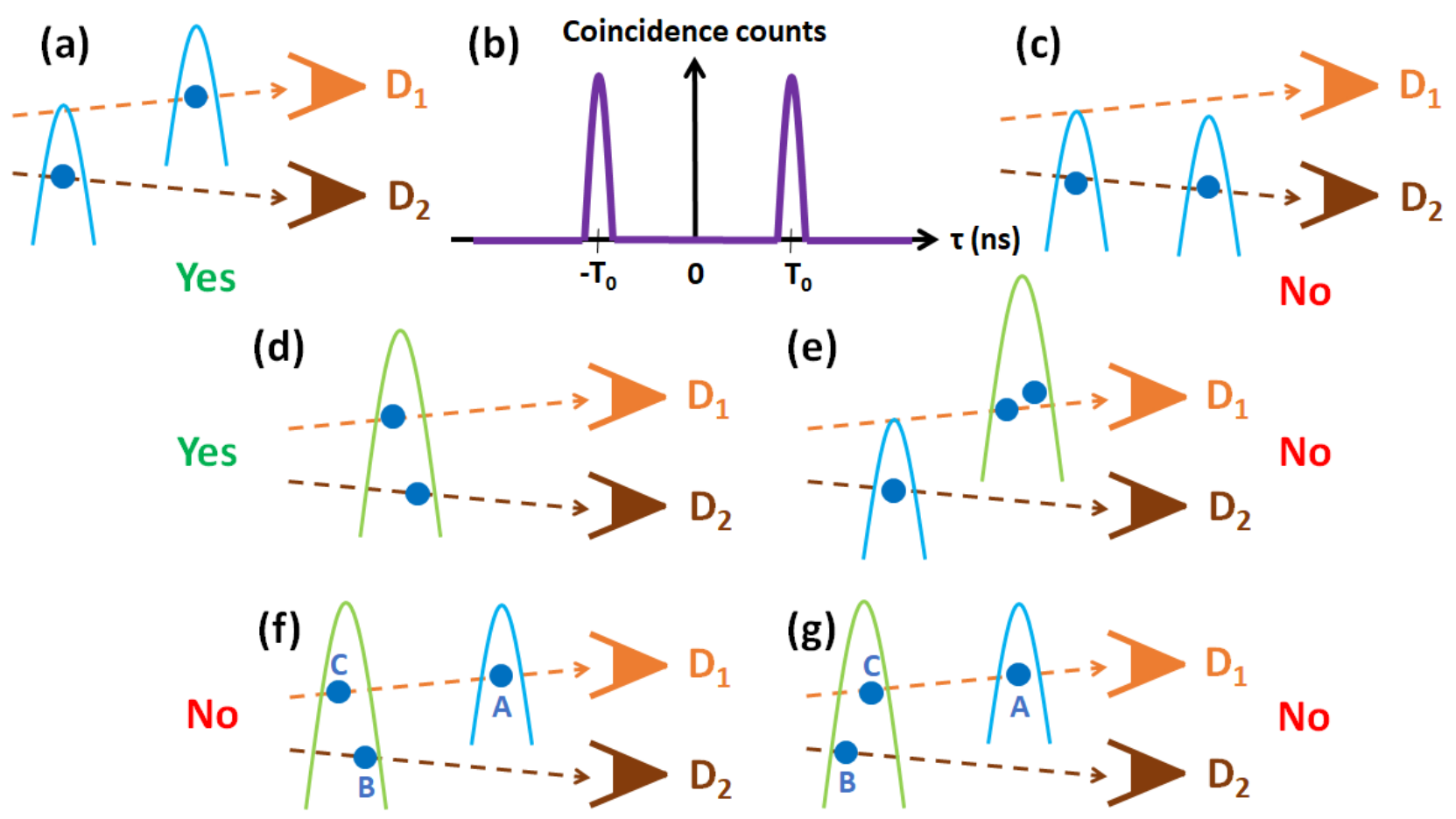

**Fig. 3-20 | Combinatoric considerations for one- and two-electron pulses.** The label *Yes* indicates the coincidence event under consideration is correctly counted, and *No* means otherwise. **(a)** Two single-electron pulses detected by the separate detectors consecutively can be resolved by the coincidence detection electronics. **(b)** Such electrons separated by one repetition time $T_0$ lead to coincidence points at $T_0$ and $-T_0$. **(c)** Both electrons strike a single detector. This alone does not give rise to any coincidence events. **(d)** For this two-electron pulse, each electron clicks at one of the two detectors. As a result, a coincidence point on the zero-delay peak about the zero time delay will follow. **(e)** Here, both electrons of the two-electron pulse strike one detector. This happens either within the dead (or resolution) time of this detector in which case the two electrons are falsely counted as one, or if outside the dead (or resolution) time, two successive clicks at $D_1$ are registered. Same as in panel (c), no coincidence event will follow, however, either the single click in the first case, or the second of the two clicks in the second case, pair with an electron detection event at the other detector. This leads to a coincidence point on one of the tall peaks centered at multiples of $T_0$ which means a fraction of the two-electron pulses also falsely contribute to the dominantly-$P_1$ peaks. **(f)** Either the electron *A* clicks with electron *B* of the two-electron pulse, or if *B* and *C* give rise to only one combined detection event at $D_2$ (similar to (e)), the electron *A* will register a coincidence count with their combination. At any

rate, the two-electron pulse escapes the $P_2$-peak. **(g)** The electrons A and C do not lead to a coincidence point just as in panel (c), however, this means that C *would* click with B similar to the panel (d). Nonetheless, this will not happen in practice. In reality, the TAC start channel opens upon arrival of electron A and *remains open* until a valid event is registered at its stop channel within the preset TAC window (similarly, for when the stop channel receives an input first). This implies that in this panel, electron A gives rise to a coincidence event in conjunction with electron C and thus the two-electron pulse will not be able to leave its trace in the final coincidence spectrum. Overall, a fraction $f_2$ of the two-electron pulses which make it to the detector pair lead to a coincidence click on the $P_2$-peak. *It is crucial to note that all of these possibilities are actually accounted for in the analytical approach of section 3.3.*

coincidence points on the zero-delay peak.

Let us first consider a couple of statistical (S) observations.

**S1.** Every coincidence point corresponds to *two* electron-detection-events, up to multi-electron registrations at a single detector within its dead (or resolution) time which cannot be resolved.

**S2.** Where no external delay is used, as in Fig. 3-14(e), for sufficiently large number of detection events, the coincidence setup only registers 50% of the actual number of total coincidences. This is because those events in which the *stop* detector clicks first are failed to be taken into account. Only when the zero time-delay is shifted to the center of the coincidence window, such as in Fig 3-14(a), shall we reliably take the total number of *observed* coincidences as the total *measured* coincidence counts across that window.

We are now prepared to analyze the results of Fig. 3-14 on combinatoric grounds. The reader must note that the $P_2$-information is lost in panel (e), and in panel (c) we do have access to it though accompanied with a lower MCA resolution. Therefore, panel (a) and its MCS counterpart are the ones we need to lay our focus on to estimate the number $N_2$ of two-electron events *on the detection plane*.

Our recipe to estimate the fraction of two-electron pulses that fail to be recorded on the $P_2$-peak is as follows. All panel labels refer to Fig. 3-14 unless stated otherwise.

1. Find $N_1$, the total number of one-electron events, from panel (b).
2. Express $N_2$, the total number of detected two-electron events from panel (a), in terms of the unknown fraction $f_2$.
3. Find the coincidence rate, denoted with *b*, from panel (e).
4. Find $N_2^P$, the total number of two-electron events, as predicted by the Poissonian distribution with the coincidence rate given by *b*, from the measured $N_1$.
5. Find the fraction $f_2$ from $N_2^P$, using Eq. (3-38) and panel (a).

We now apply this recipe to the data from Fig. 3-14.

**Step 1.** From panel (b), the total detection counts registered for each detector is

$$\begin{cases} C_1 = 130{,}143{,}497 \\ C_2 = 171{,}334{,}847 \end{cases}. \tag{3-46}$$

We can reasonably ignore the two-electron contributions to this as they are negligibly small compared with the single-electron counts. The total number of registered single-electron events, $N_1$, is therefore obtained as

$$N_1 = C_1 + C_2 = 301{,}478{,}344. \tag{3-47}$$

**Step 2.** From panel (a) we take,

$$N_2 = f_2 \times C_0^{coinc.} = f_2 \times 1.9963 \times 10^5, \tag{3-48}$$

where $C_0^{coinc.}$ is the total *coincidence counts* of the $P_2$-peak. We may note that according to S1, the corresponding number of electrons is twice as large which is a trivial observation here as we already know we are considering two-electron pulses in this step.

**Step 3.** Even though we are searching for potential deviations from a true random distribution on the detection plane, at the $N_1$-level, such deviations are marginal. Therefore, it is completely safe to assume that in panel (e), the exponential trend of the random distribution is still valid in the first approximation as we observed in Fig. 3-19. The exponential parameters in the form of $ae^{-b\tau}$ are

$$\begin{cases} a = 1.522 \times 10^5 \\ b = 2.451 \times 10^5 \ s^{-1} \end{cases}, \tag{3-49}$$

where $b$ is the sought-after coincidence rate.

**Step 4.** We can now find $N_2^P$ from Eqs. (3-47) & (3-49). Noting to the time span of the zero-delay peak which is which is one repetition time, the mean value of the Poissonian distribution for this time interval and observed rate is

$$\lambda = b\tau = 0.00323. \tag{3-50}$$

From the fact that for the Poissonian distribution,

$$\frac{N_2^P}{N_1} = \frac{\lambda}{2}, \tag{3-51}$$

the predicted Poissonian number of two-electron pulses on the detection plane is,

$$N_2^P = \frac{\lambda}{2} \cdot N_1 = 4.8769 \times 10^5. \tag{3-52}$$

**Step 5.** From Fig. 3-14(a),

$$\begin{aligned} R_{01} &= \frac{N(0\tau_0)}{N(1\tau_0)} = 0.7561 \Rightarrow N_2 = \zeta N_2^P = 3.6872 \times 10^5 \\ \Rightarrow f_2 &= \frac{N_2}{C_0^{coinc.}} = 1.874. \end{aligned} \tag{3-53}$$

Therefore, the actual number of two-electron pulses *striking the detectors* has been higher than the total coincidence counts recorded on the observed $P_2$-peak by the above factor. In other words, the situation depicted earlier in Fig. 3-20(d) seems to have been realized for $1/f_2 \simeq 54\%$ of the two-electron pulses which have made it to the detection plane in our experiment. Those of Fig. 3-20(e-g) thus constitute the remaining 46% of the possibilities – ignoring three-electron pulses and higher.

## 3.4 Some more tests

In a future experiment, we will investigate whether the antibunching contrast $V$ in this experiment will decrease or not by stretching the pulse duration. The former observation would be in support of Coulomb-induced antibunching as the corresponding repulsive force decreases by the inverse-square of the mean electron separation within each two-electron pulse. It is also instructive to consider other samples for the sake of comparison. For instance, a flat piece of copper photocathode would be of interest to see whether or not the sub-Poissonian behavior is exclusive to nanotip samples.

In another investigatory test, however, we inspected the trend in the relative reduction of the total counts in the $P_2$-peak compared with its immediate neighbors, that is, the two taller peaks at $\pm\tau_0$, with both the incident laser power and tip voltage. We know from Eq. (3-38) that this in fact directly represents the antibunching signal strength (or equivalently the sub-Poissonian contrast *V*). The rationale behind performing this experiment has to do with a crude guess; a higher laser power will increase the multiphoton absorption probability and the number of two-electron pulses $N_2$ will also increase accordingly, which is expectedly accompanied with an increase in $N_1$ (ignoring $N_3$ and higher). But if this is true, so long as the total counts on the pair of side-peaks is a good representative of the number of $N_1$ events, the absolute value of the relative reduction in the zero-delay peak $P_2$ should increase with the laser power if the antibunching signal is genuinely originated from the Coulomb interaction between the two electrons of each $N_2$ event. Put it another way, a higher laser power generates more two-electron pulses, but then the Coulomb force acts to repel them outside of one repetition time. From this follows that some of the coincidence points that would otherwise fall on the $P_2$-peak are now pushed to the taller side-peaks. The absolute value of the relative amount of reduction should then increase. The tip voltage is the other control knob at hand. Increasing this parameter makes the potential barrier narrower (see Fig. 1-9). While this leads to an increase in $N_1$, $N_2$ is not expected to be affected much by it. The relative reduction in the $P_2$-peak should therefore not change considerably by varying $V_{tip}$. In this experiment, the variation of the observed repetition time is also inspected. We know that it is approximately 13.2 ns, equal to the repetition time of the driving femtosecond laser oscillator. We inspect this as a consistency check as well as to see whether there is any distinctively observable broadening in the various coincidence peaks (for short TAC windows). Five different values of the average laser power $P_{ave}$ and seven values of $V_{tip}$ were used for this test at a fixed TAC window of 100 ns. In the former, the tip voltage was held constant at -100 V. In the latter, the laser average power was held fixed at 80 mW. Both in the first case, and in each panel of the second case, $V_f$ is set equal to half of the corresponding $V_{tip}$. During data analysis, the local maxima are identified first (red open circles). The average repetition time in each panel of Figs. (3-21) & (3-22) are then calculated as the average time interval between these consecutive peak maxima. Using the average repetition time and the known time of the maxima, the left (right) edge of each peak is defined as the time corresponding to that of its maximum minus (plus) half of the average repetition time. These are marked with green open circles. The $P_2$-peak at the center and its two side-neighbors now have well-defined intervals. Lastly, the local minima are used to calculate the average coincidence background counts per MCA channel. After background subtraction, we compute the relative reduction of the total counts on the $P_2$ peak compared with the average count of its two immediate neighbors. The result is expressed in percent and the negative sign reflects the fact that the $P_2$ peak contains fewer counts than the average of the pair of taller $N_1$-dominated peaks under consideration. Although it is not possible to draw firm conclusions from these two figures, the method can be taken advantage of in the envisioned experiment mentioned earlier which involves changing the pulse duration for data analysis. The expected increasing trend in the absolute value of the relative reduction in Fig. 3-21(g) which we had anticipated is still noteworthy.

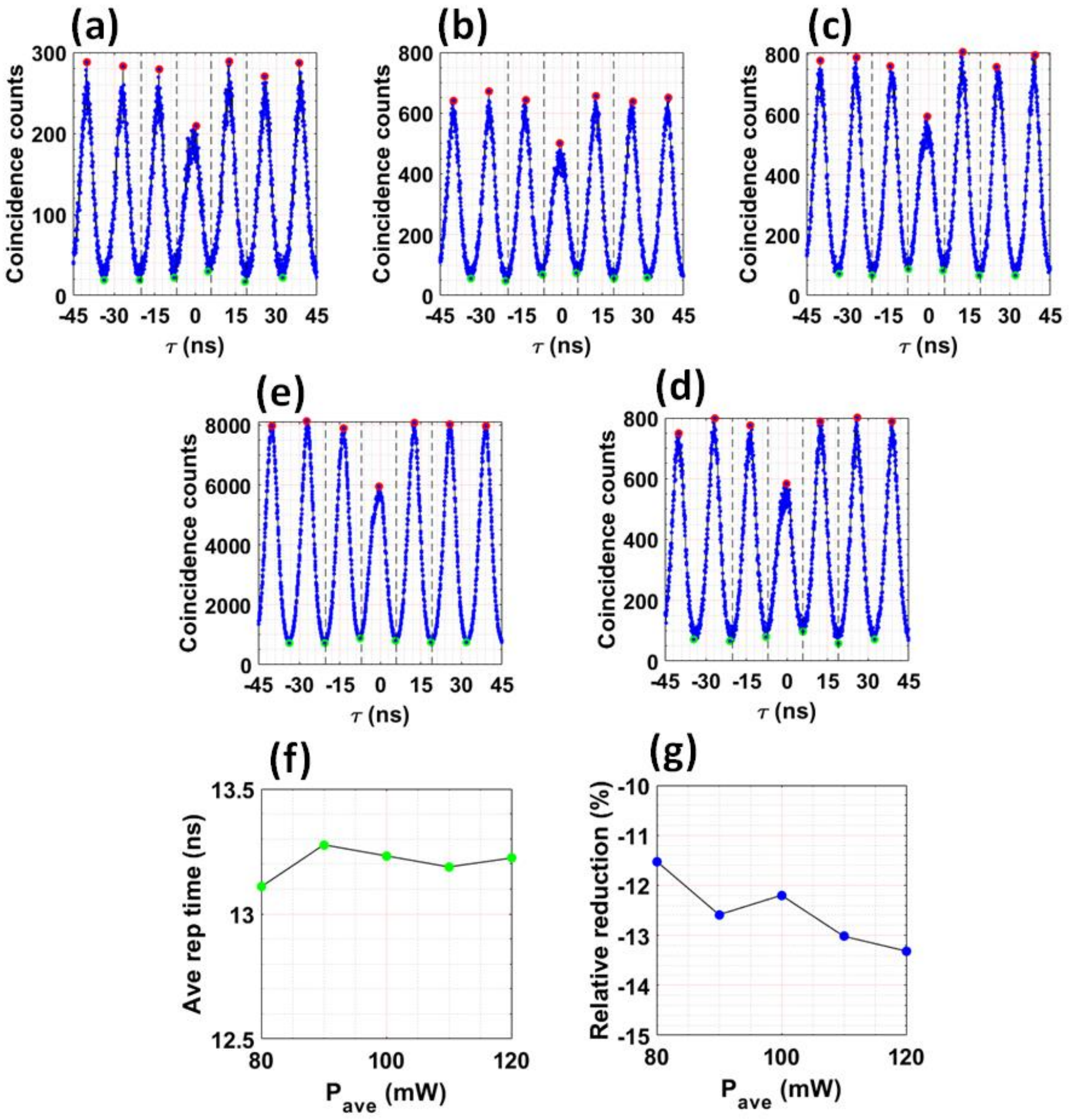

**Fig. 3-21 | Variation of the relative reduction in the $P_2$-peak counts with average laser power. (a-e)** the coincidence spectrum is shown at five laser powers and at a fixed $V_{tip}$. The measurement time is 15 min for each panel. **(f)** The corresponding average coincidence repetition time, and **(g)** the relative reduction in the $P_2$ peak counts are shown against the five laser powers used.

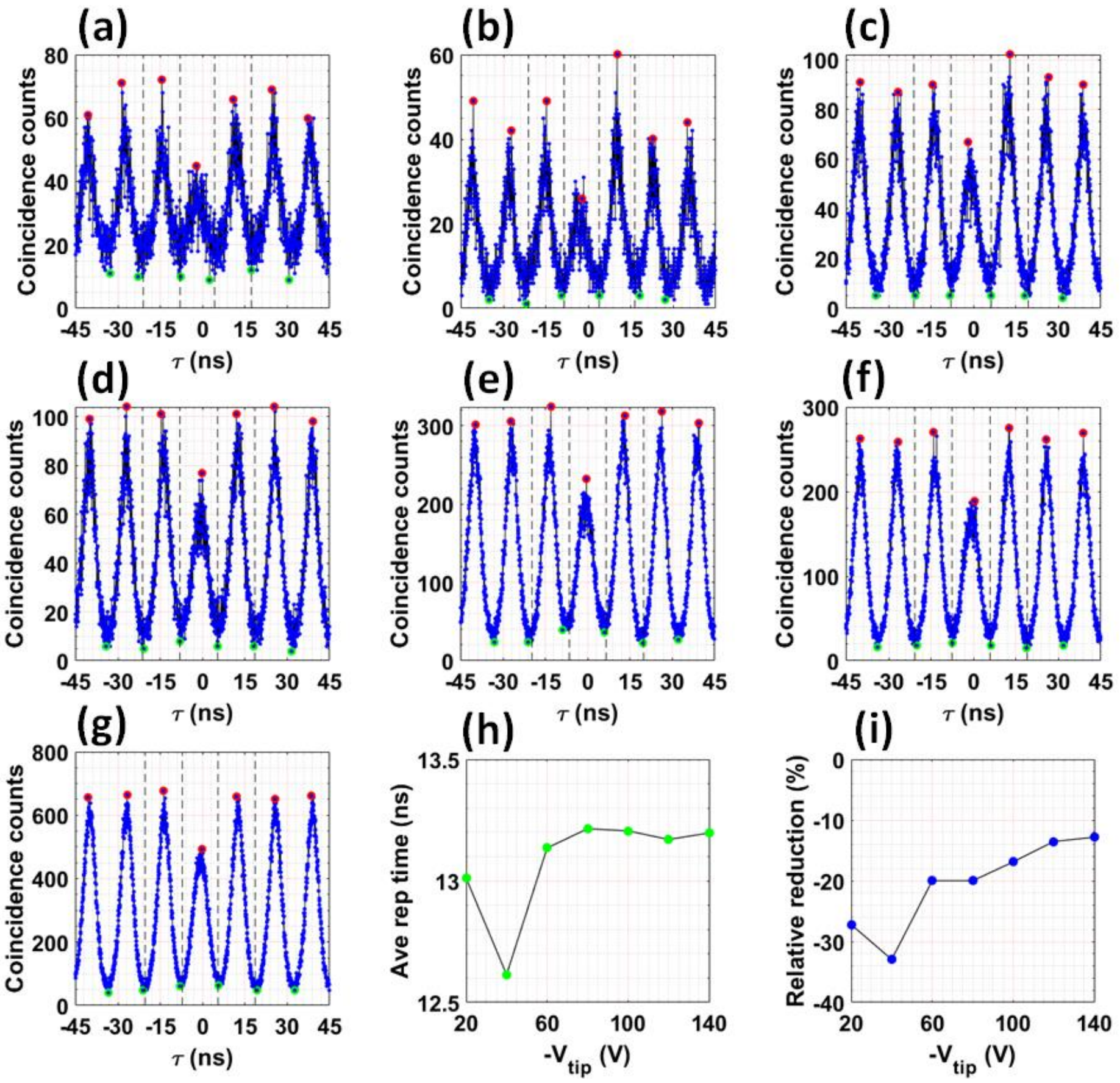

**Fig 3-22 | Variation of the relative reduction in the $P_2$-peak counts with tip voltage.** Here, the laser power is held fixed at 80 mW, contrary to Fig. 3-21. The measurement time is 15 min for each panel (a-g).

## 3.5 Next step: Quantum coherence and the HBT effect

As pointed out a few times by now, the experiment discussed here will be succeeded by a free electron HBT experiment in which a quadrupole or an einzel lens will be used to coherently illuminate the double-detector (see Fig. 1-7). The newly designed home-built chamber is ready and custom-designed magnetic shielding double-layer tubes are at hand. The latter is now a crucial component to add to our set-up as, at longer propagation distances and lower kinetic energies (see Fig. 2-6 on why lower energies are more desirable), suppression of the Earth's magnetic field is imperative. The performance of a electrostatic quadrupole lens in diverging the beam with realistic dimensions simulated with SIMION™ is shown in Fig. 3-23.

A schematic of the new home-built chamber is illustrated in Fig. 3-24, save for some details such as a number of other apertures. We calculate the transverse coherence length to compare it with the typical size of the nanotip photoemitters. This will aid in estimating the amount of magnification needed to achieve with the electrostatic lens stack – an einzel lens, in this example.

The divergence angle is $\gamma = 1\,cm/30\,cm$. From the $x$-component of the position-momentum minimum uncertainty relation $\Delta x \Delta p_x \sim \hbar/2$, the transverse coherence length $L_{coh}$ is estimated as

$$L_{coh} = \Delta x \sim \frac{\hbar}{2\Delta p_x} \approx \frac{\hbar}{2mv\gamma} = \frac{\lambda_{dB}}{4\pi\gamma} = 1.3\,nm, \tag{3-54}$$

where the de Broglie wavelength $\lambda_{dB}$ for 5 eV electrons, $5.5\times10^{-10}\ m$, is used.

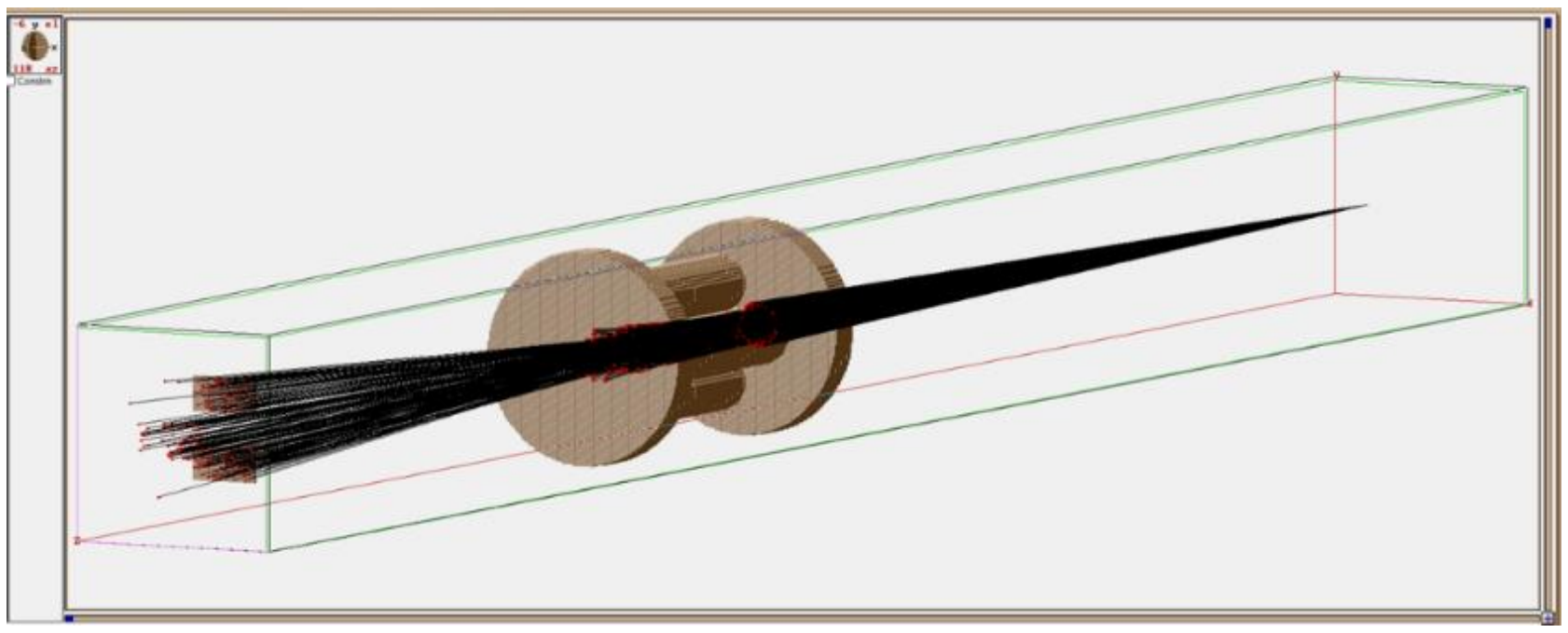

**Fig. 3-23 | The performance of a realistic quadrupole lens is simulated.** The black lines are the diverging 'classical' trajectories of the electrons originated at the point source to the right. On the left end, the two rectangular detectors are visible. The red points indicated blockade of electrons by the quadrupole and the double-detector assembly. See Fig. 1-7 for an explanation on how the size of the coherence volume can be manipulated by such an electrostatic lens. The simulation was performed by Will Brunner from Prof. T. J. Gay's group using SIMION$^{TM}$.

The coherence length is thus shorter than the nanotip radius by only a factor of 10 or so. The required magnification is therefore within reach. A few SEM images of a sharp tungsten nanotip needle that is achieved through the same electrochemical etching process of Fig. 1-4 (a) and makes it for a suitable candidate for this experiment, are shown in Fig. 3-25, with which we wrap up the present chapter.

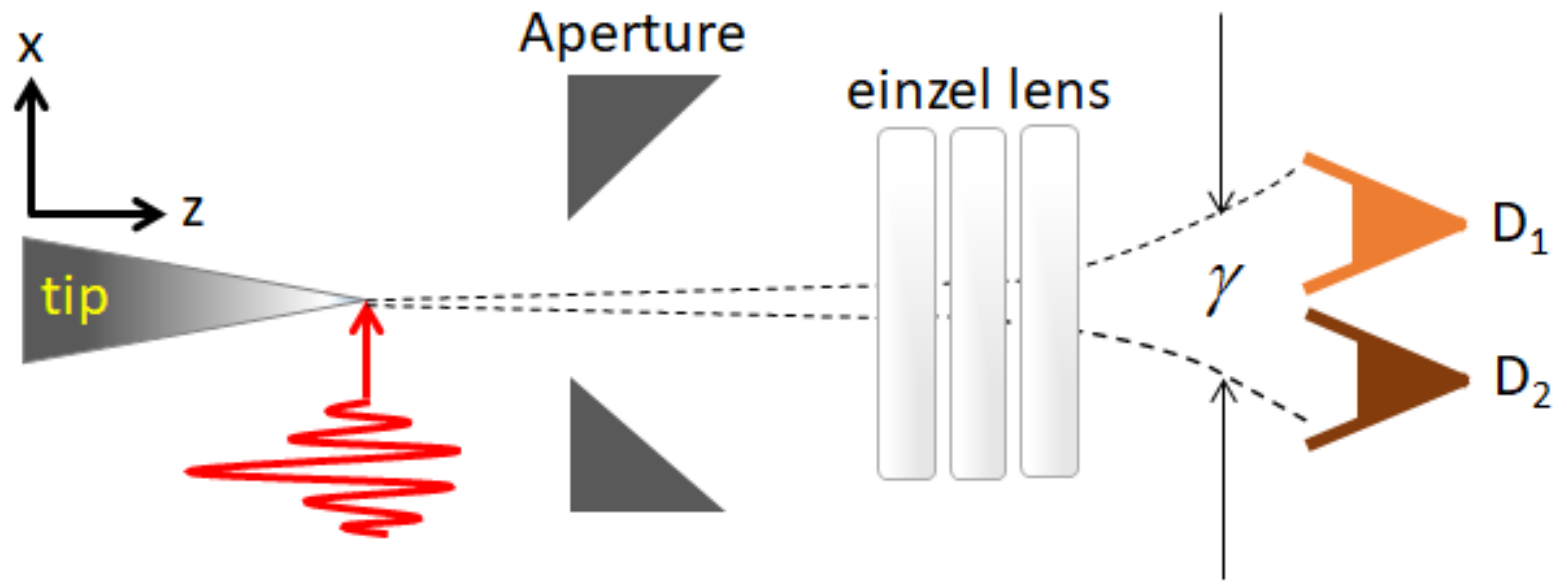

**Fig. 3-24 | Free electron HBT experiment.** A diverging electrostatic lens such as an einzel lens stack expands the coherence volume (see Fig. 1-7). The *coherent* beam divergence angle at the detection plane is $\gamma$ .

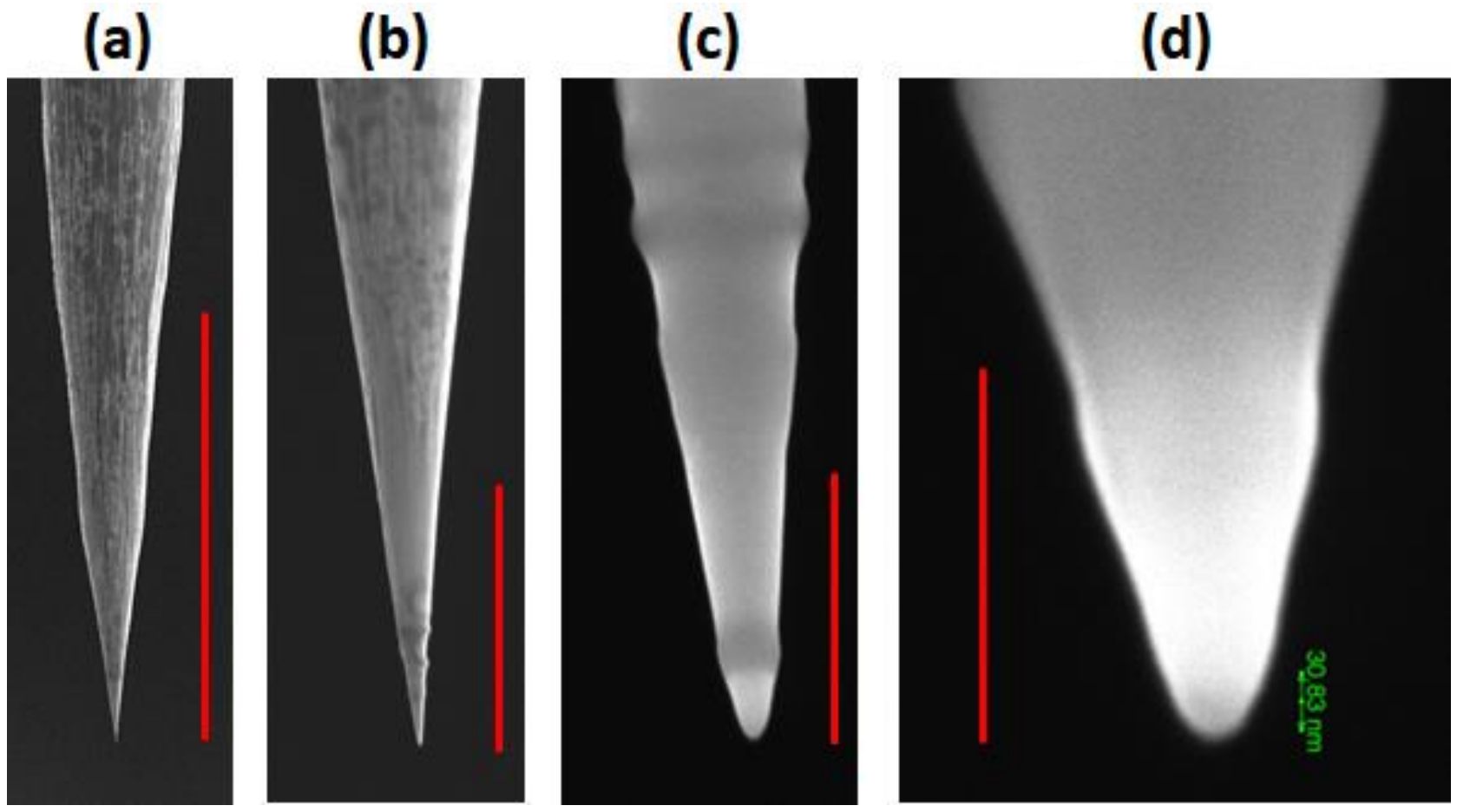

**Fig. 3-25 | SEM images of a sharp W nanotip.** The scale bars are **(a)** 100 μm, **(b)** 5 μm, **(c)** 500 nm, and **(d)** 200 nm. The geometric tip radius is seen to be around 15 nm. See the discussion surrounding Eq. (1-22) on the reported coherence measurements from nanotips. Average photoemission orders of 4-5 and emission currents of up to over 5 nA have been observed from this nanotip at fairly modest pulse energies of 1-2 nJ with linear polarization parallel to the tip shank, and an oscillator repetition rate of 76 MHz.

# CHAPTER 4: A FAST PHOTOEMISSION ELECTRON SOURCE BASED ON SURFACE PLASMON EXCITATION IN METALLIZED TAPERED FIBER OPTIC NANOTIPS

The significance of the ultrafast nanotip electron sources was overviewed in CH.1. It is assumed that the reader has read the introductory sections on Part II of this dissertation.

When an electron or photon interacts with a metal, plasmons may be excited leading to collective electronic effects. The field enhancement and confinement associated with SPR excitation has been used in recent electron photoemission experiments. Examples include SPR-enhanced electron photoemission from the corners of Ag nanocubes [62], and from the surface of a high-brightness nano-patterned Cu photocathode [63]. In both cases, ultrafast lasers, in the long-wavelength regime were used and multiphoton photoemission was identified as the electron emission mechanism. Thermionic emission through IR-absorption was ruled out as a viable channel for the observed enhanced electron emission rates at photon energies a few-fold smaller than the work function of the metals used. SPR-assisted single-photon photoemission in the short-wavelength (UV) domain from a thin film of Al has also been reported [64].

Apart from the need for enhanced emission rates in nanotip emitters, some applications require the proximity of the source to a specimen under study. Ultrafast electron point projection microscopy (ePPM) is an important example of such a source-sample configuration. In ePPM, the probe nanotip is brought into close (micrometer) proximity of a specimen, which is to be illuminated by the emitted electrons. The magnifying power of the microscope is given by $D/d$, which is the ratio of the tip-to-detector distance $D$ to the tip-to-specimen distance $d$. A shorter *d* thus yields an enhanced magnification. Since the smallest focal spot size of the ultrafast laser beams is of the same order of magnitude as the minimally desired values of *d*, it is not feasible to irradiate the nanotip apex directly by the laser beam and avoid unwanted scatterings and damage to sensitive samples. To alleviate such restrictions, grating-coupled Au nanotip emitters have been explored for ultrafast ePPM applications [65,66]. The grating is patterned on the tip shank a few microns away from the apex. The laser pulses therefore are not required to be focused onto the apex which is in the proximity of the sensitive specimen in ePPM. Instead, by irradiating the grating, surface plasmons are excited which propagate to the apex and give rise to electron emission.

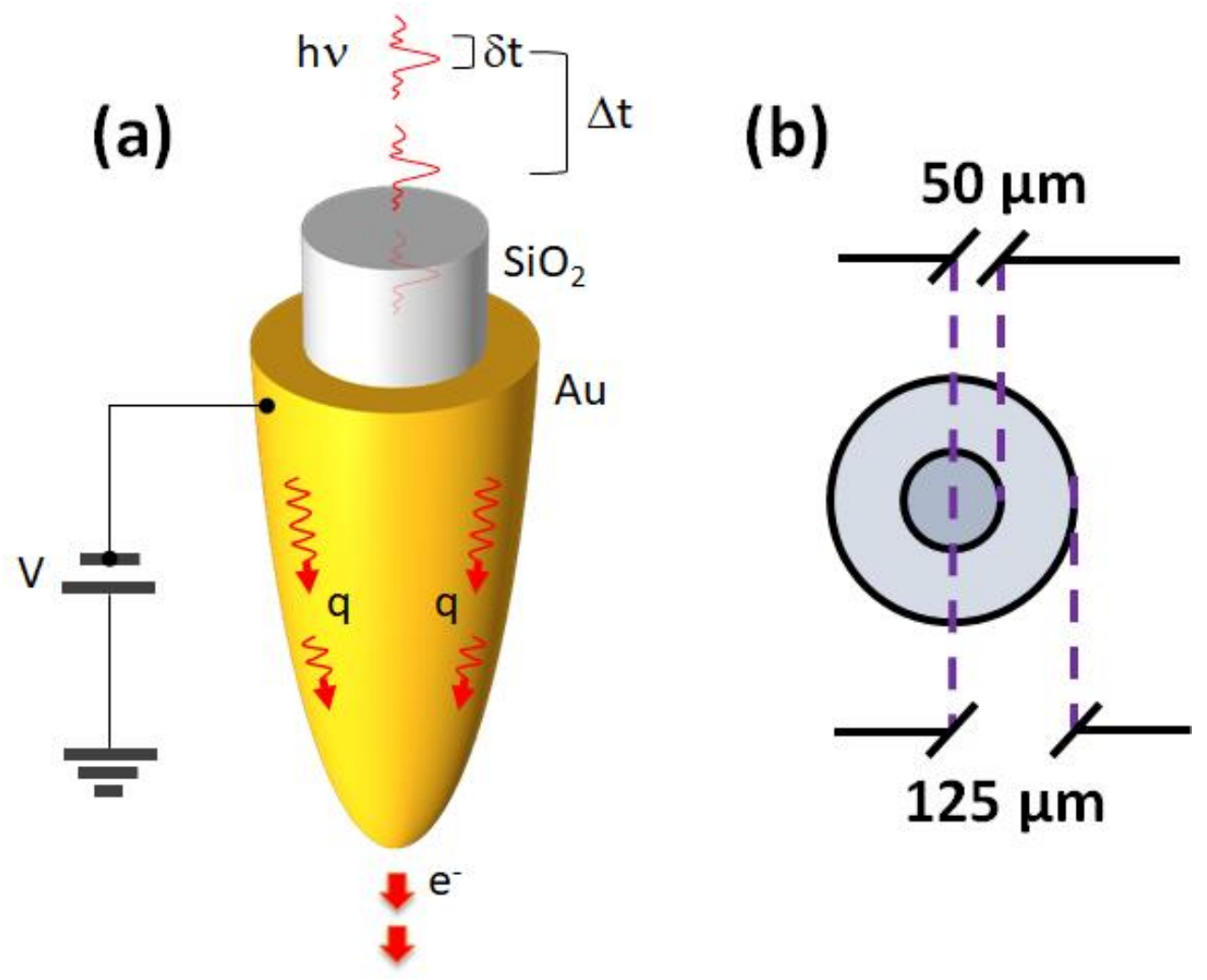

**Fig. 4-1 | Schematic depiction of our proposed nanotip source. (a)** A pulled tapered optical fiber is coated with a nanolayer of Au, 100 nm in thickness, on top of 4 nm of Cr used for robustness improvements. The light beam coupled in the silica core of the optical fiber is anticipated to excite surface plasmon resonance with momenta q (in the curved geometry) across the nanofocused region. This is expected to lead to electron emission as assisted by the enhanced electric field of the plasmons. The applied voltage of typically -100 V accelerates the emitted electrons. At higher values of the bias potential, photo-field emission (PFE) and field emission (FE) effects start to become significant as well. The figure was made by Dr. Ali Passian of ORNL. **(b)** The multimode GRIN base fibers have a core diameter of 50 μm, and a cladding diameter of 125 μm. The protective polymer coating (not shown here) is 250 μm in outer diameter.

In the present work, we introduce a new type of electron source in which the wavelength-dependent emitted current, emission time, and pulse duration can be controlled optically. Our arrangement does not require side-illumination by an external laser pulse or grating-coupled plasmon excitation. The investigated nanotip emitter is schematically depicted in Fig. 4-1. The studied system is based on a dielectric probe nanotip made by tapering a multimode (MM) graded-index (GRIN) optical fiber coated with a layer of a plasmon-supporting material, such as a noble metal like gold. By coupling a laser beam of suitable wavelength into the fiber, plasmons are excited in the curved metallic film at the tapered end. Theoretical studies [67,68] of surface plasmons excited at the metal-dielectric interface suggest the corresponding modes undergo a curvature-induced dispersion modification. As a result, useful resonant modes can be extended to the visible wavelength range in order to attain field enhancement at the apex of the probe.

Previous fiber-based electron emission studies are limited to the reports by Casandruc *et al.* [69] and Lee *et al.* [70]. The first study invoked a tungsten-coated near-field scanning optical microscope fiber probe [71] to which a 2-mW CW beam from a diode laser emitting at 405 nm was coupled. The second study involved the gold-coated *flat* end of a wide-area MM fiber core irradiated by a femtosecond laser beam at 257 nm. To achieve electron emission in the first study, external DC electric potentials larger than +1 kV were applied between the front face of a microchannel plate (MCP) detector and the fiber tip. The long switching (rise) times on the order of tens of milliseconds to a few seconds, leads the authors to indicate thermal emission as the most likely mechanism. In the second study, at 257 nm, the photon energy is larger than the Au work function of 5.1 eV [55], and the emission mechanism is reported to be the fast process of single-photon photoemission. As indicated above, the fiber was not tapered and did not reach the nanoscale.

In our work, the anticipated SPR-assisted electron emission capitalizes on the availability of resonant modes sustained by the thin nanoscale curved film. The plasmonic mechanism is corroborated by characterizing the electron emission and performing photonic measurements along with theoretical considerations including a finite difference time domain (FDTD) computational model performed by Dr. Ali Passian of the Oak Ridge National Lab (ORNL). Our aim is two-fold; firstly, we demonstrate that a multiphoton photoemission process is feasible via optical stimulation of the electron gun under study at low fs laser power levels, and secondly, we provide preliminary evidence that the observed photoemission is assisted by SPR excitation. Interestingly, we report achievement of electron emission using both low power CW and pulsed lasers. We will establish that while for the former, photo-field emission (PFE) is the driving mechanism, the latter gives rise to above-threshold emission (ATE) at the SPR wavelength and multi-photon emission (MPE) otherwise.

## 4.1 Experimental setup

Our GRIN MM optical fibers (Corning, InfiniCor™ 600) with nominal core diameter of 50 μm, cladding diameter of 125 μm, outer protective polymer coating of diameter 250 μm, and numerical aperture of NA = 0.20, were tapered using a commercial micropipette puller at ORNL. An alternative approach to make such nanotips, often for SNOM applications, is chemical etching [72]. The fiber tips were subsequently coated first with 4 nm of Cr and then with 100 nm of Au on top of Cr using electron beam deposition. Cr is used to make the interfaces more robust as Au alone would not stick well to the glass of the fiber. These steps were undertaken at ORNL. All the rest of the experimental procedures and data analyses to be discussed were performed here at UNL.

We made a home-built fiber optic vacuum feedthrough as shown in Fig. 4-2 for optimal coupling of the driving laser beam. This enables us to first optimize the lens-coupling of the laser beam into a bare auxiliary fiber. The open end is inserted into the 250-μm ferrule of a fiber chuck mounted on a 3D micro-stage to bring its coupling face into the laser focus after a convergent lens. Once the optimized light coupling is achieved, the laser beam can be end-fire-coupled into the fiber optic feedthrough using a standard mating sleeve. The feedthrough itself is constituted from a one-meter long piece of the same type of optical fiber protected by a 900-μm furcation tube. The tube is further protected by insertion into a 2.3-mm stainless steel tube. To make the feedthrough, a 2.5-mm hole was drilled through the center of a 2.75-inch vacuum flange through which the protective steel tube hosting the fiber was threaded. Both ends of the tube were subsequently connectorized through the standard procedure described in detail in Ref. [73]. The drilled hole was subsequently sealed with TorrSeal™. All of the experiments discussed in the present chapter were performed at pressures around $2\times10^{-7}$ Torr, indicating satisfactory sealing of the drilled flange hosting the feedthrough and the in-chamber mount. Standard SMA-905 fiber connectors were implemented in the feedthrough as well as to terminate the open end of the fiber tips. A high-vacuum-rated SMA-SMA mating sleeve was utilized inside the vacuum chamber to plug the terminated fiber tips onto the vacuum end of the feedthrough line.

The output voltage signals of the CEM (Dr. Sjuts™, Model KBL 510) were amplified using a BNC-terminated pre-amplifier (ORTEC, Model VT-120) with a constant gain of 200. The amplified signal was subsequently fed into a CFD (ORTEC, Model 935) in the updating mode (see Fig. 3-2(c)). The CFD output signals were registered and counted using an MCS triggered by TTL signals of an optical chopper in the experiments using CW lasers, and by those of the regenerative amplifier in the experiments with an OPA. For coincidence timing experiments, the TTL signals were given to the start input of a TAC (ORTEC,

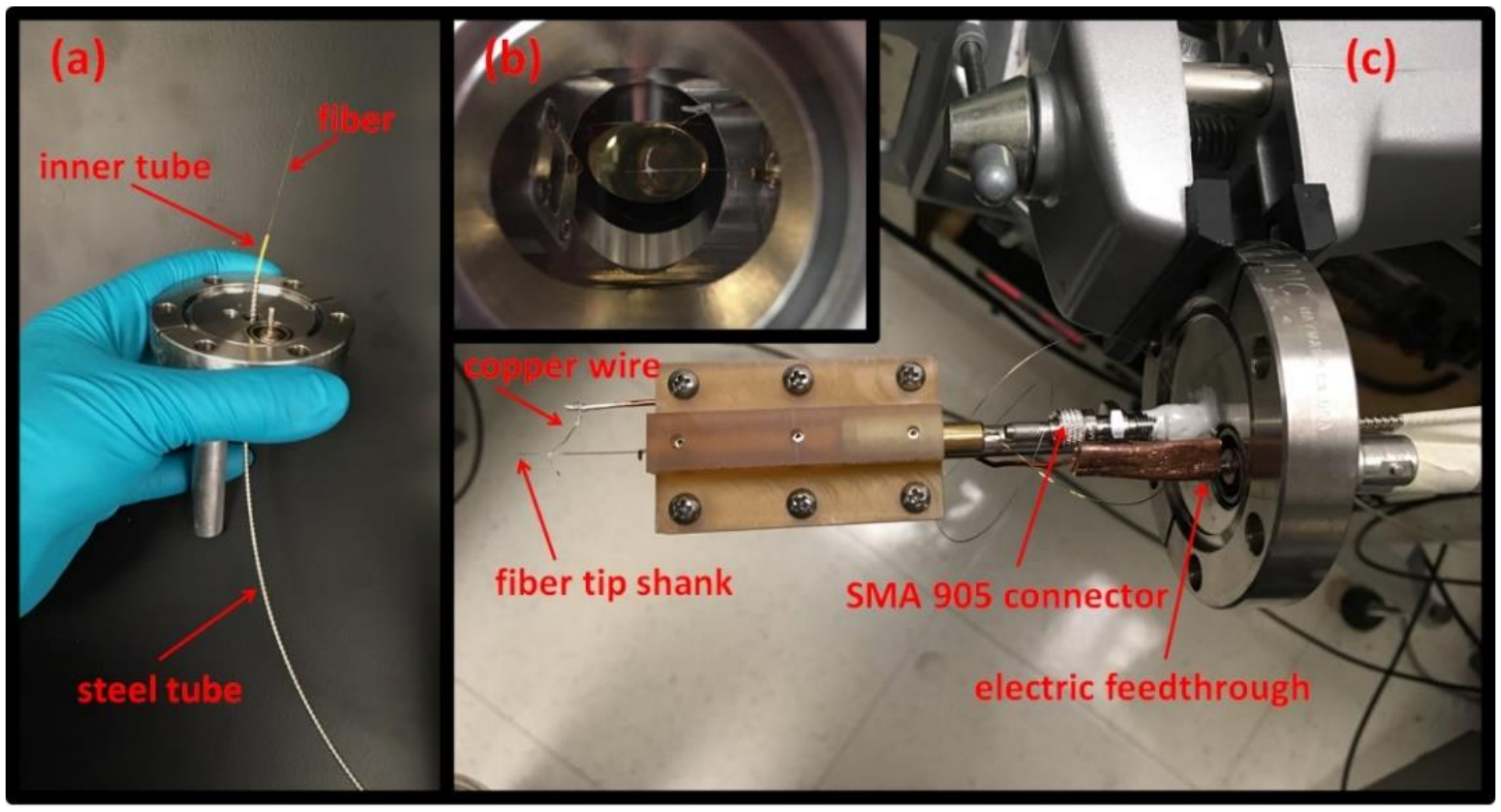

**Fig. 4-2 | A few images of the home-built fiber optic feedthrough and the mounted fiber tip. (a)** The fiber optic vacuum feedthrough under construction before connectorization and vacuum sealing. **(b)** The apex of a fiber tip source at a distance of 1 cm from the channeltron face is shown. The picture is captured through the top window of the vacuum chamber ($2\times10^{-7}$ Torr). **(c)** A mounted fiber tip source with the electrical connection provided by a fine piece of copper wire is seen here. The wire is connected to the fiber coating using silver paint. The white material at the base of the mount where the high-vacuum-rated SMA-SMA mating sleeve is attached to the 2.75-inch flange is hardened TorrSeal™.

Model 567) while the CFD output was fed into its stop channel. Using an MCA, it was possible to set an upper limit on the emission rise time as will be discussed in the following sections. For more details on the detection electronics refer to CH.3, Section 3.1.

A cartoon of the experimental setup is shown in Fig. 4-3. In panel (a) the terminated fiber tip is seen to be mounted inside a high-vacuum chamber using a side-loading fiber chuck. A fine piece of Cu wire was used to apply a negative electric bias as an accelerating potential to the shank of the metallized fiber tip using an electrical feedthrough on the host vacuum flange. The CEM, with its front electrode grounded in all the experiments of this chapter, and located at a distance of 1 cm from the apex of the nanotip, was exploited to detect the emitted electrons. A home-built fiber optic vacuum feedthrough guides the laser light into the vacuum chamber whereby end-fire-coupling the light (namely, lens-less coupling) into the 30-cm long fiber tip is accomplished. In principle, there is no restriction on the minimum length of such fiber tips other than ease of use. The feedthrough is made using the same type of fiber in order to minimize the scattering losses due to cross-sectional area mismatch during end-fire-coupling.

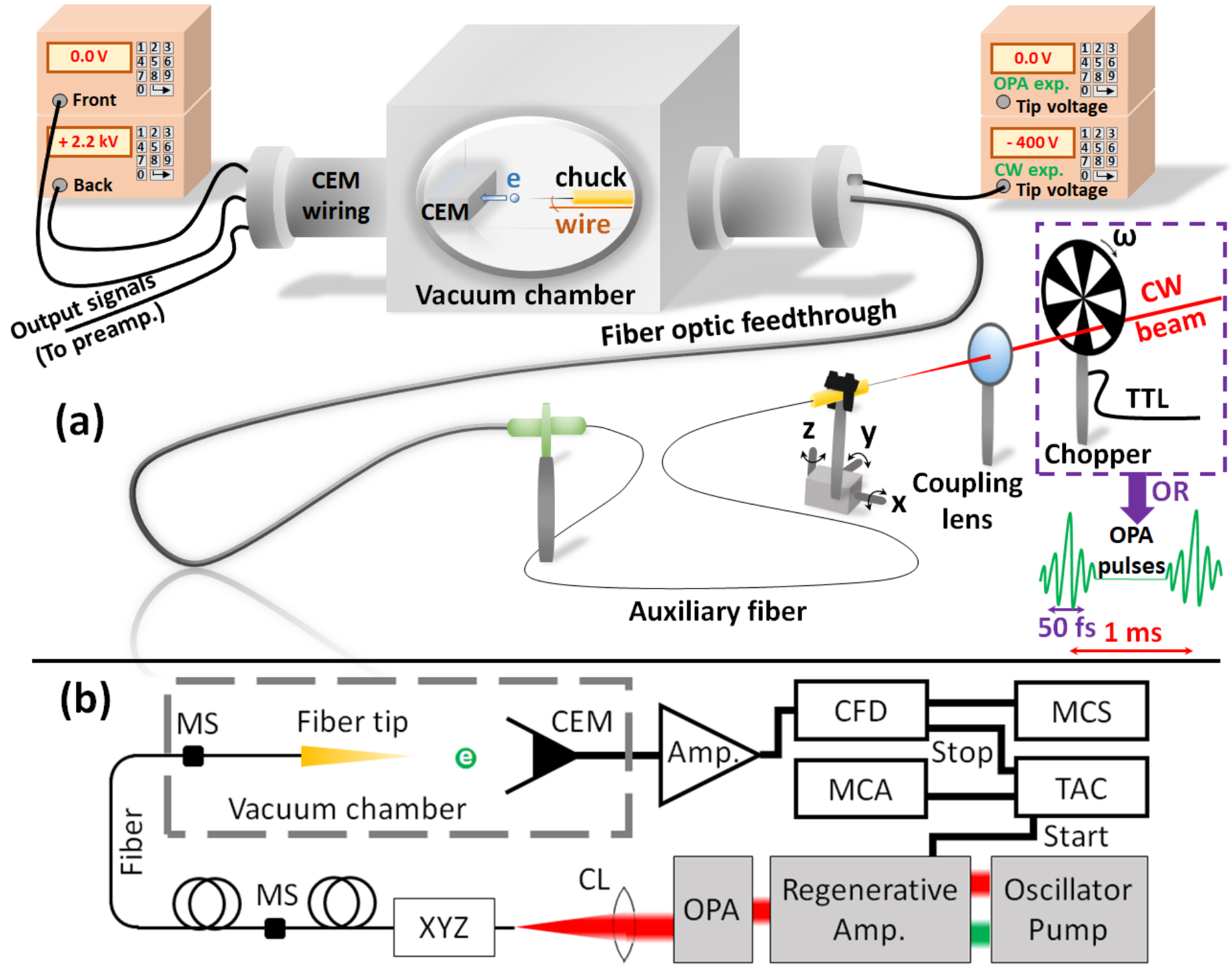

**Fig. 4-3 | A cartoon of the setup is illustrated. (a)** The connectorized fiber tip is mounted inside the vacuum chamber ($2\times10^{-7}$ Torr) using a side-loading fiber chuck. The laser light, CW or pulsed, is lens-coupled into the fiber tip through the home-built feedthrough. The optimized coupled light power can be conveniently measured after the auxiliary fiber. **(b)** A schematic diagram showing the detection electronics is depicted. The thicker black lines on the right are coaxial cables. The reader may refer to Section 3.1 for more information on the components and their functions. The CFD provides three identical outputs. The laser TTL signals fed to the TAC are either those of the regenerative amplifier in the OPA experiments, or they are identically the chopper TTL output in the CW experiments. CL is the coupling lens, and XYZ stands for 3D micro-stage. Sleeve is a standard SMA-SMA mating sleeve for end-fire (lens-less) coupling from one fiber, such as the feedthrough, to another, e.g. a short connectorized fiber tip.

In the experiments with several CW lasers, the laser beam is coupled into a home-built auxiliary fiber using a free space lens after getting chopped ON/OFF by an optical chopper which also provides the necessary timing through its output TTL signals. Throughout this chapter, all the reported laser power levels (both CW power as well as the average power of the mode-locked (ML) laser) indicate the value of this quantity optimally coupled into the auxiliary fiber measured (or calculated from a value directly measured) at its terminated end. This is done using a power-meter after plugging the end connector into a connectorized collimating lens with its antireflection (AR) coating spanning the entire visible range which completely encompasses all the wavelengths implemented in this research. Based on our observations, approximately 80% of the electromagnetic power measured after the auxiliary fiber will reach the fiber tip after propagating downstream the fiber optic line. In the experiments with a pulsed laser, the combination of the CW laser beam and chopper is replaced with the output of an optical parametric amplifier (OPA) with pulse duration of 50 fs and repetition rate of 1 kHz. The laser system driving the OPA involves a solid-state-pumped Ti:Sapphire oscillator followed by a regenerative amplifier.

A SEM image of the fiber tip used in the CW experiments presented in this chapter is given in Fig. 4-4(a), with its extracted characteristic DC tunneling

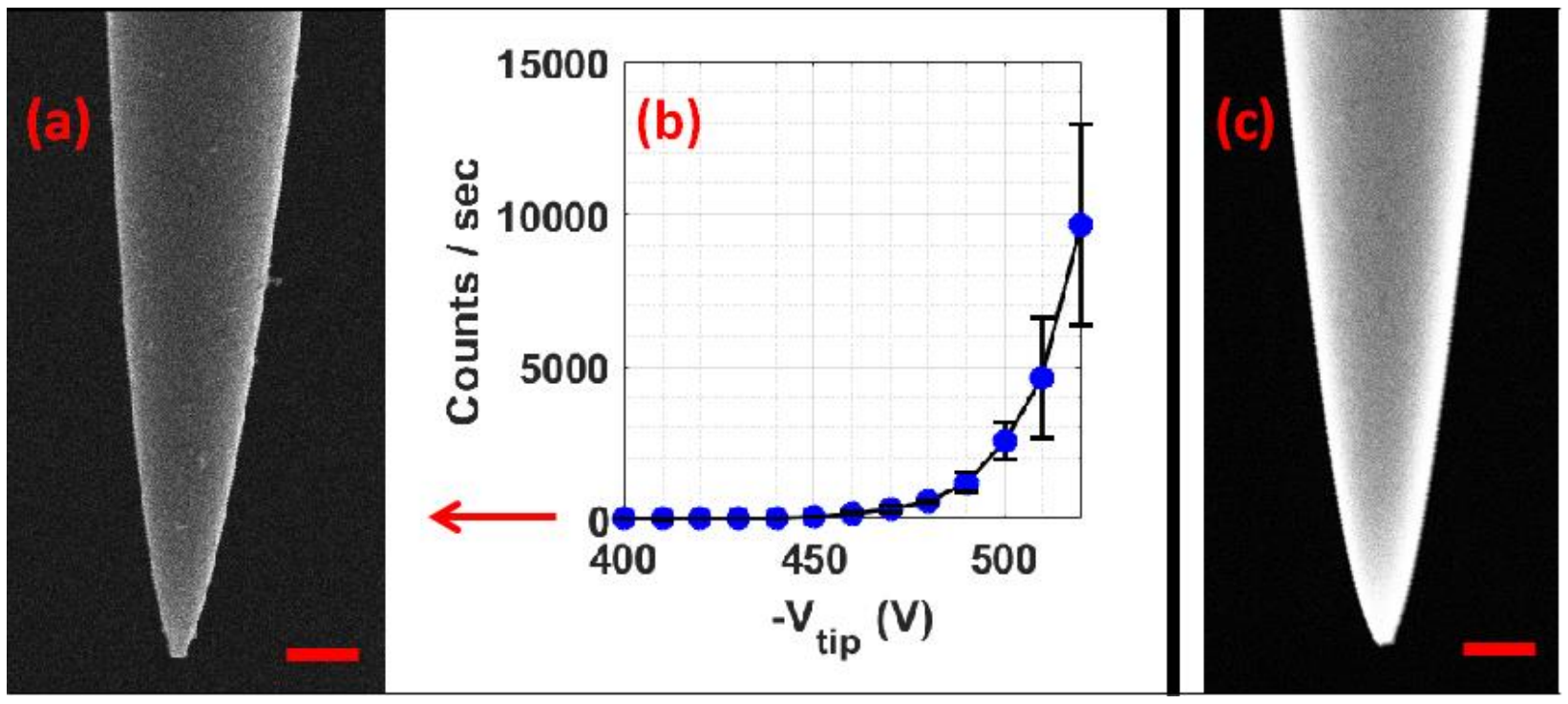

**Fig. 4-4 | SEM images taken before performing the corresponding experiments. (a)** SEM image of the tip used in the experiments with CW lasers with its **(b)** The Fowler-Nordheim (FN) curve corresponding to the tip shown in panel (a), which is consistent with an apex radius and emission area of ~ 50 nm. **(c)** SEM image of the fiber tip used in the pulsed experiments with an OPA is depicted. The red horizontal scalebars in (a) and (c) correspond to 1 μm.

Fowler-Nordheim (FN) curve shown in Fig. 4-4(b). The observed pattern is consistent with a tip radius and emission area of ~ 50 nm estimated based on the FN equation (Eq. (4-1) in the following section) [53,55]. This will be discussed in more detail in the following pages. Each data point represents the average of 6 detection counts over 5-s intervals. The errorbars correspond to one standard deviation. An SEM image of the tip used in the pulsed experiments with an OPA is also shown in panel (c). The two fiber tips in panels (a) and (c) are similarly prepared. The tip voltage $V_{tip}$ corresponding to the onset of DC field-emission was about - 450 V for both of them, consistent with the FN curve of panel (b).

## 4.2 Experiment I: Photo-field emission using CW lasers

We begin by first presenting our data set for the proposed electron emission in the CW regime, as depicted in the various panels of Figs. (4-5) & (4-6). A HeNe laser emitting at 632.8 nm (~ 633 nm) and four low power diode lasers emitting at 405 nm, 532 nm, 672 nm, and 800 nm were employed to probe the temporal and spectral response of the source in the CW regime. All of the lasers emit in the long-wavelength domain for this tip. The observed emission spectra are shown in Fig. 4-5 for the HeNe laser at 633 nm, and in Fig. 4-6 for the diode laser at 405 nm. The total data acquisition time for each combination of $V_{tip}$ and laser power $P$ is 2 minutes in all of these panels.

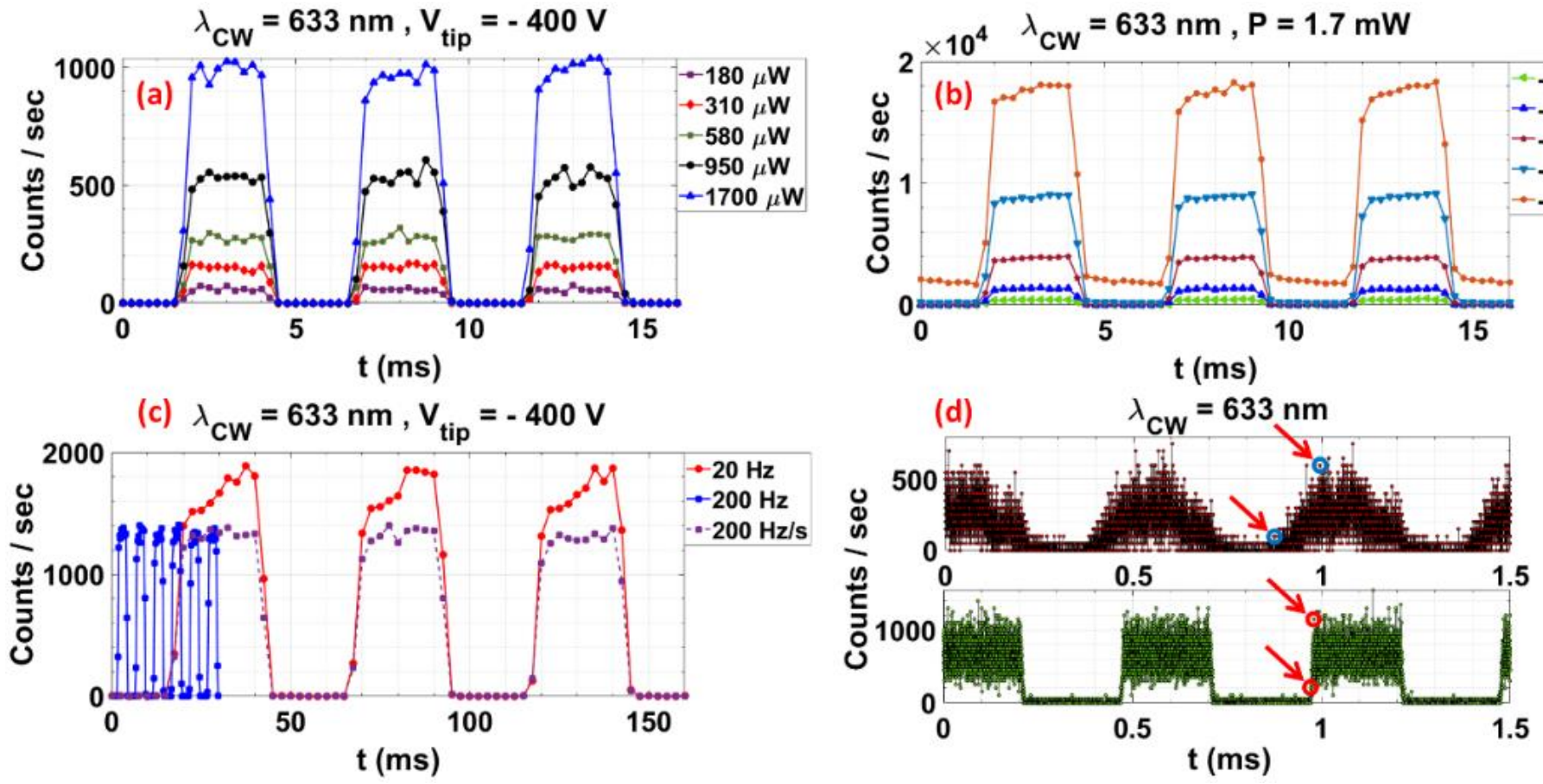

**Fig. 4-5 | CW electron emission at 633 nm.** In **(a)** the power-dependence, and in **(b)** dependence of the detection rate on the tip voltage, is demonstrated. **(c)** Two different chopping rates are used to illustrate that that electron emission can easily be modulated at different rates. The letter *s* in the legend box indicates *scaled to fit*. **(d)** With the laser beam focused onto the chopper wheel to achieve a smaller spot size the modulation rise time is increased in the bottom plot. In this panel, a 30-blade chopper wheel was operated at 2 kHz. While the electron emission rise time between the circled points indicated with the red arrows in the top plot is ~ 120 μs, that of the bottom plot is obtained as ~ 7.5 μs.

The electron emission switching using an amplitude modulated HeNe laser beam is shown in Fig. 4-5(a) for different values of $P$ at a fixed $V_{tip}$ below the onset of DC FE. The dependence of the electron emission on the input power $P$ is

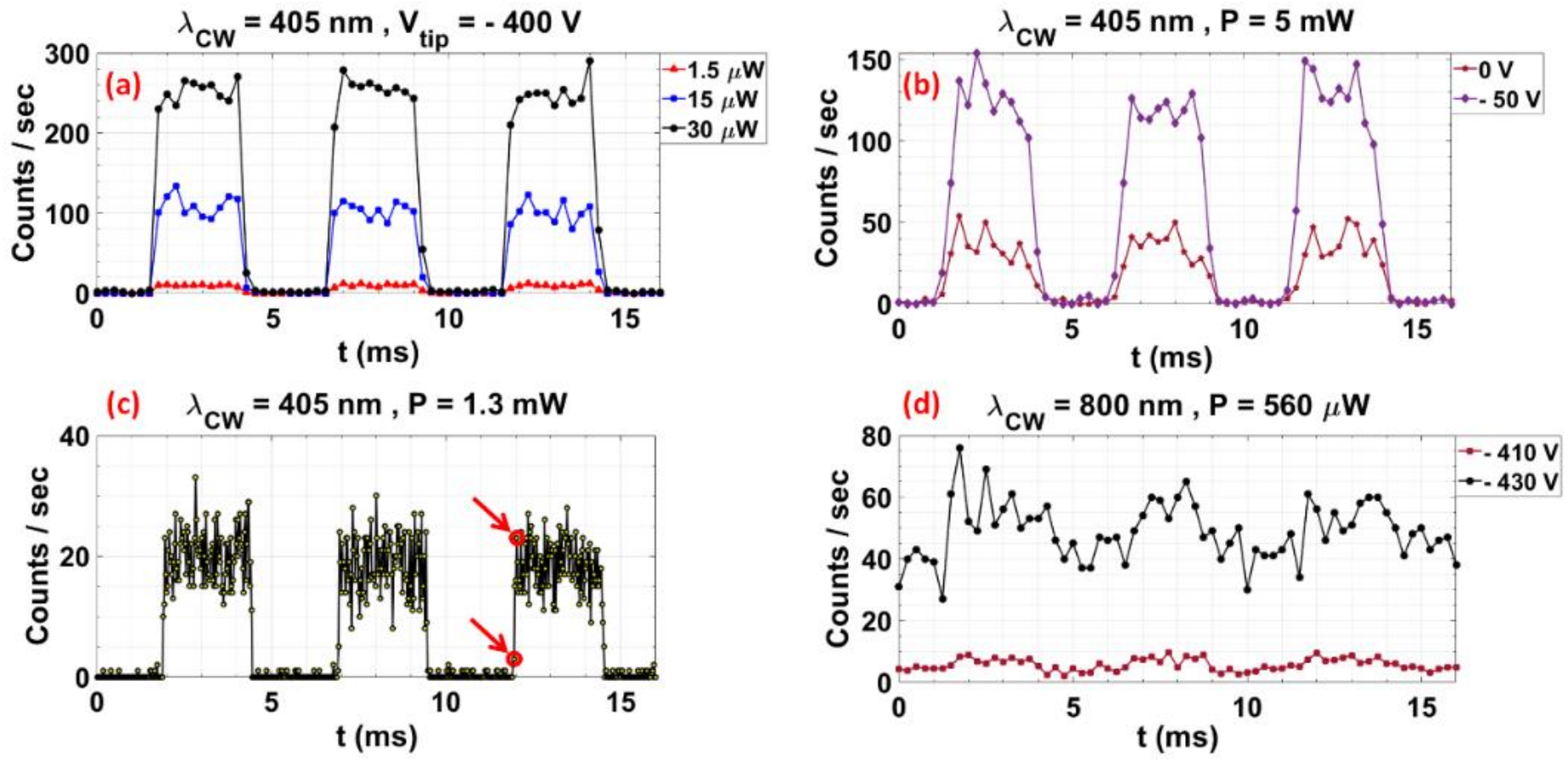

**Fig. 4-6 | CW electron emission at 405 nm and 800 nm. (a,b)** Analogues of the same panels in Fig. 4-5. The detection rates are noticeably higher at 405 nm for identical $P$ and $V_{tip}$. **(c)** The CW switching rise time with the focused beam and the chopper frequency of 200 Hz is ~ 100 μs using the circled points. **(d)** The CW beam at 800 nm does not emit electrons even at higher powers.

manifest. This is also the case for the 405-nm diode laser in Fig. 4-6 where comparable detection rates are achieved for yet lower values of $P$, notably on the order of 10 μW only. Interestingly, sub-mW CW lasers are observed to yield emission with consistent timing and switching fidelity. Figs. 4-5(b) & 4-6(b) show similar results this time at a constant $P$ and varied $V_{tip}$. For $V_{tip}$ = - 475 V in Fig. 4-5(b), which is slightly above the onset of DC FE, the background count rate in the absence of laser light is increased expectedly. In Fig. 4-6(b), it is notable to achieve laser-induced electron emission at even $V_{tip}$ = 0 for λ = 405 nm. With frequencies up to 2 kHz coupled with a shorter MCS bin width of 100 ns/bin, we could achieve a fast detection rise time of 7.5 μs as shown in Fig. 4-5(d) which is set by the input field amplitude modulation profile. Here, the laser beam was focused onto the chopper wheel plane in order to further increase the modulation speed. Fig. 4-6(c) shows fast CW switching at λ = 405 nm. Here, the chopper frequency was set to

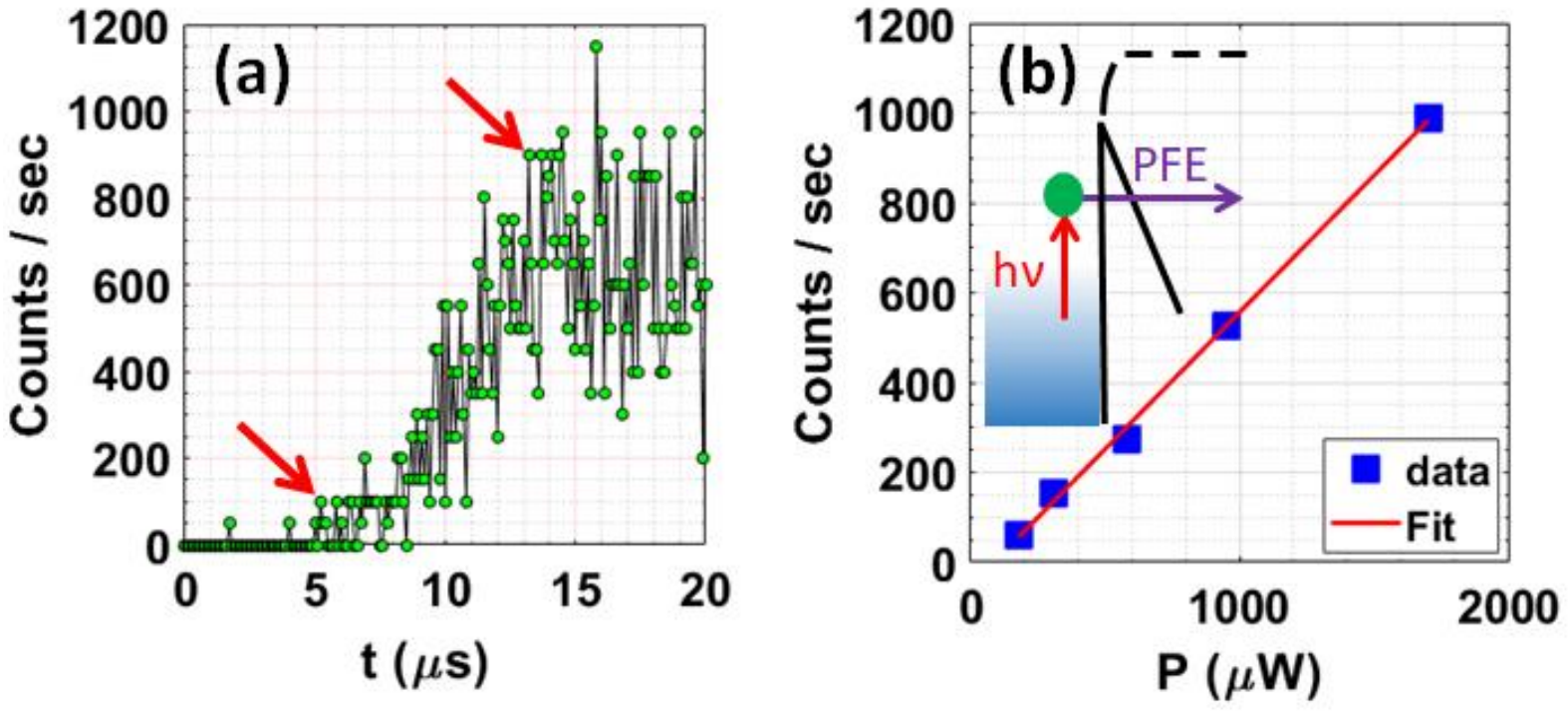

**Fig. 4-7 | Fast single-photon emission. (a)** The bottom plot of Fig. 4-5(d) is zoomed in. The fast switching on the order of 8 μs (between the red arrows) is consistent with optical processes. **(b)** Using the average count rate of the 9 data points on each of the leftmost peaks of Fig. 4-5(a), a single photon emission process is established as can be seen by the satisfactory linear fit. A PFE process is conjectured which is subsequently confirmed. See the main text. This also explains the observed increase in the emission current as the triangular potential becomes narrower progressively by increasing the $V_{tip}$. Similar results were obtained for other wavelengths including a linear photoemission fit for Fig. 4-6(a).

200 Hz and thus the observed rise time of ~ 100 μs is slower by approximately one order of magnitude.

So far, we have provided evidence for a *fast* photoemission process. Significantly, the achieved switching times are shorter than those reported in [69], also at 405 nm, by about 5 orders of magnitude. We now attempt to establish the type of photoemission mechanism. A fast single-photon photoemission process is demonstrated in Fig. 4-7 based on the leftmost peaks of Fig. 4-5(a). We thus conjecture that single-photon PFE is the emission mechanism for the implemented low power CW lasers in the long-wavelength domain. We confirmed this using the free electron field- and photo-emission theory.

The FN count rate, for a tip radius *r* and a hemispherical emission site with radius *R*, is given by [53,55]

$$C_{FN} = \left(\frac{2\pi R^2}{q}\right) \times \left(\frac{aF^2}{\phi q^2 t^2}\right) \times \exp\left(\frac{-vbq\phi^{3/2}}{F}\right), \tag{4-1}$$

in which $q$ is the elementary charge, $\phi$ is the Au work function in the absence of applied static field (~ 5.1 eV), $a = q^3 / (8\pi h)$ and $b = 8\pi\sqrt{2m} / (3qh)$ where $h$ is the Planck constant, and $F = qV_{tip} / kr$ is the force per unit charge on the tip in which $k = 5.7$ accounts for the typical geometrical shape of the metallic nanotips [46]. The parameters *t* and *v* are

$$\begin{aligned} t &= 1 + \frac{1}{9} y^2 (1 - \ln y), \\ v &= (1 - y^2) + \frac{1}{3} y^2 \ln y, \end{aligned} \tag{4-2}$$

where $y = \sqrt{q^2 F / 4\pi\varepsilon_0 \phi^2}$ is the Nordheim parameter, and $\varepsilon_0$ is the vacuum permittivity. The fitted curve to the dark FE data (red line) in Fig. 4-4(b) is obtained for $r = 42.2\ nm$ and $R = 53.0\ nm$ as the only free parameters in this model. The extrapolated tip radius $r$ is indeed consistent with the SEM image of the tip shown in Fig. 4-4(a). This is now compared with the PFE fit in Fig. 4-8. The black square data points in this figure are each the average of the 9 data points of the left-most peak (centered around t = 3 ms) in Fig. 4-5(b) for the 5 different values of $V_{tip}$. The PFE fit to this set of data is obtained as follows.

An *n*-photon absorption process is equivalent to one in which the work function is reduced by $nh\nu$ followed by the corresponding power-law term. In principle, for the (long-wavelength) photon energy in the present case, 0- to 2-photon processes are allowed based on the implemented photon energy and the gold work function, and in general, the transition probability decreases by several orders of magnitude for larger values of *n*. The count rate in the PFE model is therefore expressed as

$$C_{PFE} = a_0 C_{FN,0} + a_1 C_{FN,1} I + a_2 C_{FN,2} I^2, \tag{4-3}$$

in which $C_{FN,n}$ , $n = 0,1,2$ is the FN dark count rate of Eq. (4-1) with $\phi$ replaced with $\phi - nh\nu$ . We set $a_0 = 1$, $a_1 = 5.89\times10^{-17}\ W^{-1}m^2$, and $a_2 = 10^{-4}\times(a_1)^2$, whereupon the satisfactory PFE fit (green line) in Fig. 4-8 is achieved. The intensity at the apex is estimated at $I = 5.40\times10^8\ Wm^{-2}$ . With these parameters, the two-photon rate is negligible. To our knowledge, this is the smallest CW peak intensity in the literature for which photoemission is reported. In addition, the Schottky-reduced work functions in Fig. 4-8 fall between 3.4 eV and 3.6 eV. This is calculated from $\varphi = \phi - \sqrt{4QF}$ , where $Q = \alpha\hbar c/4$ in which $\alpha$ is the fine structure constant, $\hbar$ is the reduced Planck constant, and $c$ is the speed of light in vacuum (hence $\alpha\hbar c = 1.44\ eV \cdot nm$) [53,55].

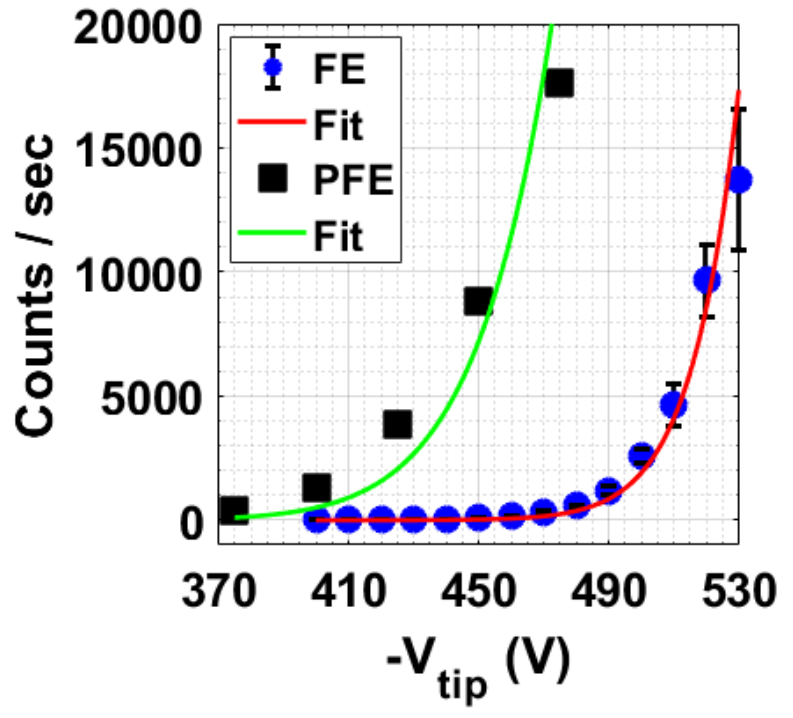

**Fig. 4-8 | CW photo-field emission mechanism established.** The Fowler-Nordheim (FN) field emission (FE) from the same fiber tip in the absence of a drive laser field is compared with the single-photon photo-field emission (PFE) process using a low-power (1.7 mW) CW HeNe laser. Blue circles display ±1σ errorbars, while black squares present the single-photon PFE, each with their corresponding fit model.

## 4.3 Experiment II: Surface-plasmon-assisted above-threshold emission using a femtosecond OPA

Following the observation and analysis of the CW emission, we employed the wavelength-tunable output beam of an optical parametric amplifier (OPA) to demonstrate the possibility of fast pulsed operation of the source. We aim to establish whether *i*) the studied nanotip emitter exhibits fast pulsed electron emission, and *ii*) the spectral response of the emission is indicative of SPR excitation as hypothesized.

The central ML wavelength of the OPA was first set to 633 nm (chosen to be identical to that of the HeNe laser used in the CW experiments of the previous section), as shown in Fig. 4-9(a), to demonstrate pulsed photoemission. Concerning laser pulse broadening in the GRIN fibers, we first note that the modal dispersion is minimal and thus its contribution to the pulse broadening may be neglected [70,74]. The stretching of the laser pulse propagating down the core in MM GRIN fibers is thus dominantly caused by chromatic dispersion. An analysis of the dispersion relation of the optical fibers from which we fabricated our probes, yields 160 fs.nm$^{-1}$.m$^{-1}$ as the upper limit for the chromatic dispersion at the operating wavelength of 750 nm. For the 3-m long fiber in our experiment, involving the auxiliary and feedthrough fibers and the fiber tip itself (which can in principle be miniaturized for device applications), an estimated 5 ps broadening is expected at 750 nm. Thus, we may safely assume a sub-10-ps broadening in the visible spectrum for the OPA pulses reaching the metallized nanotip. In spite of the pulse stretch by about 2 orders of magnitude, which leads to a reduction of the peak optical electric field at the emitting tip, we observed fast electron photoemission, as shown in Fig. 4-9(b,c). Remarkably, a strong emission is observed at the low average power of 60 μW (pulse energy of 60 nJ), without application of any DC field. To improve the temporal resolution, we employed coincidence measurements (see CH.3, Section 3.1, and Fig. 4-3(b)). Setting the coincidence window to 50 ns, a timing spectrum was obtained by correlating each delayed electron detection event with the corresponding input laser trigger signal. The observed rise time of ~ 1 ns in Fig. 4-9(d), which is only restricted by the detection resolution, thus sets a shorter experimental upper limit on the electron emission rise time and pulse duration establishing fast pulsed electron emission.

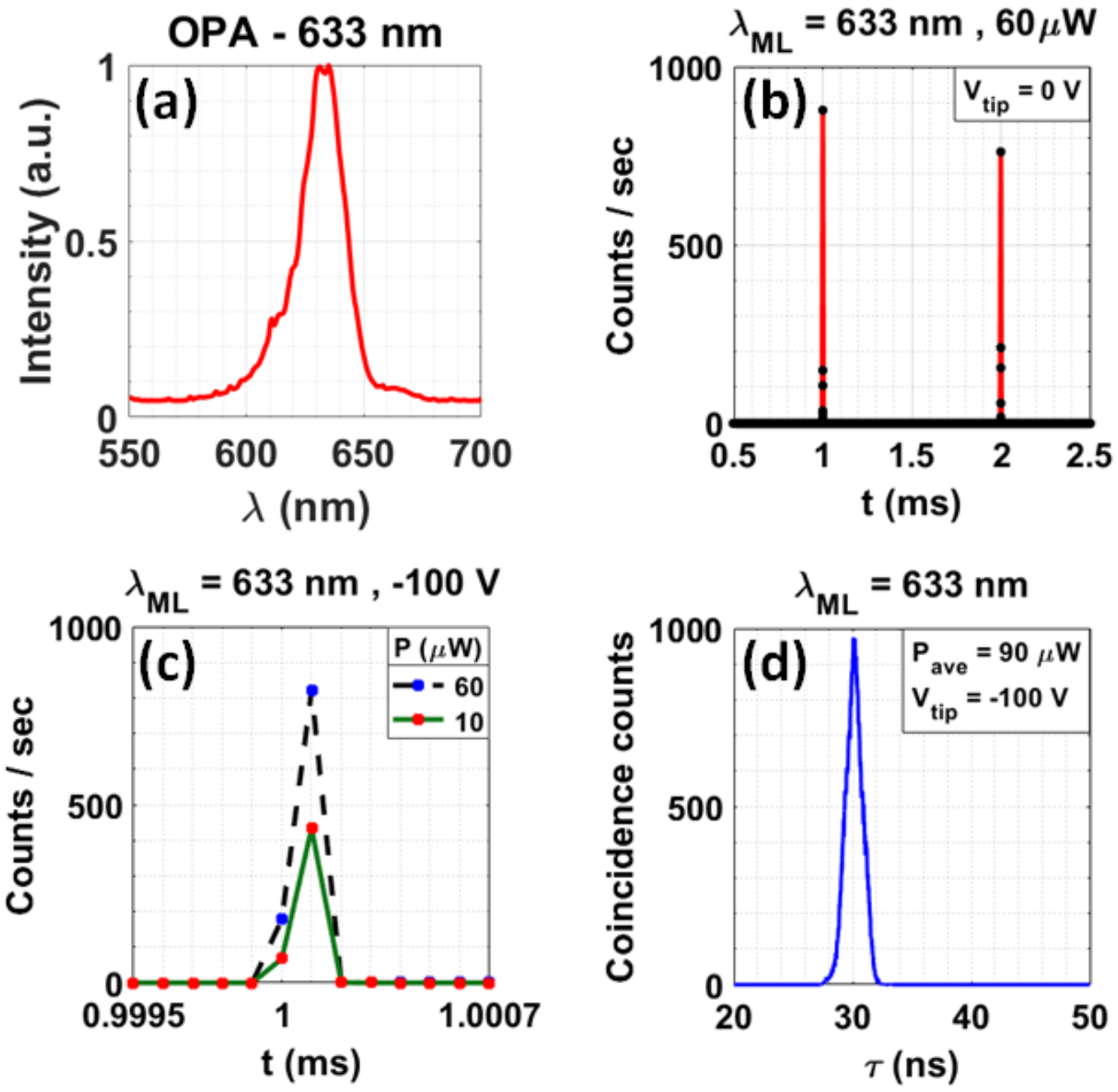

**Fig. 4-9 | Fast pulsed operation of the fiber optics based nanotip sources. (a)** The normalized output spectrum of the OPA with the central wavelength set at that of the HeNe laser (633 nm) is shown. The spectral width of the OPA at all the implemented wavelengths is about the same. **(b)** Fast electron emission driven by the OPA pulses at the same wavelength. **(c)** Electron detection peak compared for two different OPA output average power levels. **(d)** The laser-electron coincidence spectrum establishes an emission rise time of about 1 ns.

We proceed by extracting the photoemission order *n* in the visible spectrum. A rotating continuous ND wheel (THORLABS, NDC series with maximum ND = 2) was used before the coupling lens to vary the input laser intensity for this purpose. The transmission through the wheel is governed by its characteristic equation $T = 10^{-m\theta}$, in which $\theta$ is the rotation angle in degrees and *m* is a constant property of the wheel. The systematic error in calculating the transmitted power due to the finite area of the laser spot size that covers a finite range of angles is negligible. In addition, we used the regions near the rim of the wheel, about 4 cm away from the center, to minimize the angular distribution of the laser spot. The wheel was rotated at a fixed pace controlled by a computer. Consequently, $\theta$ and therefore *T* both depend on the laboratory time *t*. The direction of rotation is reversed after periods of approximately 31 s. This helped monitor statistical fluctuations and consistency of the results. The OPA output peak wavelength was changed in steps of 20 nm. Following optimization of the output beam of the OPA before coupling it to the fiber tip, the emission order was extracted. This was repeated on different days for 3 groups of wavelengths to improve the statistical significance of the results.

Before discussing the average values and the final photoemission order spectrum, let us first illustrate the power-law plots for each group separately in the following 3 figures. The slopes of the linear fits to $\log(C)$ versus $\log(P)$ in these log-log plots give the emission order in each case. $C$ is the count rate and $P$ is the input average laser power. The linear fits are weighted by the inverse of the relative error at each experimental data point. The error is defined as

$$\Delta(\log C) = \log\left(C + \sqrt{C}\right) - \log\left(C - \sqrt{C}\right), \tag{4-4}$$

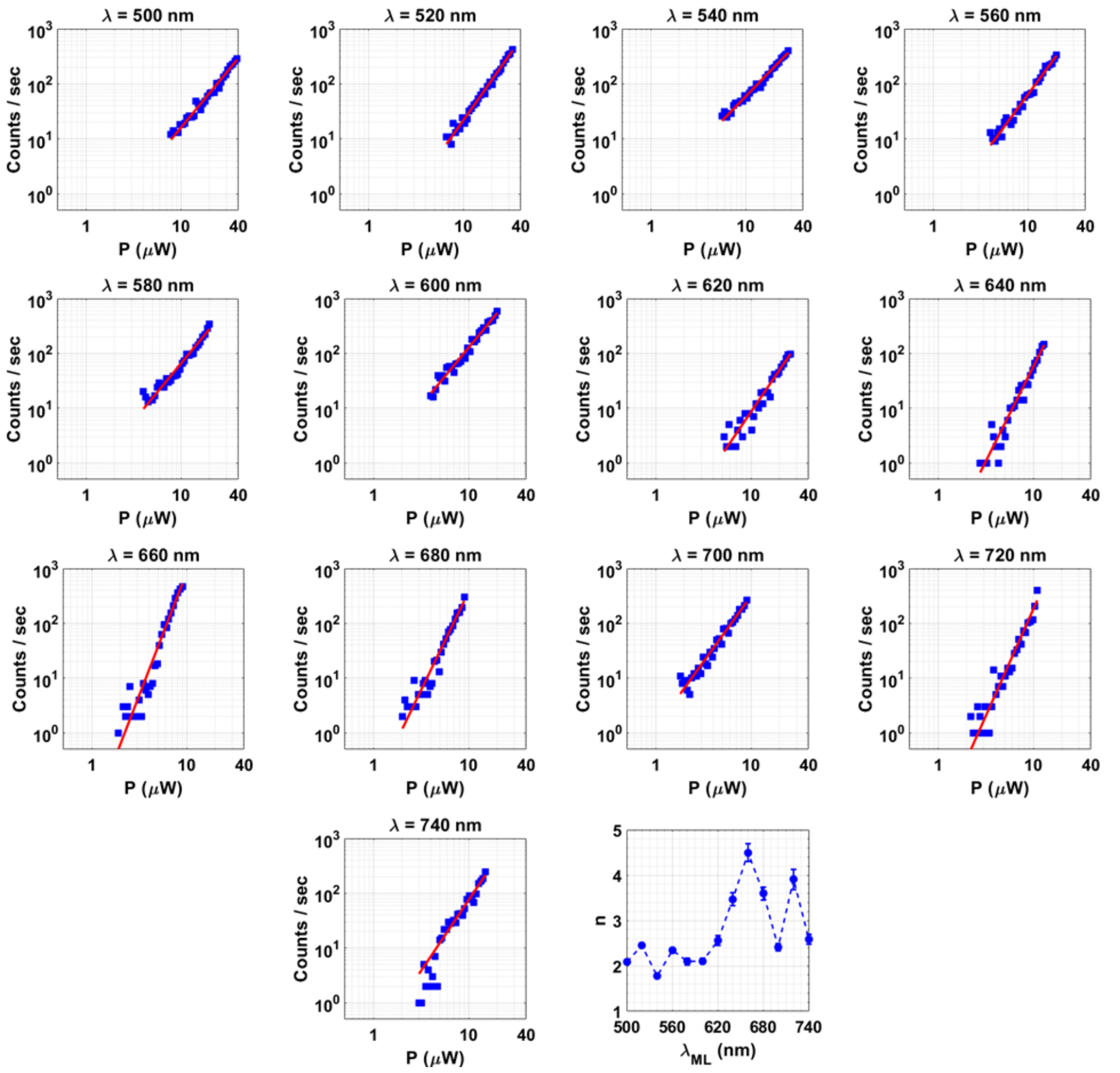

**Fig. 4-10 | Power-law fits – set 1.**

where a random counting noise per data point is assumed. The relative error is thence $\Delta\left(\log C\right)/\log C$. As can be seen in Figs. (4-10,11,12), the emission order has a peak in the 660-680 nm range. This is consistent with our initial hypothesis that at the SPR resonance region, which happens to be in the red segment of the visible spectrum for the fiber tip under study, the otherwise multiphoton photoemission (MPE) process turns into above-threshold emission (ATE) due to the extra enhancement in the localized electric field at the emitting tip, caused by the laser-induced SPR excitation. In each of these 3 figures, the last panel is a plot of the obtained emission orders at each wavelength against the corresponding peak wavelength of the mode-locked (ML) spectrum. The errorbars denote the $\pm 1\sigma$ range derived from the linear fitting process. The tip was biased at $V_{tip}$ = -50 V, and as before, the front of the CEM, 1 cm away on-axis with the fiber tip, is grounded.

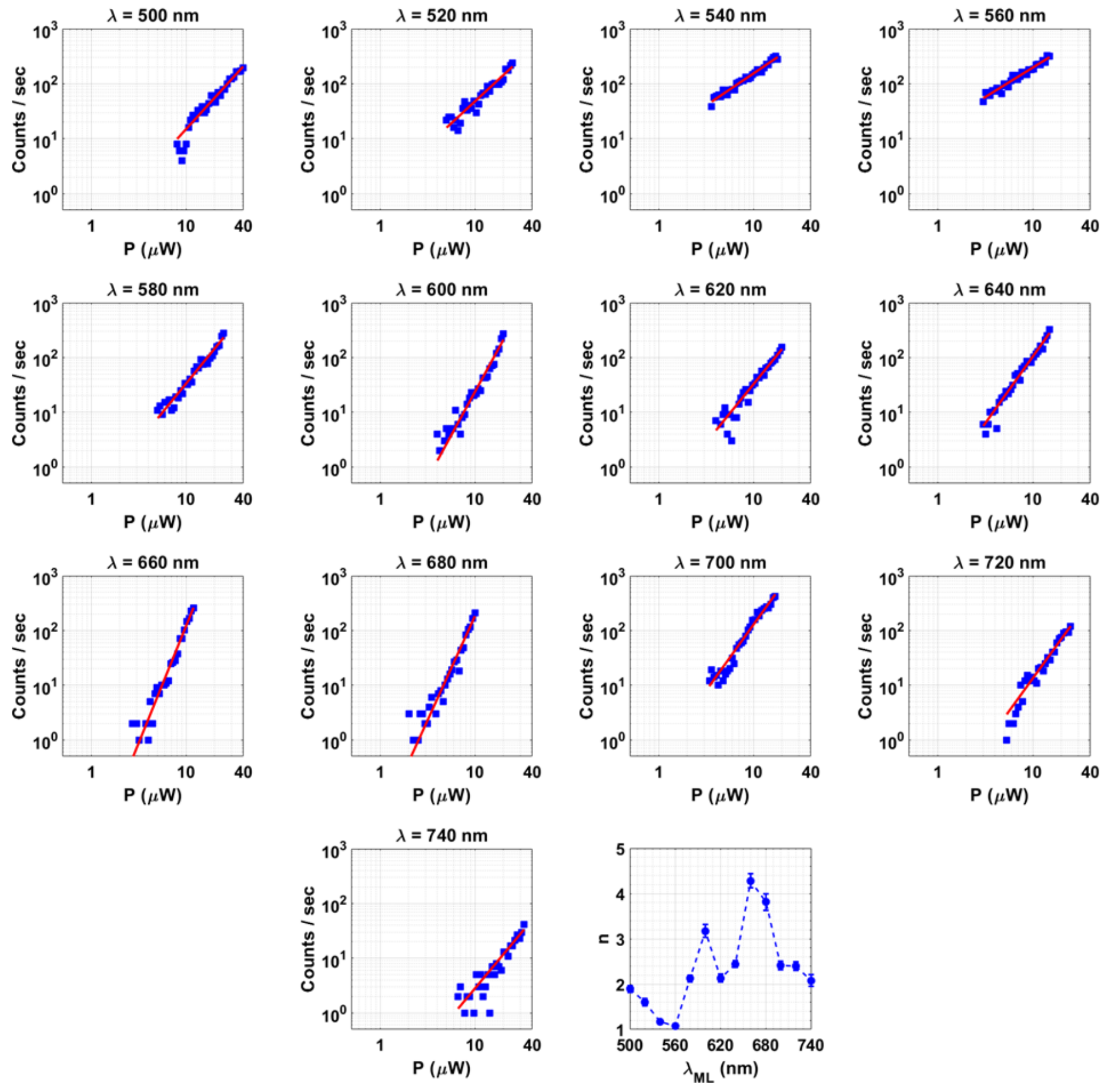

**Fig. 4-11 | Power-law fits – set 2.**

It was pointed out above that the constant *m* characterizes the transmission properties of the continuous ND wheel. The value reported by the manufacturer for this quantity is $m = 0.00741$. Upon investigation of the wavelength-dependent optical transmission through the wheel, we realized that this number might marginally deviate from the factory specifications, likely because the latter corresponds to a HeNe laser at a fixed wavelength having been used as the light source. Overall, it should be noted that while in Figs. (4-10,11,12) an approximate value of $m = 0.008$ is used, in the subsequent Fig. 4-13, the above factory value is implemented. The corresponding relative difference in the sought-after emission order $n$ is around 5-7%. This does not affect any of the conclusions about the emission mechanism based on the SPR-assisted ATE model discussed in the following.

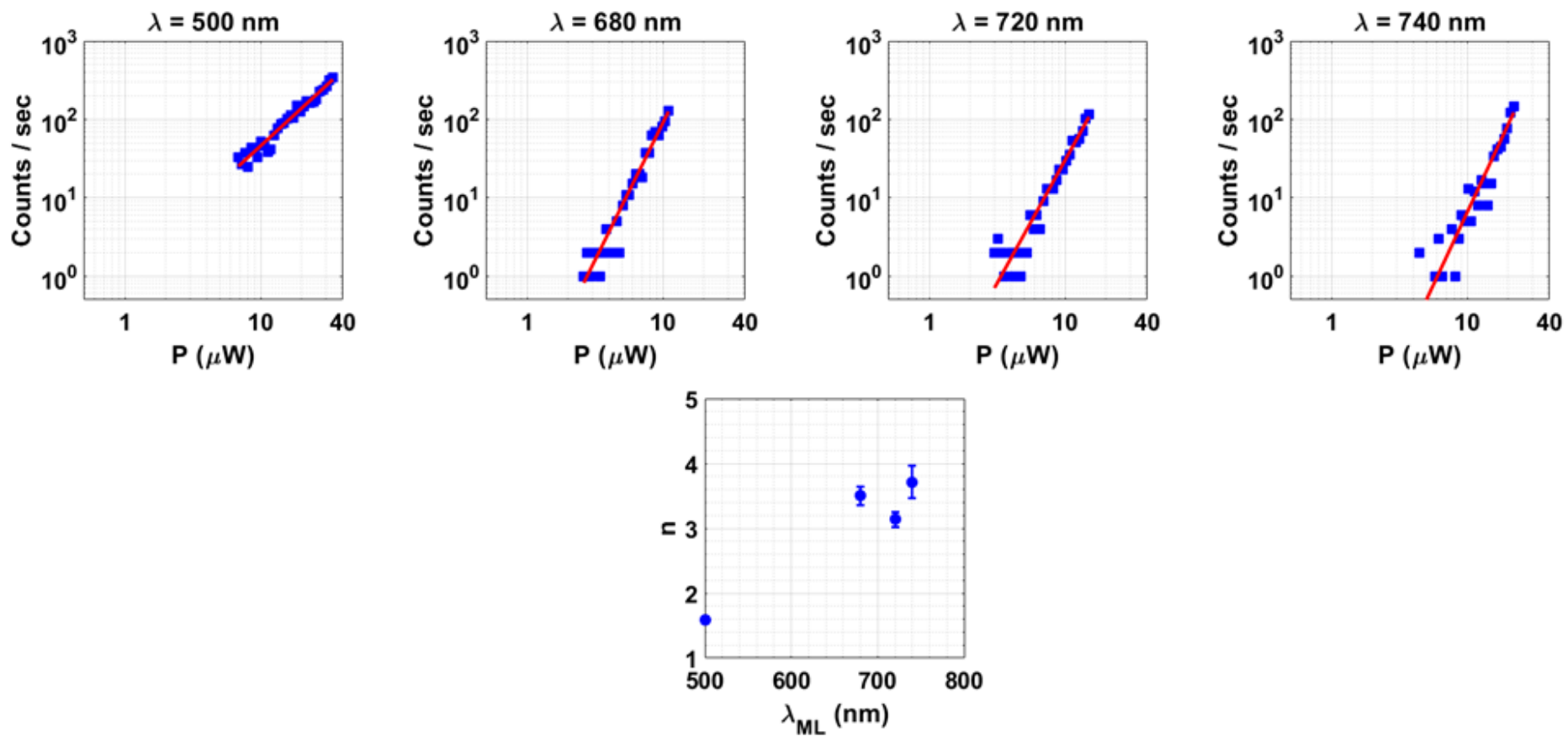

**Fig. 4-12 | Power-law fits – set 3.**

A few reversals for a resonant (660 nm) and an off-resonant (500 nm) wavelength are shown in Fig. 4-13(a) as examples to convey the observed repeatability in both the count rate and the slopes. The data in Fig. 4-13(b) correspond to the interval between the green down arrows in panel (a) which are the same experimental data points from the corresponding panels of Fig. 4-10 (as mentioned just above, the linear fits differ marginally due to the fact that slightly different values of the characteristic constant $m$ were used). This interval corresponds to the range of input powers on the horizontal axis of panel (b). The

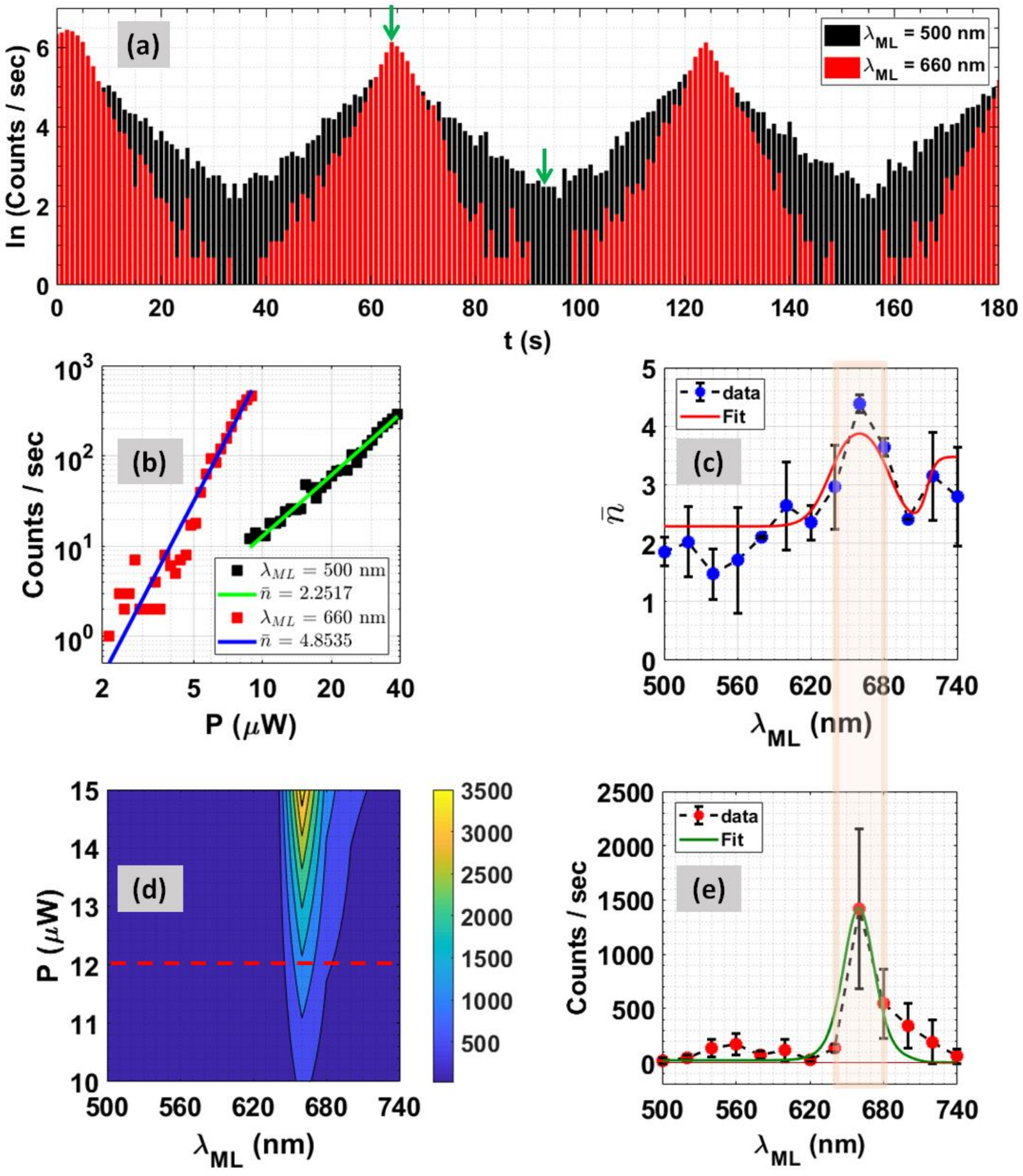

**Fig. 4-13 | OPA power-law results. (a)** Bar plot of the MCS spectrum is shown for two different wavelengths. The emergence of the peaks is due to the variable ND wheel rotating back and forth over 90-degree intervals. The computer-controlled wheel was set to perform this motion linearly in time in steps of 3 degrees per second. The horizontal axis which shows the experiment time is therefore proportional to the wheel angle, and from there, to $\log_{10}$ of the laser power. This makes it a log-log plot, and the anticipated linear trends are indeed discernible. **(b)** Linear fits to the two data sets of panel (a) over the time interval between the green arrows are shown in a log-log plot as example. In all the runs and wavelengths, the same range was used to extrapolate the

emission order $n$ to ensure consistency. **(c)** The average emission order $\bar{n}$ of the 3 runs of data – blue squares in Figs. (4-10,11,12) – is depicted. A peak at 660 nm is manifest. This is in support of a multiphoton emission process (see Fig. 1-9) turned into above-threshold emission (ATE) by SPR excitation noting to the fact that the latter is wavelength-dependent as it is originated in the spectral response of the gold nanolayer. The solid red line is the fit from the ATE model explained in the main text. **(d)** Contour plot of the average count rate corresponding to panel (c) with respect to a range of OPA average power levels and wavelengths is plotted and shown in this panel. The section at a fixed laser power of 12 μW corresponding to the red dashed line is taken to the next panel. **(e)** The extrapolated count rates at P = 12 μW from the contour plot of panel (d) is plotted against the OPA wavelength. This indicates that the *emission yield* also maximizes along with the *emission order n*, consistent with SPR dominance. The solid green line is the fit from the same ATE model and identical set of parameters that led to the red line of panel (c). The shaded segments of the spectra in panels (c) and (e) show the approximate spectral range of the SPR excitation. All error bars correspond to the ± 1σ uncertainty.

average value of the emission order $\bar{n}$ for the 3 runs of Figs. (4-10,11,12) is subsequently calculated and shown in Fig. 4-13(c). A peak around 660 nm is manifest. This is followed by a contour plot of the emission yield in Fig. 4-13(d), derived from the extrapolated power-law fits. We now discuss the SPR-assisted ATE model that generates the solid red line of Fig. 4-13 (c) and the solid green line of Fig. 4-13(e), simultaneously, demonstrating a satisfactory explanation of the observed data based on the SPR-enhancement conjecture.

The observed power law suggests that the emission mechanism relies on a linear combination of multi-photon and above-threshold processes. The latter involves emission orders larger than what is needed to overcome the barrier (see Fig. 1-9). Unlike in pure MPE processes with integer emission orders, the average emission order $\bar{n}$ in ATE is, in general, a non-integer number due to the underlying *n*-photon contributions to the process. The emission count rate is described by

$$C = N\sum_n a_n I^n p_n \equiv N\sum_n C_n, \tag{4-5}$$

where *N* is an overall scaling factor, $a_n$ is a transition probability, $p_n$ is a population distribution, and *I* is the total intensity. This equation motivates the ATE schematic in Fig. 1-9. For example, considering the 4-photon ATE process, each red arrow adds a factor *I* to the count rate. The process starts from states in the Fermi-Dirac distribution with population $p_n$, while the total process has some quantum mechanical transition probability indicated with $a_n$. For the transition probability $a_n$, $a_{n+1}/a_n \ll 1$ which indicates that higher-order processes are less likely to occur, as appropriate for a perturbative series for which $a_n \propto \xi^n$, where $\xi$ is a constant [75].

The intensity is given by

$$I = I_0\left[1 + \delta\exp\left(-\left(\lambda - \lambda_0\right)^2 / \lambda_w{}^2\right)\right], \tag{4-6}$$

where the peak laser pulse intensity, $I_0 = P\big/\left(\Delta t \pi r^2 f_{rep}\right)$, depends on the average power, *P*, the laser pulse duration, $\Delta t$, the radius of curvature of the nanotip, $r$ (~ 50 nm), and the laser repetition rate, $f_{rep}$ (= 1 kHz). The SPR peak wavelength is $\lambda_0$ = 660 nm and the resonance width is $\lambda_w$ = 30 nm. A modest field enhancement factor of $\delta$ = 3 provides reasonable agreement with the experimental data. Both the order and the count rate at the resonance restrict the magnitude of the enhancement factor. The laser linewidth is three times narrower than the SPR width, as can be seen comparing Fig. 4-9(a) with Fig. 4-13(c,e), and can thus be ignored. The population $p_n$ is that of the state from which the multi-photon excitation starts. A value approximately given by the Fermi-Dirac distribution would be expected for a metal,

$$p_n = \frac{1}{\exp\left(\frac{\phi - n\hbar\omega}{k_B T}\right)+1}, \qquad (4\text{-}7)$$

where the workfunction $\phi = 5.25\ \mathrm{eV}$ for gold is lowered by the energy of *n* photons with angular frequency $\omega$ needed for the MPE process. $k_B$ is the Boltzmann's constant and *T* is the local temperature at the nanotip. The population equals one for states below the Fermi energy, while the energy of 3 photons of wavelengths ranging from 500 to 700 nm is needed to initiate MPE from these states. A surprising observation from Fig. 4-13(c) is thus that the order parameter hovers around 2 (instead of 3) for wavelengths from 500 to 620 nm, indicating a dominant 2-photon process. Moreover, the emission rate in Fig. 4-13(e) does not drop rapidly by orders of magnitude as the wavelength increases in this same range. This is in contrast with the Fermi-Dirac distribution which would predict such a behavior above the Fermi level. We assume that the first transition in the 3-photon process saturates the distribution, which would make the population independent on the wavelength and the order value equal to 2 [76]. We model this by replacing $n \to n+1$ in Eq. (4-7). Distributions of hot electrons with a width of 1-2 eV have been predicted for a gold surface by laser excitation providing a candidate for the saturated state [77].

The average order parameter is obtained from Eq. (4-5) by determining the slope as a function of intensity on a log-log scale

$$\bar{n} = \frac{I}{C}\frac{dC}{dI} = \frac{I}{C}\sum_n n a_n p_n I^{n-1} = \frac{1}{C}\sum_n nC_n\ . \qquad (4\text{-}8)$$

Upon inspection of the right-hand side, this equation yields the average order $\bar{n}$. It has no further adjustable parameters and the model is strongly constrained by the combination of the experimentally measured order parameter and emission rate (even when the experimental data is sparse). For the model curves of Fig. 4-13(c,e), the parameters are

$$a_1 = 1\ W^{-1}m^2s^{-1},\ a_2 = 1.7\times10^{-16}\ W^{-2}m^4s^{-1},\ a_3 = 3.0\times10^{-32}\ W^{-3}m^6s^{-1},$$
$$a_4 = 1.0\times10^{-47}\ W^{-4}m^8s^{-1},\ a_5 = 7.0\times10^{-63}\ W^{-5}m^{10}s^{-1},\ \text{and}\ N = 1.3\times10^{-13}.$$

(4-9)

The objective of this study is to find a flexible electron source that lowers the amount of laser power needed, so that the source is less likely to suffer damage. The mechanism discussed above satisfies this objective. It is, however, important to look at alternative mechanisms that could explain the resonance without providing protection from damage. In particular, no significant gain has been made if a plasmonic resonance or an intermediate state resonance heats the nanotip and drives thermionic electron emission or photo-assisted thermionic emission. The strong non-linearity of thermionic emission could in principle change the order parameter as a function of wavelength in the presence of a resonance. The question then is if our data rules out thermionic emission.

The intensity-dependent data requires an assumption on how the tip temperature depends on laser power. Assuming the temperature is linearly dependent on the laser power delivered to the tip, that is $T(P)\propto P$, the Richardson-Laue-Dushman (RLD) equation [53,55],

$$C_{RLD} = \left(\frac{\pi r^2}{e}\right)\cdot A_{RLD}T(P)^2\exp\left(\frac{-\phi}{k_BT(P)}\right), \qquad (4\text{-}10)$$

predicts an electron emission that is much more strongly dependent on the power than a power-law with order 2. This would be inconsistent with our data for wavelengths in the 500-620 nm range. Moreover, a resonance would change the average order $\bar{n}$ and the emission yield much more than observed, as the laser power now enters not only in the exponent of the power, but also in the exponential. A model similar to the photon-enhanced thermionic emission (PETE) mechanism [78] would modify Eq. (4-10) by multiplication with $I^n$ and would lower the workfunction as in Eq. (4-7). These models all lead to variations in emission and order that are too large. Making the resonance much weaker can give some reasonable agreement with the electron emission yield of Fig. 4-13 (e), but only at the expense of the fact that the modeled average order parameter would lose its resonance and hence would no longer fit to the data in Fig. 4-13(c). As pointed out earlier, the simultaneous presence of a resonance in the order and emission rate strongly constrains the possible emission mechanism. The previous SPR-assisted ATE model therefore seems to provide the only viable explanation. In Eq. (4-10), $r$ is the radius of a hemispherical emission site at the nanotip, $e$ is the unit electric charge, and with $m_e$ being the electron mass, $A_{RLD} \equiv e m_e k_B^2 / 2\pi^2 \hbar^3 = 120.17\ A/cm^2 K^2$ is a constant.

Lastly in this part, we add that a previous study on gold nanowire damage predicts a cooling time of ~ 10 μs [79]. On the other hand, the silica fiber core in our gold-coated fiber tip is a thermal insulator, which hampers cooling as compared to solid metal nanotips, consistent with studies of plasmon decay in the Kretschmann configuration [80] and nanoparticles [81]. The laser pulse stretch by about 2 orders of magnitude discussed at the beginning of this section, leads to a reduction of the peak optical intensity at the emitting nanotip, whereby decreasing the chance of damage in the metal coating due to thermal effects. In relation with this, we note that the peak laser pulse intensity, $I = P / \left( \Delta t \pi r^2 f_{rep} \right)$ depends on the average laser power, $P$, the laser pulse duration $\Delta t$, the radius of curvature of the nanotip $r$ (~ 50 nm), and the laser repetition rate $f_{rep}$ (= 1 kHz). For P = 12 μW and a pulse duration of 5 ps, the peak intensity is $I_{pulsed} = 9.2 \times 10^{14}\, Wm^{-2}$ with an associated electric field of ~ 1 V $nm^{-1}$. We may compare this with the estimated intensity in the CW experiment of Section 4.2 that was pointed out after Eq. (4-3) to be $I_{CW} = 5.4 \times 10^{8}\, Wm^{-2}$.

As predicted from the plasmon dispersion relation, the finite curvature and roughness of the Au apex coating furnishes the available momentum for the plasmon-photon coupling and thus also a decay channel. Therefore, a finite transmission of the input field is expected near the tip apex. Consequently, the radiative plasmon decay detected by a photodiode detector in the far-field of the probe apex supplements additional information on the SPR spectrum and is expected to be in general agreement with the electron emission response of the fiber tip. The optical transmission which represents the photonic response of the tip was extracted at the Oak Ridge National Laboratory (ORNL) implementing the phase-sensitive detection technique. A mercury arc lamp was used as a broadband source. A high-resolution monochromator (ORIEL™ CS260™) was implemented to scan through the visible spectrum. The output signal of a photomultiplier tube (PMT) in the far-field of the fiber tip source was subsequently measured using a lock-in amplifier referenced with the optical chopper TTL signals. The result is shown in Fig. 4-14(a) along with a computational model of the absorption spectrum of the gold-coated FONTES.

To provide support for the observed photonic signal, the FDTD technique was employed to numerically compute the required fields using a Lumerical™ simulation package performed by Dr. Ali Passian at ORNL. FDTD solves the Maxwell equations over a Cartesian mesh throughout the modeling structure, where each cell is characterized by the properties of the material it occupies, here, the frequency-dependent dielectric functions of Au and $SiO_2$. From the relationship between the time increment, incident field wavelength, and nanostructure size, an appropriate mesh is created. The fields are then calculated by spatially staggering the field component on a so-called Yee-cell and updating them via leap-frog time-marching. Thus, transient fields can be effectively computed when a suitable source field is incorporated. By obtaining the fields everywhere, the absorption may be calculated from an integration of the loss function over the metallic subdomain. Within the FDTD computational domain, a 2-μm-long probe with an apex radius of curvature of 50 nm was created from a silica core and a 50 nm thin film of gold. To obtain the response of the probe, a 2-fs pulse was brought into interaction with the Au domain from within the probe achieved by incorporating a dipole radiation field near the probe on the symmetry axis. Any ensuing photon-plasmon coupling that may occur is expected to exhibit a resonance structure consistent with the normal modes of charge density oscillations on the probe surface.

The calculated absorption spectrum, shown in Fig. 4-14(a), exhibits two main features; a spectral peak at 390 nm corresponding to a resonant excitation of surface plasmons when the dipole moment of the exciting source is perpendicular to the symmetry axis of the probe, and a second peak at 626 nm corresponding to plasmon excitation with the dipole source radiation field moment parallel to the symmetry axis of the probe. The experimentally observed absorption peak (assumed to be represented by the measured transmission peak) obtained at ORNL using gold-coated fiber tip is in general agreement with the calculation results.

The simulation results of the instantaneous radiation field propagating in a tapered uncoated fiber are shown in Fig. 4-14(b-d) at three different instances of time. The radiation due to the 2-fs dipole source is transmitted out of the tapered fiber into the vacuum and the corresponding computed optical absorption is negligibly small across the visible spectrum. This is in stark contrast with the analogous simulation for a gold-coated fiber tip shown in Fig. 4-14(e-g). Reflection and resonance effects due to the presence of the Au coating are manifest here after the dipole source is excited in panel (e). Fig. 4-14(g) shows the instantaneous field distribution ~ 4 fs after the snapshot in Fig. 4-14(e) where the dipole source is almost died out while a resonant hot spot is visible near the apex due to the SPR excitation.

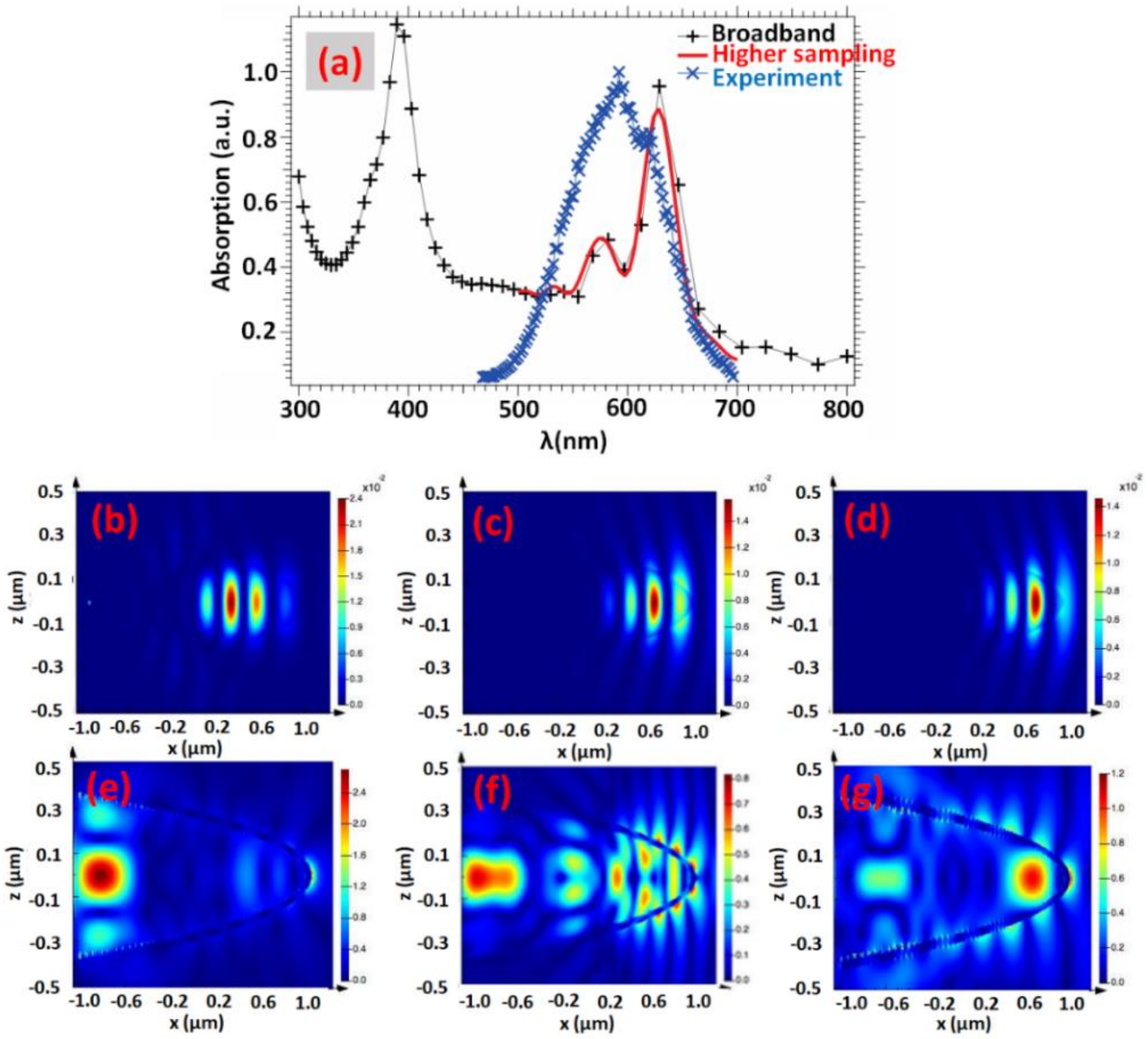

**Fig. 4-14 | FDTD simulation results. (a)** The calculated absorption spectrum shows two peaks. The one at ~630 nm is close to the region where we observed evidence of the SPR excitation in Fig. 4-13(c,e). **(b-d)** Light propagation in a tapered uncoated fiber tip is shown to lead to zero optical loss. **(e-g)** The Au-coated fiber tip supports SPR excitation. Refer to the main text for more details. The computer simulations were performed by Dr. Ali Passian at ORNL.

## 4.4 Discussion

The presented measurements demonstrate the first observation of SPR-assisted electron photoemission from an optically driven nanoscale fiber tip coated with a thin film of gold. The electron emission can be driven with a low-power femtosecond laser oscillator. The proposed photonic fiber-optics approach, as opposed to free-space scattering, is therefore a viable path to electron generation. The described fiber-based source can be readily configured to supply electrons with the same degree of 3D spatial maneuverability as in the SPM techniques. That is, fast electron pulses can be delivered from a probe with nanometer lateral and vertical resolution to specific sites of a given specimen. Similar to STM operation, it is expected that the proposed electron nanotip source can operate under ambient conditions, extending its use to a variety of nanometrology applications.

Plasmon bands in other nanostructures of similar size and curvature also yield a spectral peak associated with excitation of specific normal modes of the surface charge density in the gold thin film of a given curvature [67,68]. The computationally determined absorption spectrum of Fig. 4-14(a) exhibits two major resonance peaks associated with plasmon excitation with longitudinally and transversally polarized field components. One of the predicted peaks led to the current study. A search for the ~ 400 nm plasmonic resonance was not attempted in the present work as the employed OPA does not support laser emission in this wavelength range which could yield strong electron emission at even lower laser powers as the power-law order of photo-emission is in general smaller for more energetic photons. Higher electron emission rates with a further reduced risk of damage could be the result. Overall, a time-resolved nano-scale probe, that can deliver combinations of electron and light pulses is thus feasible. In addition, inexpensive CW diode lasers, widely available in the visible range, can be used to achieve CW photocurrents at modest applied bias potentials below the onset of dark FE as demonstrated in Section 4.2.

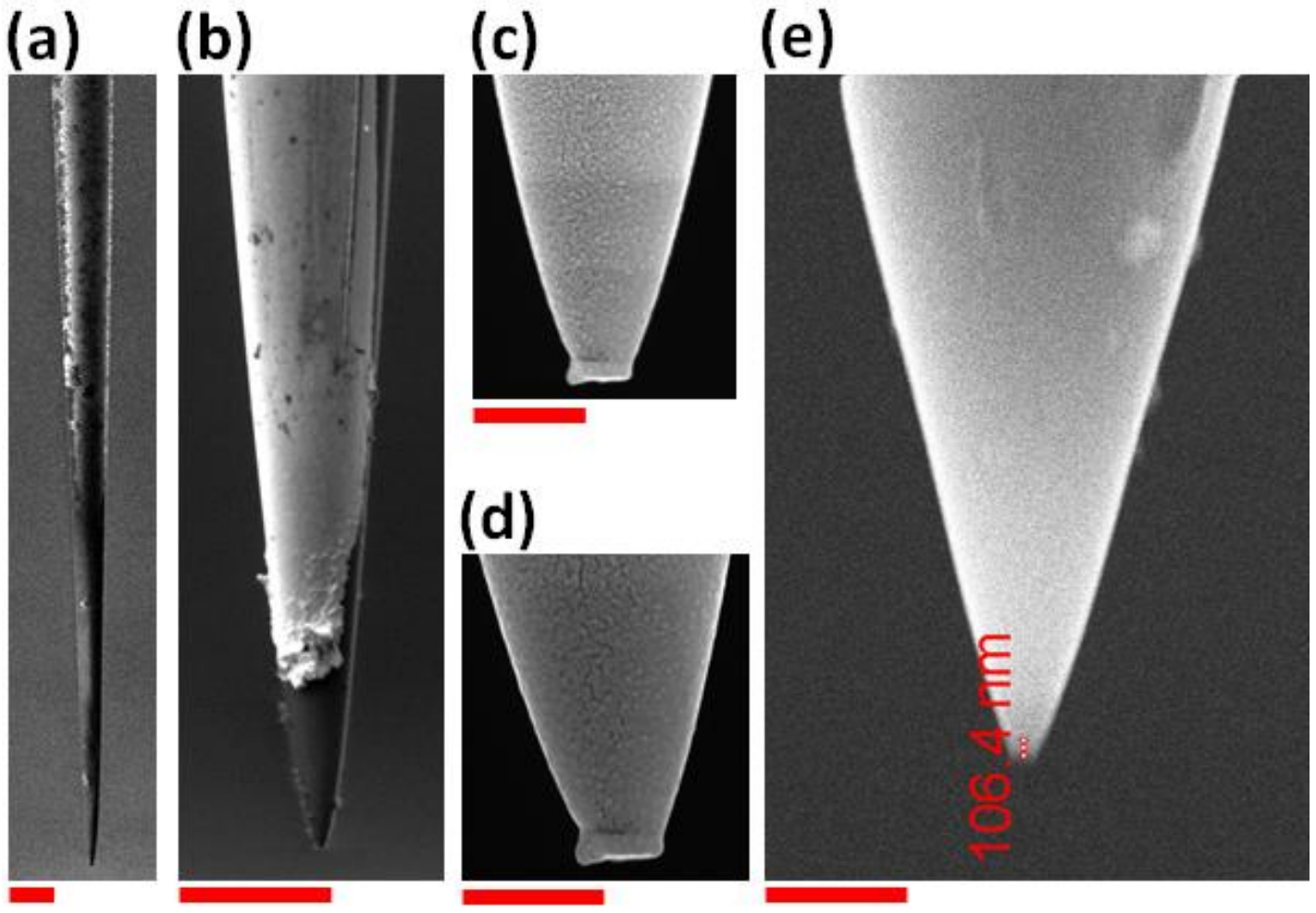

**Fig. 4-15 | SEM images after the experiment. (a)** Image of the tip used in the CW experiments (scale bar = 10 μm), and **(b)** the tip used in the OPA experiments (scale bar = 5μm). **(c)** An identically prepared tip was used in a damage test (scale bar = 500 nm). **(d)** No damage is visible after the test (scale bar = 500 nm). **(e)** An example of another sharp tip we achieved is shown (scale bar = 1 μm).

SEM images of the fiber tips used in the CW and OPA tests *after the experiments* are shown in Fig. 4-15(a) and (b), respectively. It is also important to note that the optical absorption does not necessarily damage the fiber tips. To demonstrate this, a representative fresh fiber tip shown in panel (c) was used for a damage test. The OPA output at $\lambda_{ML}$ = 500 nm and P = 14 μW was used to drive the fiber tip for 2 minutes. A constant detection rate of 150 cps was observed at $V_{tip}$ = 0. The high-resolution SEM images of this tip taken after the test in Fig. 4-15(d) shows no damage incurred to the coating. This demonstrates that for a sufficiently low range of average powers the fiber tip sources are reliably operable for extended time intervals. The survival of the two fiber tips used in the CW and OPA experiments for this work over long exposure times to the laser field can testify this. Throughout the experiments, however, high intensities and tip voltages where also tried towards the end of each set which can ultimately and abruptly disrupt the coating. We stress that the tip coatings were not damaged throughout all the experiments of this chapter. The way we know this is that the damage occurs quite abruptly and when it happens, the $V_{tip}$ corresponding to the onset of DC FE overshoots to several times its initial value (~ -450 V in this work). This is accompanied by a dramatic deterioration of the laser-induced electron yield and switching properties. From another perspective, the fact that the tips with a damaged coating do not photoemit show that the electron emission is indeed originating from the sharp metal coating near the apex and not the heated glass of the fiber core.

## 4.5 Outlook and future work

The results presented in this chapter demonstrate the first observation of dynamic electron emission from nanoscale curved thin films of gold – effectively, gold nanotips with glass cores. The detected emission current is found to unequivocally carry the multiscale time signature of the exciting laser field from CW to short pulses. The observed fast switching speeds, specifically in the CW experiment at $\lambda$ = 405 nm is particularly noteworthy as the switching speeds achieved here are 3-5 times faster than those reported in [69] using the same CW wavelength. The photonic approach, as opposed to free-space scattering, is seen to be a viable path to an electromagnetically cleaner electron generation. Interestingly, the proposed electron gun can potentially operate under ambient conditions, extending its use to a variety of nanometrology applications. In exploring wavelength tunability, our analysis revealed a resonance spectral feature around $\lambda$ = 660 nm. It is noteworthy that our finding agrees well with simulations of the optical field enhancement at a gold nanotip with the radius of curvature of 10 nm reported in [82] which gives a theoretical peak at 680 nm attributed to SPR.

As for the average pulsed laser power in the OPA experiments, the reader must note that an energy per pulse of ~10 nJ (10 μW at 1 kHz) inputted to the fiber tip in the OPA experiments at the resonant wavelength saturates the count rate at 1 kcps, namely 1 electron per laser pulse on average. The typical energy per pulse with the fs laser oscillator in the experiments with W nanotips (which do not support SPR at 800 nm) such as the ones described in CH.3, where the pulse energy can be typically 1 nJ (80 mW at 80 MHz), on the other hand, is often sufficient to give at least $10^6$ cps which is only 0.0125 electrons per laser pulse. Noting to the optical loss by a factor of 300 in far field transmission through the fiber tips, as well as the stretch in the laser pulse by a factor of at least 10 as indicated earlier, it is safe to claim that in the OPA experiments, a pulse energy of less than 10nJ/(300×10) = 3.3 pJ with the right color can generate 1 electron, while in the side-illumination of a W nanotip at 800 nm, 1-nJ 100-fs pulses can give 0.0125 electrons on average which is smaller by a factor of about 100. Therefore, it is justifiable to claim that in the OPA experiments, the average and peak powers are comparatively *low*.

We can also base the above consideration on quantum efficiency (QE), defined as (number of electrons per emitted pulse) / (number of incident photons per irradiating pulse). For fiber tips in the OPA experiments with a pulse energy of 3 pJ at 660 nm and 1 electron emission per pulse, we obtain QE = $10^{-10}$. For W tips in the oscillator experiments with a typical 1-nJ pulse energy at 800 nm and 0.0125 electrons emitted per pulse as described above, QE = $3\times10^{-15}$. The QE for the FONTESs is therefore about 5 orders of magnitude larger than that of W nanotip sources in our experiments, despite the sustained pulse stretching by an order of magnitude.

Our results presented in this chapter, support an SPR-assisted emission mechanism, and constitutes a proof-of-principle. Further investigations are warranted to establish the detailed role of plasmons in this work in order to determine and understand the many parameter-dependencies on various materials deposition, multilayer depositions, excitation laser properties, the coating thickness, size and tip curvature, exhaustively.

### 4.5.1 A few envisioned applications

Before we close this chapter, let us list and review a few potential applications of the FONTESs studied throughout this chapter.

**1. Flexible and inexpensive, suitable for UEM.** Fiber tip electron sources are flexible; once the laser beam is optimally coupled to the core of the base fiber no further alignment will be needed which makes it possible to produce alignment-free plug-and-play electron sources. In particular, the requirement for any kind of alignment will vanish entirely using laser systems with a fiber coupler on their output port, as in such cases, the fiber tip can be plugged onto the laser head directly from its coupling end through end-fire coupling. It is also notable that as we demonstrated in this chapter, inexpensive low-power CW diode lasers in the visible range of the spectrum can be used to excite electron emission assisted by surface plasmon resonance excitation. This brings down the cost of operating laser-driven electron tip sources such as W nanotip needles by at least two orders of magnitude noting to the high price of pulsed laser systems as well as the guiding optics required to irradiate the emitter held fixed inside the vacuum chamber. One can thus envision plug-and-play electron sources produced and commercialized based on an inexpensive diode laser with a replaceable fiber tip connected to its output port. Inside the vacuum chamber, on the emitting end, the tip source can be mounted on a mechanically or electronically movable stage in order to deliver the electrons anywhere in the 3D space on-demand. In addition, our fiber tip electron sources are the first bimodal emitters of their kind in the sense that one can achieve both continuous and pulsed electron emission from them using CW and pulsed lasers, respectively. The latter procures fast timing capability aside from the 3D spatial control which is desired for the so-called 4D electron microscopy applications to make real-time movies of various atomic and molecular reactions and interactions as in UED and UEM. To this goal, such fiber tip sources can be incorporated in existing SEMs and TEMs. Once again, the possibility to drive these sources alignment-free will help bypass the difficulties in transmitting the laser beam inside the microscope column to irradiate the emitter from the vacuum side.

**2. Flexible, suitable for electron nanolithography and STM.** The flexibility of the fiber tip makes of it an ideal electron source for raster scanning as in electron nanolithography applications. Specifically, such a tip source provides a higher spatial coherence, and consequently, a better focus as well. The resultant enhanced spatial resolution, along with the flexibility of the source, due to which the user no longer needs to send the laser beam into the vacuum chamber and focus it onto the tip source through sophisticated optical components, will therefore improve high-resolution near-contact electron lithography and raster scanning. Another example where such a flexible electron source will be advantageous to implement is in STM using such laser-driven fiber tip sources to probe the sample in the near-contact mode (which is imperative in this technique) without the need to irradiate the tip source from the side using conventional alignment-sensitive schemes. This will open up the route to fast time-resolved STM.

**3. Back-illuminated, suitable for ePPM.** Another application of the fiber tip electron sources is ePPM. The advantage in exploiting this technique which is demonstrated using metallic nanotips emitting through grating-coupled SPR is that the fs laser beam does not need to irradiate the apex of the nanotip directly. Rather, a nano-grating etched or patterned by ion-milling a few microns away from the apex on the shank of the emitter is shined by the laser beam to excite the surface plasmons from there. This minimizes the amount of scattered light near the apex which comes to the proximity of the sample in this technique. This is useful to avoid damaging sensitive samples like biological specimens with the fs laser pulses which would otherwise be focused to a high fluence surrounding the apex. A still more advantageous scheme will thus be to implement FONTESs which give the possibility of complete back-illumination which entirely eliminates the need to send any laser beam into the vacuum chamber where the sensitive sample is located.

**4. Addressable fiber tip bundle.** In addition to all of these, the flexibility of the fiber tips and their alignment-free operation makes it practical to use a set of them as a bundle of independent electron emitters. Such a bundle of FONTESs can be addressed by one laser and a set of beam splitters, or through multiple independent lasers, CW or pulsed, conveniently from outside the vacuum chamber. Such a bundle is envisioned to be utilized in applications such as parallel-processing in electron nanolithography.

Some of the ideas discussed above are illustrated in Fig. 4-15 with which we close the present chapter. In the next and last chapter, a potential application of FONTESs as a possible platform to create a novel spin-polarized electron source is discussed.

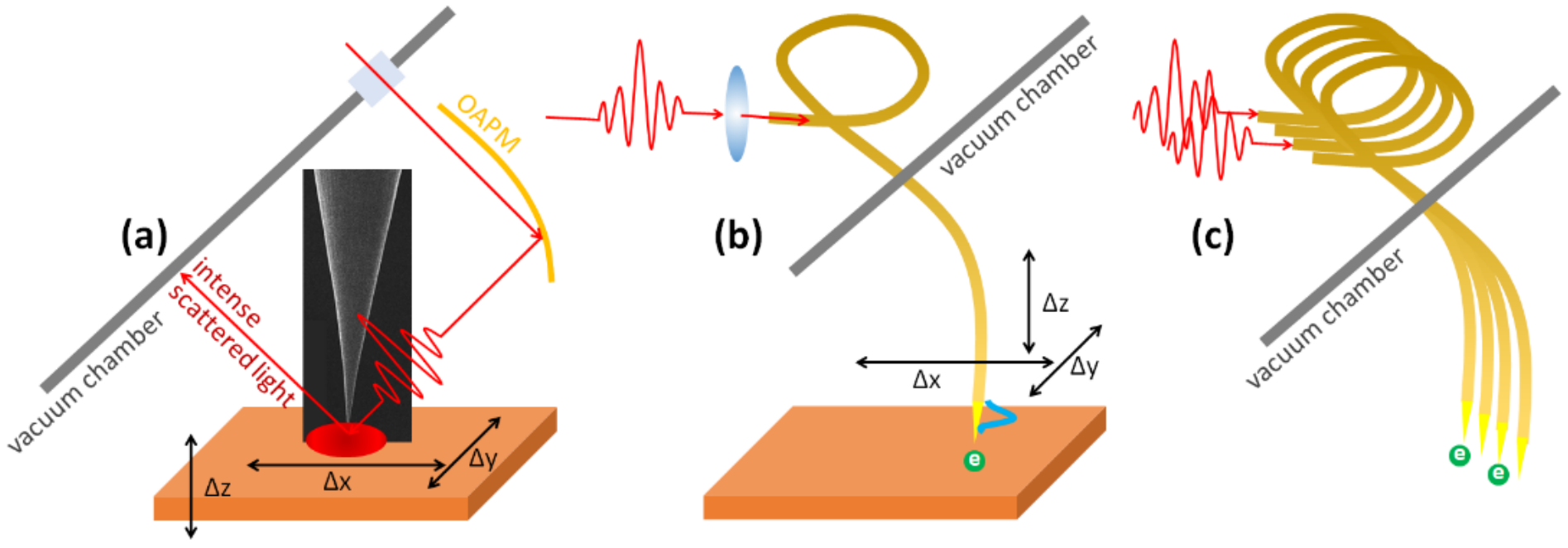

**Fig. 4-16 | A few potential applications of fiber optic nanotip electron sources. (a)** Metallic nanotip sources such as chemically etched W needles shown in the SEM image in this panel require sending intense laser pulses into the vacuum chamber. This scheme which is highly alignment-sensitive is not suitable to scan the source on top of a sample in ultrafast STM applications. Although moving the specimen itself solves this problem, the fact that the intense laser pulses need to be focused in the proximity of sensitive samples begs for better solutions. **(b)** The alignment-free fiber optic nanotip emitters characterized in this chapter seem to be ideal to employ for this and similar purposes. In this scheme, no laser light needs to be sent into the vacuum chamber in free space. The blue pulse at the fiber taper represents an excited plasmon. **(c)** Another potential advantage of fiber-based sources is that it is, in principle, possible to bundle them. This way, the various *point-like* electron sources may conveniently be addressed from outside the vacuum chamber using CW or pulsed lasers as needed.

# CHAPTER 5: PROPOSAL FOR A LASER-DRIVEN SPIN-POLARIZED ELECTRON SOURCE BASED ON THE SPIN HALL EFFECT USING FIBER OPTIC NANOTIPS

In CH.2, we theoretically demonstrated that compared with unpolarized spatially coherent sources of electrons, exploiting spin-polarized electron sources (SPESs) will enhance the HBT antibunching signal. Moreover, a switchable SPES provides a *spin-knob* that can be implemented to observe the change in the antibunching signal amplitude by turning it on and off for a conclusive demonstration of the effect of the PEP on ultrashort free electron pulses. This is indeed the case due to the fact that the repulsive Coulomb force is independent of spin.

The SPES normally taken advantage of in our group is based on GaAs, often shards of GaAs substrate. It is shown that circularly polarized femtosecond pulses of a Ti:Sapphire laser oscillator centered near 800 nm can give rise to ultrashort electron pulses with over 10% spin polarization; the electron spin direction is set by the irradiating circularly polarized light helicity [4,5]. Appropriate wave-plates then act as the sought-after *spin-knob* by switching the laser polarization from circular to linear, and vice versa. Fabricating GaAs nanotips will therefore give rise to spatially coherent optically controllable SPESs which can be accommodated to study the effects of the PEP in ultrashort electron pulses. A few SEM images of a typical GaAs shard are shown in Fig. 5-1.

Currently, there is ongoing research in Prof. T. J. Gay's group here at UNL to make GaAs nanotips by electrochemical etching. In a proof-of-concept effort, a wedge of GaAs that was thinned through etching was used by the author of the present dissertation to experiment with making tip-like structures at its thinned end by focused ion beam (FIB) milling using FEI™ Helios-660 dual-beam microscope.

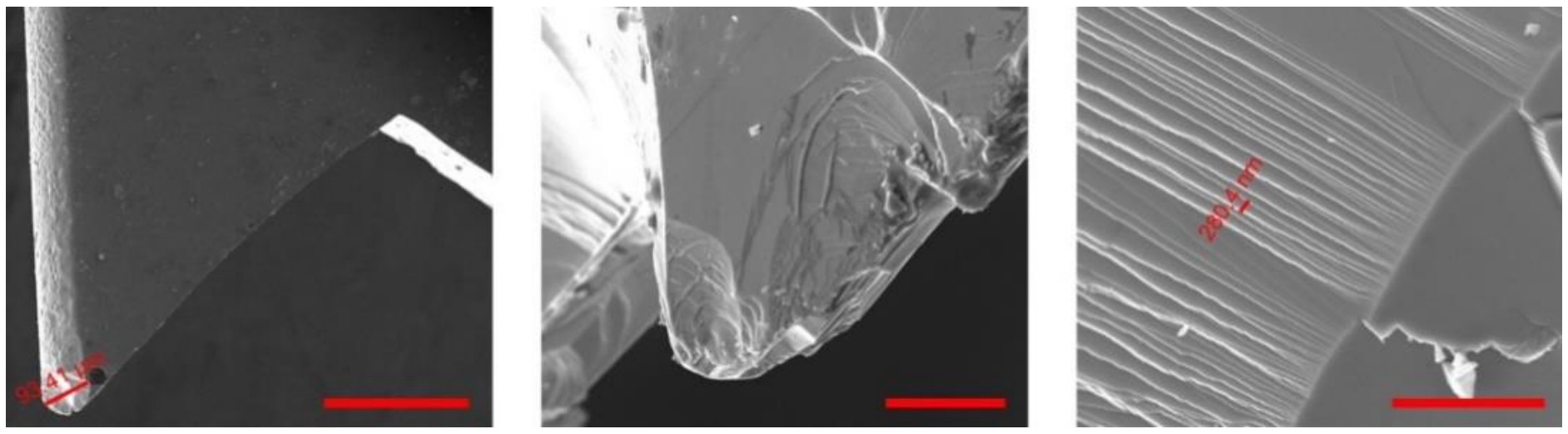

**Fig. 5-1 | SEM images of a typical GaAs shard.** From left to right, the horizontal scale bars represent a length of 300 μm, 10 μm, and 4 μm. In the experiment with such a shard, a femtosecond near-IR laser beam with circular polarization is focused on the tip of the shard, where tiny structures, some of them nanotip-like, exist.

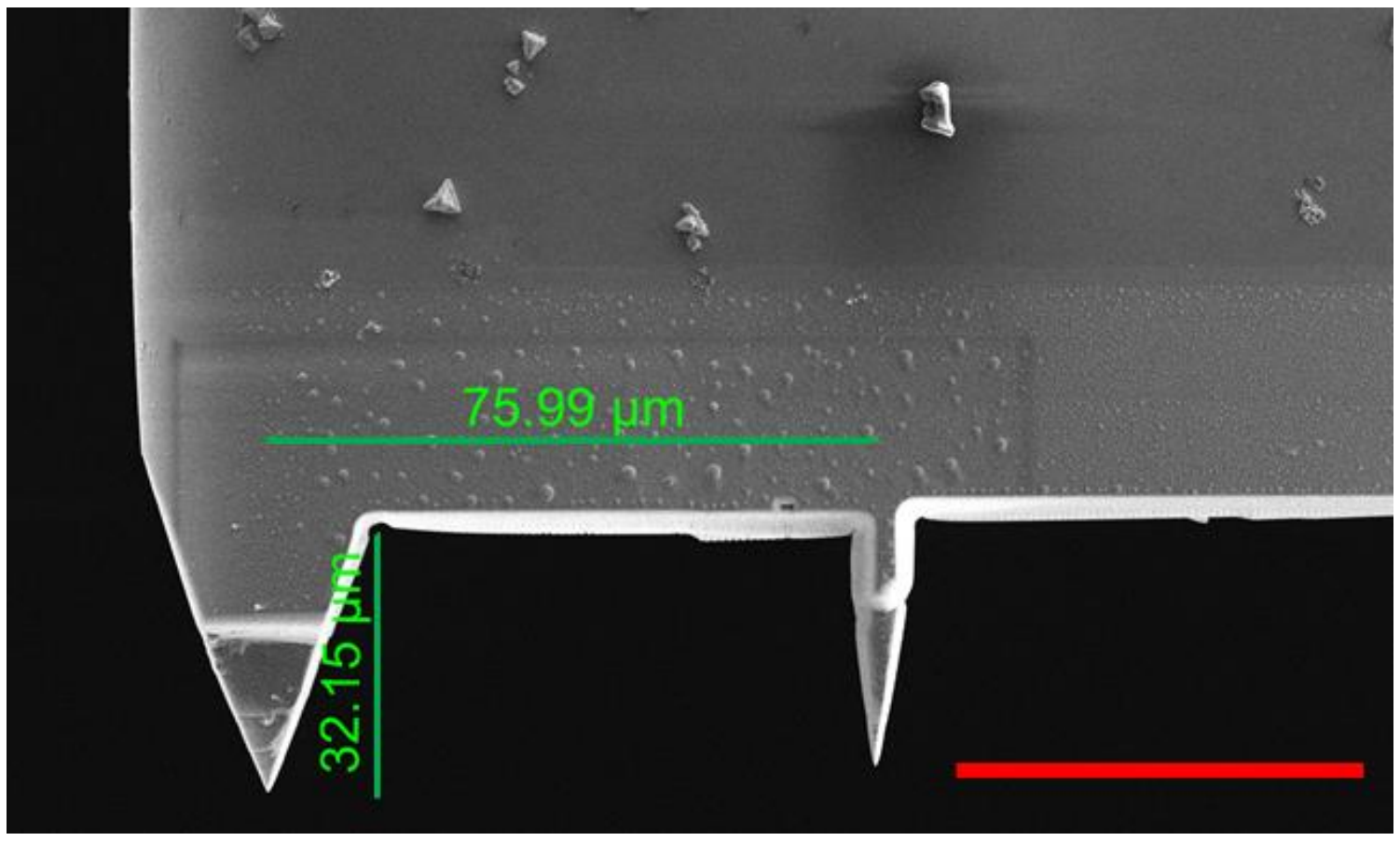

**Fig. 5-2 | SEM image of a focused-ion-beam-milled GaAs wedge.** Two tips were milled out of a thinned GaAs wedge. The horizontal scale bar represents a length of 50 μm.

A SEM image of the modified wedge is shown in Fig. 5-2. In the future, *annular* FIB milling when possible should ideally give rise to more well-defined tip radii and overall shape as in Ref. [83] where a silver nanotip is reported to have been made by this method.

The SPESs are not restricted to GaAs. As another example, magnetic samples were shown to be capable of giving rise to spin polarized emission. For instance, spin polarized field-emission from iron- and cobalt-coated tungsten tips have been reported [84]. In contrast with the GaAs source, however, these tips are not expected to lead to fast optical switching of the spin polarization noting to the fact that typically, an external magnetic field would need to be applied to switch the magnetization. This also suffers from technical difficulties including generation of the magnetic field inside the high-vacuum chamber and in the proximity of electron-optical components. Nonetheless, in a different context aimed to study the magnetization precession phenomenon in a thin film of iron, the photoemitted electrons from the specimen following its excitation with a near-IR fs laser beam that triggers the precession dynamics, where shown to carry spin-polarization [85]. In this work, a thin 8-monolayer film of iron to which an in-plane saturating magnetic field has been applied, was illuminated by a fs pump laser at 800 nm and duration $< 120$ fs. A burst of approximately 300 soft X-ray probe pulses from a free electron laser at 182 eV and 10 kHz supersedes each pump pulse at controlled time delays. Subsequently, the cascade of ejected photoelectrons which are liberated from the iron sample by the X-ray probe pulses is sent to a Mott polarimeter for detection of their transient spin polarization. Subsequently, by controlling the time delay between consecutive pump and probe pulses, the magnetization dynamics of the iron sample under the effect of the fs pump laser, encoded in the emitted photoelectrons, was reconstructed. This suggests that instead of an external magnetic field, an all-optical pump-probe scheme [86] may be used in the future to generate spin polarized photoelectrons from magnetized samples – nanotips in particular – upon altering their surface magnetization by incident near-IR pump laser pulses. The average degree of spin polarization of the emitted electrons is then expected to be adjustable with the pump-probe time delay within a sample-dependent dynamic range. In the present chapter, our goal is to design a new electron source based on another spin-dependent effect, namely, the spin Hall effect (SHE).

In the following sections, we first explain two key concepts; the *Mott scattering* and the notion of *spin current*. We will then move on to an overview of the family of Hall effects, most notably for our purpose, the SHE. Lastly, our SPES design based on a fiber optic nanotip electron source (FONTES) which was the topic of CH.4 will be given in the final section. This is a confluence of the two parts of this dissertation; a FONTES-based SPES that is of potential to be used in the future to study the effects of the PEP in pulses of free electrons with applications in electron microscopy, nanophotonics, and beyond.

## 5.1 Mott scattering

The present section provides an overview of the Mott scattering (MS) based on the Ref.'s [87,88]. This is a key concept in the context of spin-polarized electron sources and beams. The reason is that the electron beam spin-polarization is generally measured through the MS which is a type of electron-nucleus collision process. In fact, to characterize our spin-polarized sources, we take advantage of a home-built instrument called a Mott polarimeter which functions based on the MS of the incident (polarized) electrons off the heavy nuclei of a gold rod [5]. In the following, the physics of MS is reviewed in brief.

For an electron moving with velocity $\vec{v}$ in the electric field $\vec{E}$ of a nucleus, the induced magnetic field in the electron rest frame is given by

$$\vec{B} = -\frac{1}{c}\vec{v} \times \vec{E}, \tag{5-1}$$

in *cgs* units. Taking $\vec{r}$ as the position vector of the electron seen by a nucleus with atomic number $Z$, the magnetic field becomes

$$\vec{B} = \frac{Ze}{cr^3}\vec{r} \times \vec{v} = \frac{Ze}{m_e cr^3}\vec{L}, \tag{5-2}$$

where $\vec{L}$ is the orbital angular momentum of the electron, $m_e$ is its mass, and $e$ is the unit electric charge. The corresponding spin-orbit (SO) interaction potential is given by

$$V_{SO} = -\vec{\mu}_s \cdot \vec{B}, \tag{5-3}$$

where $\vec{\mu}_s$ is the spin magnetic moment of the electron related to its spin operator $\vec{S}$ through the equation,

$$\vec{\mu}_s = -\left(\frac{ge}{2m_e c}\right)\vec{S}, \tag{5-4}$$

where $g \simeq 2$ is the electron spin g-factor. Combining Eqs. (5-2) to (5-4), the SO interaction potential is obtained as

$$V_{SO} = \frac{Ze^2}{2m_e^{\,2}c^2r^3}\vec{L}\cdot\vec{S}. \tag{5-5}$$

It is precisely due to this SO interaction that the electron scattering cross section,

$$\sigma(\theta) = I(\theta)\left[1 + \vec{P}\cdot\hat{n}\right], \tag{5-6}$$

becomes spin-dependent. Here, $I(\theta)$ is the spin-averaged scattering intensity, $\vec{P}$ is the spin polarization of the incident electron, and $\hat{n}$ is the unit vector perpendicular to the scattering plane defined by the linear momenta of the incident and scattered electrons. Noting that an unpolarized incident electron beam can be pictured as having an identical number $N_\uparrow$ of spin-up as that of spin-down electrons, $N_\downarrow$, all moving in the same direction, the net polarization of the scattered beam at angle $\theta$ is defined as

$$P(\theta)=\frac{N_\uparrow - N_\downarrow}{N_\uparrow + N_\downarrow}. \tag{5-7}$$

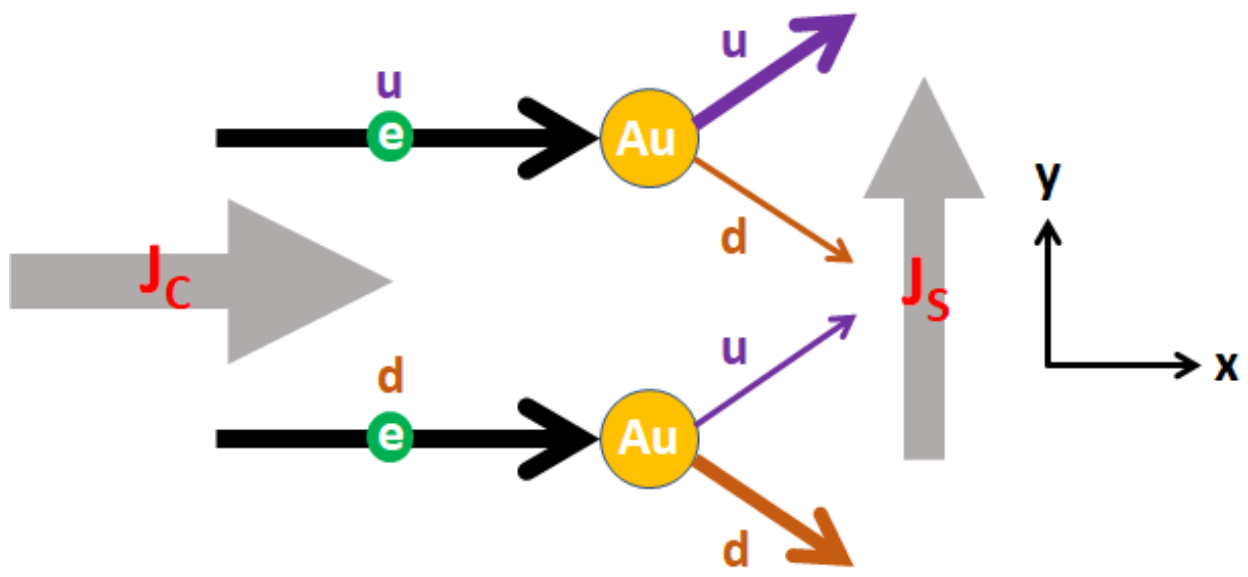

**Fig. 5-3 | Schematic representation of the Mott scattering (MS).** In this example, the majority of spin-up (u) electrons of the incident unpolarized charge current (CC) density $J_C$ are scattered from the Au nuclei in the +y direction, while most of the spin-down (d) electrons are scattered in the opposite direction, that is, the -y direction. Whereupon these scattering events, a non-zero spin polarization is formed which is carried by the scattered electrons in the direction transverse to that of the flow of the incident unpolarized CC, giving rise to a spin current (SC) in the +y direction. (Figure adopted from Ref. [89].)

The utility of this definition is therefore manifest; for an unpolarized beam $N_\uparrow = N_\downarrow$ and the polarization is zero as expected. For a partially polarized beam with more spin-up or spin-down electrons the numerator is non-zero. The ratio is 1 (-1) for a completely polarized spin-up (-down) beam. In addition, as can be seen from Eq. (5-5), the effect is more pronounced for high-Z nuclei such as those of gold. A schematic diagram showing the MS of electrons is depicted in Fig. 5-3.

## 5.2 Spin current

As discussed in the previous section and specifically in Fig. 5-3, the MS is a mechanism that can give rise to a spin-polarized current of electrons commonly referred to as spin current (SC). It is straightforward to quantify this core concept of the present chapter by noting to its similarities with the regular charge current (CC) that the reader is likely to be more familiar with from classical electrodynamics.

We know that CC is nothing but a flow of net electric charge. This can happen, say, in a metal where the ionic cores with a net positive charge form the crystal lattice. In their rest frame, the positive charges stand still while the negatively charged free electrons flow against the direction of an applied electric field. This generates a net current of electric charge which is known as CC, namely, a net current of spin-unpolarized electrons. It is important to note that for a CC to form, the particle flow through a cross-section of the beam must be non-zero. In the above example, a net flow of electric charge was due to the collective motion of electrons in one direction. However, with 50% of the free electrons in a conductive specimen moving, say, in the $+x$-direction and the remaining 50% moving in the $-x$-direction there is no net flow of charge across any cross-section of the current and hence no CC exists in this situation. It is clear that this is the case, for instance in the conductors in the absence of any external electric field, where the free electrons undergo thermal motion in random directions. On the contrary, when an external electric field is applied, all of the free electrons tend to move against the field forming a non-zero CC. In contrast, as discussed in the previous section, the MS is indeed a type of interaction which introduces a spin-dependent asymmetry in the scattering angle. With the incident beam being partially spin-polarized, the scattered total CC will be non-zero but will also be spin-polarized. If such electrons are evenly divided into two groups flowing in opposite directions but with opposite spin orientations, the resultant current will be a *pure* SC, namely one that carries no net electric charge but with a net spin polarization. The reason why we should consider the difference between the two spin-up and spin-down current densities can be easily comprehended noting to the fact that for a spin-up electron crossing a beam section from left to right, there is a net transfer of one unit of up spin to the right. Now if a spin-down electron crosses the identical section again to the right and at the same time, the net transferred spin to the right is zero. However, if the spin-down electron crosses the section from right to left instead, an up spin is added to the right volume while a down spin is subtracted from it which raises its total spin by 2 units of up-spin. Taking into account the CC, for a spin-up electron moving in the +*x*-direction, there is a net non-zero charge ($1.6\times10^{-19}C$) and a net non-zero spin ($\hbar/2$) transferred in the same direction. For a second spin-up electron joining the first one in moving in the +*x*-direction, both the net charge and the net spin transfer are doubled. But what if we add a spin-down electron to the first one instead of a spin-up electron? The total charge still doubles, however, the net spin that crosses a beam section is now zero. This is a *pure* CC which carries no spin polarization. Now consider the scenario in which two spin-parallel electrons move in the opposite directions. Here the net CC is zero while the net SC is also zero. Lastly, for a spin-up electron crossing a beam section towards the +*x*-direction and a spin-down electron crossing the same beam section towards the -*x*-direction at the same time, the net charge transferred through the section is zero same as in the previous case, however, this time the net spin transfer is non-zero which makes it a *pure* SC. This is illustrated in Fig. 5-4.

From classical electrodynamics, we know the current density is defined as

$$\vec{j} = nq\vec{v}, \tag{5-8}$$

with $n$ denoting the number density of particles of charge $q$ and velocity $\vec{v}$. Similarly, for the spin-up electrons moving in one direction after MS from a group of scattering centers such as Au nuclei, the CC density is

$$\vec{j}_{\uparrow} = n_{\uparrow} e \vec{v}_{\uparrow}, \tag{5-9}$$

while for the spin-down electrons which flow in the direction opposite to their spin-up counterparts in the same scattering process, the CC density is given by

$$\vec{j}_{\downarrow} = n_{\downarrow} e \vec{v}_{\downarrow}. \tag{5-10}$$

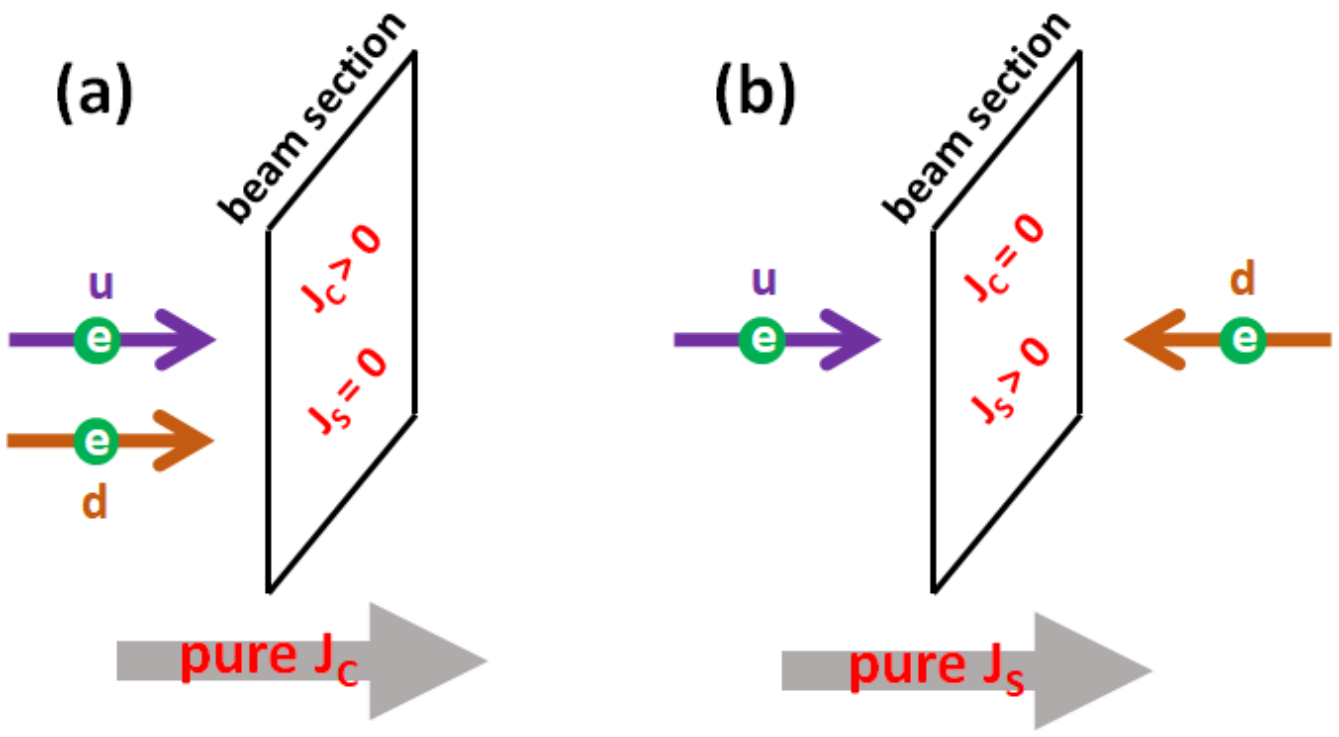

**Fig. 5-4 | Charge and spin currents are compared. (a)** A pure charge current (CC) crosses the representative beam cross-section shown in this panel. The incident current density from the left is spin-unpolarized. Consequently, no net spin is transferred across this section unlike a net electric charge, making it a pure CC. **(b)** A pure SC is formed when no net electric charge crosses an arbitrary beam section but there is a non-zero net spin transferred to each side of the section. In the example shown here, the net electric charge remains unchanged both to the left and to the right side of the section. In contrast, the net spin increases by two positive (up-spin) units on the right side of the section while it decreases (down-spin) by the same amount in the left side upon passage of the two electrons through the representative section shown here. As a result, there is no net transfer of electric charge in the electron beam, while a pure SC is present. For the same reason, pure SCs cannot be detected by a voltmeter alone.

The up (down) arrow indicates that the respective quantity belongs to spin-up (-

down) electrons. Consequently, the net charge current $\vec{j}_c$ on the axis on which the scattered spin-up and spin-down electrons travel is

$$\vec{j}_c = \vec{j}_\uparrow + \vec{j}_\downarrow = e\left(n_\uparrow \vec{v}_\uparrow + n_\downarrow \vec{v}_\downarrow\right). \tag{5-11}$$

In the MS of an unpolarized current of electrons, $n_\uparrow = n_\downarrow = n$ and $\vec{v}_\uparrow = -\vec{v}_\downarrow$. Plugging these into Eq. (5-11), we readily observe that the total CC density $\vec{j}_c$ for the scattered electrons is zero. This quantifies the qualitative description given earlier for the MS.

The SC can also be quantified on the same footing. Noting to the discussion that led to Fig. 5-4, the SC density $\vec{j}_s$ is defined as the difference between the spin-up and spin-down current densities, say, for the transversely scattered electrons in the MS, which gives

$$\vec{j}_s = \vec{j}_\uparrow - \vec{j}_\downarrow. \tag{5-12}$$

Plugging the Eqs. (5-9) & (5-10) in Eq. (5-12) we arrive at

$$\vec{j}_s = \vec{j}_\uparrow - \vec{j}_\downarrow = e\left(n_\uparrow \vec{v}_\uparrow - n_\downarrow \vec{v}_\downarrow\right) = e\vec{v}_\uparrow\left(n_\uparrow + n_\downarrow\right) = -e\vec{v}_\downarrow\left(n_\uparrow + n_\downarrow\right), \tag{5-13}$$

for the SC density vector. While for a pure CC, $j_c \neq 0$ and $j_s = 0$, for a pure SC the opposite is the case, that is, $j_c = 0$ and $j_s \neq 0$.

The examples discussed before Fig. 5-4 for different directions of propagation of spin-up and spin-down electrons are direct consequences of the continuity equation,

$$\nabla.\vec{j}_c + \frac{\partial \rho_c}{\partial t} = 0, \quad (5\text{-}14)$$

noting to the fact that the electric charge density $\rho_c$ is the source for CC. When the charge density is zero, the net current through a beam section perpendicular to the direction of the flow of the electrons is zero which can be easily verified using the Stoke's theorem. It is worth mentioning here that the spin-flip mechanisms provide examples in which the source term for a similar equation with $\vec{j}_s$ replacing $\vec{j}_c$ is non-zero. The MS can also be put on the same footing.

## 5.3 The spin Hall effects

The Hall effect (HE) was first discovered by Edwin Hall in 1879, but that was not the end of the story. Later findings revealed several other effects which share a number of their core features with those of the HE. Due to their notable similarities, we can now categorize all of them in one group labeled HEs. One of the members of this family of physical phenomena is the spin Hall effect (SHE) which is at the focus of the present chapter. Here, we first give an overview of the ordinary HE before concentrating on the main topic of this final chapter.

With an external electric potential applied between the two ends of a flat *non-magnetic* conductor or semiconductor sample, and in the presence of an applied magnetic field $\vec{B}$ perpendicular to both the surface of the specimen and the applied electric field $\vec{E}$ the Lorentz force on the flowing electrons with charge *e* and drift velocity $\vec{v}$,

$$\vec{F} = -e\left(\vec{E} + \vec{v} \times \vec{B}\right), \quad (5\text{-}15)$$

gives rise to the Hall field. This is identically the induced electric field in the direction $\vec{j}_c \times \vec{B}$. In this expression, $\vec{j}_c$ is the electric CC density due to the applied electric potential. As a result, a charge accumulation with opposite signs forms on the opposite edges of the sample in the direction perpendicular to $\vec{j}_c$ [47, pp.153 156]. The induced transverse electric potential is called the Hall voltage denoted with $V_H$. The standard experimental set-up to measure this quantity, which is one of the important characteristic properties of conductors and semiconductors, is illustrated in Fig. 5-5.

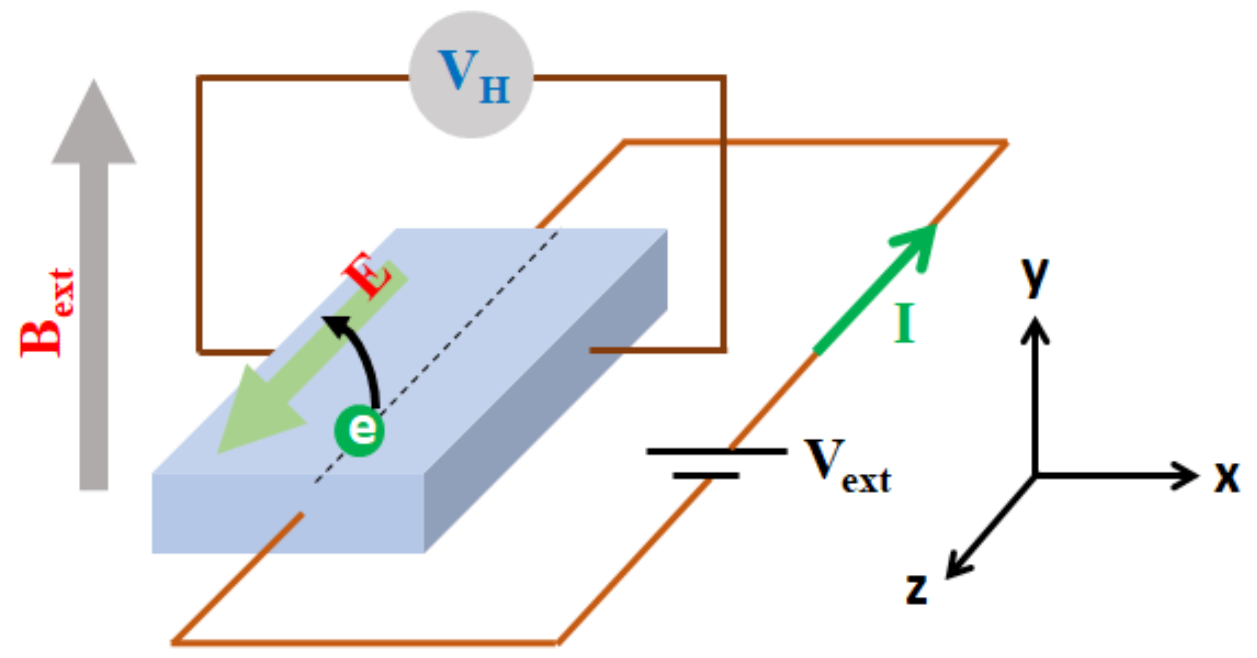

**Fig. 5-5 | Schematic illustration of the Hall effect (HE) set-up.** In the HE, the voltmeter measures the induced Hall voltage $V_H$ generated as a result of the potential difference given to the slab by the applied voltage $V_{ext}$. The flowing charge current experiences a Lorentz force in the presence of the perpendicular magnetic field $B_{ext}$ applied to the non-magnetic slab of the conducting or semiconducting sample. This gives rise to the Hall current in the transverse direction which in turn gives rise to the measurable Hall voltage $V_H$. As a result, two layers of charge accumulation with opposite signs form at the opposite edges of the specimen separated by the slab width. The Hall field induced in the example considered here is in the $+x$-direction. (Figure adopted from Ref. [90].)

The spin-dependent group of HEs, which include the SHE, the inverse spin Hall effect (ISHE), the anomalous Hall effect (AHE), and the quantum spin Hall effect (QSHE), are essentially governed by the MS overviewed in section 5.1. We are now well-positioned to categorize the family of spin-dependent HEs. The main resemblance of these phenomena to the ordinary HE is that in each of them a current, either a CC or a SC, is applied to the sample by an externally controlled agent (e.g. an external electric field to drive a CC), and a SC or a CC is resultantly induced in the specimen in the transverse direction. In the ordinary HE, both currents are CCs. For spin-dependent HEs, at least one of the currents is spin-polarized and is either a polarized CC or a pure SC. We first review the main features of the members of this group before focusing on the SHE.

- **Anomalous Hall effect (AHE):** A CC flowing through a ferromagnetic sample induces a transverse *polarized* CC, i.e. a *non-pure* SC. The effect is attributed to spin-dependent MS events in the sample and not the contribution of the intrinsic magnetization to the applied magnetic field. We have already seen that the MS tends to give rise to a transverse SC. Here, the net magnetization of the sample favors parallel spin orientation. As a result, there are more electrons of one spin orientation (parallel to the magnetization) than the oppositely oriented ones. Since the electrons with opposite spin orientations tend to move in opposite directions in the MS as explained before, the imbalance in their number gives the induced polarized current a non-zero electric charge as well. The AHE thus pertains to the ferromagnetic conductors only [91].

- **Spin Hall effect (SHE):** The passage of a CC through a non-magnetic sample in the absence of any external magnetic field induces a pure in-plane SC perpendicular to the CC as a result of the MS of the unpolarized electrons from the lattice nuclei and crystal impurities. The polarization of the induced SC is perpendicular to the scattering plane defined by the applied CC and the induced SC. Consequently, spin accumulations with opposite polarizations pile up on the opposing edges of the sample transverse to the induced SC. Unlike in the AHE, there is no net magnetization or applied magnetic field in this case. This makes the number of spin-up and spin-down electrons which flow transversely to the applied CC to be equal to each other. As a result, the induced CC density is zero in this case while the pure SC density is non-zero. It is important to note that application of an external magnetic field will diminish and ruin the SHE by causing magnetization precession. Therefore, the SHE is strictly pertinent to *non-magnetic* specimens. In addition, comparing the SHE with the ordinary HE, the *spin diffusion length* can be much longer in the SHE compared with the *Debye screening length* which is the equivalent quantity in the ordinary HE. While the latter is typically on the order of 1-10 nm, the former, depending on the specimen, can be from a few nanometers to above 1 μm [92].

- **Inverse spin Hall effect (ISHE):** This effect is similar to the SHE with the roles of the CC and the SC reversed. A pure SC generated in a *non-magnetic* sample induces a transverse in-plane CC. Due to the MS events, if the spin-up electrons deflect to their right, say, toward the -*y*-direction, the spin down electrons scatter to their left. However, since they are flowing against the direction of the spin-up electrons in a pure SC, their left side is again in the -*y*-direction. This means that all of the scattered spin-up and spin-down electrons flow towards the -*y*-direction in this example. As discussed in the previous section, this corresponds to a pure CC which carries no net spin polarization [92].

- **Quantum spin Hall effect (QSHE):** In certain 2D semiconductors, the spin Hall conductance is quantized while the charge Hall conductance is negligible. Application of a magnetic field is not required in this case [92]. Noting to the fact that the SHE is much more favorable to implement in fabrication of polarized electron tip sources – since it does not suffer from geometrical restrictions for design purposes – we forgo a more detailed discussion of the QSHE in this chapter.

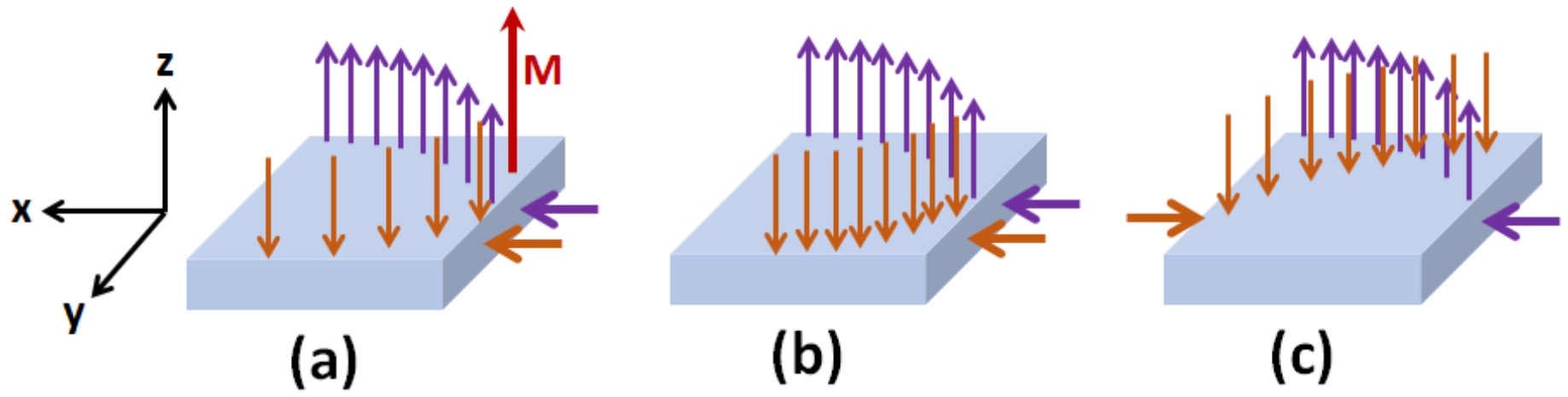

**Fig. 5-6 | Schematic comparison of the anomalous Hall effect (AHE), spin Hall effect (SHE), and the inverse spin Hall effect (ISHE).** The spin-up (-down) electrons are represented by the purple (brown) arrows pointing in the +*z* (-*z*) direction, respectively. The purple (brown) arrow on the *x*-axis shows the direction of the *x*-component of the linear momenta of the spin-up (-down) electrons which enter the thin film. **(a)** The *AHE* in a ferromagnetic sample with magnetization vector M in the +*z*-direction is shown. A CC starts to flow inside the sample in the +*x*-direction while the magnetization vector makes the spin-up orientation more favorable in the ongoing MSs. As a result, the net spin polarization of the induced transverse CC is non-zero making it a polarized CC that can be detected using a voltmeter. **(b)** The *SHE* which can arise in non-magnetic conductors is depicted in this panel. A CC is generated inside the sample in the +*x*-direction. Because of the MS from the atomic nuclei and impurities the scattered spin-up and spin-down electrons are deflected in opposite directions across the *y*-axis. On this transverse direction, the net electric charge is zero since the same number of electrons flow in the +*y*- and -*y*-directions. However, the net spin is not zero. A pure CC thus leads to a pure SC in the SHE, which, as pointed out in Fig. 5-4, cannot be detected by a voltmeter. **(c)** In the *ISHE* a pure SC flowing on the *x*-axis induces a pure CC in the transverse direction. The MS in this example scatters the spin-up electrons toward the -*y*-direction which is to the right of the *x*-component of their linear momenta. The same scattering mechanism tends to scatter the spin-down electrons to the left of their linear momenta, which, noting to the fact that the spin-down electrons are flowing against their spin-up counterparts in the incoming pure SC, also pushes them toward the -*y*-direction. Consequently, the transversely induced current is a pure CC, one

that carries no net spin polarization and can be detected by a voltmeter. (Figure adopted from Ref. [92].)

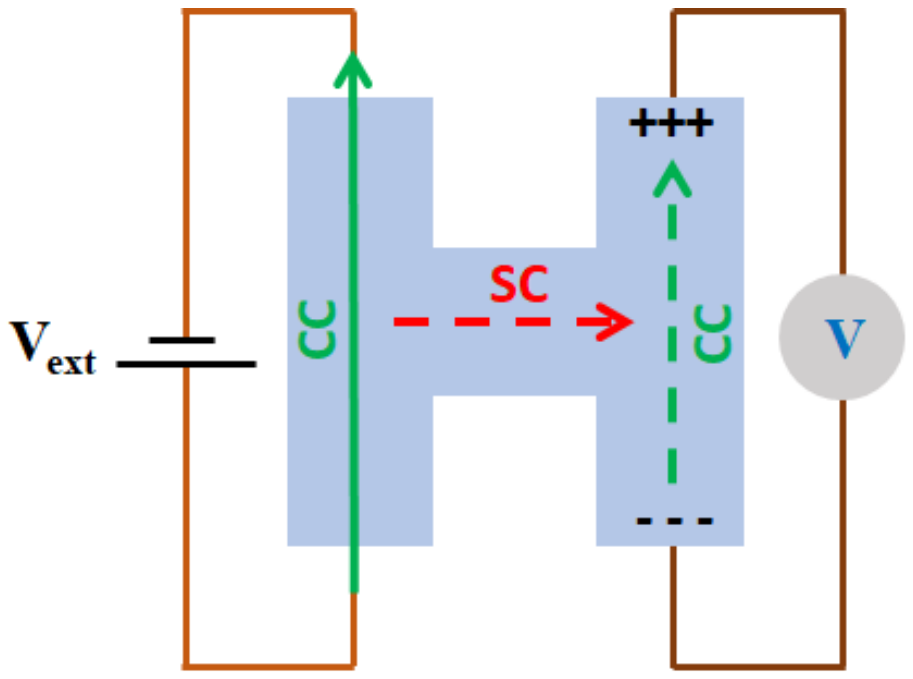

**Fig. 5-7 | The H-bar device designed to detect the spin Hall effect is shown schematically.** The CC in the left vertical bar induces a transverse pure SC through SHE in the horizontal bar in the middle. This SC in turn induces a CC in the right bar through ISHE. A non-zero voltage measured across the length of the right bar then implies the existence of the transverse SC. (Figure adopted from Ref. [93].)

An illustrative comparison is made between AHE, SHE, and ISHE, in Fig. 5-6.

As discussed above, the induced transverse current in the SHE is a pure SC. Since there is no net transfer of electric charge in this case it is not possible to measure this induced current using a voltmeter. A standard way to verify the existence of the pure SC in the SHE is through exploitation of the combined effects of SHE and the ISHE in a device which looks like the letter H and is commonly known as Hall bar or H-bar. The operation principle of this tool is explained in Fig. 5-7.

One of the characteristic properties pertinent to the SHE is called the Hall angle and is denoted with $\theta_H$ . This is defined through the equation,

$$\vec{j}_s = \theta_H \left( \hat{z} \times \vec{j}_c \right), \qquad (5\text{-}16)$$

which relates the induced SC density $\vec{j}_s$ with the applied CC density $\vec{j}_c$ where the unit vector $\hat{z}$ is perpendicular to the sample and defines the spin-up direction. Consequently,

$$\theta_H = \frac{j_s}{j_c}. \qquad (5\text{-}17)$$

The reader must note that $j_s$ and $j_c$ are the magnitudes of the induced pure SC density and the applied pure CC density, respectively, in arbitrary units, and that these vectors are perpendicular to each other [92,94].

Table 5-1 summarizes the two main properties of a few representative materials which are known to be suitable to manifest the SHE. One of these quantities is the Hall angle defined by Eq. (5-17) which gives the conversion efficiency of the SHE; the larger the induced SC density, the larger the Hall angle. The other important quantity is the spin diffusion length $L_s$. For practical applications, it is highly important to have a sufficiently thick layer of spin accumulation at each opposing edge of the sample. In particular, for the purpose of fabricating a laser-driven polarized electron source, where laser pulses are focused on a sharp nanotip, the penetration depth of the laser must be less than $L_s$ to make sure that only spin polarized electrons will be emitted from the tip into the free space.

| | $\theta_H$ | $L_s$ |
|---|---|---|
| **W** | 0.33 | <10 nm |
| **Pt** | 0.07 | <10 nm |
| **Ta** | -0.15 | <10 nm |
| **$Cu_{99.5}Bi_{0.5}$** * | -0.24 | 80 nm |

* The calculated and measured values for CuBi in the literature are given at 10 K.

**Table 5-1 | The reported characteristic values of Hall angle $\theta_H$ and spin diffusion length $L_s$ for four materials suitable to demonstrate the spin Hall effect are compared.** The high-Z elements tungsten (W), platinum (Pt), and tantalum (Ta) are known to procure a relatively high SO coupling in scattering of incident electrons. These are called materials with *intrinsic* SO coupling. In the final row, CuBi, the Bi ions act as scattering centers with a high SO coupling whereby giving rise to a strong SHE. Since this is due to doping external scattering agents, this compound is said to show *extrinsic* SO couplings [94-96].

## 5.4 Proposals to fabricate spin-polarized electron nanotip emitters based on the spin Hall effect

**1. Polarized nanotip photoemitter based on the STM tip design of Ref. [93]**

The only work I encountered in the literature in which a SPES is designed based on the SHE is reported in Ref. [93] which is aimed for STM applications. It is pointed out in Ref. [93] that the proposed design is inspired by a type of superconducting quantum interference device (SQUID) Josephson junction which is made of a sharp insulator with a triangular section coated on its two opposite edges with a superconductor. The coating is instead taken to be CuBi which can manifest the SHE (see Table 5-1). The design projected in Ref. [93] is illustrated in Fig. 5-8 schematically. We can now readily envision incorporation of such a nanotip as a photocathode with a fs laser beam focused onto its apex as in the experiments of CH. 3 (see Fig. 3-12). Considering the practical values of the spin diffusion length at room temperature given in Table 5-1 on the one hand, and a typical penetration depth of a few tens of nanometers in metals on the other hand, extraction of a fraction of the spin polarized electrons from the proximity of the outer boundary of the tip without flipping of their spins is anticipated. Equally importantly, the spin direction of the output beam can be reversed straightforwardly by inverting the direction of the flow of the applied CC that drives the SHE in the nanolayer. This greatly facilitates the Mott electron polarimetry of the source to characterize its spin polarization as frequent switching of the spin direction is required in that case for statistical analysis and in order to account for the various sources of instrumental asymmetry. For instance, the spin direction of pulsed-laser-driven GaAs sources is optically reversible by reversing the helicity of the irradiating circularly polarized laser pulses. Here, reversing the direction of the CC will have the same effect by changing the spin orientation beneath the outermost surface of the nanotip. Additionally, given the metallic nature of the nanotip, say, using W or Pt as the SHE-supporting material, dependence of the emission current and spin-polarization on various emission mechanisms, including DC field-emission, photo-field emission, and single- to multi-photon and above-threshold emission, is of interest to investigate in detail. The same technique that we have used to produce the fiber optic nanotips of CH. 4 can be implemented to fabricate the insulating core of the proposed tip discussed here.

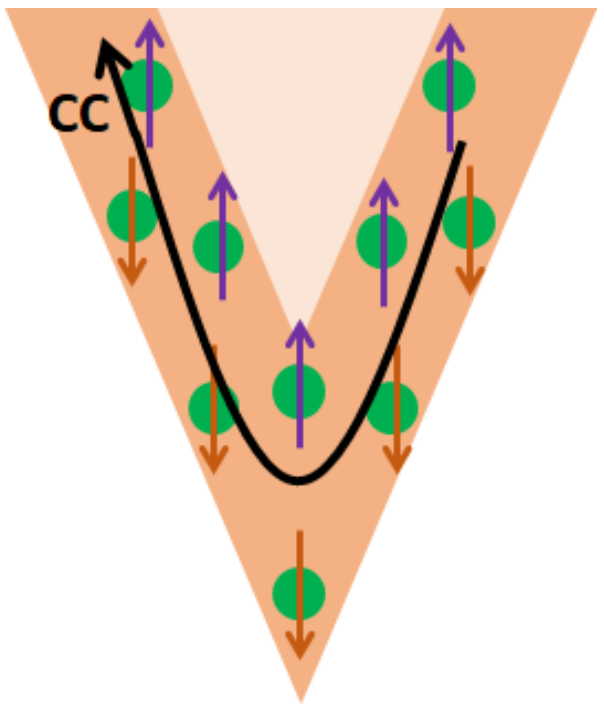

**Fig. 5-8 | The SHE-based STM tip design from Ref. [93] is depicted schematically.** A 1-mm thick rod of insulating quartz or borosilicate glass is pulled and tapered under heating using micropipette puller (same technique as that was used to make our fiber nanotip sources of CH. 4). The tip apex diameter is reported to be controllable down to 100 nm. The insulating tip is coated with CuBi whose relevant properties were indicated in Table 5-1 (at 10 K). The coated part is V-shaped and thus can carry a linear CC as denoted by the black curved arrow. In the SHE, the induced SC is perpendicular to the CC. As a result, a spin accumulation forms at the boundary of the insulator borosilicate and the conductor CuBi, while another spin accumulation with the opposite spin orientation is produced at the vacuum boundary of CuBi. (Figure and details adopted from Ref. [93, p.25].) – Compared with the cited work, the direction of the flow of the CC is reversed here to correctly reflect the negative Hall angle of CuBi according to Table 5-1.

**2. SHE-supporting metallized optical fiber nanotip**

The second design of a SPES which functions on the basis of the SHE exploits a FONTES introduced and discussed in detail in CH.4. We take advantage of the proposed design of Ref. [93] discussed above. The idea is to make the (V-shape) nanotip by metallizing the FONTES with an SHE-supporting material. The first open question is whether the coating can also support SPR at the available laser wavelengths. The second open question would be if the electron emission into the vacuum is accompanied by spin-flip processes, and if yes, whether or not these processes constitute the dominant trend. It was, however, pointed out earlier in this chapter that in Ref. [85] a non-zero spin-polarization is reported for the emitted photoelectrons in a pump-probe experiment using a thin film of ferromagnetic iron. Nonetheless, we leave out further scrutiny of these basic problems for a future work. For the time being, we focus on our qualitative proof-of-concept design in the remainder of this final chapter.

We choose tungsten for the metal coating due to its reported giant Hall angle at room temperature as depicted in Table 5-1, and in spite of its shorter spin diffusion length. Any metal with appropriate properties to support SHE can replace it in practice. The reader must note that the fiber core and cladding are insulators often made of differently doped glass, hence, when the fiber end is pulled or chemically etched, the tapered segment will look similar to the proposal of Ref. [93] shown in Fig. 5-8. The V-shape W coating of the FONTES can cover a suitable length of the fiber jacket as well. This way, two electrodes which can simply be two fine pieces of copper wire can be attached to the coated part, say, a few centimeters away from the apex, using silver paint, in order to supply the driving CC through vacuum flange feedthroughs. The CC which flows in the V-shape metal coating beneath the outer surface of the tip will thus induce a SC through the SHE which is everywhere perpendicular to the CC. This gives rise to two oppositely oriented spin accumulations at the boundaries of the metal coating with the outer surface of the tapered fiber core, and with the vacuum in which the fiber tip is mounted, respectively. In addition, it is important to note that by simply reversing the direction of the flow of the CC by reversing the polarity of its driving current source, the polarization of the accumulated spin layers can be reversed. As mentioned in the discussion of the previous design, this property is sought after in Mott polarimetry, where switching of the spin direction in quasi-periodic cycles is needed for the sake of statistical post-analysis. The same fiber nanotip, of course, can still be exploited in a side-illumination set-up to achieve a cleaner spin-polarized output than in back-illumination as anticipated from the possibility of an avalanche of random scatterings in the latter case. Our design is depicted in Fig. 5-9 schematically which wraps up this final chapter of the present dissertation.

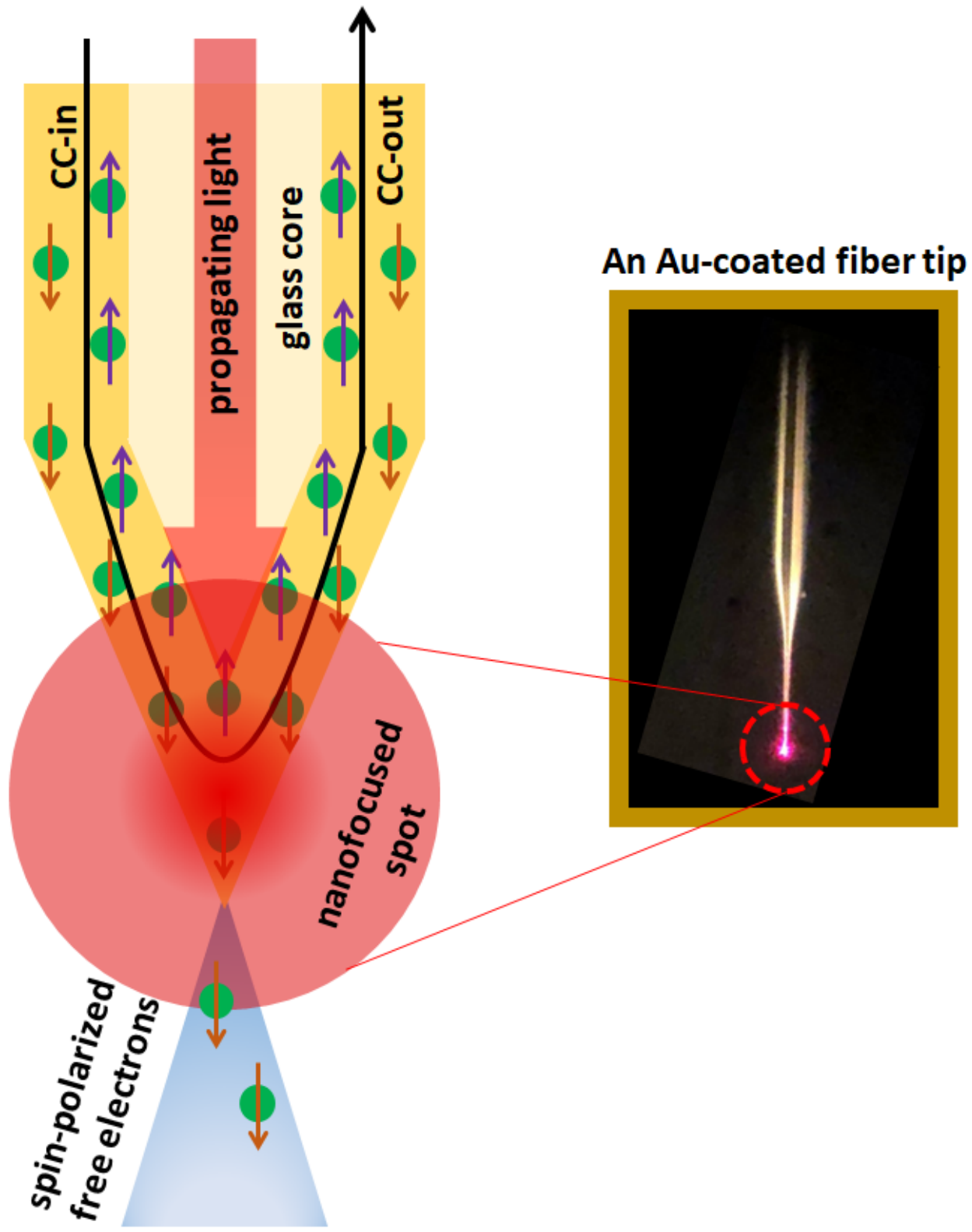

**Fig. 5-9 | Schematic illustration of the proposed SHE-based fiber optic nanotip source of spin-polarized free electrons.** A CC flows inside the V-shape tungsten layer by means of two conductive connections sufficiently far from the tip apex which are not shown here. This can be achieved using a pair of fine copper wires attached to two electrical feedthrough pins on a nearby vacuum flange so that an external current source can be used to supply this CC. Because of the SHE, a SC which is perpendicular to the CC at every section of the flowing CC density is induced. Consequently, at the two boundaries of the metal coating, one with the outer layer of the tapered fiber core, and the other with the surrounding vacuum, spin accumulations with opposite orientations form which are depicted with the purple and brown up and down arrows. The laser light coupled into the base fiber core, either CW or pulsed, is provided through a fiber optic vacuum feedthrough in an alignment-free back-illumination

scheme as explained in CH.4. In particular, with a fs pulsed laser beam, the thermal effects are expected to be minimal, and in the absence of spin flip mechanisms, the emitted electrons would maintain their well-defined spin. The spin-polarization of the photoelectrons is controlled by switching the supplied CC polarity. The inset to the right shows a gold-coated fiber optic nanotip under an optical microscope with a CW diode laser beam at $\lambda$ = 672 nm coupled into its core. The observed light which is understood to have been scattered out of the nanofocused area surrounding the tip apex exhibits the red blob at the tapered segment which is schematically magnified on the left illustration.

# APPENDIX: MATLAB™ PROGRAMS

In this appendix, the main programs I have written in the MATLAB™ language for the projects covered in the present dissertation are provided to the readers. In multiple situations, variations of the same code were compiled for different purposes such as producing different but related figures. In such cases, only the main version of the program is given here for brevity. To copy all or part of the programs of this Appendix in any programming language, please inform the author by email at: sam.keramati@huskers.unl.edu

## 1. Matter-wave single-slit far-field diffraction in space

Three figures are generated running this program: 1) The diffraction pattern, 2) the relative phase at the main peak, and 3) the relative phase at the first zero.

```
% 1D far-field diffraction in space - SK
clear
clc
hbar = 1.054e-34;    % in J.s
m = 9.1e-31;         % electron mass in Kg
D = 1e-1;            % peak position on the screen
T = 1e-6;            % flight time
alpha = m*D/(hbar*T);
lambda_dB = 2*pi/abs(alpha);
V = D/T;
a = 1e-7;        % spatial size of the source (a<<D)
b = abs(D/2);
stepsperlambda = 100;
M = round(stepsperlambda*a/lambda_dB);
N = 2e3;         % number of steps in the detection range
if M > 1e4  % conditional to make sure M is not too large.
    disp('Error: M > 1e4')
    return;
end
d_a = linspace(0,a,M+1);
d_b = linspace((D-b),(D+b),N+1);
K = zeros(M+1,N+1);               % PI Kernel
phi_i = 1*exp(1i*alpha.*d_a);     % the initial wavefunction
for v = 1:N+1
    K(:,v) = sqrt(m/(1i*2*pi*hbar*T))*exp((1i*m*((d_a -...
          (d_b(1,v))).^2))/(2*hbar*T));
```

```
end
probi = phi_i.*conj(phi_i);
P_i = trapz(d_a,probi);
prob_i = real(probi/P_i);
phi_f = phi_i*K;
probf = phi_f.*conj(phi_f);
P_f = trapz(d_b,probf);
prob_f = real(probf/P_f);
firstmin = (2*pi*hbar*T/(m*a)) + D;
relphase0 = ((m/(2*hbar*T))*(((firstmin - d_a).^2) -...
(firstmin - 0)^2) + alpha.*d_a)/pi;
plot(d_a,relphase0,'linewidth',2), grid
title('Relative phase of the source points at...
the first zero')
xlabel('Source point (m)','fontsize',20)
ylabel('Relative phase/\pi','fontsize',20)
xlim([min(d_a) max(d_a)]);
ax = gca;
ax.FontSize = 18;
ax.XColor = 'k';
ax.YColor = 'k';
ax.XMinorTick = 'on';
ax.YMinorTick = 'on';
ax.XMinorGrid = 'on';
ax.YMinorGrid = 'on';
box on
figure
relphaseD = ((m/(2*hbar*T))*(((D - d_a).^2) - (D - 0)^2)...
+ alpha.*d_a)/pi;
plot(d_a,relphaseD,'linewidth',3), grid
title('Relative phase of the source points at...
the main peak')
xlabel('Source point (m)','fontsize',20)
ylabel('Relative phase/\pi','fontsize',20)
xlim([min(d_a) max(d_a)]);
ax = gca;
ax.FontSize = 20;
ax.FontWeight = 'bold';
ax.XColor = 'k';
ax.YColor = 'k';
ax.FontWeight = 'bold';
ax.XMinorTick = 'on';
ax.YMinorTick = 'on';
ax.XMinorGrid = 'on';
ax.YMinorGrid = 'on';
box on
figure
plot(d_b,prob_f,'linewidth',3), grid
title('Single-slit far-field diffraction in 1D space')
```

```
xlabel('Detection point (m)','fontsize',20)
ylabel('Normalized probability density','fontsize',20)
xlim([min(d_b) max(d_b)]);
ax = gca;
ax.FontSize = 20;
ax.FontWeight = 'bold';
ax.XColor = 'k';
ax.YColor = 'k';
ax.FontWeight = 'bold';
ax.XMinorTick = 'on';
ax.YMinorTick = 'on';
ax.XMinorGrid = 'on';
ax.YMinorGrid = 'on';
box on
hold on;
plot(firstmin,min(prob_f),'xr','linewidth',3);
hold off;
% End
```

## 2. Matter-wave double-slit far-field interference in space

This program outputs a figure showing both double-slit coherent and incoherent superpositions.

```
% 1D far-field 2-slit interference in space - SK
clear
clc
hbar = 1.054e-34;    % in J.s
m = 9.1e-31;         % electron mass in Kg
D = 1e-1;            % peak position on the screen
T = 1e-6;            % flight time
alpha = m*D/(hbar*T);
lambda_dB = 2*pi/abs(alpha);
V = D/T;
disp('Velocity (m/s) = '),disp(V)
a = 1e-7;        % spatial size of the source (a<<D)
disp('Slit width (m) = '),disp(a)
stepsperlambda = 50;
M = round(stepsperlambda*a/lambda_dB);
N = 2e3;         % number of steps in the detection range
if M > 1e4
    disp('Error: M > 1e4')
    return;
end
d_a1 = linspace(-3*a/2,-a/2,M+1);
d_a2 = linspace(a/2,3*a/2,M+1);
b = abs(D/30);
```

```
d_b = linspace((D-b),(D+b),N+1);
K1 = zeros(M+1,N+1);    % Kernel from the 1st source
K2 = zeros(M+1,N+1);    % Kernel from the 2nd source
phi_i1 = 1*exp(1i*alpha.*d_a1);    % initial wavefunction
phi_i2 = 1*exp(1i*alpha.*d_a2);    % initial wavefunction
for v = 1:N+1
   K1(:,v) = sqrt(m/(1i*2*pi*hbar*T))*exp((1i*m*((d_a1 -...
          (d_b(1,v))).^2))/(2*hbar*T));
   K2(:,v) = sqrt(m/(1i*2*pi*hbar*T))*exp((1i*m*((d_a2 -...
          (d_b(1,v))).^2))/(2*hbar*T));
end
probi1 = phi_i1.*conj(phi_i1);
P_i1 = trapz(d_a1,probi1);
prob_i1 = real(probi1/P_i1);
probi2 = phi_i2.*conj(phi_i2);
P_i2 = trapz(d_a2,probi2);
prob_i2 = real(probi2/P_i2);
% Coherent interference
phi_f = (phi_i1*K1) + (phi_i2*K2);
probf = phi_f.*conj(phi_f);
P_f = trapz(d_b,probf);
prob_f = real(probf/P_f);
% Incoherent sum
phi_f1 = phi_i1*K1;
probf1 = phi_f1.*conj(phi_f1);
P_f1 = trapz(d_b,probf1);
prob_f1 = real(probf1/P_f1);
phi_f2 = phi_i2*K2;
probf2 = phi_f2.*conj(phi_f2);
P_f2 = trapz(d_b,probf2);
prob_f2 = real(probf2/P_f2);
prob_fIncoh = (1/2)*(prob_f1 + prob_f2);
plot(d_b,prob_f,'linewidth',3), grid
title('Double-slit far-field interference in 1D space')
xlabel('x (m)','fontsize',20)
ylabel('P (m^-^1)','fontsize',20)
xlim([min(d_b) max(d_b)]);
ax = gca;
ax.FontSize = 20;
ax.FontWeight = 'bold';
ax.XColor = 'k';
ax.YColor = 'k';
ax.FontWeight = 'bold';
ax.XMinorTick = 'on';
ax.YMinorTick = 'on';
ax.XMinorGrid = 'on';
ax.YMinorGrid = 'on';
box on
hold on
```

```
plot(d_b,prob_fIncoh,'--r','linewidth',3)
legend('Coherent','Incoherent')
hold off
% End
```

## 3. Matter-wave single-slit far-field diffraction in time

This program leads to Fig. 2-3(a), including its inset as a separate figure, and the plot of the relative phase at the main peak.

```
% 1D far-field diffraction in time - SK
clear
clc
hbar = 1.054e-34;  % in J.s
m = 9.1e-31;       % electron mass in Kg
D = 5e-2;          % distance from the screen to the source
T = 50e-9;         % peak time on the screen
alpha = -1*m*(D^2)/(2*hbar*(T^2));
period = 2*pi/abs(alpha);
V = D/T;           % average velocity (not used below)
a = 5e-15;         % temporal size of the source (a<<T)
b = T/2;
disp('Source-to-screen distance (m) = '), disp(D)
disp('Temporal size of the source (s) = '), disp(a)
stepsperperiod = 100;
M = round(stepsperperiod*a/period);
N = 2e3;        % number of steps in the detection range
if M > 1e4
    disp('Error: M > 1e4')
    return;
end
t_a = linspace(0,a,M+1);
t_b = linspace((T-b),(T+b),N+1);
K = zeros(M+1,N+1);    % PI Kernel
phi_i = 1*exp(1i*alpha.*t_a);
for v = 1:N+1
      K(:,v) = sqrt(m./(1i*2*pi*hbar.*(t_b(1,v) - t_a)))...
          .*exp((1i*m*(D^2))./(2*hbar.*(t_b(1,v) - t_a)));
end
phi_f = phi_i*K;
probf = real(phi_f.*conj(phi_f));
P_f = trapz(t_b,probf);
prob_f = real(probf/P_f);
relphase0 = ((m*(D^2)/(2*hbar))*(1./(firstmin - t_a) -...
(1/(firstmin - 0))) + alpha*t_a)/pi;
plot(t_a,relphase0,'linewidth',3), grid
```

```
title('Relative phase of the source points at the...
first zero')
xlabel('Source time (s)','fontsize',20)
ylabel('Relative phase/\pi','fontsize',20)
xlim([min(t_a) max(t_a)]);
axis square
ax = gca;
ax.FontSize = 20;
ax.FontWeight = 'bold';
ax.XColor = 'k';
ax.YColor = 'k';
ax.GridColor = 'k';
ax.XMinorTick = 'on';
ax.YMinorTick = 'on';
ax.XMinorGrid = 'on';
ax.YMinorGrid = 'on';
box on
figure
relphaseT = ((m*(D^2)/(2*hbar))*(1./(T - t_a) -...
(1/(T - 0))) + alpha*t_a)/pi;
plot(t_a,relphaseT,'linewidth',3), grid
title('Relative phase of the source points at the...
main peak')
xlabel('Source time (s)','fontsize',20)
ylabel('Relative phase/\pi','fontsize',20)
xlim([min(t_a) max(t_a)]);
ax = gca;
ax.FontSize = 20;
ax.FontWeight = 'bold';
ax.XColor = 'k';
ax.YColor = 'k';
ax.GridColor = 'k';
ax.XMinorTick = 'on';
ax.YMinorTick = 'on';
ax.XMinorGrid = 'on';
ax.YMinorGrid = 'on';
box on
axis square
figure
plot(t_b,prob_f,'linewidth',3), grid
title('Single-slit far-field diffraction in time')
xlabel('t (s)','fontsize',20)
ylabel('P (s^-^1)','fontsize',20)
xlim([min(t_b) max(t_b)]);
ax = gca;
ax.FontSize = 20;
box on
axis square
ax.FontWeight = 'bold';
```

```
ax.XColor = 'k';
ax.YColor = 'k';
ax.GridColor = 'k';
ax.XMinorTick = 'on';
ax.YMinorTick = 'on';
ax.XMinorGrid = 'on';
ax.YMinorGrid = 'on';
hold on;
plot(firstmin,min(prob_f),'xr','linewidth',3);
hold off;
% End
```

## 4. Matter-wave double-slit far-field interference in time

This program generates Fig. 2-3(d).

```
% 1D far-field 2-slit interference in time - SK
clear
clc
hbar = 1.054e-34;  % in J.s
m = 9.1e-31;       % electron mass in Kg
D = 5e-2;          % distance from the screen to the source
T = 50e-9;         % peak time on the screen
alpha = -1*m*(D^2)/(2*hbar*(T^2));
period = 2*pi/abs(alpha);
V = D/T;           % average velocity (not used below)
a = 5e-15;         % temporal size of the source (a<<T)
b = T/2;
stepsperperiod = 100;
M = round(stepsperperiod*a/period);
N = 2e3;        % number of steps in the detection range
if M > 1e4
    disp('Error: M > 1e4')
    return;
end
t_a1 = linspace(0,a,M+1);
tau_emission = 2*a;
t_a2 = t_a1 + tau_emission;
t_b = linspace((T-b),(T+b),N+1);
disp('Source-to-screen distance (m) = '), disp(D)
disp('Slit duration (s) = '), disp(a)
disp('Second source delay (s) = '), disp(tau_emission)
K1 = zeros(M+1,N+1);    % Kernel from the 1st source
K2 = zeros(M+1,N+1);    % Kernel from the 2nd source
phi_i1 = exp(1i*alpha.*t_a1);
phi_i2 = exp(1i*alpha.*t_a2);
for v = 1:N+1
  K1(:,v) = sqrt(m./(1i*2*pi*hbar.*(t_b(1,v) - t_a1)))...
```

```
          .*exp((1i*m*(D^2))./(2*hbar.*(t_b(1,v) - t_a1)));
  K2(:,v) = sqrt(m./(1i*2*pi*hbar.*(t_b(1,v) - t_a2)))...
          .*exp((1i*m*(D^2))./(2*hbar.*(t_b(1,v) - t_a2)));
end
% Coherent interference
phi_f = (phi_i1*K1) + (phi_i2*K2);
probf = phi_f.*conj(phi_f);
P_f = trapz(t_b,probf);
prob_f = real(probf/P_f);
% Incoherent sum
phi_f1 = phi_i1*K1;
probf1 = real(phi_f1.*conj(phi_f1));
P_f1 = trapz(t_b,probf1);
prob_f1 = real(probf1/P_f1);
phi_f2 = phi_i2*K2;
probf2 = phi_f2.*conj(phi_f2);
P_f2 = trapz(t_b,probf2);
prob_f2 = real(probf2/P_f2);
prob_fIncoh = (1/2)*(prob_f1 + prob_f2);
plot(t_b,prob_f,'linewidth',3), grid
title('Double-slit far-field interference in time')
xlabel('t (ns)','fontsize',20)
ylabel('P (s^-^1)','fontsize',20)
xlim([min(t_b) max(t_b)]);
xticks([2.5e-8 3e-8 3.5e-8 4e-8 4.5e-8 5e-8 5.5e-8...
6e-8 6.5e-8 7e-8 7.5e-8]);
xticklabels({'25','30','35','40','45','50',...
'55','60','65','70','75'});
ax = gca;
ax.FontSize = 20;
ax.FontWeight = 'bold';
ax.XColor = 'k';
ax.YColor = 'k';
ax.GridColor = 'k';
ax.XMinorTick = 'on';
ax.YMinorTick = 'on';
ax.XMinorGrid = 'on';
ax.YMinorGrid = 'on';
box on
hold on
plot(t_b,prob_fIncoh,'--r','linewidth',3)
legend('Coherent','Incoherent')
hold off
% End
```

# 5. The fermionic HBT effect

This programs outputs the antisymmetric or the symmetric component of the 1D fermionic HBT joint probability density. In the case below, it generates the red solid or dashed lines of Fig. 2-5(a) along with the black incoherent diffraction pattern.

```
% Partially coherent HBT - SK
% S/AS components with 2 particles and N temporal slits.
% The coherence time is 2*(slit duration). Pulse duration
% is N*(slit duration).
clear
clc
tic;
hbar = 1.054e-34;  % in J.s
m = 9.1e-31;       % electron mass in Kg
D = 5e-2;          % distance from the screen to the source
a = 5e-15;         % temporal size of the source (a<<T)
T = 50e-9;         % peak time on the screen (the first
detection time)
alpha = -1*m*(D^2)/(2*hbar*(T^2));
period = 2*pi/abs(alpha);
stepsperperiod = 100;
M = round(stepsperperiod*a/period);
N = 2e3;           % number of steps on the screen
if M > 1e4
    disp('Error: M > 1e4')
    return;
end
b = T/2;
t_b = linspace(T-b,T+b,N+1);
alphaDel = -1*m*(D^2)./(2*hbar.*(t_b.^2));
t_a1 = linspace(0,a,M+1);
tau_emission = 1*a;
t_a2 = t_a1 + tau_emission;
t_a3 = t_a2 + tau_emission;
disp('Source-to-screen distance (m) = '), disp(D)
disp('One-electron detection time (s) = '), disp(T)
disp('Coherence time (s) = '), disp(2*a)
K1 = zeros(M+1,N+1);    % Kernel for source1
K2 = zeros(M+1,N+1);    % Kernel for source2
K3 = zeros(M+1,N+1);    % Kernel for source3
phi_i11 = zeros(N+1,M+1);
phi_i12 = zeros(N+1,M+1);
phi_i21 = zeros(N+1,M+1);
phi_i22 = zeros(N+1,M+1);
```

```
phi_i31 = zeros(N+1,M+1);
phi_i32 = zeros(N+1,M+1);
for u = 1:M+1
    phi_i11(:,u) = exp(1i*alpha*t_a1(1,u));
    phi_i12(:,u) = exp(1i*alphaDel*t_a1(1,u));
    phi_i21(:,u) = exp(1i*alpha*t_a2(1,u));
    phi_i22(:,u) = exp(1i*alphaDel*t_a2(1,u));
    phi_i31(:,u) = exp(1i*alpha*t_a3(1,u));
    phi_i32(:,u) = exp(1i*alphaDel*t_a3(1,u));
    K1(u,:) = sqrt(m./(1i*2*pi*hbar* ...
(t_b(1,:) - t_a1(1,u)))).*exp((1i*m*(D^2))./(2*hbar*...
(t_b(1,:) - t_a1(1,u))));
    K2(u,:) = sqrt(m./(1i*2*pi*hbar* ...
(t_b(1,:) - t_a2(1,u)))).*exp((1i*m*(D^2))./(2*hbar*...
(t_b(1,:) - t_a2(1,u))));
    K3(u,:) = sqrt(m./(1i*2*pi*hbar* ...
(t_b(1,:) - t_a3(1,u)))).*exp((1i*m*(D^2))./(2*hbar*...
(t_b(1,:) - t_a3(1,u))));
end
phi_f11 = phi_i11*K1;
phi_f12 = phi_i12*K1;
phi_f21 = phi_i21*K2;
phi_f22 = phi_i22*K2;
phi_f31 = phi_i31*K3;
phi_f32 = phi_i32*K3;
psi1 = (1/sqrt(2))*(phi_f11.*phi_f22 - phi_f12.*phi_f21);
prob_psi1f = real(psi1.*conj(psi1));
P_f1 = trapz(t_b,prob_psi1f);
prob_psi1_f = bsxfun(@rdivide,prob_psi1f,P_f1);
psi1_B = (1/sqrt(2))*(phi_f11.*phi_f22 + phi_f12.*phi_f21);
prob_psi1_Bf = real(psi1_B.*conj(psi1_B));
P_f1_B = trapz(t_b,prob_psi1_Bf);
prob_psi1_B_f = bsxfun(@rdivide,prob_psi1_Bf,P_f1_B);
probf11 = real(phi_f11.*conj(phi_f11));
P_f11 = trapz(t_b,probf11);
prob_f11 = bsxfun(@rdivide,probf11,P_f11);
probf12 = real(phi_f12.*conj(phi_f12));
P_f12 = trapz(t_b,probf12);
prob_f12 = bsxfun(@rdivide,probf11,P_f12);
probf31 = real(phi_f31.*conj(phi_f31));
P_f31 = trapz(t_b,probf31);
prob_f31 = bsxfun(@rdivide,probf31,P_f31);
probf32 = real(phi_f32.*conj(phi_f32));
P_f32 = trapz(t_b,probf32);
prob_f32 = bsxfun(@rdivide,probf32,P_f32);
prob_fIncoh = probf11.*probf32 + probf12.*probf31;
P_Incoh = trapz(t_b,prob_fIncoh);
prob_f_Incoh = bsxfun(@rdivide,prob_fIncoh,P_Incoh);
Num = 3;    % N > 2
```

```
q_Coh = 2/Num;
q_Incoh = (Num - 2)/Num;
prob_f = q_Coh*prob_psi1_f + q_Incoh*prob_f_Incoh;
prob_f_B = q_Coh*prob_psi1_B_f + q_Incoh*prob_f_Incoh;
disp('Pulse length (s) = '), disp(Num*a)
A = ((prob_f_Incoh(((N/2)+1),((N/2)+1))) -...
prob_f(((N/2)+1),((N/2)+1)))/...
((prob_f_Incoh(((N/2)+1),((N/2)+1))) +...
prob_f(((N/2)+1),((N/2)+1)));
disp('A (contrast) = '), disp(A)
Tdelay = t_b - T;
plot(Tdelay,prob_f(:,((N/2)+1)),'linewidth',3)
xlim([min(Tdelay) max(Tdelay)]);
title('Matter wave partially coherent HBT effect')
xlabel('\tau (ns)','fontsize',20)
ylabel('P(s^-^1)','fontsize',20)
xticks([-2.5e-8 -2e-8 -1.5e-8 -1e-8 -.5e-8...
0 .5e-8 1e-8 1.5e-8 2e-8 2.5e-8]);
xticklabels({'-25','-20','-15','-10',...
'-5','0','5','10','15','20','25'});
ax = gca;
ax.FontSize = 20;
ax.FontWeight = 'bold';
ax.XColor = 'k';
ax.YColor = 'k';
ax.GridColor = 'k';
ax.XMinorTick = 'on';
ax.YMinorTick = 'on';
ax.XMinorGrid = 'on';
ax.YMinorGrid = 'on';
box on
hold on
plot(Tdelay,prob_f_B(:,((N/2)+1)),'--r','linewidth',3)
grid
hold off
hold on
plot(Tdelay,prob_f_Incoh(:,((N/2)+1)),...
'-black','linewidth',3)
legend('AS','S','Incoh.')
hold off
toc
% End
```

## 6. Degenerate antisymmetric joint probability vs time and energy

This program generates a surface plot of the antisymmetric degenerate joint probability density in the fermionic HBT effect against the mutual detection time and electron kinetic energy. The output is a figure similar to Fig. 2-6.

```
% Degenerate AS component of HBT vs time and energy - SK
clear
clc
hbar = 1.054e-34;  % in J.s
m = 9.1e-31;       % electron mass in Kg
D = 5e-2;          % distance from the screen to the source
a = 5e-15;         % temporal size of the source (a<<T)
sizew = 100;
T = zeros(1,sizew);
KE = zeros(1,sizew);
for w = 1:sizew
T(1,w) = 25e-9 + (.25e-9)*w;
alpha = -1*m*(D^2)/(2*hbar*(T(1,w)^2));
KE(1,w) = (.5*m*(D/T(1,w))^2)/(1.6e-19);
period = 2*pi/abs(alpha);
stepsperperiod = 100;
M = round(stepsperperiod*a/period);
N = 2e2;
if M > 1e4
    disp('Error: M > 1e4')
    return;
end
t_a1 = linspace(0,a,M+1);
tau_emission = 1*a;
t_a2 = t_a1 + tau_emission;
b = 10e-9;
t_b = linspace(T(1,w)-b,T(1,w)+b,N+1);
alphaDel = -1*m*(D^2)./(2*hbar.*(t_b.^2));
K1 = zeros(M+1,N+1);    % Kernel for source1
K2 = zeros(M+1,N+1);    % Kernel for source2
phi_i11 = zeros(N+1,M+1);
phi_i12 = zeros(N+1,M+1);
phi_i21 = zeros(N+1,M+1);
phi_i22 = zeros(N+1,M+1);
for u = 1:M+1
    phi_i11(:,u) = exp(1i*alpha*t_a1(1,u));
    phi_i12(:,u) = exp(1i*alphaDel*t_a1(1,u));
    phi_i21(:,u) = exp(1i*alpha*t_a2(1,u));
```

```
    phi_i22(:,u) = exp(1i*alphaDel*t_a2(1,u));
    K1(u,:) = sqrt(m./(1i*2*pi*hbar*(t_b - t_a1(1,u))))...
        .*exp((1i*m*(D^2))./(2*hbar*(t_b - t_a1(1,u))));
    K2(u,:) = sqrt(m./(1i*2*pi*hbar*(t_b - t_a2(1,u))))...
        .*exp((1i*m*(D^2))./(2*hbar*(t_b - t_a2(1,u))));
end
phi_f11 = phi_i11*K1;
phi_f12 = phi_i12*K1;
phi_f21 = phi_i21*K2;
phi_f22 = phi_i22*K2;
psi1 = (1/sqrt(2))*(phi_f11.*phi_f22 - phi_f12.*phi_f21);
prob_psi1f = real(psi1.*conj(psi1));
P_f1 = trapz(t_b,prob_psi1f);
prob_psi1_f(w,:,:) = bsxfun(@rdivide,prob_psi1f,P_f1);
prob(w,:) = prob_psi1_f(w,:,((N/2)+1));
t(:,w) = t_b;
end
surf(KE,t,prob','linewidth',1)
xlabel('KE (eV)','fontsize',20)
ylabel('t (s)','fontsize',20)
zlabel('P (s^-^1)','fontsize',20)
ax = gca;
ax.FontSize = 20;
ax.FontWeight = 'bold';
ax.Color = 'k';
box on
% End
```

## 7. Antibunching signal analysis

This program outputs figures similar to the various panels of Figs. 3-21 & 22.

```
% Program goal: Find the relative reduction of the zero-
% delay peak compared with the average of its 2 side
% peaks after finding the rep time using the maxima and
% the average BG using the minima. - SK
clear
clc
% import the 1st column of MCA data file, label it y
yy = y(160:1860);
x = (147:1847)';
ChW = 0.053;   % channel width
xx = ChW*x;
plot(xx,yy,'-o','MarkerSize',3,'MarkerFaceColor','b',...
    'MarkerEdgeColor','b','Linewidth',1,'Color','k'), grid
xlabel('\tau (ns)','FontSize',20,'FontWeight','bold')
ylabel('Coincidence counts','FontSize', ...
20,'FontWeight','bold')
```

```
axis square
xlim([7.79 97.79]);
ax = gca;
ax.FontSize = 20;
ax.FontWeight = 'bold';
ax.XColor = 'k';
ax.YColor = 'k';
ax.GridColor = 'r';
ax.XMinorTick = 'on';
ax.YMinorTick = 'on';
ax.XMinorGrid = 'on';
ax.YMinorGrid = 'on';
xticks([7.79 22.79 37.79 52.79 67.79 82.79 97.79]);
xticklabels({'-45','-30','-15','0','15','30','45'});
ax.XAxis.MinorTickValues = 7.79:3:97.79;
[peaks, times] = findpeaks(yy,xx,'MinPeakDistance',12);
rept = sum(diff(times))/length(diff(times));
disp('Ave rep time (ns) = '), disp(rept)
mint = islocalmin(yy,'MaxNumExtrema',6,...
'MinSeparation',150);
BG = sum(yy(mint))/length(yy(mint));
disp('Ave BG counts per Ch = '), disp(BG)
hold on
plot(times,peaks,'or','linewidth',2)
plot(xx(mint),yy(mint),'og','linewidth',2)
LL = times(3) - rept/2;
LR = times(3) + rept/2;
RL = times(5) - rept/2;
RR = times(5) + rept/2;
nLL = round(LL/ChW) - 147;
nLR = round(LR/ChW) - 147;
nRL = round(RL/ChW) - 147;
nRR = round(RR/ChW) - 147;
yrange = get(gca,'ylim');
plot([xx(nLL) xx(nLL)],yrange,'--k')
plot([xx(nLR) xx(nLR)],yrange,'--k')
plot([xx(nRL) xx(nRL)],yrange,'--k')
plot([xx(nRR) xx(nRR)],yrange,'--k')
yyBG = yy - BG;
t0 = RL - LR;
deltaT = (t0 - rept)*1e3;   % ps
disp('Zero-peak broadening (ps) = '), disp(deltaT)
sumL = sum(yyBG(nLL:nLR));
sum0 = sum(yyBG(nLR:nRL));
sumR = sum(yyBG(nRL:nRR));
deltaC = 100*(sum0 - sumR)/sumR;
disp('Sub-Poissonian contrast (%) = '), disp(deltaC)
disp('t0 (ns) = '), disp(t0)
disp('sum0 = '), disp(sum0)
```

```
disp('sumR = '), disp(sumR)
disp('sumL = '), disp(sumL)
% End
```

## 8. Coincidence histograms from the SIMION™ raw data

This is the program written to generate various coincidence histograms from the SIMION™ raw data after broadening of the peaks by fitting using a set of Gaussian temporal profiles. Refer to the discussions surrounding Table 3-3 for more information.

```
% Coincidence histograms from the SIMION raw data - SK
clear
clc
T = load('filename.txt');
Tfull = T;
tic
T = sortrows(T,2);
w1 = normrnd(0,2e-3,length(T),1);
w2 = normrnd(0,2e-3,length(T),1);
w3 = normrnd(0,1.1e-3,length(T),1);
w4 = normrnd(0,1.1e-3,length(T),1);
t1 = T(:,2) + w1;
t2 = T(:,2) + w2;
t3 = T(:,2) + w3;
t4 = T(:,2) + w4;
y = T(:,1);
detector = sign(y);
TACsign_raw = -diff(detector);
TACsign = TACsign_raw;
for n = 1:length(TACsign)-1
    if (abs(TACsign(n+1))-abs(TACsign(n)))==0
        TACsign(n+1) = 0;
    end
end
posclicks = find(TACsign==2);
negclicks = find(TACsign==-2);
postau1 = t1(posclicks + 1) - t1(posclicks);
negtau1 = -(t1(negclicks + 1) - t1(negclicks));
tau1 = [negtau1;postau1];
postau2 = t2(posclicks + 1) - t2(posclicks);
negtau2 = -(t2(negclicks + 1) - t2(negclicks));
tau2 = [negtau2;postau2];
postau3 = t3(posclicks + 1) - t3(posclicks);
negtau3 = -(t3(negclicks + 1) - t3(negclicks));
```

```
tau3 = [negtau3;postau3];
postau4 = t4(posclicks + 1) - t4(posclicks);
negtau4 = -(t4(negclicks + 1) - t4(negclicks));
tau4 = [negtau4;postau4];
toc
edges = ((-.8*((13.1617e-4)/2)-3.5*13.1617e-3)...
    :.8*13.1617e-4:(.8*((13.1617e-4)/2)+3.5*13.1617e-3));
h1 = histogram(tau1,edges);
C1 = h1.Values;
h2 = histogram(tau2,edges);
C2 = h2.Values;
h3 = histogram(tau3,edges);
C3 = h3.Values;
h4 = histogram(tau4,edges);
C4 = h4.Values;
C = C1 + C2 + C3 + C4;
edges(length(edges)) = [];
edges = (edges + ((.8*13.1617e-4)/2))*1e3 + 51.3;
plot(edges,C,'--','Linewidth',2,'Color',[0.07 .62 1]),grid
axis square
xlabel('\tau (ns)','FontSize',20,'FontWeight','bold')
ylabel('Coincidence counts','FontSize',...
20,'FontWeight','bold')
box on
grid
ax = gca;
ax.FontSize = 20;
ax.FontWeight = 'bold';
ax.XColor = 'k';
ax.YColor = 'k';
ax.GridColor = 'k';
ax.XMinorTick = 'on';
ax.YMinorTick = 'on';
ax.XMinorGrid = 'on';
ax.YMinorGrid = 'on';
% End
```

## 9. Multiphoton emission order

This program outputs the power-law plot along with the photoemission order and its standard deviation as explained in CH. 4. A continuous ND wheel was used to vary the laser power. The data were recorded with an MCS.

```
% Power-law plot - SK
clear
clc
```

```
MCS = load('filename.Asc');
x = MCS(:,1);
y = MCS(:,2);
t = x(1:30);
C = y(65:94);
% Initial (max) power in muW (at 1 kHz)
P_0 = xxx; % enter the initial power
m = 0.00741;
P = P_0*10.^(-m*3*t);
[u,v] = find(~C);
C(u) = [];
P(u) = [];
Plog = log(P);
Clog = log(C);
sdClog = log(C + sqrt(C)) - log(C - sqrt(C));
relerrClog = sdClog./Clog;
W = 1./relerrClog;
LinearModelTerms = {'x','1'};
linearfittype = fittype({'x','1'});
f = fit(Plog,Clog,linearfittype,'Weight',W)
error = confint(f);             % 95% confidence interval
n_onesigma = (1/4)*(error(2,1) - error(1,1));   % 1 sigma
loglog(P,C,'s','MarkerSize',10,'MarkerFaceColor',...
'b','MarkerEdgeColor','b')
grid, axis square, box on
xlabel('P (\muW)','FontSize',20,'FontWeight','bold')
ylabel('Counts / sec','FontSize',20,'FontWeight','bold')
title('\lambda = xxx nm')
ax = gca;
ax.FontSize = 20;
ax.FontWeight = 'bold';
ax.XColor = 'k';
ax.YColor = 'k';
ax.GridColor = 'k';
ax.XMinorTick = 'on';
ax.YMinorTick = 'on';
ax.XMinorGrid = 'on';
ax.YMinorGrid = 'on';
hold on
fittedline = exp(f.b)*P.^(f.a);
loglog(P,fittedline,'LineWidth',3,'Color','r')
disp('a = '), disp(exp(f.b))
disp('n = '), disp(f.a)
disp('one sigma of n = '), disp(n_onesigma)
% End
```

# 10. Photo-field emission and Fowler-Nordheim models

This program outputs the PFE model of Fig. 4-8 along with its separate constituent terms in two panels. The minimum and maximum values of the reduced work function are also displayed in the Command Window for reference. The FN model is a special case as explained in CH. 4.

```
% PFE model - SK
clear
clc
Vtip = 375:.5:475;
VtipExp = 375:25:475;
C = [415,1321,3827,8805,17603];
q = 1.6e-19;
h = 2*pi*1.054e-34;
c = 3e8;
m = 9.1e-31;
phi = 5.1*q;
lambda = 633e-9;
photonE = h*c/lambda;
epsilon = 8.85e-12;
aa = (q^3)/(8*pi*h);
bb = (8*pi*sqrt(2*m))/(3*q*h);
k = 5.7;
r = 42.2e-9;    % tip radius (free parameter)
F = q*Vtip/(k*r);
reducedphi = (phi - sqrt(q*1.44e-9*F))/q;
disp('min reduced phi (eV) = '),disp(min(reducedphi))
disp('max reduced phi (eV) = '),disp(max(reducedphi))
disp('photon energy (eV) = '),disp(photonE/q)
R = 53e-9;      % emission site radius  (free parameter)
y0 = (1/phi)*sqrt((q^2)*F/(4*pi*epsilon));
v0 = (1 - y0.^2) + (1/3)*(y0.^2).*(log(y0));
t0 = 1 + (1/9)*(y0.^2).*(1 - log(y0));
C_FN0 = (2*pi*(R^2)/q)*((aa*F.^2)./...
(phi*(q^2)*(t0.^2))).*...
    exp(-(v0*bb*q*(phi^1.5))./F);
y1 = (1/(phi-photonE))*sqrt((q^2)*F/(4*pi*epsilon));
v1 = (1 - y1.^2) + (1/3)*(y1.^2).*(log(y1));
t1 = 1 + (1/9)*(y1.^2).*(1 - log(y1));
C_FN1 = (2*pi*(R^2)/q)*((aa*F.^2)./...
((phi-photonE)*(q^2)*(t1.^2))).*...
    exp(-(v1*bb*q*((phi-photonE)^1.5))./F);
y2 = (1/(phi-2*photonE))*sqrt((q^2)*F/(4*pi*epsilon));
v2 = (1 - y2.^2) + (1/3)*(y2.^2).*(log(y2));
t2 = 1 + (1/9)*(y2.^2).*(1 - log(y2));
```

```
C_FN2 = (2*pi*(R^2)/q)*((aa*F.^2)./...
((phi-2*photonE)*(q^2)*(t2.^2))).*...
    exp(-(v2*bb*q*((phi-2*photonE)^1.5))./F);
a0 = 1;
a1 = 5.89e-17;
a2 = (1e-4)*a1^2;
I = 5.4e8;
C_FN = a0*C_FN0 + a1*C_FN1*I + a2*C_FN2*I^2;
subplot(1,2,1)
plot(VtipExp,C,'bs','MarkerSize',10,'MarkerFaceColor','b')
grid, axis square, box on
xlabel('-V_t_i_p (V)','FontSize',20,'FontWeight','bold')
ylabel('Counts / sec','FontSize',20,'FontWeight','bold')
ax = gca;
ax.FontSize = 20;
ax.FontWeight = 'bold';
hold on
plot(Vtip,C_FN,'-r','linewidth',2)
subplot(1,2,2)
loglog(Vtip,a0*C_FN0,Vtip,a1*C_FN1*I,Vtip,a2*C_FN2*I^2,...
    'Linewidth',2,'Linewidth',2), grid
xlabel('-V_t_i_p (V)','FontSize',20,'FontWeight','bold')
ylabel('Counts / sec','FontSize',20,'FontWeight','bold')
axis square
ax = gca;
ax.FontSize = 20;
ax.FontWeight = 'bold';
ax.XColor = 'k';
ax.YColor = 'k';
ax.GridColor = 'k';
ax.XMinorTick = 'on';
ax.YMinorTick = 'on';
ax.XMinorGrid = 'on';
legend('n=0','n=1','n=2')
ax.YMinorGrid = 'on';
% End
```

## 11. Above-threshold photoemission model

This program outputs panels (c) and (e) of Fig. 4-13 separately.

```
% ATE model - SK
clear
clc
Cexp = (1e3)*[.0222 .0489 .1361 .1691 .0724 .1143...
.0308 .1354 1.4196 .5446 .3436 .1942 .0620];
```

```
Cerr = (1e3)*[0.0279 0.0211 0.0770 0.0994 0.0330...
0.1025 0.0244 0.0384 .7348 0.3185 0.2047 0.2001 0.0674];
nexp = [1.8566 2.0221 1.4728 1.7083 2.1090 2.6410...
2.3480 2.9561 4.3872 3.6377 2.4105 3.1481 2.7952];
nerr = [0.2491 0.6003 0.4347 0.8991 0.0181 0.7542...
0.3015 0.7259 0.1531 0.1578 0.0047 0.7536 0.8373];
lambda = 500:1:740;
llambda = 500:20:740;
lambda0 = 660;
q = 1.6e-19;
c = 3e8;
h = 6.626e-34;
P = 1e-5;
r = 1e-5;
delta = 10;
I_0 = delta*(P/(2*pi*r^2));
ef = 3;          % enhancement factor due to SPR excitation
w = 3e1;         % nm
I = I_0*(1 + ef*exp(-((lambda - lambda0).^2)./(w^2)));
phi = 5.25;      % eV
E = (h*c./((1e-9)*lambda))/q;
k_B = 8.617e-5; % eV/K
T = 300;         % K
FD1 = 1./(exp((phi - (2*E))./(k_B*T)) + 1);
FD2 = 1./(exp((phi - (3*E))./(k_B*T)) + 1);
FD3 = 1./(exp((phi - (4*E))./(k_B*T)) + 1);
FD4 = 1./(exp((phi - (5*E))./(k_B*T)) + 1);
FD5 = 1./(exp((phi - (6*E))./(k_B*T)) + 1);
a1 = 1e6;
a2 = 1;
a3 = 1e-6;
a4 = 2e-12;
a5 = 8e-18;
C = (a1.*FD1.*I + a2.*FD2.*I.^2 + a3.*FD3.*I.^3...
+ a4.*FD4.*I.^4 + a5.*FD5.*I.^5);
n = (I./C).*(a1.*FD1 + 2*a2.*FD2.*I + 3*a3.*FD3.*I.^2 +...
    4*a4.*FD4.*I.^3 + 5*a5.*FD5.*I.^4);
subplot(1,2,1)
errorbar(llambda,Cexp,Cerr,...
'--o','MarkerSize',8,'MarkerFaceColor','r',...
    'MarkerEdgeColor','r','Linewidth',2,'Color','k')
xlabel('\lambda_M_L(nm)','FontSize',20,'FontWeight','bold')
ylabel('Counts / sec','FontSize',20,'FontWeight','bold')
axis square
box on
grid
ax = gca;
ax.FontSize = 20;
ax.FontWeight = 'bold';
```

```
ax.XColor = 'k';
ax.YColor = 'k';
ax.GridColor = 'k';
ax.XMinorTick = 'on';
ax.YMinorTick = 'on';
ax.XMinorGrid = 'on';
ax.YMinorGrid = 'on';
xlim([500 740]);
xticks([500 560 620 680 740]);
xticklabels({'500','560','620','680','740'});
hold on
plot(lambda,(.78e-9)*C,'Linewidth',2,'Linewidth',2,...
'Color','[0 .5 0]'), grid
xlabel('\lambda_M_L (nm)','FontSize',20,...
'FontWeight','bold')
ylabel('Counts / sec','FontSize',20,...
'FontWeight','bold')
xlim([500 740]);
subplot(1,2,2)
errorbar(llambda,nexp,nerr,'-- o',...
'MarkerSize',8,'MarkerFaceColor','b',...
    'MarkerEdgeColor','b','Linewidth',2,'Color','k')
hold on
plot(lambda,n,'Linewidth',2,'Linewidth',2,'Color','r')
grid
xlabel('\lambda_M_L (nm)','FontSize',20,...
'FontWeight','bold')
ylabel('n','FontSize',20,'FontWeight','bold')
xlim([500 740]);
xticks([500 560 620 680 740]);
xticklabels({'500','560','620','680','740'});
axis square
ax = gca;
ax.FontSize = 20;
ax.FontWeight = 'bold';
ax.XColor = 'k';
ax.YColor = 'k';
ax.GridColor = 'k';
ax.XMinorTick = 'on';
ax.YMinorTick = 'on';
ax.XMinorGrid = 'on';
ax.YMinorGrid = 'on';
% End
```